\def\cm{cm$^{-1}$}
\def\clo4{(TMTSF)$_2$\-ClO$_4$}
\def\pf6{(TMTSF)$_2$\-PF$_6$}
\def\asf6{(TMTSF)$_2$\-AsF$_6$}
\def\etbr{$\kappa$-(BE\-DT\--TTF)$_2$\-Cu\-[N\-(CN)$_{2}$]Br}
\def\etcl{$\kappa$-(BE\-DT\--TTF)$_2$\-Cu\-[N\-(CN)$_{2}$]Cl}
\def\etbrcl{$\kappa$-(BE\-DT\--TTF)$_2$\-Cu\-[N\-(CN)$_{2}$]\-Br$_{x}$Cl$_{1-x}$}
\def\aeti{$\alpha$-(BEDT\--TTF)$_2$I$_3$}
\def\etcn{$\kappa$-(BE\-DT\--TTF)$_2$\-Cu$_2$(CN)$_3$}
\def\stfcn{$\kappa$-[(BE\-DT\--TTF)$_{1-x}$(BE\-DT\--STF)$_x$]$_2$\-Cu$_2$(CN)$_3$}
\def\agcn{$\kappa$-(BE\-DT\--TTF)$_2$\-Ag$_2$(CN)$_3$}
\def\etscn{$\kappa$-(BE\-DT\--TTF)$_2$\-Cu\-(SCN)$_2$}
\def\tetcz{$\theta$-(BEDT-TTF)$_2$\-CsZn(SCN)$_4$}
\def\tetrz{$\theta$-(BEDT-TTF)$_2$\-RbZn(SCN)$_4$}
\def\dmit{$\beta^{\prime}$-EtMe$_3$\-Sb[Pd(dmit)$_2$]$_2$}
\def\cat{$\kappa$-H$_3$(Cat\--EDT-TTF)$_2$}
\def\dcat{$\kappa$-D$_3$(Cat\--EDT-TTF)$_2$}
\def\hgbr{$\kappa$-(BE\-DT\--TTF)$_2$\-Hg\-(SCN)$_{2}$Br}
\def\hgcl{$\kappa$-(BE\-DT\--TTF)$_2$\-Hg\-(SCN)$_{2}$Cl}

\documentclass[]{tandf2}

\usepackage{epstopdf}
\usepackage{subfigure}

\usepackage[numbers,sort&compress]{natbib}
\bibpunct[, ]{[}{]}{,}{n}{,}{,}
\makeatletter
\def\NAT@def@citea{\def\@citea{\NAT@separator}}
\makeatother

\begin{document}

\articletype{Advances in Physics}

\title{Molecular Quantum Materials: \\
Electronic Phases and Charge Dynamics \\
in Two-Dimensional Organic Solids}

\author{
\name{Martin Dressel\textsuperscript{a}
and Silvia Tomi{\'c}\textsuperscript{b}
}
\affil{\textsuperscript{a}1.~Physikalisches Institut, Universit\"at Stuttgart, Stuttgart, Germany; \\ \textsuperscript{b}Institut za fiziku, Zagreb, Croatia}
}

\maketitle

\begin{abstract}
This review provides a perspective on recent developments and their implications for our understanding of novel quantum phenomena in the physics of two-dimensional organic solids. We concentrate on the phase transitions and collective response in the charge sector, the importance of coupling of electronic and lattice degrees of freedom and stress an intriguing role of disorder. After a brief introduction to low-dimensional organic solids and their crystallographic structures, we focus on the dimensionality and interactions and emergent quantum phenomena. Important topics of current research in organic matter with sizeable electronic correlations are Mott metal-insulator phase transitions, charge order and ferroelectricity. Highly frustrated two-dimensional systems are established model compounds for studying the quantum spin liquid state and the competition with magnetic long-range order. There are also unique examples of quantum disordered state of magnetic and electric dipoles. Representative experimental results are complemented by current theoretical approaches.
\end{abstract}


\tableofcontents

\section{Introduction}
    \label{sec:introduction}
When molecular solids and in particular organic conductors started flourishing in the 1970s and 1980s, the spotlight was directed on their strong electronic anisotropy down to one dimension and subsequently the superconducting phases in the TMTSF and BEDT-TTF salts \cite{Keller75,Schuster75,Pal77,Keller77,Miller78,Hatfield79,Barisic79,Devreese79,Alcacer80,Bernasconi81,Miller82,
Tanaka86,Jerome87,Delaes87,Metzger90,Saito90,Kresin90,Jerome82,KagoshimaBook,IshiguroBook,Mori06,Jerome08, Little16,Gutfreund16}.
While these are still thrilling topics with many unresolved questions,
for the last ten years or so the focus of interest shifted
towards novel electronic properties that are related to the strong electron-electron correlations present in these compounds. By now, organic conductors  are recognized as versatile molecular quantum materials, which exhibit numerous exciting and unprecedented features. In comparison to inorganic compounds, molecular conductors are cleaner and easier to tune by chemical and physical means; hence they are well established as model compounds for the investigation of several fundamental problems. Since the field has been steadily covered by monographies and compilations
\cite{IshiguroBook,GrajaBook,WilliamsBook,FargesBook,GiamarchiBook,ChemRev04,JPSJ06,ToyotaBook,LebedBook,Crystals12,MoriBook,Crystals18},
we restrain ourselves from presenting a comprehensive and balanced overview on the entire field. Instead we select
the outstanding themes and issues: those which --~from our point of view~-- are of paramount importance to and have the largest impact on condensed matter physics in general.

\label{sec:outline}

In prototypical strongly correlated materials, such as heavy fermions or transition metal compounds, $f$- and $d$-electrons with their rather narrow bands govern the electronic properties. Often the Fermi surface is very complex and multiple bands compete. Molecular materials, on the other side, are characterized by delocalized  $\pi$-electrons that are distributed over the extended organic molecule.
If the almost planar molecules are stacked or assembled face-to-face in certain patterns constituting bulk crystals, the protruding orbitals of adjacent molecules overlap slightly, forming narrow electronic bands.
Although the unit cell of molecular crystals contains many atoms, it consists of a small number of molecules.
In most cases, only one electronic band cuts the Fermi energy, leading to rather simple Fermi surfaces
\cite{WosnitzaBook}.
Typical bands consist only of lowest unoccupied molecular orbital (LUMO) and highest occupied molecular orbital (HOMO), which often are isolated from other bands. Therefore, the effective Coulomb interaction between electrons on the HOMO and LUMO orbitals is poorly screened by other bands.
These two factors (small bandwidth and ineffective screening) make the molecular conductors mostly strongly correlated electron systems.

Since heavy fermions are intermetallic compounds with delocalized electrons, strong electronic correlations are adequately captured by simple renormalization of the Fermi-liquid parameters \cite{ColemanBook}. Transition metal compounds are most successfully described by the Hubbard model, {\it i.e.} a lattice fermion model including onsite and intersite interaction as required. Here strong electronic repulsion causes a metal insulator transition that takes place either by diverging mass or -- more commonly -- by vanishing carrier number \cite{Imada98}.
In many transition metal oxides, hybridization with the oxygen $p$ orbitals takes place and the character of low-energy charge excitations changes; these Mott insulators are actually charge transfer insulators.
In this regard organics represent the best examples of Mott-Hubbard insulators with the lowest energy excitation between the lower and upper Hubbard bands. Here the quasi-particle mass increases significantly as the Mott transition is approached.

By now it became clear that molecular conductors are an original class of strongly correlated electron systems.
They possess several features that makes them a diverse playground for the study of quantum many body physics and the properties of quantum materials. Organic solids are available as extremely pure single crystals of millimeter size and stable under ambient conditions. The energy scales fall in an easily accessible range, as far as temperature, magnetic field and pressure is concerned: superconductivity occurs around 10~K; but can be destroyed by applying less than 20~T; the typical bandwidth is less than 1~eV; due to the large compressibility of organic compounds, pressure of less than 1~GPa already leads to significant alterations of the physical properties.
Due to different organic molecules that can be combined with a number of counterions and the numerous stacking pattern possible in two-dimensions, a variety of prototypical behaviors of correlated electron systems can be achieved and easily modified by chemical means.

But there are some caveats that have to be mentioned: Several of the methods, which have been advanced over the last decades to investigate the electronic and magnetic properties of inorganic materials cannot easily be applied to molecular compounds.
Limited crystal size and the presence of hydrogen essentially prevent neutron scattering experiments, the most powerful tool for exploration of magnetic properties and dispersion. Furthermore, scanning tunneling methods as well as photoemission spectroscopy are challenging due to the surface properties and ionic structure. Nevertheless, the study of organic crystals has significantly profited from the enormous progress in instrumentation seen over the last years and spectacular achievements are reported on a regular basis.

We start in Chapter~\ref{sec:ChargeOrder} with a survey given on charge order and ferroelectricity in two-dimensional molecular solids, in particular covering the experimental achievements and theoretical insight gained over the last decade or so.
By now, dimerized BEDT-TTF salts are widely recognized as the prime examples for studying
the metal-insulator transition driven by onsite Coulomb repulsion.
Hence in Chapter~\ref{sec:MottTransition} we cover various aspects of the Mott transition, starting from
the scaling behavior around the critical endpoint and followed by the spatial coexistence of metallic and insulating regions at the first-order Mott transition. In the following Chapter we consider the highly-frustrated magnetic state leading to the quantum spin liquid ground state.
In Chapter~\ref{sec:quantumdisorder} we turn our attention to charge degrees of freedom and explore their interplay with magnetic degrees of freedom and how varying dimerization and geometrical frustration results in the particular ordered and liquid ground states. The Chapters should be self-contained with numerous cross-references that allow the reader to select certain topics of particular interest.


\section{Structural Aspects}
    \label{sec:structure}
Before proceeding to the detailed discussion of the electronic properties of molecular quantum materials,
in this Chapter we want to summarize briefly some structural aspects of those compounds most prominent in this regards.

\subsection{BEDT-TTF salts}
\label{sec:BEDT-TTF}
\begin{figure}
    \centering\includegraphics[clip,width=0.5\columnwidth]{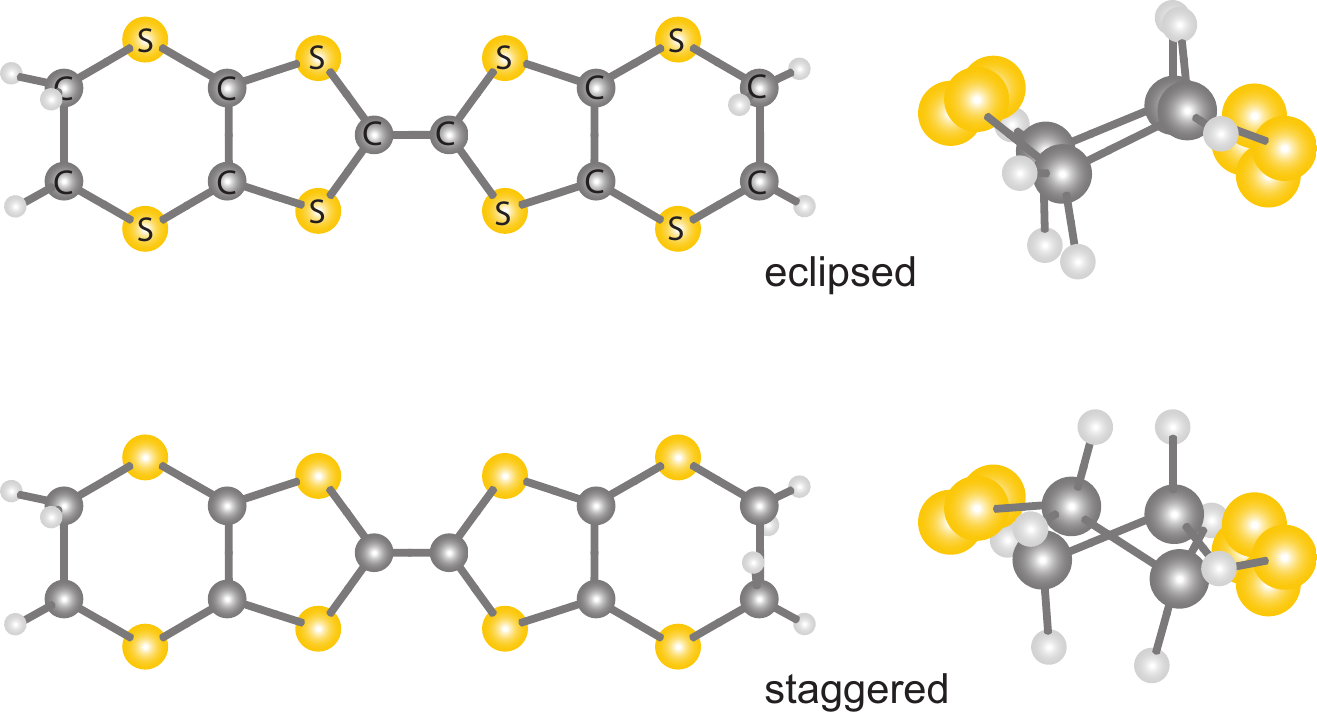}
	\caption{Molecular structure of BEDT-TTF or ET, which stands for bis(ethylenedithio)tetrathiafulvalene. The eight sulfur atoms are indicated by yellow, the carbon atoms by dark grey balls, the hydrogens are depicted by light gray. The molecule is not absolutely flat, but the endgroups are tilted either in eclipsed or staggered fashion, as indicated.}
	\label{fig:BEDT-TTF}
\end{figure}

In succession of the Fabre salts based on TMTTF (tetramethyl-tetrathiafulvalene) and Bechgaard salts based on TMTSF (tetramethyl-tetraselenafulvalene) molecule, the BEDT-TTF molecules become the widely used building block for the two-dimensional organic superconductors and molecular quantum materials. Here BEDT-TTF or sometimes simply ET, stands for bis(ethylenedithio)tetrathiafulvalene (C$_{10}$H$_8$S$_8$)
and was first synthesized by Saito {\it et al.} in 1982 \cite{Saito82}; as depicted in Figure~\ref{fig:BEDT-TTF}, the core is still the TTF-unit, but now extended by two sulfur and two ethylene groups on each side. Owed to the extended size, the molecule is not completely flat anymore but the two terminal groups are slightly twisted. This can be done in the same direction, called eclipsed, or opposite directions (staggered); opening the possibility of intrinsic disorder being present. In the case of the organic superconductor \etbr, for instance, the crystals exhibit ordering of the ethylene endgroups upon cooling: While at room temperature 70\%\ of the molecules are in the eclipsed configuration, it becomes 29\%\ at $T=100$~K \cite{Strack05}.

As pointed out by Girlando \cite{Girlando11a}, the symmetry of the BEDT-TTF molecule is often assumed
D$_{2h}$, implying a completely planar molecule \cite{Kozlov87,Kozlov89,Eldridge95}, in order to reduce the 72 independent vibrational degrees of freedom when calculating the
molecular vibrations. The actual symmetry, however, is D$_2$ in the case of a staggered molecule, and C$_{2h}$ in the case of eclipsed. As a matter of fact, the neutral molecule acquires a boat conformation with C$_2$ symmetry \cite{Kobayashi86,Demiralp95}, but this aspect is commonly neglected in the ionic crystal.

In order to vary the electronic orbitals, some of the sulfur atoms can be substituted. Replacing the four central S by Se, for instance, leads to BEDT-TSF or BETS, {\it i.e.} bis(ethylenedithio)tetraselenafulvalene. Alternatively, the substitution can be performed only on one side, resulting in the asymmetric BEDT-STF, {\it i.e.} bis(ethylenedithio)selenathiafulvalene. This will be utilized in several studies presented in Section~\ref{sec:QSLMott}.

Typically the compounds grow as (BEDT-TTF)$_2$$X$, where $X$ stands for a monovalent anion. The electronic charge of half a hole per BEDT-TTF molecule is distributed over the entire molecule, with the highest density around the central C=C, followed by the other two carbon double bonds.
In most of the (BEDT-TTF)$_2$$X$ salts, the organic donor molecules are packed more-or-less upright
in layers; they are held together by strong in-plane covalent bonds but weak out-of-plane van der Waals forces.
The molecular layers alternate along the third direction with sheets of monovalent anions or polymeric networks.
The interface between the donor and acceptor layers is determined by a hydrogen-bond network
constituted by the terminal ethylene groups. Commonly, anions are considered to serve as spacer and as charge reservoir, but the role of donor-anion interaction in stabilization of diverse electronic phases has been emphasized \cite{Pouget18}.

\begin{figure}
    \centering\includegraphics[clip,width=0.9\columnwidth]{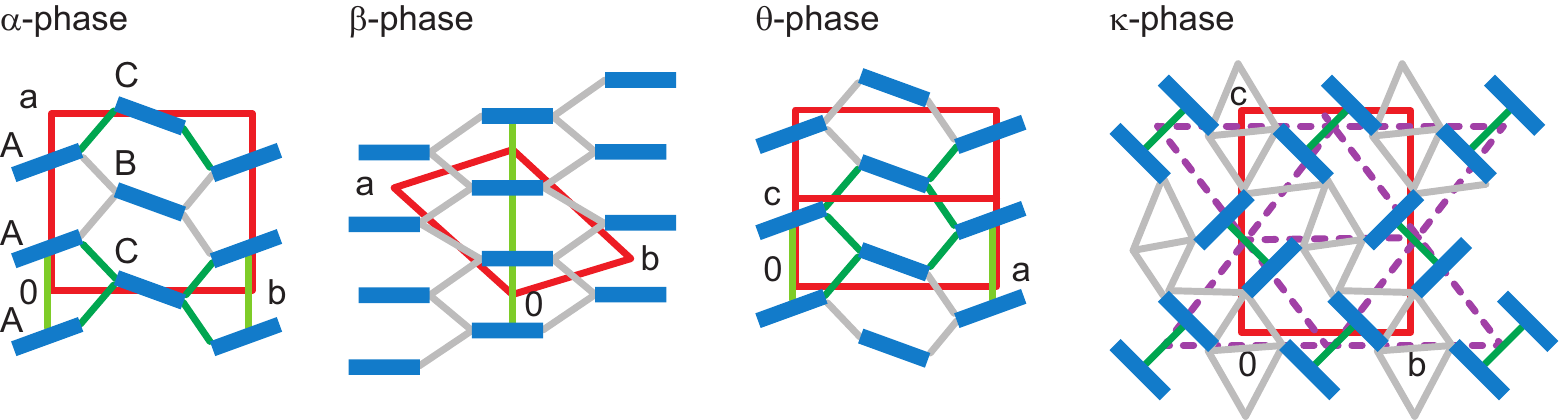}
	\caption{Schematic representation of some of the (BEDT-TTF)$_2$$X$ packing motives by looking at the quasi-two-dimensional layer.
The pattern labeled as $\alpha$-, $\beta$-, $\theta$- and $\kappa$-phase possesses a different
degree of dimerization. Only the strongest inter-donor interactions are indicated.
While in the $\alpha$- $\beta$- and $\theta$-polymorphs chains of BEDT-TTF molecules can be identified,
the lattice of the $\kappa$-phases consists of pairs of molecules so-called dimers arranged almost perpendicular to each other.
Here an effective triangular lattice of dimeric units can be identified as indicated by the dashed purple lines (suggested by \cite{Pouget18}).}
	\label{fig:structure_general}
\end{figure}

Due to the larger number of orbitals and extended size compared to the TMTSF-molecule,
multiple polymorphs can be found for most BEDT-TTF compounds, labeled by Greek letters \cite{WilliamsBook,Mori84,Mori98a,Mori99b,Mori99c}. Figure~\ref{fig:structure_general} displays some of the packing motives of relevance here. In the case of the $\alpha$- $\beta$- and $\theta$-phase, the
organic molecules are arranged in stacks, while in $\kappa$-salts pairs of molecules so-called dimers are formed that are almost orthogonal.
Within the stacks of the $\beta$-phase, neighboring molecules are slightly shifted with respect to each other; for the $\alpha$- and $\theta$-phase they are alternatively tilted. As indicated, neighboring stacks are coupled, resulting in two-dimensional, almost isotropic properties. The  dimers of the $\kappa$-lattice compose an effective triangular lattice giving way to frustration, as we will see in Chapter~\ref{sec:Frustration}.

\subsubsection{\rm $\alpha$-(BEDT-TTF)$_2$I$_3$}
\begin{figure}[h]
    \centering\includegraphics[clip,width=1.0\columnwidth]{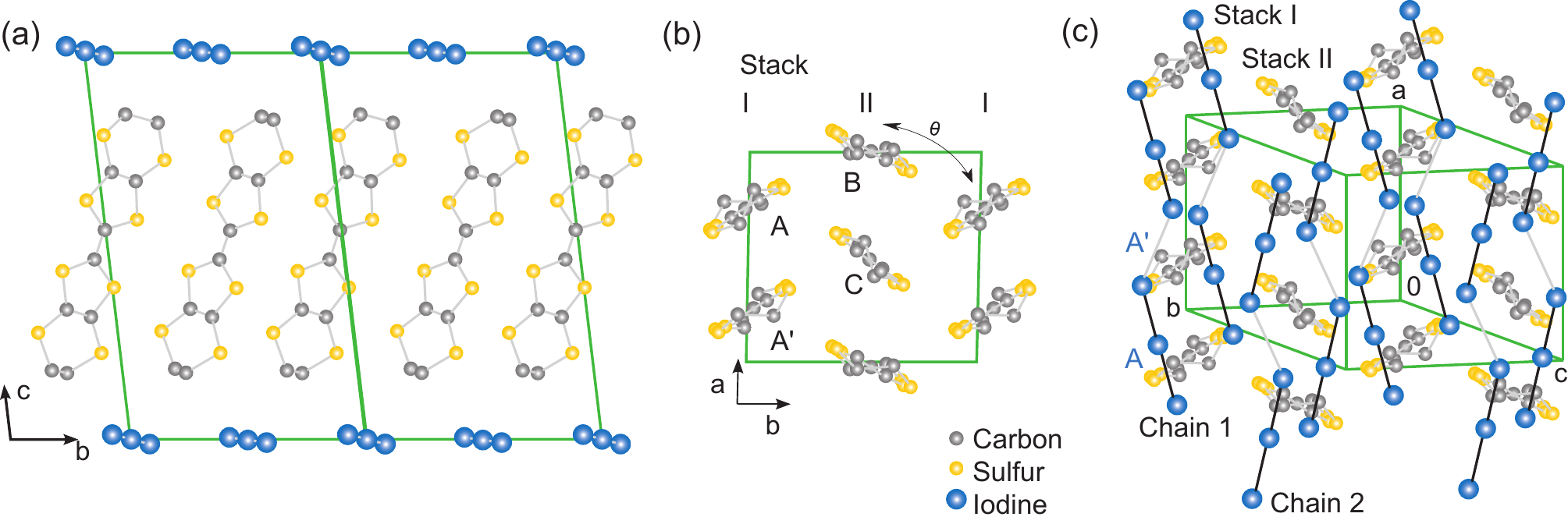}
	\caption{Crystal structure of \aeti.
(a)~The BEDT-TTF molecules are arranged in the $ab$-layers with the molecular axis
not normal but slightly tilted to this layer. Sheets of I$_3^-$-anions separate these layers
in $c$-direction.
(b)~View along the molecular axis reveals two distinct stacks. Stack I contains the BEDT-TTF molecules A and A$^{\prime}$, which are identical by symmetry. The molecules B and C in stack II are distinct to the rest. A dihedral angle between BEDT-TTF molecules in neighboring stacks is labeled by $\theta$.
(c)~Axiometric view of \aeti\ with the unit cell indicated in green.
There are two arrangements of the I$_3^-$ anions, labelled as chain 1 and chain 2.
	\label{fig:structure_alpha}}
\end{figure}

In the case of \aeti, the triclinic crystal structure (space group P$\bar{1}$) is an alternation of insulating I$_3^-$ anion layers and conducting layers of donor molecules BEDT-TTF$^{0.5+}$, displayed in Figure~\ref{fig:structure_alpha}(a) \cite{Bender84a}.
The BEDT-TTF molecules form a herringbone pattern and are organized in a
triangular lattice with two types of stacks.
At room temperature, stack I is
weakly dimerized and composed of crystallographically equivalent molecules A and
A$^\prime$ related by an inversion center, while the stack II is a uniform chain composed of distinct B and C molecules, shown in Figure~\ref{fig:structure_alpha}(b).
These two types of stacks are interconnected by many S$\cdots$S short contacts that provide the electronic delocalization within the $ab$-layer.
The I$_3^-$ anions form two distinct chains, labeled by chain 1 and chain 2, illustrated in panel (c). Thus the unit cell contains four BEDT-TTF molecules.
At high temperatures, the system is a semimetal with small electron
and hole pockets in the Fermi surface \cite{Bender84b,Mori84}.
There is a slight charge disproportionation
(among B and C molecules, while there is none among A and
A$^\prime$) already at ambient condition that gets considerable when cooled below the charge-ordering phase transition temperature at $T_{\rm CO} = 135$~K,
as discussed in Section~\ref{sec:COweaklydimerized}.

\subsubsection{\rm $\beta^{\prime\prime}$-(BEDT-TTF)$_2$SF$_5$$R$SO$_3$}
The  $\beta^{\prime\prime}$-(BEDT-TTF)$_2$SF$_5$$R$SO$_3$ compounds with different groups $R$
are isostructural, crystallizing in the P$\bar{1}$ triclinic system, with two formula units per
unit cell \cite{Ward00,Geiser96}. The structure is characterized by layers
of BEDT-TTF in the $ab$-crystal plane, separated by the all-organic anions, shown in Figure~\ref{fig:structure_beta}.
Within the cation layer, the BEDT-TTF are arranged in tilted stacks, typical
of the $\beta^{\prime\prime}$-packing motif \cite{Mori98a} with the strongest
interaction along the crystallographic $b$-axis,
{\it i.e.} perpendicular to the stacks.
Along the stacking direction $a$, two BEDT-TTF molecules
are related by inversion symmetry (A and A$^{\prime}$; B and B$^{\prime}$);
and the two pairs in neighboring stacks, AA$^{\prime}$ and BB$^{\prime}$, are
crystallographically independent.
Figure~\ref{fig:structure_beta}(c) presents four different anions constituting systems that have been subject of intense investigations discussed in Section~\ref{sec:COweaklydimerized}:
\begin{figure}[h]
    \centering\includegraphics[clip,width=0.9\columnwidth]{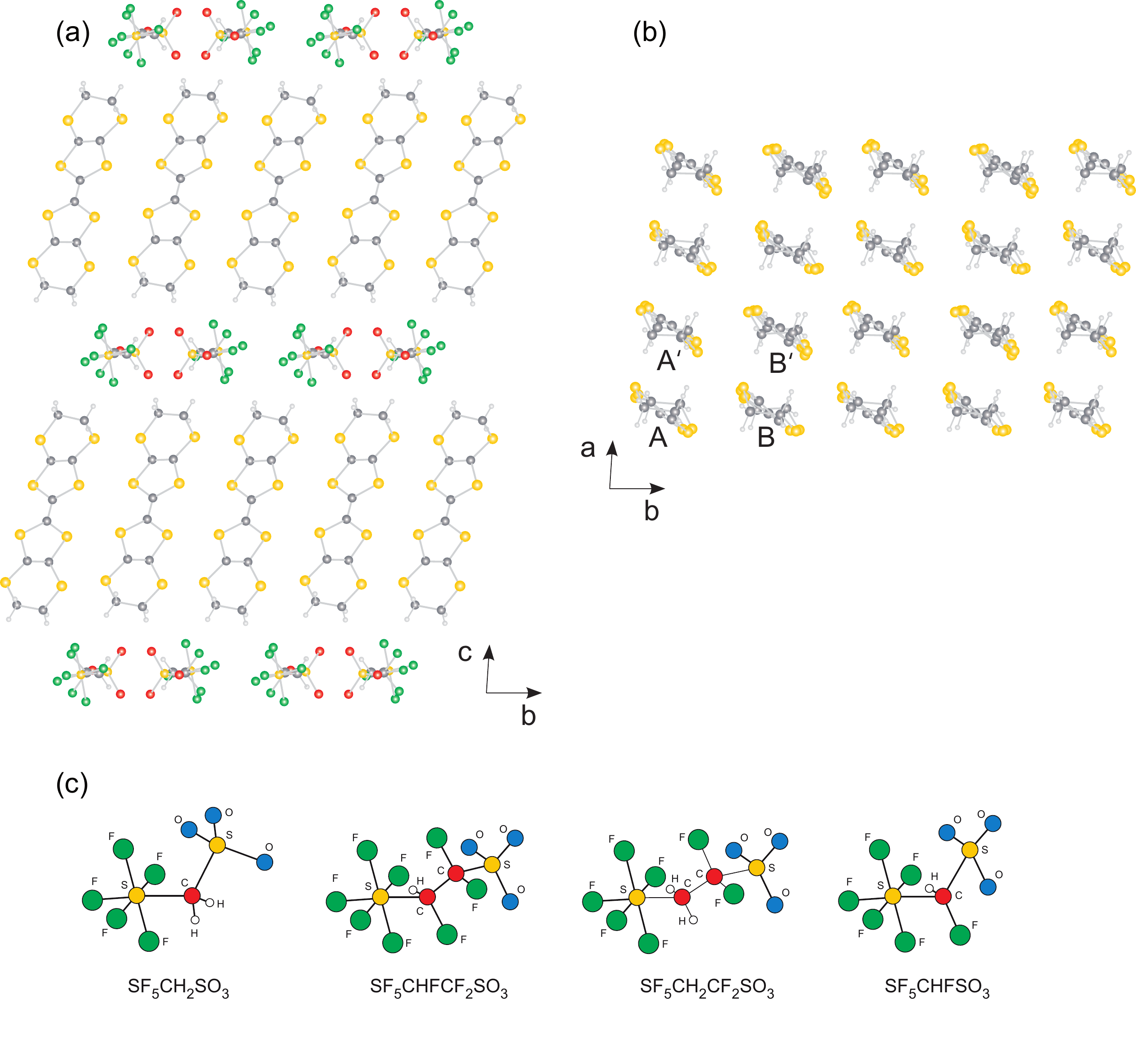}
	\caption{Crystal structure of $\beta^{\prime\prime}$-(BEDT-TTF)$_2$SF$_5$$R$SO$_3$.
(a)~View onto the $bc$-plane illustrates how the BEDT-TTF molecules form layers,
separated in $c$-direction by sheets of the all-organic anions. (b)~Within the $ab$-plane, the BEDT-TTF molecules are arranged in two slightly distinct stacks along the $a$-direction with uniform distance.
(c)~The four anions SF$_5$$R$SO$_3^-$ differ by the central entity:
$R$ equal CH$_2$, CHFCF$_2$, CH$_2$CF$_2$ and CHF, respectively.
		\label{fig:structure_beta}}
\end{figure}
$\beta^{\prime\prime}$-(BEDT-TTF)$_2$SF$_5$CH$_2$SO$_3$ (denoted $\beta^{\prime\prime}$-I)
is a charge-ordered insulator,  $\beta^{\prime\prime}$-(BEDT-TTF)$_2$SF$_5$CHFCF$_2$SO$_3$
($\beta^{\prime\prime}$-MI) undergoes a metal-insulator transition at 180~K,
$\beta^{\prime\prime}$-(BEDT-TTF)$_2$SF$_5$CH$_2$CF$_2$SO$_3$  ($\beta^{\prime\prime}$-SC)
is a superconductor at $T_c=5$~K driven by charge fluctuations, and
$\beta^{\prime\prime}$-(BEDT-TTF)$_2$SF$_5$CHFSO$_3$ ($\beta^{\prime\prime}$-M) remains metallic
down to low temperatures. The phase diagram of Figure~\ref{fig:betaphasediagram} summarizes the ground states of this family.
At $T=300$~K the crystal structure of $\beta^{\prime\prime}$-I exhibits some degree of disorder in the terminal ethylene groups \cite{Ward00}. No disorder is observed in the structure of $\beta^{\prime\prime}$-SC, which, however, has been collected at $T=123$~K.

\subsubsection{\rm $\theta$-(BEDT-TTF)$_2$RbZn(SCN)$_4$}
\begin{figure}[h]
    \centering\includegraphics[clip,width=0.7\columnwidth]{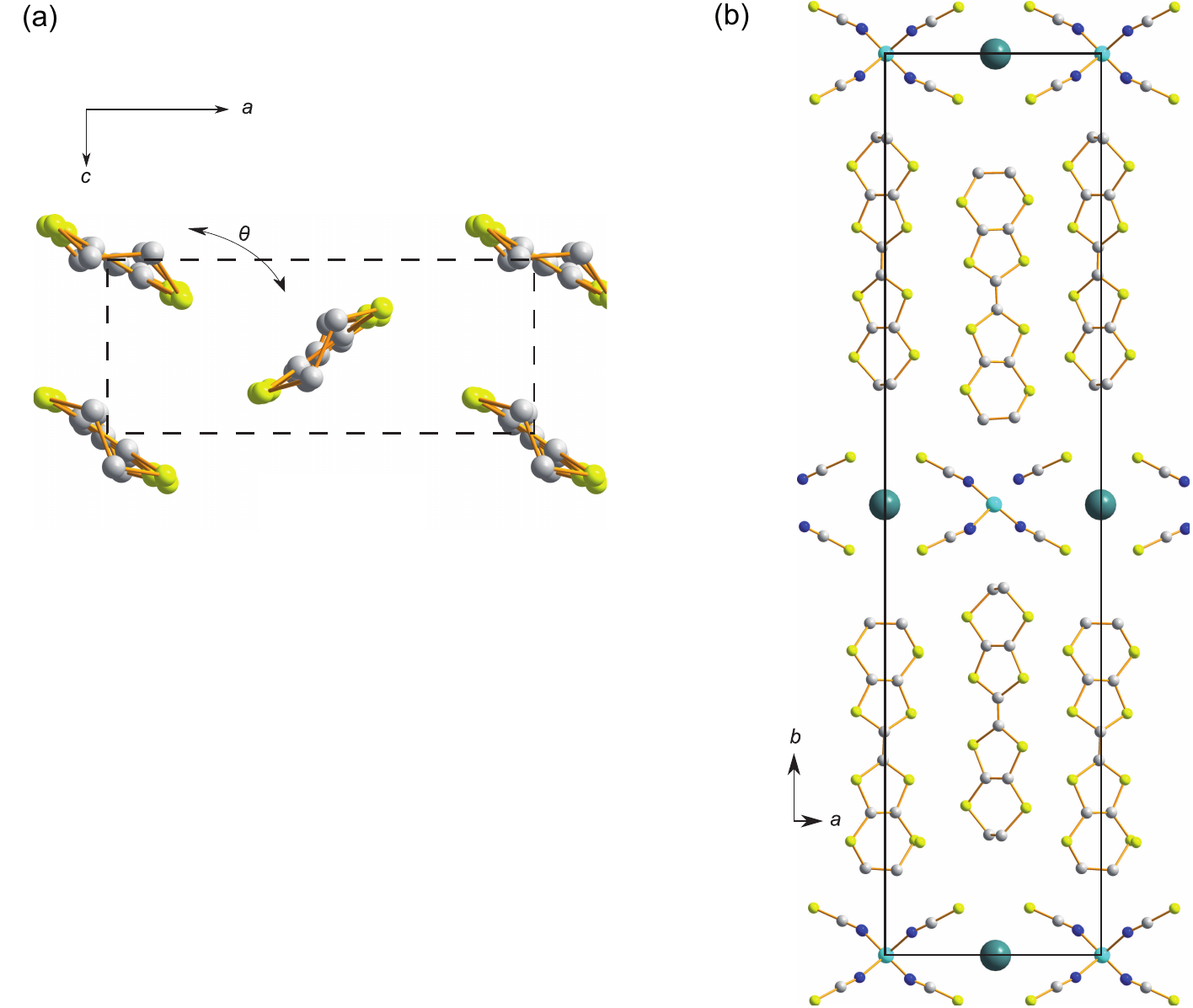}
	\caption{Crystallographic  structure of $\theta$-(BEDT-TTF)$_2$RbZn(SCN)$_4$.
(a)~View of BEDT-TTF molecules from one out of two cation layers in the $ac$-plane showing their triangular arrangement. They are tilted by a dihedral angle $\theta$ with respect to each other. (b)~Unit cell of
\tetrz\ containing four BEDT-TTF molecules and two RbZn(SCN)$_4$ entities.}
	\label{fig:structure_theta}
\end{figure}

$\theta$-(BEDT-TTF)$_2$RbZn(SCN)$_4$ crystallizes in the higher symmetry orthorhombic structure (space group I222) with four {BEDT-TTF}$^{0.5+}$ molecules and two [RbZn(SCN)$_4]^-$ anions per unit cell,
shown in Figure~\ref{fig:structure_theta}.
All BEDT-TTF molecules are crystallographically equivalent and stack along the $a$-axis. Since the molecules are strongly tilted with respect to each other, there is a large orbital overlap between neighboring stacks leading to a two-dimensional conductivity in the $ac$-plane \cite{Mori99b};  in $b$-direction the layers are separated by the anions as shown in panel(b). This two-dimensional RbZn(SCN)$_4^-$ network is built from two (SCN)$_2$ chains, connected by a Rb$^+$, in which Zn$^{2+}$ is tetrahedrally coordinated to the SCN$^-$ groups. At high temperatures the system is characterized by degenerated bands and a two-dimensional closed Fermi surface with three-quarter filling,
resulting in a metallic conductivity behavior within the molecular planes \cite{HMoriPRB98}.
Similar to \aeti, charge disproportionation develops in the metallic state and when cooled slowly a phase transition into the long-range charge-order state sets in at $T_{\rm CO} = 190$~K, as discussed in Section~\ref{sec:COweaklydimerized}. Conversely,
rapid cooling inhibits order and a charge glass state is formed, as discussed in Section~\ref{sec:ChargeGlass}.

\subsubsection{\rm $\kappa$-(BEDT-TTF)$_2$$X$ salts}
The $\kappa$-(BEDT-TTF)$_2$$X$ salts are characterized by the BEDT-TTF dimers, which are almost orthogonal to each other. Due to the strong dimerization, the $\kappa$-salts are prime examples of half filled Mott systems. The intradimer coupling $t_d$ is a fair estimate of the on-site Coulomb repulsion $U$ \cite{Saito95,McKenzie98}. Advancing previous calculations \cite{Emge86,Kobayashi87,Campos96,Fortunelli97a,Fortunelli97b,Rahal97,Mori98a,
Schlueter02,Nakamura09,Kandpal09},
Scriven and Powell computed the effective Coulomb interaction
within the BEDT-TTF dimers by density functional theory (DFT)  \cite{Scriven09b}.

\begin{figure}[h]
    \centering\includegraphics[clip,width=0.8\columnwidth]{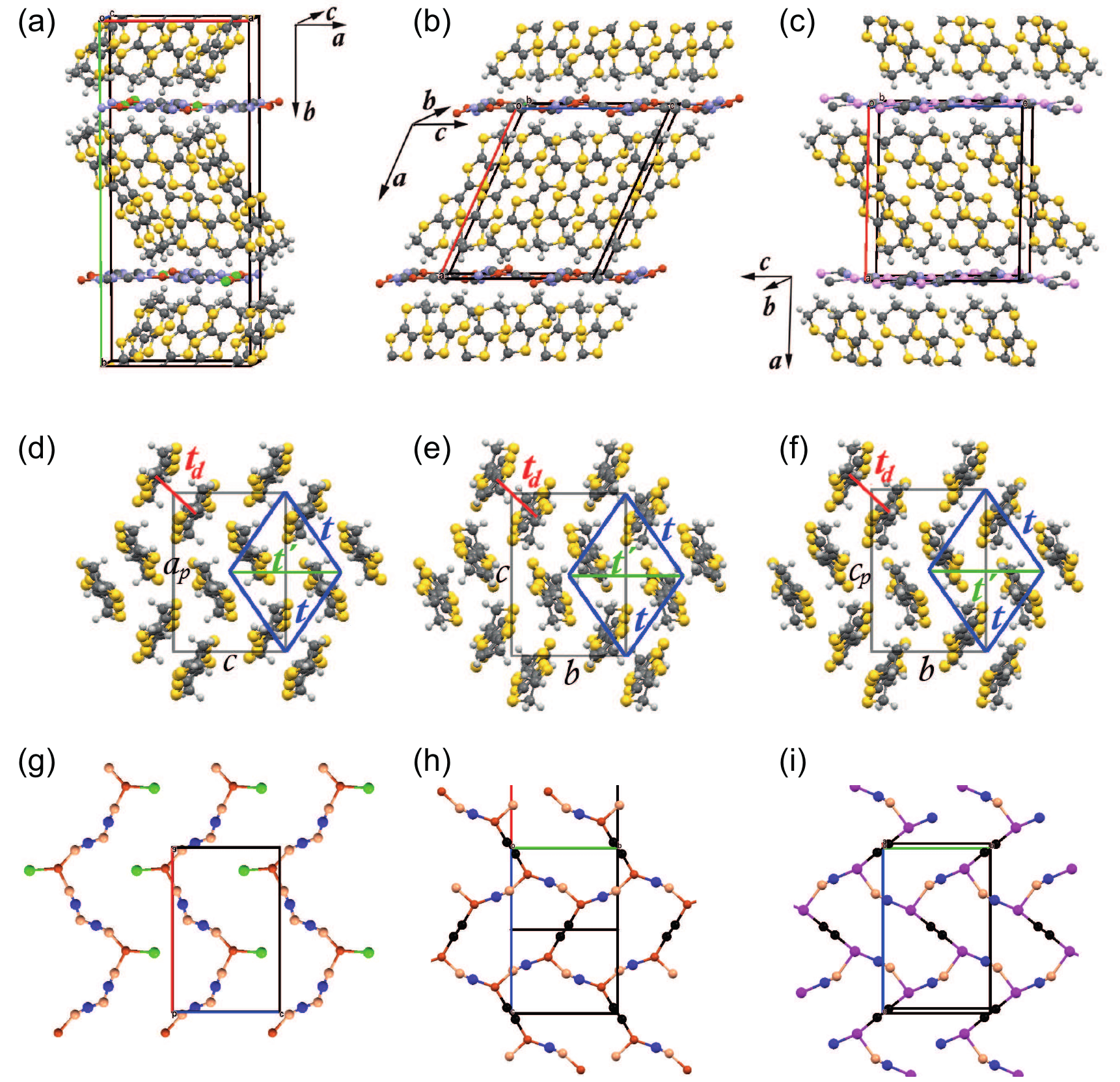}
	\caption{Crystal structure of three $\kappa$-phase salts: (a,d,g) \etcl, (b,e,h) \etcn\ and (c,f,i) \agcn. The lines mark the unit cell. In panels (d,g) for clarity reasons only one out of two cation and anion layers constituting the unit cell is shown.
(d,e,f) View of BEDT-TTF dimers arranged in anisotropic triangles in the two-dimensional planes. Carbon, sulfur and hydrogen atoms of the BEDT-TTF molecule are colored in dark gray, yellow and light gray, respectively. The interdimer transfer integrals are denoted by $t$ and $t^{\prime}$, and the intradimer transfer integral by $t_\mathrm{d}$. The ratio $t^{\prime}/t$ measures the degree of frustration. (g,h,i) View of the anion network in the two-dimensional planes. Chlorine, cooper, silver, carbon and nitrogen are colored in green, red, pink, blue and orange, respectively. Ordered cyanide (CN)$^-$ groups exist in all three systems, while  CN$^-$ groups (labeled by black) located at inversion centers are present only in \etcn\ and \agcn\ and are source of intrinsic disorder. Note that for \etcl\ the anion network reveals a kind of linear bonding scheme, while for \etcn\ and \agcn\ the anion network displays two-dimensional bonding arrangement.}
	\label{fig:structure_kappa}
\end{figure}

As usual, the organic molecules are arranged in layers separated by the anion sheets. In the case of \etcl\ and \etbr\ the BEDT-TTF dimers are tilted in alternating directions
[Figure~\ref{fig:structure_kappa}(a)]. In each of two organic layers, related by mirror symmetry, all four BEDT-TTF molecules are equivalent. Hence, \etcl\ crystallizes in the space group Pnma.
The BEDT-TTF layers are separated by polymeric Cu[N(CN)$_2$]Cl$^{-}$ anions along the $b$-direction.
Figure~\ref{fig:structure_kappa}(b) and (c) display the extended unit cell of \etcn\ and \agcn, respectively. The space group is commonly solved in monoclinic P2$_1$/c, in which all four BEDT-TTF molecules are equivalent \cite{Geiser91, Hiramatsu17}. However, the P2$_1$/c is only the average structure, while the exact structure presents a triclinic symmetry with two non-equivalent crystallographic sites \cite{Foury18,Foury20}.
In the panel (d) to (f) the projection of one of the layers is shown. Neighboring dimers are rotated by about $90^{\circ}$ with respect to each other. The ratio $t^{\prime}/t$ of next-nearest-neighbor ($t^{\prime}$) and nearest-neighbor ($t$) coupling between the dimers measures the degree of frustration.
For \etcl\ $t^{\prime}/t \approx 0.5$ is rather small, while for the spin liquid candidates \etcn\ and \agcn\ the ratio $t^{\prime}/t \approx 0.85$ indicates high frustration on the anisotropic triangle. Using tight-binding analysis, it was demonstrated that also the molecular conformation of the ethylene endgroups has some influence on the electronic structure \cite{Guterding15}.
Panels (g) to (i) show the anion networks. In \etcl\
all cyanide (CN) groups positioned between the copper atoms are ordered in a zigzag line along the $a$-axis so that the anion layer consists of one-dimensional chains [panel (g)]. In contrast to that, in \etcn\ and \agcn\
in addition to ordered CN groups (so-called chain CN) between Cu/Ag atoms in the chains along the $b$-axis, there are CN groups (so-called bridging CN) located at the inversion centers. These bridging CN groups connect Cu/Ag atoms along the $c$-axis so that the anion network is formed in two-dimensions [panels (h) and (i)]. The triangular coordination of Cu and Ag implies frustration since each Cu/Ag atom can be linked either to two N and one C atom, or to one N atom and two C atoms introducing intrinsic disorder.
This ambiguity was eliminated in the recently synthesized salt $\kappa$-(BEDT-TTF)$_2$Cu[Au(CN)$_2$]Cl that is highly frustrated $t^{\prime}/t = 1.19$ but possesses no disorder in the polymeric anions \cite{Tomeno20}.
The properties of \etcl, \etcn\ and \agcn\ are discussed in Chapters~\ref{sec:MottTransition} and \ref{sec:Frustration}, and Section~\ref{sec:QSL+afm}.

\begin{figure}
    \centering\includegraphics[width=0.8\columnwidth]{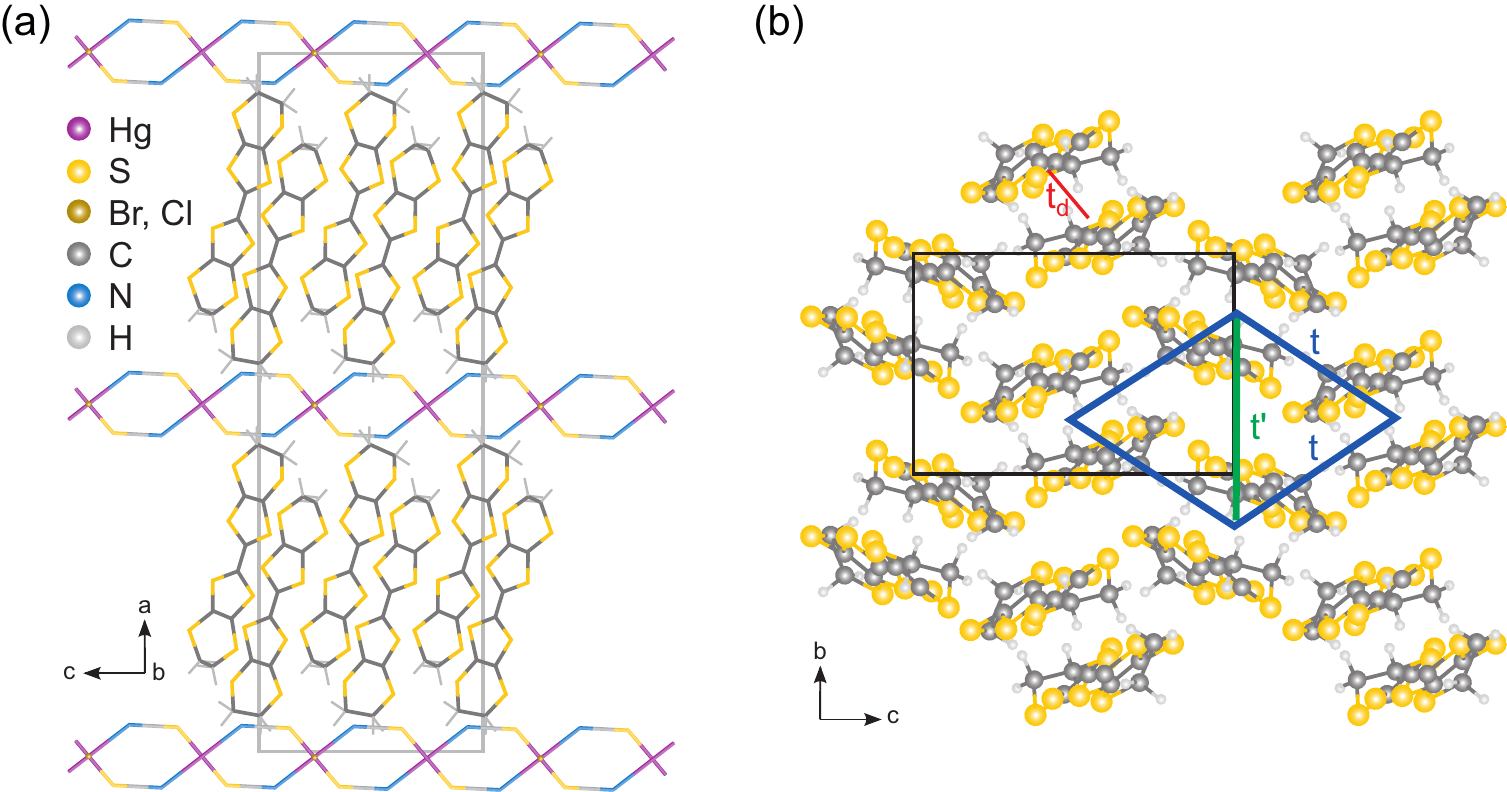}
    \caption{Crystal structure of \hgcl\ and \hgbr.
(a)~The molecular layers along the conducting $bc$-plane are separated by the
[Hg(SCN)$_2$Cl]$_{\infty}^{-}$ anions and [Hg(SCN)$_2$Br]$_{\infty}^{-}$ in the case of \hgcl\ and \hgbr, respectively. Note that within a dimer the two molecules are displaced along the molecular axis, leading to a reduced transfer integral $t_d$.
(b)~Two face-to-face BEDT-TTF molecules form dimer with intra-dimer transfer integral $t_d$.
The dimers are arranged on an anisotropic triangular
lattice with effective transfer integrals $t$ and $t^{\prime}$.}
\label{fig:structure_HgBrCl}
\end{figure}

The crystal structure of  \hgcl\ and \hgbr\ corresponds to space group C2/c \cite{Drichko14}.
It consists of alternating layers of BEDT-TTF radical cations and anions along the crystallographic $a$-axis, as shown in Figure~\ref{fig:structure_HgBrCl}(a).
The anionic layer contains [Hg(SCN)$_2$Cl]$_{\infty}^-$ and [Hg(SCN)$_2$Br]$_{\infty}^-$ chains in the case of \hgcl\ and \hgbr, respectively. It is worth to note that in contrast to many other BEDT-TTF salts, in the present family, {\it i.e.} in \hgcl\ and the bromine-containing sister compound, the anions are not completely flat, but have a sizable width. This implies that the BEDT-TTF layers are separated more than usually.

As illustrated in panel (b), in \hgcl\ as well as in \hgbr\ the dimers form a triangular lattice with rather large frustration $t^{\prime}/t \approx 0.80$ \cite{Drichko14,Gati18a}.
The $\kappa$-type packing motif is slightly distorted compared to the previous ones in \etcl\ or \etcn. The large number of S$\cdots$S intermolecular interactions between dimer units leads to a pretty strong $V$.
For that reason, for  \hgcl\ and \hgbr\ the interdimer interaction becomes so important, that the system cannot be treated by the simple Hubbard model with a half-filled conduction band. Another distinction is that within the dimer, the BEDT-TTF molecules are slightly shifted with respect to each other, leading to a reduced intradimer transfer integral $t_d$ and thus smaller $U$.

For \hgcl\ and \hgbr\ there is one crystallographically unique BEDT-TTF molecule in the unit cell.
In the case of \hgcl\ at room temperature, both ethylene end-groups are disordered, but below $T=100$~K a staggered conformation prevails. Interestingly, the unit cell is slightly larger than in the sister compound \hgbr\ \cite{Aldoshina93}; this trend is just opposite than expected from the larger Br ion compared to Cl.
However, there seems to be no possibility to transform the physical properties of \hgbr\ towards those of \hgcl, but also not {\it vice versa}. The two structurally similar compounds display different low-temperature behavior. \hgcl\
undergoes a pronounced metal-insulator transition due to charge ordering at $T_{\rm CO} = 30$\,K,
as discussed in Section~\ref{sec:COdimerized}, whereas \hgbr\ when cooled below metal-insulator transition at $T_{\rm CO} = 80$\,K develops quantum dipole liquid state with glassy signatures, as discussed in Section \ref{sec:DipoleLiquid}.

\subsection{Other molecular compounds}
During the last decades there have been numerous alternative attempts towards organic superconductors, most of them with rather limited success.
The molecular conductors based on the anion radicals [$M$(dmit)$_2$] ($M$ = Ni and Pd) synthesized by R. Kato \cite{Kato14} is probably the most prominent family as it exhibits  several interesting properties. More
recently H. Mori suggested an even other approach via a H-bonded molecular unit-based organic conductor with a fused structure of short hydrogen bonds and stacked $\pi$-electron systems.

\subsubsection{\rm $\beta^{\prime}$-EtMe$_3$\-Sb[Pd(dmit)$_2$]$_2$}
\begin{figure}
    \centering\includegraphics[clip,width=0.6\columnwidth]{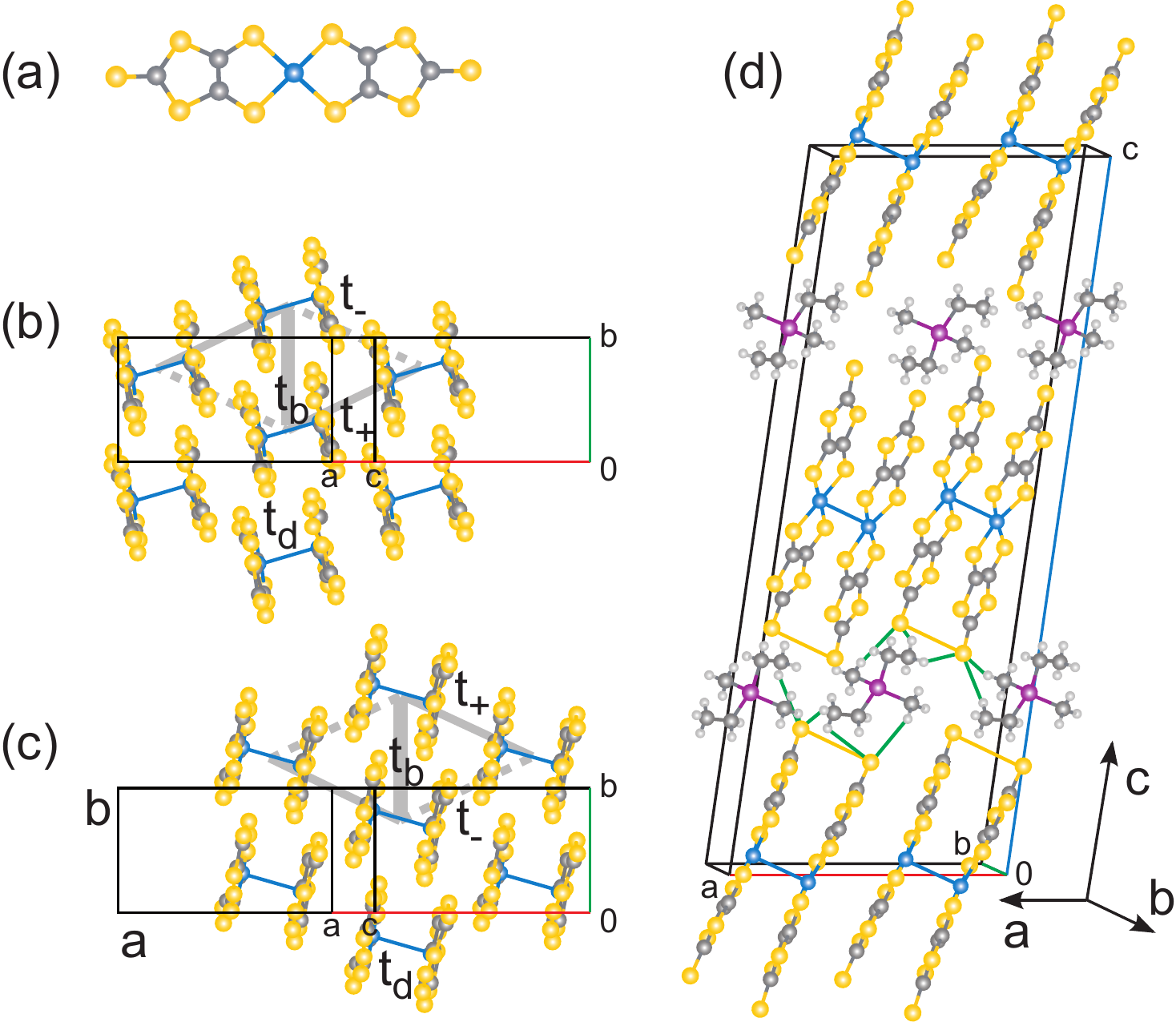}
	\caption{Crystal structure of \dmit.
(a)~Sketch of the molecule (1,3-dithiole-2-thione-4,5-dithiolate), Pd(dmit)$_2$: yellow, grey and green circles denote sulfur, carbon, and palladium atoms, respectively.
(b)~and (c) View of Pd(dmit)$_2$ dimers in the $ab$-plane projected along the direction tilted
$17^{\circ}$  away from the $c$-axis. The panels show dimers in two neighboring
layers which have ($a+b$) and ($a - b$) stacking directions, respectively. An almost isotropic triangular lattice is denoted by gray lines, the interdimer transfer integrals are labeled by $t_b$ (thick gray) and $t_{+}$ (thin gray), and $t_{-}$ (dashed gray), while the intradimer transfer integral is labeled by $t_d$. (d)~Side view of the extended unit cell. Antimony and hydrogen atoms are denoted by violet and light grey circles. Possible hydrogen bonds
between the end S ions of the Pd(dmit)$_2$ molecules and H of Et (CH$_2$-CH$_3$) or Me (CH$_3$) groups of the cations are indicated by full green lines.}
	\label{fig:structure_DMIT}
\end{figure}

\dmit\ radical anion salt is based on the metal dithiolene complex Pd(dmit)$_2$ \cite{Kato12a}. The crystal has a layered structure along the $c$-axis with symmetry space group C2/c (Figure~\ref{fig:structure_DMIT}). Two Pd(dmit)$_2$ molecules form a dimer with one negative charge, and the dimers are arranged to form an almost isotropic triangular lattice with the ratio $t^{\prime}/t \approx 0.8$.
There are two equivalent anion layers consisting of four molecular dimers (two in the central and two in the side layers), where dimers stack face-to-face along two diagonal directions $[110]$ and $[1\bar{1}0]$.
There are four monovalent cations EtMe$_3$Sb; each cation consists of three methyl CH$_3$ groups labeled as Me$_3$ and one ethyl CH$_2$-CH$_3$ group labeled as Et,
which can occupy one of two different equally probable orientations. The latter indicates that the formation of several crystallographic configurations is possible. Quantum spin liquid is suggested as the ground state of \dmit; its properties are discussed in Chapters~\ref{sec:MottTransition} and \ref{sec:Frustration}.

\subsubsection{\rm $\kappa$-H$_3$(Cat\--EDT-TTF)$_2$}
\begin{figure}
    \centering\includegraphics[clip,width=0.9\columnwidth]{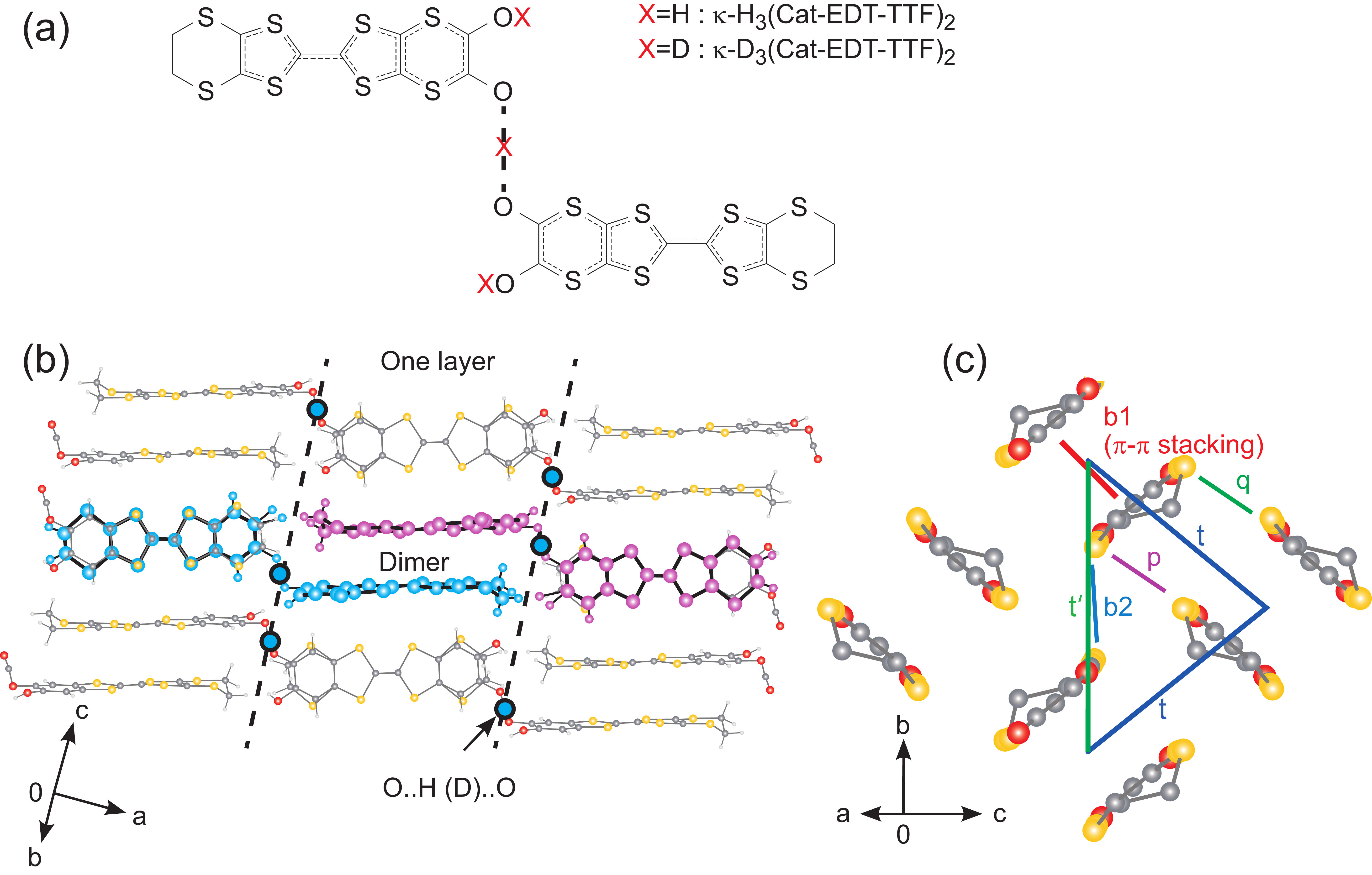}
	\caption{Crystal structure of \cat and \dcat.
(a)~Hydrogen-bonded molecular unit of the $\kappa$-$X$$_3$(Cat\--EDT-TTF)$_2$,
where $X$ stands for either a hydrogen or deuterium atom, in the case of \cat\ and \dcat, respectively. Two Cat-EDT-TTF molecules are related by twofold rotational symmetry with respect to the
central hydrogen atom, and are thus crystallographically equivalent to each other.
(b)~Packing arrangement of the molecular \cat\ and \dcat\ units.
Two face-to-face Cat-EDT-TTF molecules, colored by pink and light blue, form a dimer.
(c) View of the two-dimensional crystal structure in the layer shown in panel (b).
There are four different kinds of inter-molecular interactions in the two-dimensional layer; the corresponding transfer integrals are labeled as
$b1$ (face-to-face $\pi –\pi$ stacking), $b2$, $p$, and $q$.
The dimers are arranged on an anisotropic triangular lattice in the $(b,c)$ plane.
The interdimer transfer integrals are denoted by $t$ and $t^{\prime}$, and the intradimer transfer integral by $t_\mathrm{d}$. The ratio $t^{\prime}/t$ measures the degree of frustration.
All hydrogen atoms are
omitted for the simplicity reasons (after \cite{Yamamoto16}).}
	\label{fig:structure_CAT}
\end{figure}

In comparison to the previous radical cation (BEDT-TTF) and anion [Pd(dmit)$_2$] salts, \cat\ and its deuterated isotopologue \dcat\ do not contain counter-ion species.
\cat\ consists of two crystallographically equivalent catechol-fused ethylenedithio-tetrathiafulvalene
(Cat-EDT-TTF) molecules linked by a symmetric anionic [O$\cdots$H$\cdots$O]$^{-}$-type strong hydrogen-bond as shown in
Figure~\ref{fig:structure_CAT}(a) \cite{Isono13,Ueda14,Ueda15,Yamamoto16}.
H-bonded hydrogen is located at the central position between two oxygen atoms and these H-bonds result in the generation of holes (+0.5) on both Cat-EDT-TTF molecuels. The crystal holds C2/c symmetry.
These conductors enabled the construction of an unprecedented packing structure where the two-dimensional
$\pi$-electron conducting layers shown in Figure~\ref{fig:structure_CAT}(c) are connected
by the H-bonds [panel (b)].
In these layers there are four kinds of inter-molecular interactions ($b1$, $b2$, $p$, and $q$); among them the strongest interaction is $b1$ from the face-to-face $\pi –\pi$ stacking. Two face-to-face Cat-EDT-TTF molecules are paired in a dimer and the dimers form a two-dimensional triangular lattice in the $bc$-plane. The degree of frustration measured by the ratio of the couplings
between dimers $t^{\prime}/t \approx 1.25$ is slightly one-dimensional unlike $t^{\prime}/t$ values found in $\kappa$-(BEDT-TTF)$_2$$X$.
\cat\ is suggested as a quantum spin liquid candidate with simultaneously developed quantum disordered state of electric dipoles, while \dcat\ undergoes a metal-insulator transition to a charge-ordered state at $T_{\rm CO} = 185$~K. We discuss their properties in Sections~\ref{sec:propertiesQSL} and \ref{sec:cat}.

\section{Charge Order and Ferroelectricity}
    \label{sec:ChargeOrder}
Among the quantum phenomena emerging in organic low-dimensional solids with strong electron correlations,
ferroelectricity driven by charge order is probably the most prominent one.
Ferroelectricity was discovered 100 years ago \cite{Valasek1921};
nowadays it is of great fundamental and technological importance, widely exploited in modern devices such as memory storage elements, filters, high-performance insulators, and most recently in the energy harvesting technologies \cite{Rabe07}. Ferroelectrics exhibit a spontaneous electric polarization that can be reversed by applying an external electric field. Conventional ferroelectrics are commonly classified into two groups termed as displacive and order-disorder ferroelectrics \cite{LinesGlassBook}. In the former, the polarization originates in the relative shift of positive and negative ions due to lattice distortion, while the polarization in the latter case arises from the collective alignment of permanent electric dipoles. The most important developments in electronic technology, {\it viz.} the size and switching time reduction, and functional improvement of mobile electronic components, require novel ferroelectric materials with properties that  enable overcoming the present limitations in designing new generations of energy efficient electronic devices. A promising route may lie in materials with strongly correlated electrons \cite{Dagotto05, Hu20}. These materials undergo various phase transitions into broken-symmetry electronic ground states, including electronic ferroelectricity and multiferroicity, due to a fine interplay of charge, spin, lattice and orbital interactions \cite{Dagotto01, Seo00, Yoshioka12, Hotta12, Ikeda05, Milward05, Hemberger05, Kimura08}.

Electronic ferroelectricity is basically charge-driven, meaning that the macroscopic electric polarization primarily depends on electron correlations and is closely related to charge ordering. In contrast to conventional displacive ferroelectrics, the mass of charge particles responsible for polarization is therefore very small and large lattice distortions are not expected. This implies the existence of strong dielectric fluctuations and polar domains of small size. Novel collective modes of excitation are also expected to provide ultrafast optical responses. Another distinction from conventional ferroelectric materials arises from the interplay of several competing electronic interactions involved in the charge ordering phase transition resulting in paraelectric-ferroelectric phase separation and causing relaxor-like dispersion in the dielectric behavior. A universal signature of the relaxor is a broad frequency-dependent peak in the real part of the temperature-dependent dielectric susceptibility.
The latter is commonly observed in relaxor ferroelectrics, which exists in a random field state due to internal heterogeneities and disorder \cite{Cross87, Vugmeister90, Crowley11}. Notably, signatures of anomalous relaxor-like behavior and inherent randomness have also been observed in organic solids hosting charge ordering and electronic ferroelectricity, however, a full understanding is still missing.

\begin{figure}
	\centering\includegraphics[clip,width=0.7\columnwidth]{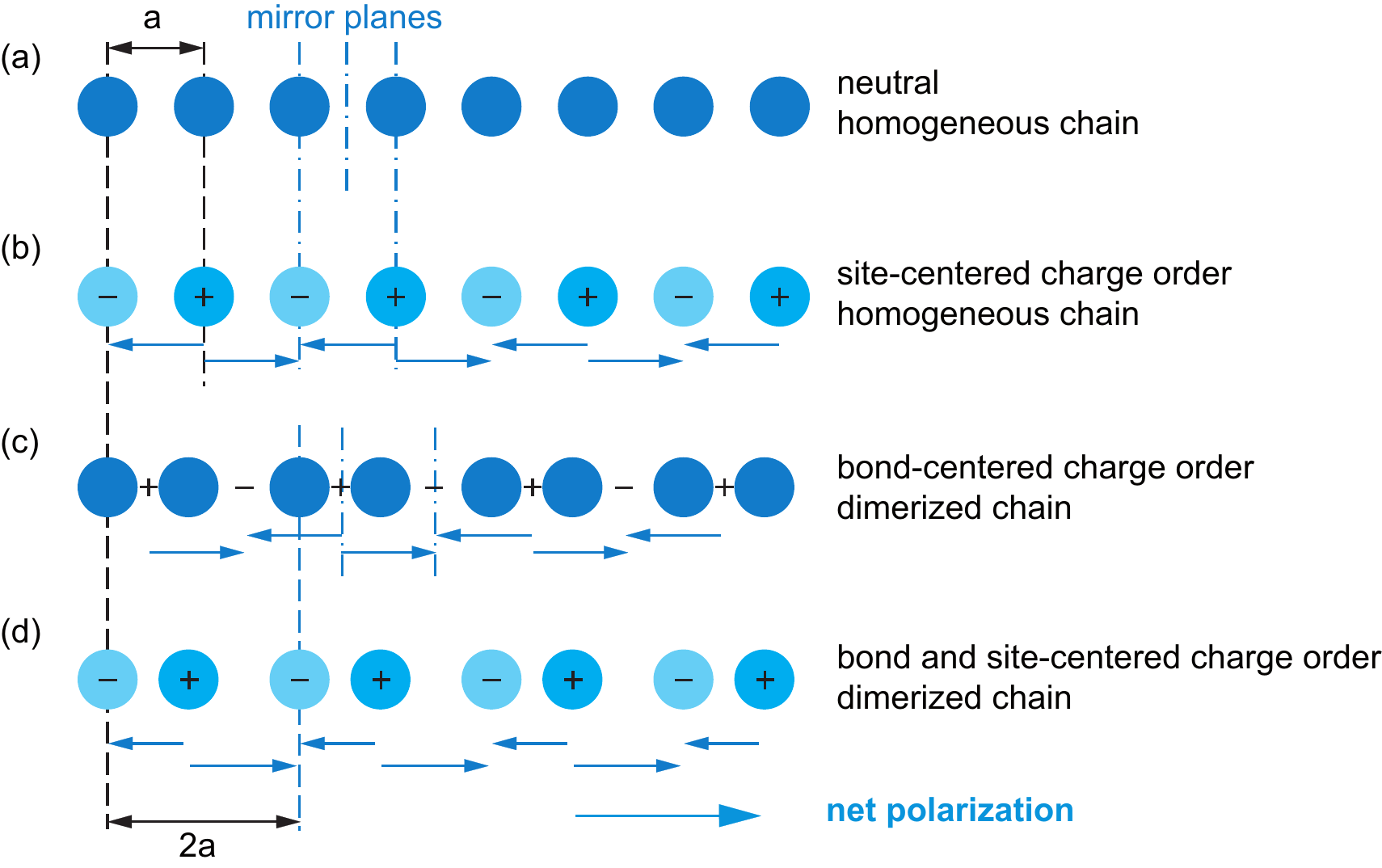}
\caption{Schematic explanation of the basic mechanism by which charge ordering in a one-dimensional chain gives rise to electronic ferroelectricity (following \cite{VandenBrinkKhomskii08}). The blue arrows denote dipolar moments. Site-centered charge order breaks the inversion symmetry between the sites (b), but only together with bond-centered charge order resulting in a lattice dimerization (c), charge order leads to a complete loss of inversion symmetry and a net polarization is established (d).
	\label{fig:modelCOFE}}
\end{figure}

The charge-driven ferroelectricity originating from an electronic phase transition was first shown theoretically  to occur in the strongly correlated insulating phase of the Falicov-Kimball model \cite{Portengen96}.
More recently, the basic mechanism by which the charge ordering can lead to ferroelectricity was elucidated by Efremov {\it et al.} \cite{Efremov04} and van den Brink and Khomskii \cite{VandenBrinkKhomskii08}: it depends on the simultaneous presence of non-equivalent sites and bonds as depicted in Figure~\ref{fig:modelCOFE}. It is worth noting that non-equivalent sites and bonds are not necessarily associated with the site-centered and bond-centered charge ordering; rather they may be simply due to the presence of charge sites with different valences and native dimerized structure, respectively.

Electronic ferroelectricity can be also spin-driven, implying that the macroscopic electric polarization is primarily due to spin frustration and closely related to magnetic ordering \cite{Hill00,Kimura03}. The materials with magneto-electric coupling, in which ferroelectricity and magnetism coexist, commonly known as multiferroics, open prospects of charge control by applied magnetic fields (and {\it vice versa}). In the $21^{\rm st}$ century the flury of activity in the subject of multiferroicity is considerably rising worldwide due to both the intriguing fundamental physics problems and the technological importance in designing new generation of spintronics devices \cite{Cheong07, Chandra18, Zhao18, Gong19}. There exist several extensive review articles on multiferroicity, e.g.\ \cite{VandenBrinkKhomskii08, Wang09, Ishihara10} to name only a few. Multiferroicity and magnetoelectric coupling of electrical and magnetic properties, which is a more common phenomenon \cite{Ederer04, DongDagotto19}, have been identified mostly among magnetic oxides and chalcogenides. However, ferroelectric and magnetic properties may also be interwined in some low-dimensional organic solids.
Notably, magneto-dielectric effects have been detected in several two-dimensional organic solids albeit no sign of the charge-ordering phase transition has been found. These materials show signatures of relaxor ferroelectricity, and multiferroicity was suggested involving the coupled order-disorder ferroelectric and magnetic order.
Several theoretical approaches assuming the presence of quantum electric dipoles due to inherent frustration of molecular dimer units have been developed to explain the experimental observations. They all proposed the important role of spin-charge coupling in the formation of low-temperature phases in charge and spin sectors. This intriguing topic will be addressed in Chapter~\ref{sec:quantumdisorder}.

Charge-driven ferroelectricity has been proposed to be realized in a number low-dimensional molecular organic solids based on the structural dimer units and charge ordering phase transitions (see Figure~\ref{fig:modelCOFE}). Quasi-one-dimensional (TMTTF)$_2$$X$ and TTF-CA materials are considered as prototype examples of electronic ferroelectrics whose properties are rather well understood after the years of extensive research. Here TMTTF stands for tetramethyltetrathiafulvalene which forms a charge-transfer salt with centrosymmetric anions $X^{-}$ such as PF$_6^{-}$, AsF$_6^{-}$, SbF$_6^{-}$, Br$^{-}$  and noncentrosymmetric anions $X^{-}$ such as ReO$_4^{-}$, BF$_4^{-}$. The Fabre salts (TMTTF)$_2$$X$ are dimerized Mott insulators consisting of the dimer units of two neighboring TMTTF molecules, and of anions along the segregated quasi-one-dimensional chains with one hole per TMTTF dimer at high temperatures. A charge-ordering  phase transition, in which TMTTF molecules within a dimer become non-equivalent, involves coupling of the electronic TMTTF subsystem with the anion lattice. The charge ordering phase transition and the low-temperature state show some typical fingerprints of ferroelectricity including structural changes and charge disproportionation due to symmetry breaking, and dielectric response. Several extensive review papers cover experimental and theoretical aspects of electronic ferroelectricity in these systems intensively studied during the first decade  of the present century \cite{Brazovskii08,Monceau12,TomicDressel15,LunkenLoidl15}.

Another family of organic ferroelectrics are mixed-stack compounds, the most prominent being TTF-CA. Here TTF stands for tetrathiafulvalene and CA for tetrachloro-p-benzoquinone called chloranil. The planar TTF or its derivatives, together with chloranil or its variants form a single chain with alternating $\pi$-electron donor and acceptor molecules. At ambient conditions, the material is basically neutral but under elevated pressure or temperature or photoexcitation the TTF-CA becomes ionic. At the first-order phase transition the change of ionicity and dimerization appear simultaneously, resulting in symmetry breaking and a sudden increase of the dielectric constant. Ferroelectricity is also manifested in polarization hysteresis, strong nonlinear effects and ultra-fast optical response. Several papers provide an overview of ferroelectricity in these systems \cite{HoriuchiChPhys06, HoriuchiJPSJ06, TomicDressel15}. In addition, development of novel routes for design and applications of organic ferroelectrics based on single-component polar organic molecules, hydrogen-bonded supramolecules and charge-transfer complexes with TTF, has been presented by an extensive review and articles by Horiuchi and Tokura \cite{HoriuchiTokura08,HoricuhiTokura15}.

Although the signatures of electronic ferroelectricity, such as large dielectric constant, characteristic dielectric dispersion, hysteresis, second-harmonic generation, ultra-fast dynamics, switching, bistable resistance and domain-wall dynamics have also been observed in two-dimensional organic solids,
a complete characterization and full understanding as in conventional ferroelectrics is often missing. Considerable efforts have been devoted to elucidate the nature of collective charge excitations in the charge-driven ferroelectric phases as well as their coupling to applied dc and ac fields. And while some of their features --~such as screened Arrhenius-like dispersion~-- resemble the well-established electrodynamics of charge-density waves in one- and two-dimensions \cite{Gruener88, GrunerBook, VuleticPR06}, others --~like giant nonlinear conductivity and negative differential resistance~-- appear different and have not been encountered until now.
Distinct mechanisms to explain these remarkable effects have been suggested ranging from melting of charge order to dielectric breakdown and sliding.
Diverse approaches have been proposed concerning the mechanism behind charge-ordering phase transition and associated ferroelectricity, ranging from the purely Coulomb-driven to cation-anion hydrogen bonding interactions. Most important, a vital role of geometrical frustration in the formation of ferroelectric phase, related to a triangular arrangement of organic molecules in two-dimensional organic solids, has been recently reviewed in two publications \cite{Ishihara10, Hotta12}.

In this Chapter we present the current state of research and give a survey of the relevant work performed in the studies of charge order and ferroelectricity in two-dimensional molecular solids during the last decade. For further details we refer the reader to three review papers published a few years ago \cite{Ishihara14, TomicDressel15, LunkenLoidl15}.
We focus on three major quarter-filled systems that have drawn most attention during this period:
\aeti, $\theta$-(BEDT-TTF)$_2$\-RbZn(SCN)$_4$ and $\beta^{\prime\prime}$-(BEDT-TTF)$_2$\-SF$_5$$R$SO$_3$ (Sections \ref{sec:COPhaseDiagram}, \ref{sec:COweaklydimerized} and \ref{sec:COFerroelectricity}).
In the final Section~\ref{sec:COdimerized}, we will have a look at \hgcl, where charge order with ferroelectric signatures occurs in a dimerized compound with an effectively half-filled band.

    \subsection{Phase diagrams of quarter-filled systems}
\label{sec:COPhaseDiagram}
Two-dimensional organic materials based on BEDT-TTF molecule hosting electronic ferroelectricity driven by charge order possess extremely rich phase diagrams; their origin lies in the competition between the tendency of electrons to delocalize and strong interactions between charge, spin and lattice. In the conducting layers, BEDT-TTF molecules
form a geometrically frustrated triangular lattice and each molecule accommodates half a hole so that the conduction bands are quarter filled. In contrast to dimer Mott insulators at half filling, which are mainly governed by the on-site Coulomb repulsion $U$, here the combined action of a $U$ and a sizeable inter-site Coulomb repulsion $V$, analyzed within an extended Hubbard model, is expected to lead to the formation of a charge-ordered ground state \cite{Kino95,Seo00, McKenzie01,Seo04,Kino96}. Indeed, it turns out that $V$ is strong enough so that charge-ordered states are observed in numerous materials based on the BEDT-TTF molecule possessing  different types of crystal morphologies, labelled by $\alpha$, $\theta$ and $\beta^{\prime\prime}$ \cite{WilliamsBook,Mori98a,Mori99b,Mori99c}, as depicted in Figure~\ref{fig:structure_general}. On the other hand, in materials with $\kappa$-pattern, in which molecules are paired in dimers (s.c. dimerized materials), the Mott physics plays important role (see Chapters~\ref{sec:MottTransition} and \ref{sec:Frustration}), so that no tendency towards charge order has been expected. However, this common view has been recently challenged by a discovery of charge ordered ferroelectricity in \hgcl\ \cite{Drichko14,Hassan18,Gati18a}. The charge ordered state is quickly suppressed by a rather low pressure, however without any trace of superconductivity \cite{Lohle17,Lohle18}.

\begin{figure}
	\centering\includegraphics[clip,width=0.6\columnwidth]{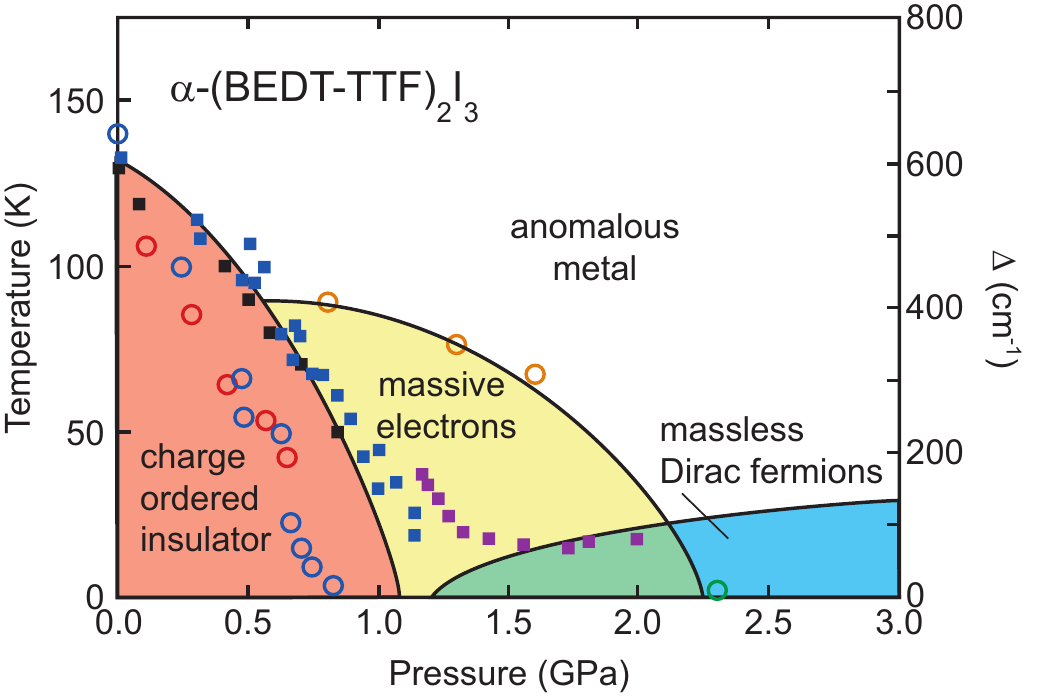}
	\caption{The phase diagram of \aeti\ shows how with varying pressure and temperature different phases and boundaries develop. The charge-ordered transition $T_{\rm CO}$ is suppressed by pressure as probed by various methods (squares, left scale); concomitantly the charge gap $\Delta$ closes as extracted from thermal activated conductivity and optical properties (circles, right axis).
The charge-ordered insulator and an anomalous metal coexist at high temperatures and low pressures, whereas they share the common boundary with the massless Dirac fermionic state at lower temperature and higher pressure (after \cite{Uykur19}).
		\label{fig:alphaphasediagram}}
\end{figure}
The charge-ordered ferroelectric \aeti\ is certainly of utmost importance among the class of two-dimensional molecular materials with quarter filling, thanks to its rich phase diagram
with a number of exotic quantum phenomena ranging from electronic ferroelectricity to superconductivity, from non-linear transport to zero-gap semiconductivity and ferrimagnetism characterized by massless Dirac fermions \cite{Dressel94, Tajima06, Hirata16, Uykur19}. Superconductivity with $T_c$ $\approx$ 7\,K is reported to occur under 0.2 GPa of uniaxial strain applied along the in-plane stacking $a$-axis \cite{Tajima02}.
A stable superconducting phase with $T_c=8$~K can be achieved by tempering the crystals at $70^{\circ}$C for a few days \cite{Schweitzer87}.
The most outstanding feature of the \aeti\ phase diagram is that a charge-ordered ferroelectricity lies nearby a massless Dirac fermions state, as displayed in Figure~\ref{fig:alphaphasediagram}. At ambient pressure, charge-ordering develops below the metal-insulator phase transition $T_\mathrm{CO} = 135$~K with inversion symmetry broken and charge disproportionated between neighboring sites. Increasing pressure suppresses charge order and a massless Dirac fermions phase emerges, thus making $\alpha$-(BEDT-TTF)$_2$I$_3$ the first bulk material, in which the impact of electron correlations on the Dirac-point conductance can be studied \cite{Alemany12, Liu16, Uykur19} (see Section~\ref{sec:DiracElectrons}).

\begin{figure}[h]
	\centering\includegraphics[clip,width=0.4\columnwidth]{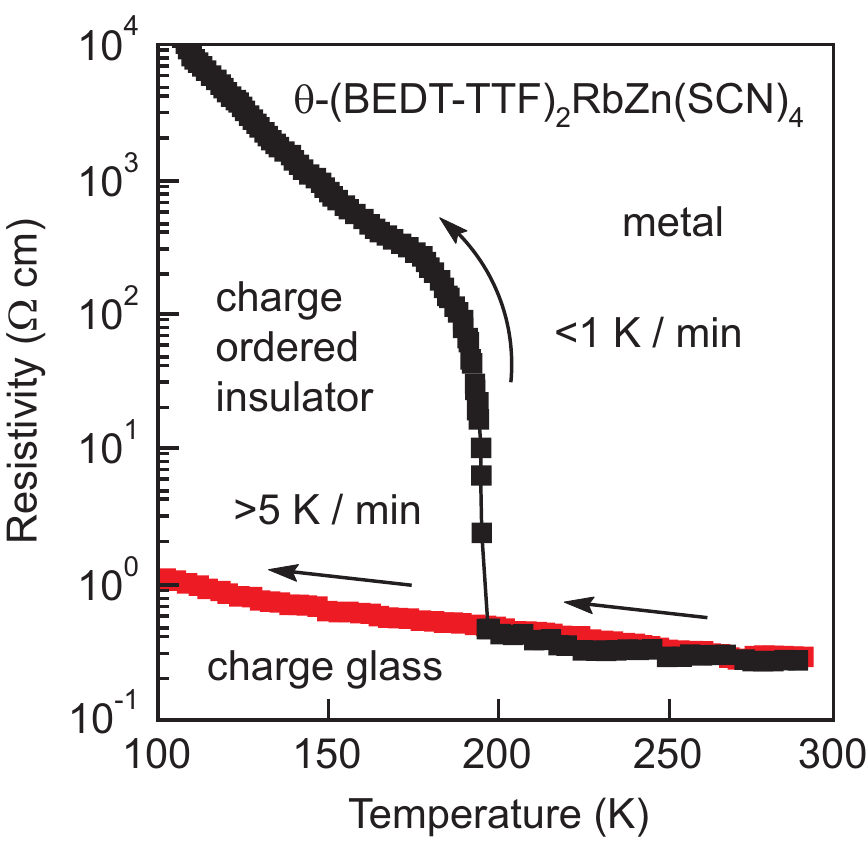}
	\caption{Temperature-dependent resistivity of $\theta$-(BEDT-TTF)$_2$RbZn(SCN)$_4$ measured with different cooling rates. Slow cooling leads to a sharp metal-insulator transition at $T_\mathrm{CO} = 190$~K into the charge-ordered state. Rapid cooling (less than 1~K per minute) prevents this order leaving space to charge glass formation (after \cite{Kagawa13}).
		\label{fig:thetaphasediagram}}
\end{figure}
Another prominent example of electronic ferroelectricity due to charge order is $\theta$-(BEDT-TTF)$_2$RbZn(SCN)$_4$. A remarkable feature of this material is that charge order sets in  right at the boundary of the charge-glass in the phase diagram (Figure \ref{fig:thetaphasediagram}). The competition between charge-ordered phase and charge glass is governed by geometrical frustration.
Increasing the degree of frustration by replacing the RbZn(SCN)$_4^-$ anions by CsZn(SCN)$_4^-$, which form nearly isotropic triangular lattice, completely suppresses the formation of charge order \cite{Sato14}. In $\theta$-(BEDT-TTF)$_2$RbZn(SCN)$_4$ charge order  develops at long-range scale only under sufficiently slow cooling conditions (cooling speed slower than 1 K/min). Similarly to \aeti, a charge-ordered state builds up below the metal-insulator transition $T_\mathrm{CO} = 190$~K with the inversion symmetry broken and charge imbalance between neighboring sites. On the other hand, when cooled rapidly, the charge ordering is avoided and a charge glass with slow dynamics is formed below $T_\mathrm{g} = 170$~K \cite{Kagawa13} (see Section~\ref{sec:ChargeGlass}).

Among other weakly dimerized compounds with charge order, the BEDT-TTF family of  $\beta^{\prime\prime}$-(BEDT-TTF)$_2$SF$_5$$R$SO$_3$ has attracted a lot of attention
because they constitute an all-organic superconductor with large polyfluorinated anions [Figure~\ref{fig:structure_beta}(c)] built with H$\cdots$F hydrogen bonding \cite{Geiser96}.
Some years ago, Merino and McKenzie suggested these materials for investigating the interplay between charge order and superconductivity \cite{Merino01, Kaiser10}. However, only recently, a generalized phase diagram has been established showing that both the charge disproportionation and the temperature of the charge-ordering phase transition scale with intersite repulsion $V$,
\begin{figure}[h]
    \centering\includegraphics[clip,width=0.7\columnwidth]{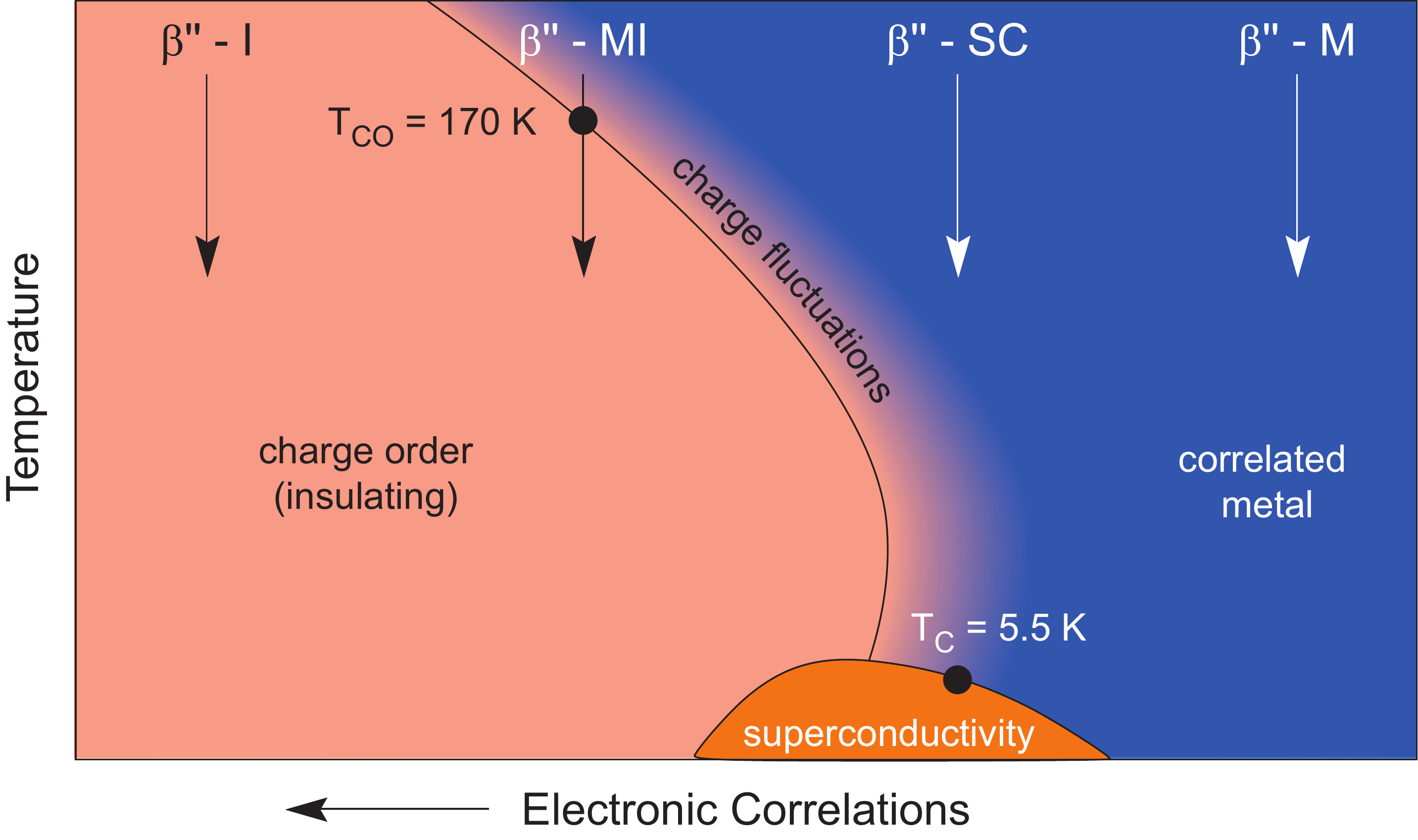}
	\caption{Phase diagram of $\beta^{\prime\prime}$-(BEDT-TTF)$_2$SF$_5$$R$SO$_3$:
as electronic correlations get weaker, the charge order is gradually suppressed and superconductivity sets in. $\beta^{\prime\prime}$-I, $\beta^{\prime\prime}$-MI, $\beta^{\prime\prime}$-SC and $\beta^{\prime\prime}$-M stand for materials with $R$ equal CH$_2$, CHFCF$_2$, CH$_2$CF$_2$ and CHF, respectively; cf. Figure~\ref{fig:structure_beta}(c)
(adopted from  \cite{PustogowPRB19}).
		\label{fig:betaphasediagram}}
\end{figure}
as theoretically expected \cite{PustogowRC19, PustogowPRB19, Drichko09, Girlando11u, Girlando12, Girlando14}.  Starting at the strongly correlated side $\beta^{\prime\prime}$-I ($R$ = CH$_2$) is an insulator up to the room temperature, followed by $\beta^{\prime\prime}$-MI ($R$ = CHFCF$_2$) which undergoes a metal-to-insulator phase transition at 170\,K, $\beta^{\prime\prime}$-SC ($R$ = CH$_2$CF$_2$) becomes a superconductor at $T_c=5.5$~K, while $\beta^{\prime\prime}$-M ($R$ = CHF) is metal at all temperatures. Most remarkably, a superconducting state occurs close to the charge-ordered insulating state as plotted in the phase diagram of Figure \ref{fig:betaphasediagram}; an enhanced charge disproportionation leads to an increase of superconducting transition temperature $T_c$ indicating a critical role of charge fluctuations in its formation.

    \subsection{Ferroelectricity driven by charge order in weakly dimerized solids}
\label{sec:COweaklydimerized}
Ferroelectricity driven by a charge-order phase transition is predominantly established in weakly dimerized solids with quarter-filled bands.
Among them \aeti\ represents the most outstanding example, which has attracted strong interest among researchers worldwide since it was introduced to the community by Schweitzer and collaborators in 1984 \cite{Bender84a,Bender84b}. At ambient pressure, \aeti\ undergoes a metal-insulator phase transition from a high-temperature semimetallic state to a charge-ordered at $T_\mathrm{CO}$ = 135\,K. Below this temperature, a striped pattern of charge disproportionation sets in. The inversion symmetry is broken, the charge is disproportionated between the molecular sites within the dimerized stack of the unit cell and consequently a net polarization is created resulting in ferroelectricity.

In the present and the subsequent Section~\ref{sec:COFerroelectricity}, we give a comprehensive presentation of the relevant findings in \aeti\ and compare them with the results obtained in the studies of ferroelectricity in slowly cooled $\theta$-(BEDT-TTF)$_2$RbZn(SCN)$_4$ wherever appropriate. For completeness reasons, we briefly address $\beta^{\prime\prime}$-(BEDT-TTF)$_2$SF$_5$CHFCF$_2$SO$_3$, in which charge order does not lead to ferroelectricity; inversion symmetry persists at all temperatures and ferroelectricity cannot be established.

\subsubsection{Semi-metallic state at high temperatures}
\label{sec:COhightemperatures}
\begin{figure}[h]
	\centering\includegraphics[clip,width=0.5\columnwidth]{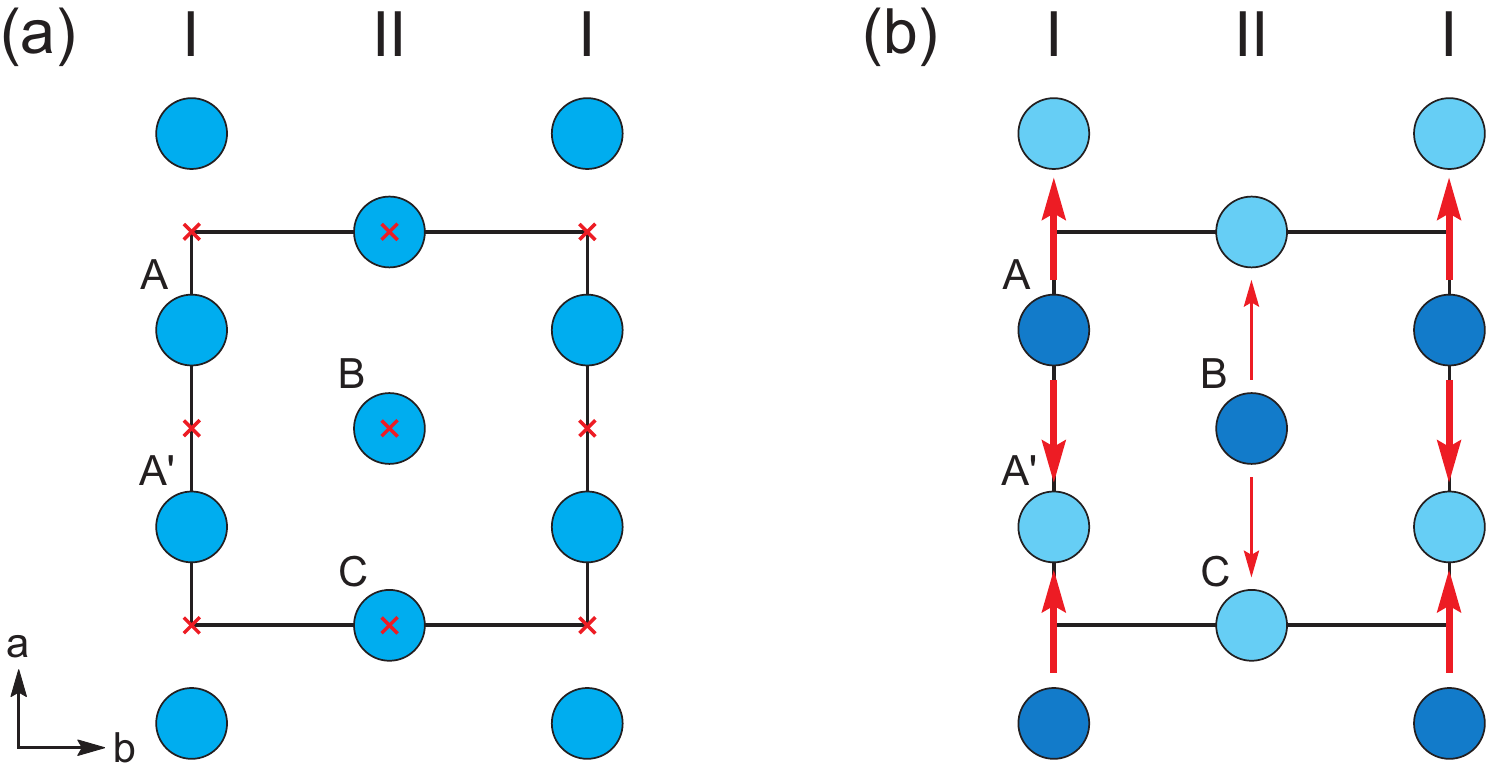}
	\caption{(a) Schematic representation of the in-plane molecular structure of \aeti\ in the high-temperature semi-metallic state. Within the $(ab)$-plane the BEDT-TTF molecules form stacks along the $a$-direction in a herringbone way and exhibit a triangular arrangement with adjacent stacks. Molecules belonging to weakly dimerized stack I and non-dimerized stack II are denoted as A, A$^\prime$, and B, C, respectively. Crosses denote the inversion center relating A and A$^\prime$ molecules. (b)~Charge-density distribution in the in-plane molecular layer of \aeti\ in the low-temperature charge-ordered state. Molecules with higher (charge-rich) and lower (charge-poor) charge density are shown by dark and light blue circles, respectively. Electric dipoles are indicated by arrows: importantly, only dipoles in stack I contribute to a net polarization, while those in stack II cancel each other out.}
		\label{fig:alphamolecularstructure}
\end{figure}

We first address the semi-metallic state at high temperatures. Here, $\alpha$-(BEDT-TTF)$_2$I$_3$ \cite{Bender84a, Bender84b} crystallizes in non-polar space group ${\rm P}\bar{1}$. As is described in Section~\ref{sec:structure}, the unit cell contains four {BEDT-TTF}$^{0.5+}$ molecules and two  I$_3^-$ anions. Commonly, two stacks are distinguished: within column I, which is weakly dimerized, BEDT-TTF molecules, denoted as A and A$^\prime$,  are related through an inversion center and thus are equivalent; column II consists of two different types of molecules, denoted as B and C, which are uniformly spaced and lie at the crystallographic inversion centers [Figure \ref{fig:alphamolecularstructure} (a)].

According to the density functional calculations \cite{Alemany12}, as well as extended H\"{u}ckel molecular-orbital calculations \cite{Mori84}, the system is a semi-metal with very small electron and hole pockets; this results in an experimentally observed weakly metal-like conductivity within the molecular plane \cite{Bender84a, Bender84b,Dressel94}. The direct experimental proof for semimetallicity is provided by recent Hall effect and magnetoresistance measurements, which show that dc transport is governed by the high mobility of electrons and holes resulting in an almost temperature-independent conductivity. The value of the Hall coefficient confirms the idea of a quarter-filled band;
the dominantly interpocket scattering equalizes the mobilities of the two types of charge carriers \cite{IvekCulo17}.

\begin{figure}
	\centering\includegraphics[clip,width=0.7\columnwidth]{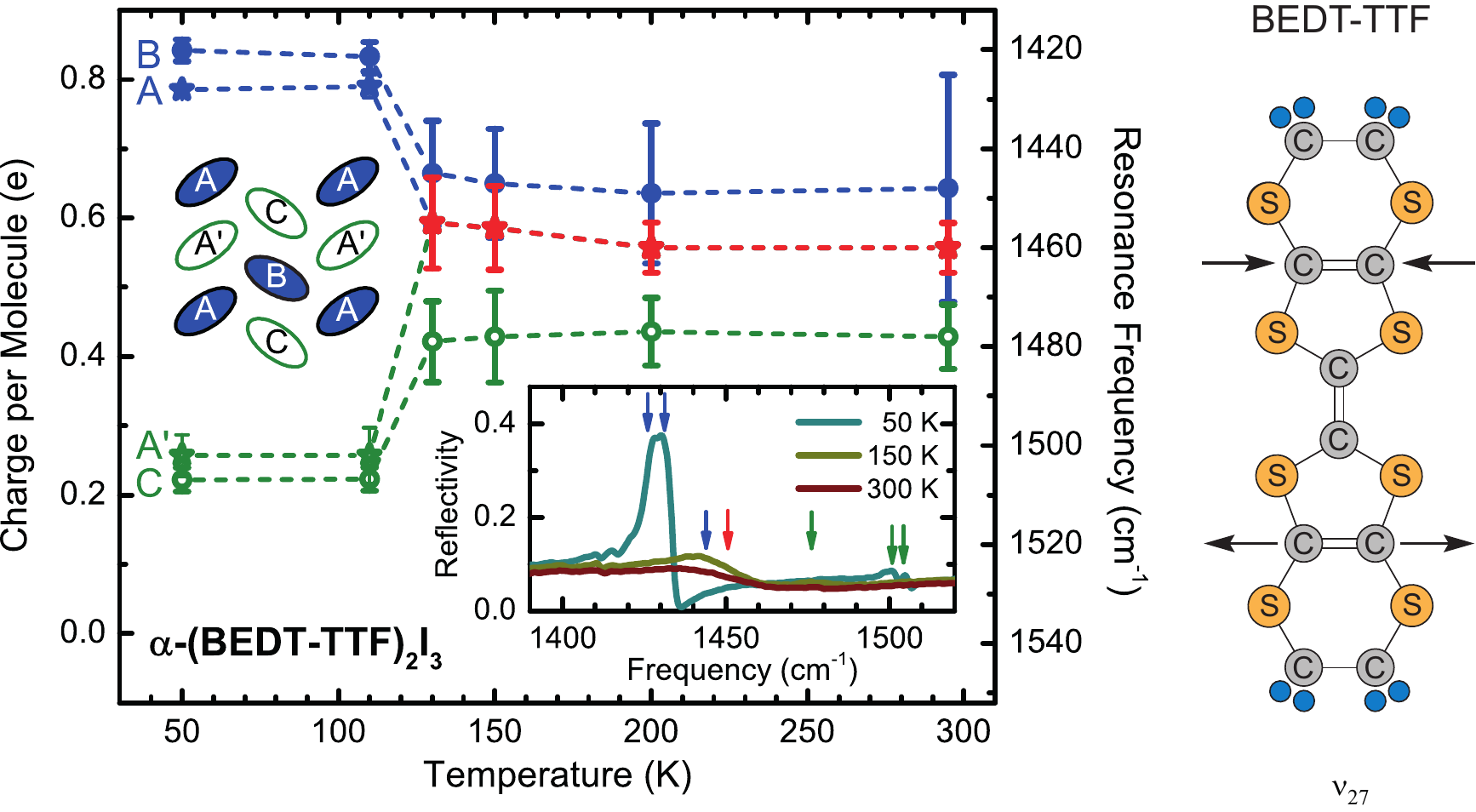}
	\caption{Temperature dependence of the charge distribution (left axis) and the frequency position of $\nu_{27}(b_{1u})$ stretching mode (right axis) of $\alpha$-(BEDT-TTF)$_{2}$I$_{3}$. Below the phase transition temperature of
		$T_{\rm CO}=135$~\,K, the charge per molecule changes drastically for all four sites. The inset shows reflectivity in the frequency range of the $\nu_{27}(b_{1u})$ vibrations. The positions of the vibrational modes are illustrated by the
		colored arrows, corresponding to the dots in the main frame. At 300\,K, blue color indicates the low-frequency vibration of the B molecule, green the C molecule at high frequencies, and the red color the molecules A and A$^\prime$ located in between. The charge ordering phase transition is visible by the splitting of the peak: now two features are present each one with a double-peak structure (from \cite{Beyer16}). The sketch on the right illustrates the structure of the BEDT-TTF molecule and the antisymmetric vibrations of the outer C=C double bond involved in the $\nu_{27}(b_{1u})$ mode leading to an alternating charge flow along the molecular axis.}
		\label{fig:alphanu27}
\end{figure}

Notably, a striped charge pattern is observed at room temperature by scanning tunneling microscopy \cite{Katano15}. Partial charge disproportionation and charge fluctuations already present at $T=300$~K persist all the way down to the charge-ordering phase transition. The charge imbalance was first suggested based on nuclear-magnetic-resonance (NMR) measurements by Takahashi and collaborators \cite{Takano01b, Takahashi06} and was later deduced from x-ray diffraction measurements \cite{Kakiuchi07} using empirical relationship between intramolecular bond lengths and charge density \cite{Guionneau97} and confirmed by optical vibrational measurements \cite{Wojciechowski03,Yue10, IvekPRB11}.
The best local probes utilized in these experiments are
the highly charge-sensitive stretching modes of the BEDT-TTF molecules, such as the symmetric Raman-active $\nu_{3}$ mode, and the asymmetric infrared-active $\nu_{27}$ mode of the outer C=C bonds
\cite{Maksimuk01,Dressel04a,Yamamoto05,Girlando11a,Yakushi12}. The $\nu_{27}(b_{1u})$ band at high temperatures is shown in the inset in Figure \ref{fig:alphanu27} {}\cite{Beyer16}. The result corresponds well to the one reported previously \cite{IvekPRB11,TomicDressel15}. With an -- in a first approximation -- linear shift of 140~cm$^{-1}$ per unit charge, the charge disproportionation $2\delta_\mathrm{\rho}$ is calculated from
\begin{equation}
2\delta_{\rho} = \frac{\delta\nu_{27}(b_{1u})}{140~{\rm cm}^{-1}/e} \quad ,
\end{equation}
where $\delta\nu_{27}$ is the difference in frequency positions between the two $\nu_{27}$ vibration bands associated with two non-equal BEDT-TTF molecules, i.e 2$\delta_{\rho}$ = $\rho_{\rm rich}$ - $ \rho_{\rm poor}$ \cite{Yamamoto05}. For the BEDT-TTF cations, an increase in positive charge loosens the bonds, {\it i.e.} the vibrational features are redshifted.

The broad absorption band observed in the infrared spectra at about 1445~\cm, corresponding to an average charge of $+0.5e$ per molecule hides a non-uniform site charge distribution. Resolving the associated splitting indicates that charge imbalance is present only within stack II and is rather small $2\delta_\mathrm{\rho} < 0.2e$ (Figure \ref{fig:alphanu27}); a similar result is deduced from x-ray scattering measurements \cite{Kakiuchi07,TomicDressel15}. The values of charge density at different molecular sites A, A$^\prime$, B and C agree well with results obtained by density functional theory (DFT) calculations: $+0.52e$ (A, A$^{\prime}$), $+0.55e$ (B) and $+0.38e$ (C) \cite{Ishibashi06, Alemany12}.

The question remains how to explain the charge disproportionation at elevated temperatures. Kakiuchi {\it et al.} discard the local electric potential distribution originating from the iodine ion in the anionic layer and suggest that the charge disproportionation may be due to distribution of transfer integrals among the BEDT-TTF molecules \cite{Kakiuchi07}. Their view was challenged by the findings of Alemany {\it et al.} \cite{Alemany12}, who argue that the hole concentration in the highest-occupied molecular orbitals is intrinsically related to the strength of hydrogen bonding between the teminal hydrogens of BEDT-TTF molecules and the anions. They conclude that molecules A and B participate in a number of very short H-bonds with anions, whereas molecules C are not involved in those; instead they make the strongest sulfur-sulfur contacts insuring in this way the stability and electronic delocalization within molecular layers.

At the end, we note that short-range charge disproportionation and charge fluctuations at high temperatures also develop in slowly cooled $\theta$-(BEDT-TTF)$_2$RbZn(SCN)$_4$ as evidenced by NMR line broadening and diffuse x-ray scattering \cite{Miyagawa00, Chiba04, Watanabe04}.

\subsubsection{Metal-insulator transition into the charge-ordered state}
\label{sec:MITchargeorder}
In the following, we address the charge-ordering phase transition and the charge-ordered state which develops upon cooling.
When the temperature reaches $T_\mathrm{CO}=135$\,K, several striking changes in the physical properties, electronic as well as in structural, take place.

\begin{figure}[h]
	\centering\includegraphics[clip,width=0.4\columnwidth]{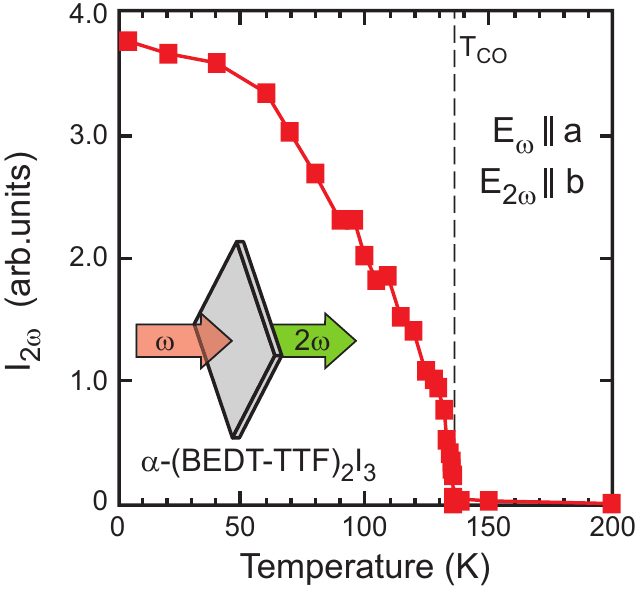}
	\caption{Temperature dependence of the optical second-harmonic-generation (SHG) signal in \aeti. The SHG signal, recognized as a most reliable fingerprint of inversion symmetry breaking, is generated right at the charge-ordering phase transition $T_{\rm CO} = 135$~K, and steadily increases further as the temperature is reduced (after \cite{YamamotoJPSJ08}).
		\label{fig:alphaSHG}}
\end{figure}
Optical second-harmonic generation provides the definite proof that inversion symmetry is lost. As displayed in Figure~\ref{fig:alphaSHG}, the SHG signal sets in abruptly at $T_\mathrm{CO}$ and develops gradually with further decreasing temperature \cite{YamamotoJPSJ08, Denev11}. In addition, large polar domains
develop with several 100 $\mu$m in size and opposite polarization as detected by interferometric experiments of the second-harmonic signal \cite{Yamamoto10}. These findings agree perfectly with the results of x-ray diffraction measurements: At $T_\mathrm{CO}$ lattice deformation breaks the inversion symmetry; the symmetry is reduced from ${\rm P}\bar{1}$ space group at high temperatures to polar space group P1 and twin, right-handed and left-handed domains are formed due to the acentric structure \cite{Kakiuchi07}.

\begin{figure}[b]
	\centering\includegraphics[clip,width=0.9\columnwidth]{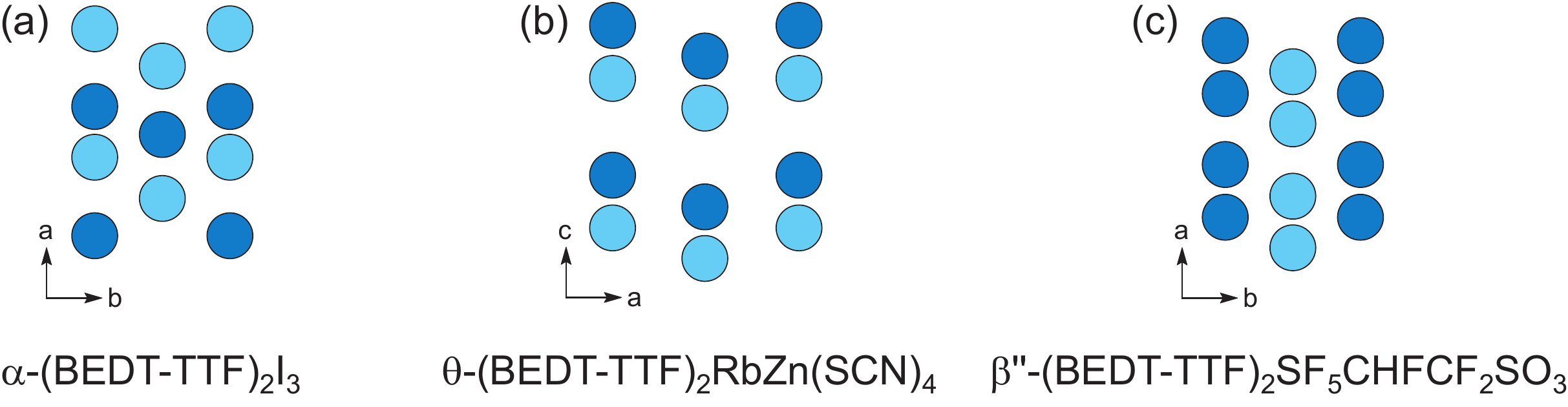}
	\caption{Schematic representation of the horizontal charge stripe arrangement observed in (a) $\alpha$-(BEDT-TTF)$_2$I$_3$, (b) $\theta$-(BEDT-TTF)$_2$RbZn(SCN)$_4$ and in (c) $\beta^{\prime\prime}$-(BEDT-TTF)$_2$SF$_5$CHFCF$_2$SO$_3$. Dark and light shaded circles denote charge-rich and charge-poor molecules, respectively. Dimerized molecular chains run along the $a$-axis in \aeti\ and $\beta^{\prime\prime}$-(BEDT-TTF)$_2$SF$_5$CHFCF$_2$SO$_3$, and along the $c$-axis in $\theta$-(BEDT-TTF)$_2$RbZn(SCN)$_4$, due to labelling convention. Charge order is accompanied by bond order in \aeti\ and in $\theta$-(BEDT-TTF)$_2$RbZn(SCN)$_4$, while there is no bond order in $\beta^{\prime\prime}$-(BEDT-TTF)$_2$\-SF$_5$CHFCF$_2$SO$_3$.}
	\label{fig:stripes}
\end{figure}

Remarkably, upon cooling the charge distribution changes drastically right at $T_{\rm CO}$
as deduced from the temperature-dependent intramolecular deformations. Raman and infrared vibrational spectroscopy as well as $^{13}$C-NMR measurements unanimously confirm these results \cite{Takano01, Moldenhauer93, Wojciechowski03, Kakiuchi07, Yue10, Clauss10, Hirata11, IvekPRB11, Beyer16}.
Figure \ref{fig:alphanu27} displays the abrupt splitting of the charge-sensitive $\nu_{27}(b_{1u})$ mode at $T_\mathrm{CO}= 135$~K, resembling a first-order phase transition;
further cooling leads to only minor changes.
Eventually two pairs of bands are observed: a stronger one around 1425~\cm\
and a weaker one slightly above 1500~\cm, signaling the formation of four different molecular sites
in the unit cell. The lower-frequency bands correspond to a charge of approximately $+0.79e$ and $+0.84e$
per molecule, and the upper-frequency bands to $+0.25e$ and $+0.22e$, respectively.
The overall results reveal the formation of horizontal stripes of charge-rich and charge-poor molecular sites as illustrated in Figure \ref{fig:stripes}(a).
\begin{figure}
	\centering\includegraphics[clip,width=0.35\columnwidth]{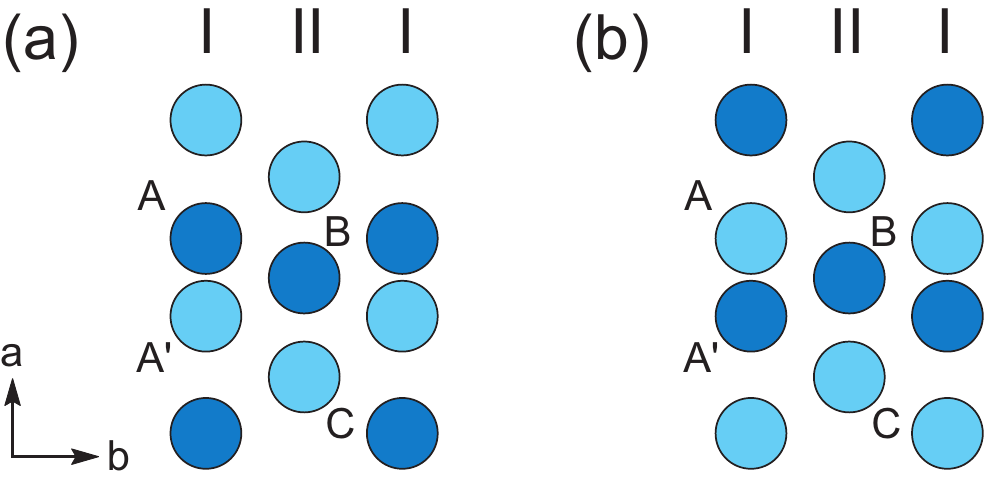}
	\caption{Twin domains in $\alpha$-(BEDT-TTF)$_2$I$_3$. The stripes are formed either (a) by the charge-rich molecules A und B or (b) by
		the charge-rich molecules A$^\prime$ and B.}
	\label{fig:alphatwin}
\end{figure}
In stack II the molecules are equally spaced, but B and C differ in charge.
On the other column, the  molecules A and A$^\prime$
are crystallographically equivalent at temperatures above $T_\mathrm{CO}$,
but posses different charge for $T<T_{\rm CO}$. Now a net-dipole moment is generated
as sketched in Figure \ref{fig:alphamolecularstructure}(b). Conversely, the equivalent bonds within stack II result in a cancellation of the electric dipoles between the charge-rich and charge-poor neighboring molecules. Importantly, there are two ways of rearranging stack I below $T_\mathrm{CO}$:
(A, A$^\prime$)(A, A$^\prime$) and (A$^\prime$, A)(A$^\prime$, A).
The two possible valence arrangements along the $a$-axis lead to twin polar domains,
as illustrated in Figure~\ref{fig:alphatwin}.

When $\theta$-(BEDT-TTF)$_2$RbZn(SCN)$_4$ crystals are slowly cooled, a similar charge order is identified below the metal-insulator transition at $T_\mathrm{CO} = 190$~K \cite{HMoriPRB98,Chiba01,Chiba01b,Watanabe04}. $^{13}$C-NMR, Raman, infrared spectroscopy and x-ray structural studies, all have concordantly
revealed the formation of horizontal charge stripes similar as illustrated in
Figure~\ref{fig:stripes}(b); the stripes are formed  in-plane along the $a$-axis with a charge-rich and charge-poor pattern along the $c$-axis: $+0.8e:+0.2e$ on two non-equal BEDT-TTF molecules within dimers stacked in two columns in the unit cell \cite{Takahashi06,Miyagawa00,Chiba01,Wang01,Yamamoto02,Drichko09}.
Therefore, the net-dipole moment is generated by the electric dipoles on the molecular dimers within both BEDT-TTF columns,
in contrast to the ferroelectricity established in $\alpha$-(BEDT-TTF)$_2$I$_3$.
Another distinction concerns the bonds between molecules along the two diagonal directions in the plane:
for \tetrz\ the bonds are identical, yielding an uniform Heisenberg coupling between spins. Consequently, no gap opens in the spin sector at $T_\mathrm{CO}$ and the system behaves as a paramagnetic insulator down to the spin-Peierls phase transition at $T_{\rm SP}=30$~K, where it enters into the spin-singlet phase \cite{HMoriPRB98,Seo06,Miyagawa00}. Notably, the bonds become inequivalent in the singlet phase,
resembling the charge-ordered state of \aeti.
Finally, we note that structural changes associated with a loss of inversion centers between the BEDT-TTF molecules and the doubling of unit cell along the in-plane $c$-axis
are much larger than in \aeti \cite{HMoriBCSJ98,HMoriPRB98,Watanabe04}.

\subsubsection{Character of the metal-insulator phase transition}
A close inspection of the presented results infers
that the phase transition between the high-temperature paraelectric and
low-temperature ferroelectric states is of first-order:
the charge distribution among the BEDT-TTF molecules within the dimerized chains,
the SHG signal, as well as the reduced space symmetry evidence a very rapid variation at $T_\mathrm{CO}=135$~K (cf.\ Figures~\ref{fig:alphanu27} and \ref{fig:alphaSHG}).
Similar abrupt changes are also noticed in other quantities.
At $T_\mathrm{CO}$ the dc and ac conductivity suddenly drops by several orders of magnitude along all three crystallographic directions, optical spectroscopy and measurements of the static susceptibility
evidence that charge and spin gaps suddenly open,  revealing the insulating and diamagnetic nature of the ferroelectric low-temperature state in \aeti\ \cite{Bender84a, Bender84b, RothaemelPRB86, IvekPRB11, IvekCulo17, Dressel94, Lunkenheimer15}.

\begin{figure}[h]
	\centering\includegraphics[clip,width=0.5\columnwidth]{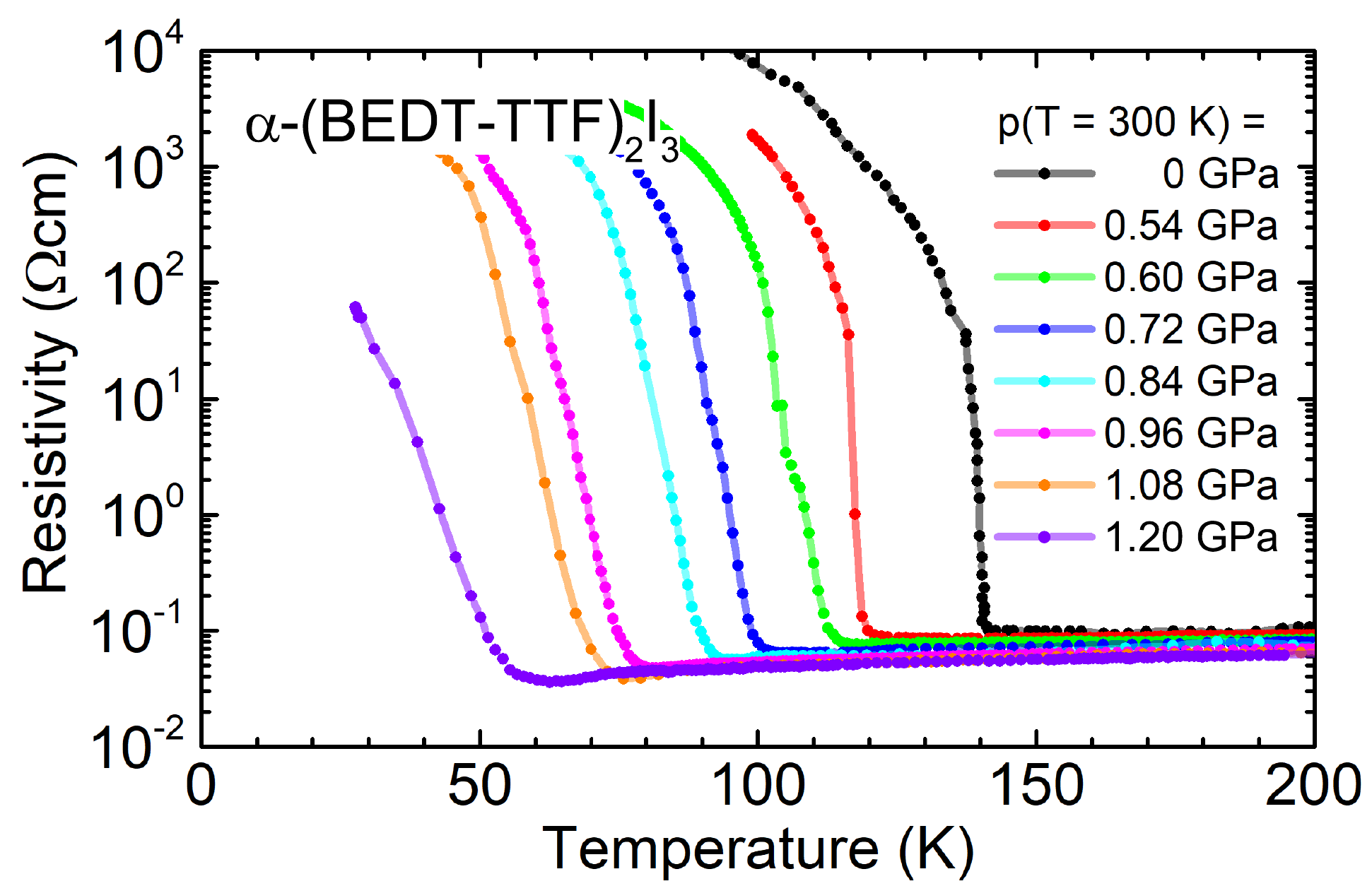}
	\caption{Temperature dependence of the in-plane dc resistivity of $\alpha$-(BEDT-TTF)$_{2}$I$_{3}$ under different pressure values as determined at room temperature (extracted from \cite{Beyer16}). The behavior for higher pressure and lower temperatures is plotted in Figure~\ref{fig:dcpressure}.}
	\label{fig:alphapressure}
\end{figure}
By applying hydrostatic pressure, the charge-order phase transition shifts to lower temperatures at the rate 80-90~K/GPa and is fully suppressed above 1.5~GPa. The charge disproportionation is concomitantly reduced by $0.17~e$/GPa and the optical gap decreases by 470~\cm/GPa \cite{Wojciechowski03, Beyer16}. Notably, a rather abrupt change in charge dispropotionation at $T_\mathrm{CO}$ persists at all applied pressures, whereas steepness of the conductivity drop is gradually diminished, as seen from Figure~\ref{fig:alphapressure}. For pressure values above 1.5~GPa and at low temperatures, the BEDT-TTF molecules carry nearly the same amount of charge as they do at ambient conditions in the semi-metallic state. The observed behavior may be ascribed to the pressure-induced enlarged bandwidth and the resulting decrease of effective intersite Coulomb repulsion \cite{Wojciechowski03, Beyer16}.
In Section~\ref{sec:DiracElectrons} we come back to the low-temperature behavior when higher pressure is applied, and continue with a detailed discussion of correlation effects and Dirac electrons.

Let us just mention here that in slowly cooled $\theta$-(BEDT-TTF)$_2$\-RbZn(SCN)$_4$, unlike the latter compound~-- the charge-order phase transition rises with hydrostatic pressure, which may be due to the enhancement of electronic correlations caused by an increase of diheadral angle between neighboring BEDT-TTF molecules \cite{HMoriPRB98}. Ref.\ \cite{Dressel04a, IvekPRB11, Yakushi12, TomicDressel15} give extensive account on charge-ordering phase transition and charge-sensitive Raman and infrared vibrational modes in the metallic phase of the charge-ordered state in $\theta$-(BEDT-TTF)$_2$\-RbZn(SCN)$_4$ as well as in \aeti.

On the one hand, the observations summarized above indicate that only the ferroelectric order parameter, which is associated with the symmetry breaking at $T_\mathrm{CO}$ and the appearance of charge disproportionation at the dimerized BEDT-TTF molecular sites, are indeed of first-order \cite{Kittel}.
On the other hand, the dc conductivity reveals no hysteresis and the temperature behavior of the transport gap indicates a dominantly mean-field character \cite{IvekPRB11, IvekCulo17}. Thus, the situation does not appear simple; we recall similar controversies and dilemmas in numerous strongly correlated electron systems \cite{Dagotto05}.

\begin{figure}
	\centering\includegraphics[clip,width=0.7\columnwidth]{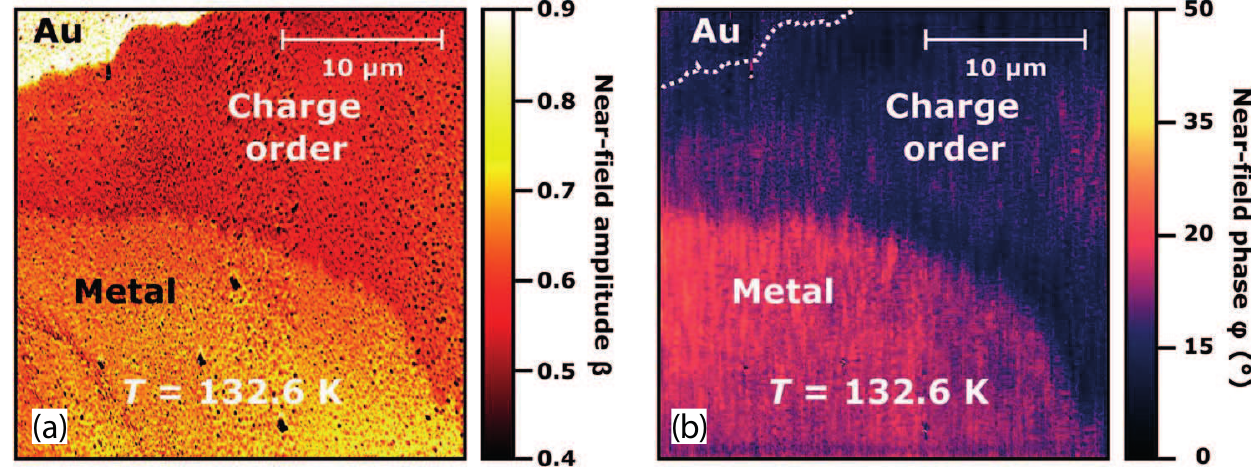}
	\caption{Near-field image of a 28$\mu$m $\times$ 28$\mu$m area recorded with a spatial resolution of 25~nm slightly below the charge ordering phase transition in \aeti.
The microscopic images are obtained by the (a)~amplitude and (b)~phase of the scattered CO$_2$ laser beam ($\lambda = 11~\mu$m) at an oscillating afm tip.
There is a well-defined phase boundary between the metallic (light) and insulating (dark) regions proving its first-order character. The evaporated gold layer (upper left corner) serves as a reference (after \cite{PustogowSciAdv18}).}
	\label{fig:alphanearfield}
\end{figure}

The recent cryogenic scanning optical near-field microscopy study sheds light on this intriguing issue by revealing the spatial evolution of this phase transition,
which evidences its first-order nature \cite{PustogowSciAdv18}.
Recorded images around $T_{\rm CO}=135$~K demonstrate a pronounced phase segregation with a sharp boundary between metallic and insulating regions within a few hundred mK temperature range (Figure~\ref{fig:alphanearfield}). The narrow coexistence range explains why the dc conductivity experiments failed to detect any hysteresis. Remarkably, such a sharp transition occurs only in a homogeneous single crystal. Conversely, when the sample is subject to appreciable strain --~a situation that may easily arise due to improper mounting~--  metallic and insulating regions spatially coexist in a wide temperature range. As a consequence of the pressure dependence presented in Figure~\ref{fig:alphapressure}, the phase transition is suppressed to lower temperatures.

\subsubsection[Origin of the charge-order phase transition]{Origin of the charge-order phase transition and ferroelectric ground state}
\begin{figure}[h]
	\centering\includegraphics[clip,width=0.9\columnwidth]{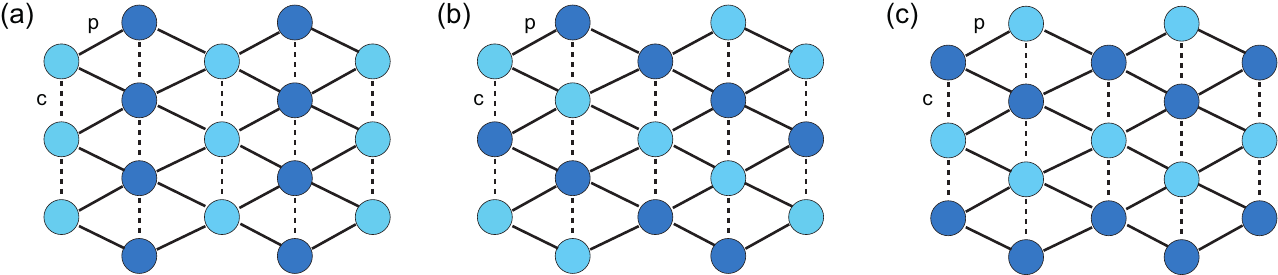}
	\caption{Schematic presentation of charge order patterns: (a) vertical stripe, (b) diagonal stripe, (c) horizontal stripe. All patterns have identical energies within the classical limit, assuming that intersite Coulomb interactions are identical in all three crystallographic directions. According to Clay {\it et. al.} \cite{Clay19} the horizontal stripe is the most stable charge order thanks to the ``1010'' charge pattern in the direction of weakest intermolecular hopping and the ``1100'' along the two diagonal directions with strong intermolecular hopping.}
	\label{fig:COpatterns}
\end{figure}

In an attempt to explain the metal-insulator phase transition observed in \aeti, Fukuyama and collaborators  suggested that electron-electron interactions play a crucial role in this regard \cite{Kino95a, Kino95b}.
Including the intersite Coulomb repulsion $V$, they
identified the importance of charge order in low-dimensional quarter-filled compounds \cite{Seo97,Fukuyama00,Seo03,Seo04}, opening a completely new chapter for these molecular quantum materials.
Seo {\it et al.} also worked out a solely electronic mechanism \cite{Seo00,Seo06} including on-site and inter-site Coulomb interactions, $U$ and $V$ in an extended Hubbard model; it results in a Wigner-crystal-type phase with three possible charge-order patterns as depicted in Figure~\ref{fig:COpatterns}. Importantly, in this mean-field approach the horizontal stripe pattern is only found if the dimerization is explicitly included.

The explanation on purely electronic grounds, however, was challenged subsequently; we now understand that a coupling to the lattice cannot be neglected. In a charge transfer crystal electron-electron and electron-phonon coupling are always present simultaneously. Thus, even if the ferroelectric phase transition is primarily an electronic instability, {\it i.e.} driven by charge ordering, the lattice modifications remain an important side effect due to a finite coupling of electrons to the lattice. The prevailing electronic origin of ferroelectricity is supported by strong non-linear effects and ultra-fast photoresponse; this will be discussed further in Section~\ref{sec:nonlinear+ultafastresponse}.

Along this avenue Clay, Mazumdar and collaborators proposed a model beyond the mean-field theory,
termed ``paired electron crystal'' that includes electron-electron and electron-phonon interactions \cite{Clay02,Gomes16,Clay19}. Numerical many-body calculations successfully reproduce the experimentally established charge pattern shown in Figures~\ref{fig:alphanu27} and \ref{fig:stripes}(a,b).
Remarkably, only a relatively weak electron-phonon interaction has to be incorporated for forming horizontal stripes;
in addition, the bond dimerization perpendicular to them appears as a consequence of the charge ordering.
The authors suggest that the simultaneous formation of the paired electron crystal and the coexisting spin-singlet state
in a sole phase transition occurs only due to lower symmetry in \aeti\ as compared
to $\theta$-(BEDT-TTF)$_2$\-RbZn(SCN)$_4$ where charge ordering and spin-singlet transition take place at different temperatures.

In another approach, Alemany {\it et al.}
argue that electronic interactions between molecular sites are not the driving force of the phase transition; rather it is the interplay between coupled BEDT-TTF molecular and anion subsystems that stabilizes charge order and ferroelectricity \cite{Alemany12, Alemany15}.
Notably, density-functional theory (DFT) calculations
yield  no significant changes in the wave-vector dispersion of the band structure below $T_\mathrm{CO}$, suggesting that only modifications of the local anion-molecular interactions occur at the ordering. In that case even small displacements of anions toward (away) the charge-rich (charge-poor) molecules would increase (decrease) the hydrogen bonding between anions and BEDT-TTF molecules, and thus induce charge redistribution; the resulting striped charge pattern qualitatively corresponds to the experimental findings.

A close look at the phase transition of \aeti\ by Pouget {\it et al.} provides additional support to the relevance of this mechanism \cite{PougetMH18}.
They also point out that the paradigmatic metal-insulator phase transition at 135~K is actually a semimetal-semiconductor phase transition. The presence of two types of charge carriers, electrons and holes, allow at high temperatures
that the electron-hole interaction generates excitonic effects and plays a role in the semimetal-semiconductor phase transition \cite{Rossnagel11}. Finally, the existence of two types of carriers questions the basic assumption of the extended Hubbard model, which considers a quarter-filled system with one type of carriers only.
Further work is needed to consistently explain the complex nature and origin of the phase transition in \aeti.


\subsubsection{Charge-ordered states with no ferroelectricity}

It is instructive to compare the charge order in \aeti\ and $\theta$-(BEDT-TTF)$_2$\-RbZn(SCN)$_4$, on the one hand, with the one observed in $\beta^{\prime\prime}$-(BEDT-TTF)$_2$SF$_5$CHFCF$_2$SO$_3$, on the other hand.
In all three systems, the charge density is arranged in horizontal stripes, however, other properties of the charge order in
$\beta^{\prime\prime}$-(BEDT-TTF)$_2$SF$_5$CHFCF$_2$SO$_3$ are strikingly different \cite{PustogowRC19, PustogowPRB19}.
In the first two systems, below the metal-insulator phase transition
differently charged molecules arrange themselves along dimerized stacks; thus allow the formation of ferroelectricity.
Conversely, in $\beta^{\prime\prime}$-(BEDT-TTF)$_2$SF$_5$CHFCF$_2$SO$_3$,
the BEDT-TTF molecules have equal charge densities within dimerized stacks,
while a charge disproportionation of $2\delta_{\rho} \approx 0.10e$ exists between crystallographically nonequivalent BEDT-TTF molecules belonging to two neighboring stacks [Figure \ref{fig:stripes}(c)].
Thus, charge order is present without bond order; and consequently no net polarization can be established (see Figure~\ref{fig:modelCOFE}).
Another important distinction lies in the fact that charge order persists at all temperatures without any noticeable temperature dependence of the charge imbalance; hence it is not associated with the metal-insulator phase transition at 190~K,
below which the high-temperature non-polar space group ${\rm P}\bar{1}$ does not change.
The origin of the temperature-independent charge order, as well as the metal-insulator phase transition remains to be clarified.
The phase transition may be related to the pronounced one-dimensional character of dimer chains running along the interstack direction at low temperatures, or it may be due to anion ordering \cite{Schlueter01, PustogowPRB19}. It is remarkable that the particular structure in $\beta^{\prime\prime}$-(BEDT-TTF)$_2$\-SF$_5$CHFCF$_2$SO$_3$
enables several short hydrogen bonds between the BEDT-TTF molecules and the oxygen and fluorine atoms of the anions; at present, however, it is not established whether these short hydrogen bonds differ for two non-equivalent BEDT-TTF molecules.
In any case, the anions are disordered at room temperature as well as at $150$~K,
{\it i.e.} below the metal-insulator phase transition. On the other hand, the ethylene end groups of the BEDT-TTF molecules,
which are disordered at room temperature, are ordered below the phase transition at $T=150$~K.
This change may contribute to the mechanism of the phase transition, as it does in $\theta$-(BEDT-TTF)$_2$\-RbZn(SCN)$_4$ \cite{Alemany15}. These hypotheses should be verified in the future.


\subsection{States nearby charge order}
\label{sec:NearChargeOrder}
\subsubsection{Charge glass}
\label{sec:ChargeGlass}

\begin{figure}[h]
	\centering\includegraphics[clip,width=0.4\columnwidth]{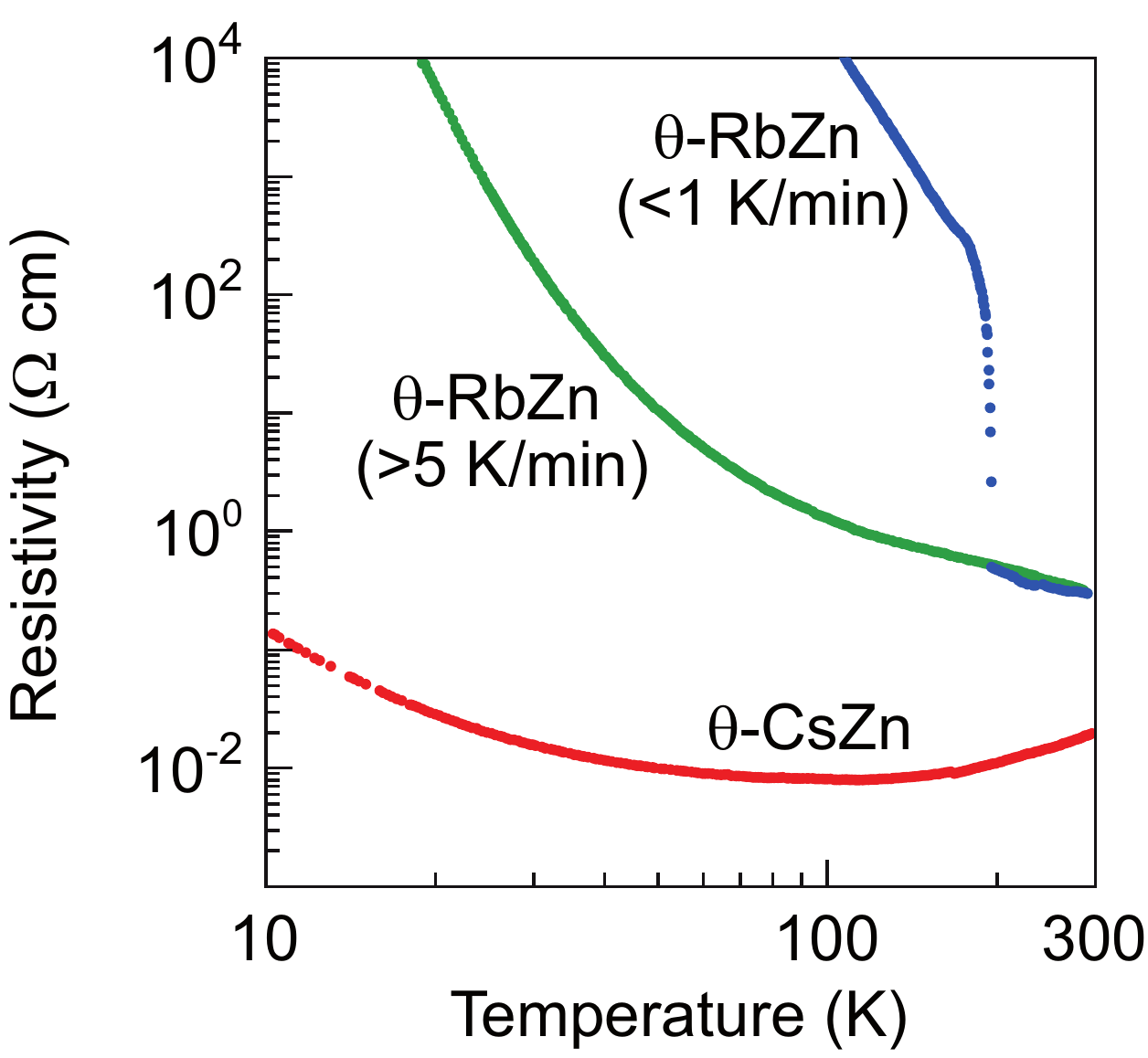}
	\caption{Resistivity behavior $\rho(T)$ of \tetrz\ drastically depends on the cooling rate: the
charge-order phase transition is suppressed by rapid cooling leaving space to charge-glass formation.
The transition is also suppressed by replacing Rb by Cs in \tetcz:
the obtained resistivity behavior does not change with cooling rate in the range of $0.1 - 10$~K/min (after \cite{Sato14}).
		\label{fig:thetaresistivity}}
\end{figure}

Glassy phases and related phenomena such as metastability and slow relaxation are found in a variety of systems when long-range order is prevented by randomness, rapid cooling or competing interactions \cite{Dagotto05,Dyre06};
these effects are commonly attributed to disorder.
A more intriguing case is the disorder-free glassy behavior established in geometrically-frustrated spin systems \cite{Anderson56,Bouchaud98}, while the influence of geometric frustration in the charge sector is less explored \cite{Merino05}.
Glassy phenomena have been demonstrated in the compound \tetrz\ with triangular arrangement of molecular units, where a high degree of charge frustration is present \cite{Kagawa13,Sato14,Sato17}.
As discussed in the previous Section~\ref{sec:COweaklydimerized}, the material displays a metal-to-insulator transition at $T_{\rm CO} = 190$~K
when cooled slowly, leading to a charge-order-driven ferroelectric ground state.
The transition can be avoided either by rapid cooling or by replacing the RbZn(SCN)$_4^-$ anions with CsZn(SCN)$_4^-$ as shown in Figure~\ref{fig:thetaresistivity}.
In the absence of long-range charge order, the resistivity remains low, the dielectric peak is suppressed \cite{Nad07} and the rapidly-cooled phase bears similarities with the high-temperature phase, e.g.\ it exhibits slow dynamics of the order of kHz \cite{Chiba04}.

\begin{figure}[h]
	\centering\includegraphics[clip,width=1\columnwidth]{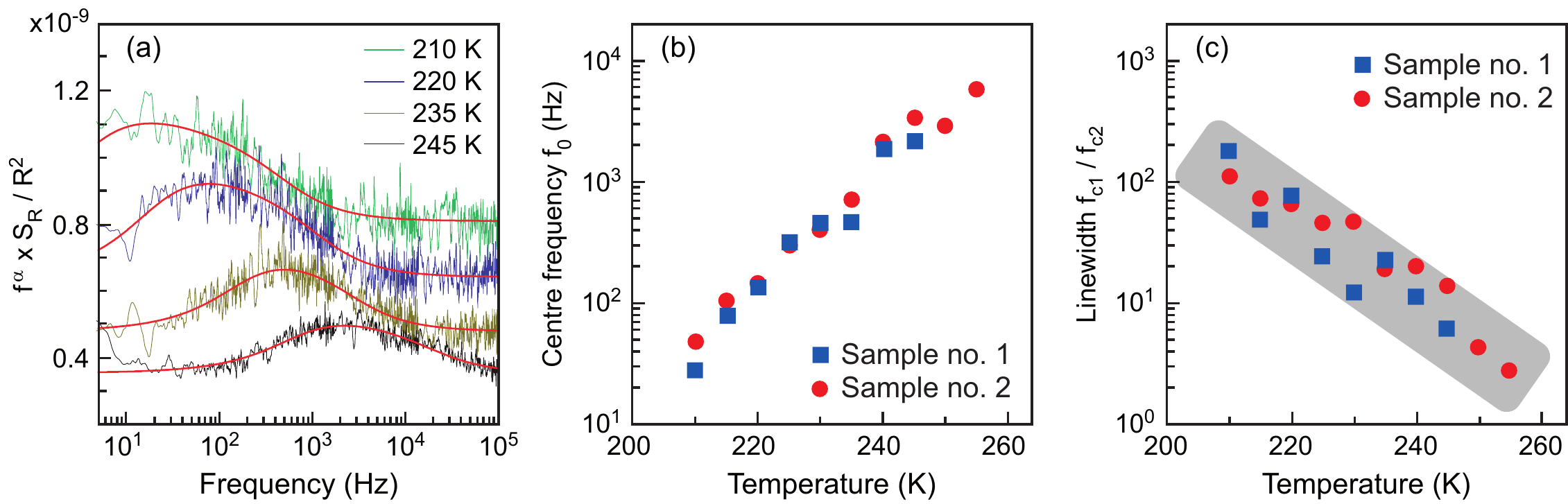}
	\caption{Resistance fluctuations in the rapidly cooled \tetrz. (a) Resistance power spectrum density multiplied by frequency at representative temperatures.  Full lines are fits to continously distributed Lorentzians plus 1/$f^\alpha$.  (b) Slowing down of the characteristic frequency and (c) growth of the dynamic heterogeneity (after \cite{Kagawa13}).
		\label{fig:thetaglass}}
\end{figure}

\paragraph{Slow charge dynamics}
\label{sec:chargedynamics}
The temperature evolution of the slow charge dynamics in organic charge transfer salts can be revealed
by resistance fluctuation spectroscopy \cite{Mueller11}, as demonstrated in Figure~\ref{fig:thetaglass}.
One observes resistance fluctuations with a characteristic frequency superimposed on the background $1/f$
noise for both compounds: \tetrz\ when cooled rapidly \cite{Kagawa13}, and \tetcz\ \cite{Sato14}.
When plotting $f^{\alpha}\times S_R$ as a function of frequency, a broad peak is uncovered that rises out of a constant background [Figure~\ref{fig:thetaglass}(a)]. The spectral shape is well described by a superposition of continuously distributed Lorentzians with high-frequency $f_1$ and low-frequency $f_2$ cutoffs. When $T$ is reduced, the peak position shifts to lower frequency and its linewidth strongly increases [panels (b) and (c)], indicating that the dynamics becomes slower and more heterogeneous. Both features are fingerprints of glassy freezing in supercooled liquids \cite{Ediger00}. It is interesting to note that the growth of slow dynamics is correlated with the evolution of two-dimensional charge clusters, recorded by x-ray diffuse scattering.
These charge clusters are present already at high temperatures and
\begin{figure}[b]
	\centering\includegraphics[clip,width=0.7\columnwidth]{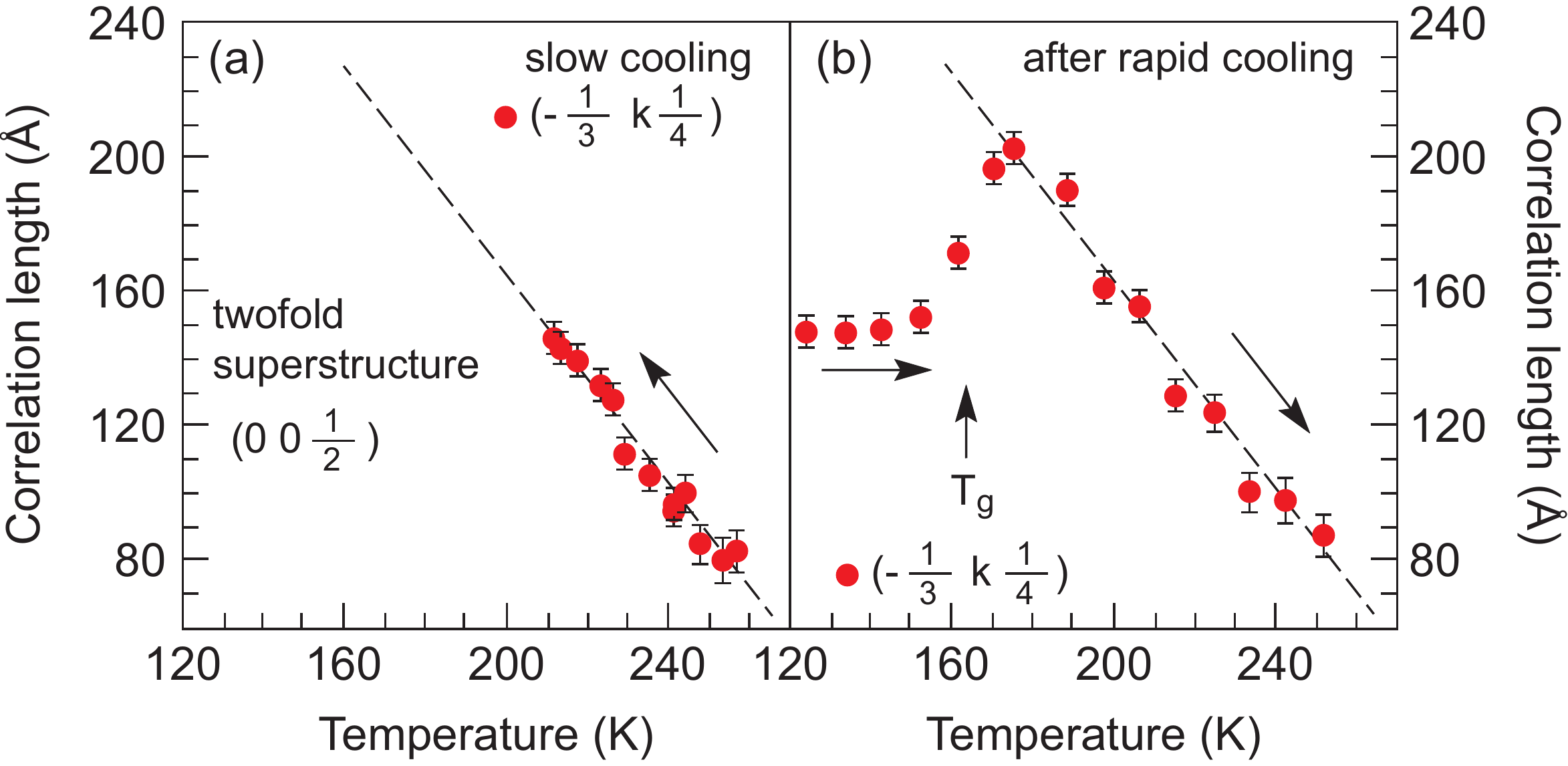}
	\caption{Temperature evolution of the charge cluster correlation length $\xi(T)$ during (a) slow cooling and (b) after rapid cooling of \tetrz\ (after \cite{Kagawa13}).
		\label{fig:thetaglassxray}}
\end{figure}
their intensities and correlation lengths increase on cooling down to $T_{\rm CO}$, as illustrated in Figure~\ref{fig:thetaglassxray}.
After rapid cooling, when the charge ordering is suppressed, charge clusters are only recorded at $T=120$~K. The correlation length $\xi = 140$~\AA\ corresponds to about 25 triangular units and is temperature independent up to $T=150$~K, indicating a frozen metastable state with no long-range order. With further warming, $\xi$ increases and at $T_g$ $\approx$ 165\,K attains the value of about 200~\AA\ in accord with data recorded on cooling.
The temperature behavior of $\xi$ indicates that $T_g$ can be identified as the glass transition temperature of charge cluster dynamics.

\paragraph{Evolution of the electronic crystal in time}
\label{sec:timeevolution}

\begin{figure}[h]
	\centering\includegraphics[clip,width=0.4\columnwidth]{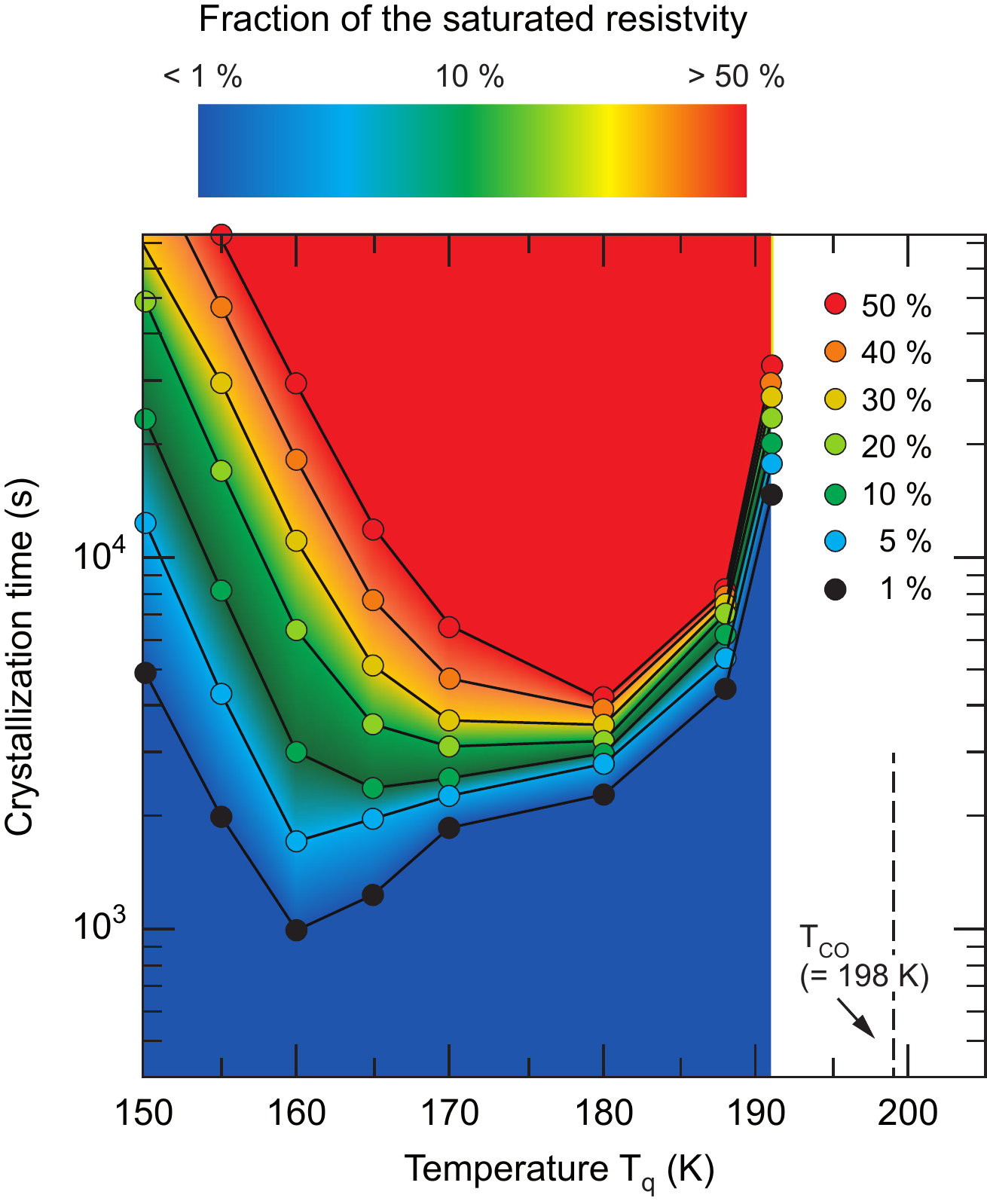}
	\caption{Time-temperature-transformation (TTT) curves of $\theta$-(BEDT-TTF)$_2$RbZn(SCN)$_4$ deduced from the time evolution of resistivity after rapid cooling at different temperatures $T_q$. The color scale denotes the percentage of resistivity recovery (after \cite{Sato17}).
		\label{fig:thetacrystallization}}
\end{figure}

\begin{figure}[h]
	\centering\includegraphics[clip,width=0.6\columnwidth]{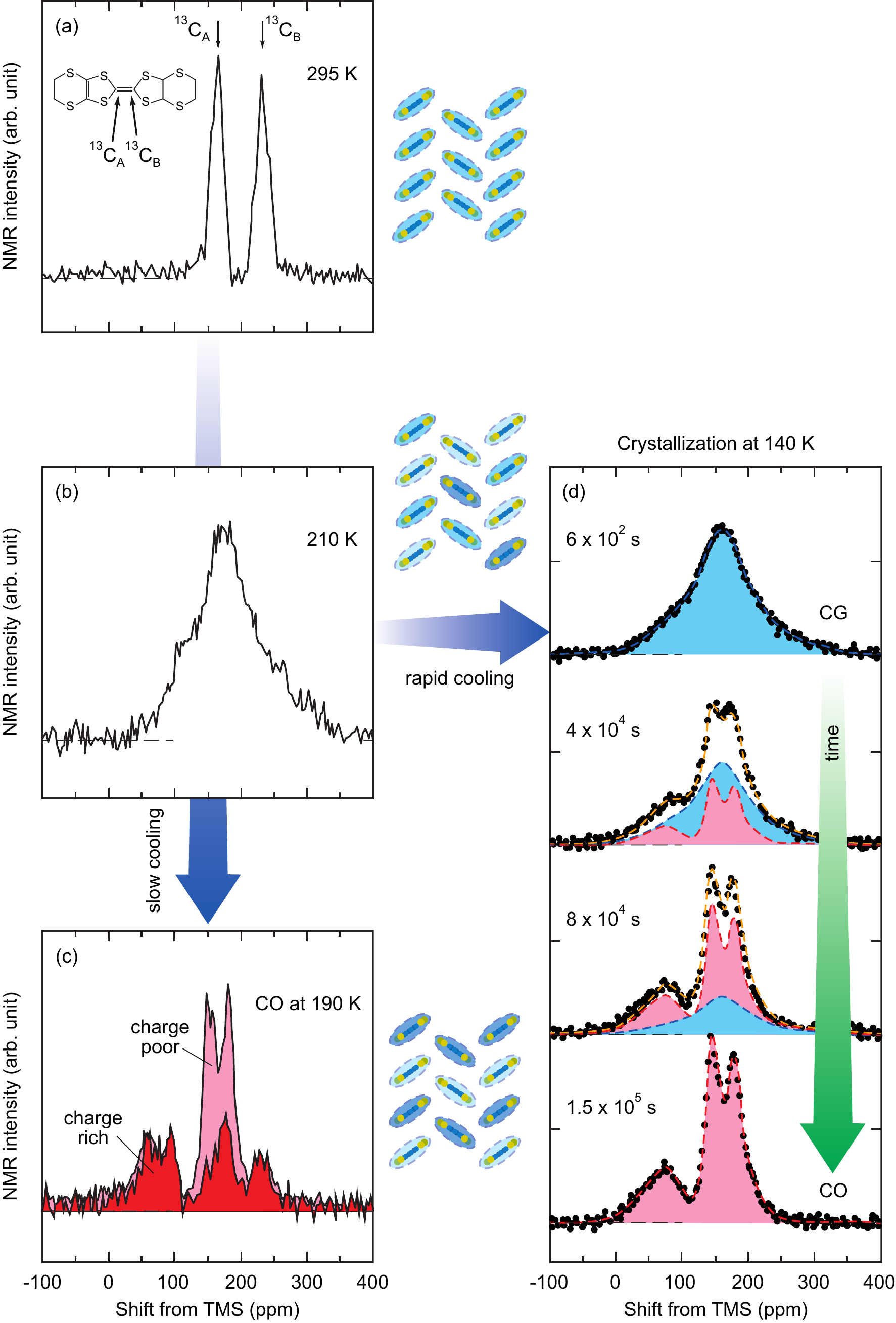}
	\caption{Temperature and time evolution of the $^{13}$C-NMR spectra of \tetrz. Slow cooling results in the charge-ordered electronic crystal, while after rapid cooling the charge glass is established. (a) At room temperature the spectrum consists of two narrow lines due to two non-equivalent $^{13}$C sites in the BEDT-TTF molecule indicating homogeneous charge density. (b) Upon cooling the spectrum broadens due to the slowing down of charge dynamics. (c) The spectrum indicates that charge poor and charge rich sites are formed in the charge ordered stated at $T=190$~K. (d) Time evolution of $^{13}$C-NMR spectra after rapid cooling during the crystallization process at $T=140$~K. The blue and red components are attributed to the charge glass and charge order state (after \cite{Sato17}).
		\label{fig:thetaNMRcrystallization}}
\end{figure}

Having established the charge glass in rapidly cooled \tetrz\ as a metastable state, the time evolution of the electronic crystal state can be followed by monitoring at a  fixed temperature $T_q < T_{\rm CO}$ how the resistivity and NMR spectra evolve in time after the crystal was cooled down rapidly.
The resistivity is much lower in the glass state compared to the electronic-crystal charge-ordered state
and thus characterizes the electronic state macroscopically.
In Figure~\ref{fig:thetacrystallization}, the crystallization time is plotted versus $T_q$:
the dome-like structure is known as time-temperature-transformation (TTT) curve,
commonly observed for crystallization of structural, ionic glasses, or metallic alloys \cite{Loeffler00}.
Analyzing the TTT curve within classical nucleation theory allows the determination of
nucleation and growth dominated regimes.
Microscopic evidence for the electronic crystal growth from a glass state was obtained
by $^{13}$C-NMR investigations. Figure~\ref{fig:thetaNMRcrystallization} compares the
NMR spectra obtained during the slow cooling process with the time evolution of spectra obtained at $T=140$~K after rapid cooling. Remarkably, the broad spectrum observed initially at 140\,K changes its structure in time and eventually adopts the shape characteristic for a charge-ordered state.

\paragraph{Charge vitrification and aging}
\label{sec:aging}

Eventually charge-order prevents further investigations of the glassy phenomena in \tetrz\ under slow cooling.
This problem can be circumvented by studying \tetcz, where the higher degree of frustration
fully prevents long-range charge order.
Indeed, additional fingerprints of glassy states such as cooling-rate-dependent charge vitrification and
non-equilibrium aging behavior were successfully  demonstrated in \tetcz\ \cite{Sato14}.

\begin{figure}
	\centering\includegraphics[clip,width=0.7\columnwidth]{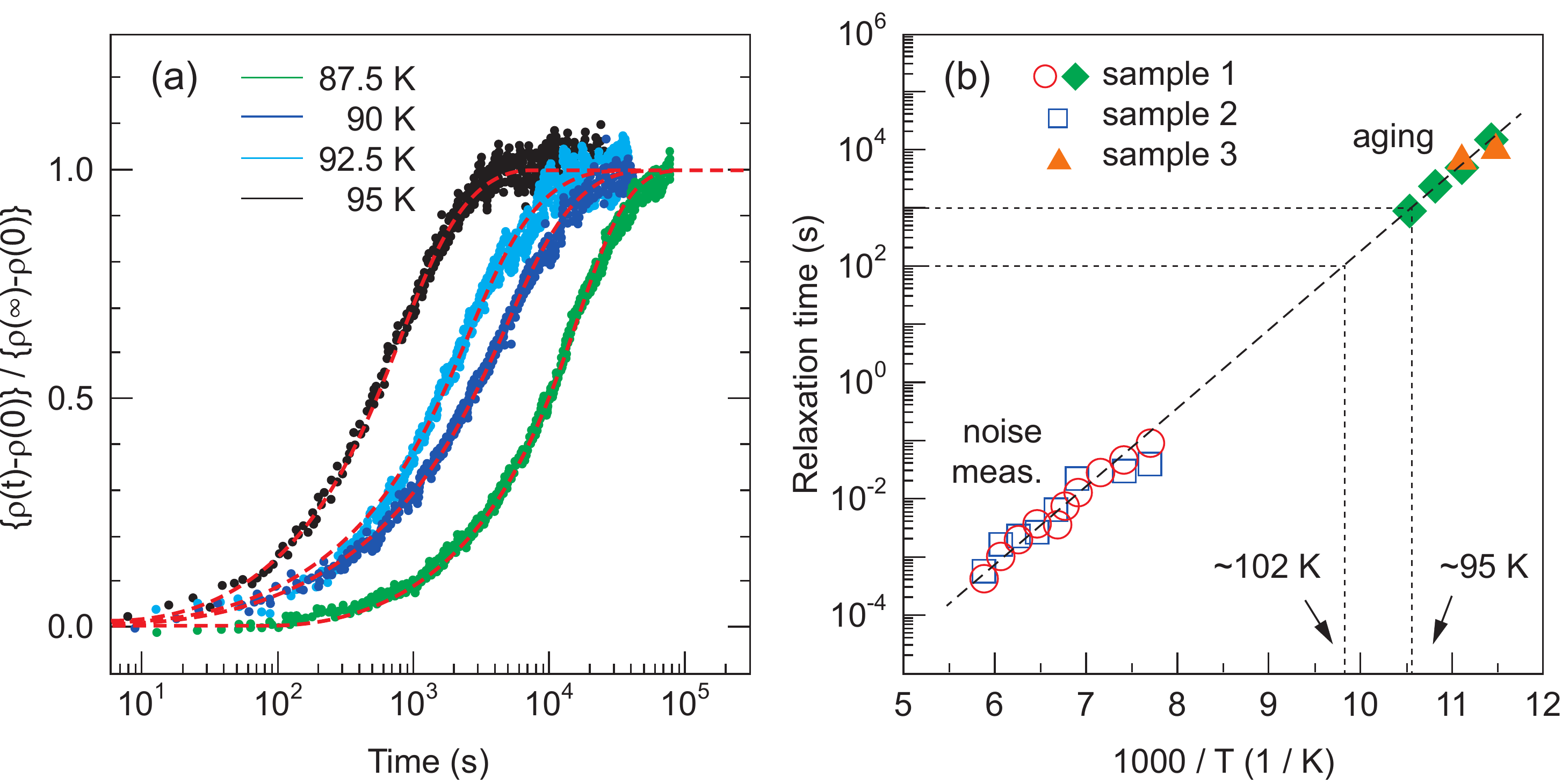}
	\caption{Aging behavior of $\theta$-(BEDT-TTF)$_2$CsZn(SCN)$_4$. (a) Evolution of the resistivity in time at representative temperatures below the glass transition $T_g$. The lines are fits to the Kohlrausch-Willimas-Watts (KWW) relaxation law. (b) Aging relaxation time extracted from the Kohlrausch-Willimas-Watts behavior $\tau_{\rm aging}$ and $\tau_{\rm noise}$ = 1/$f_0$ extracted from the resistance fluctuations peak as a function of inverse temeprature. The line is a fit to Arrheniuos law giving the activation energy of about 2600 K. The dotted lines denote the temperature range where $\tau_{\rm aging}$ reaches values 100-1000\,s widely recognized to define  $T_g$ (after \cite{Sato14}).
		\label{fig:thetaglassaging}}
\end{figure}

(i) From the resistivity {\it vs.} temperature curves under different sweeping rates, we can conclude that faster  cooling leads to a higher glass transition temperature $T_g$, as expected in glass phenomenology.
(ii) The charge-fluctuation broad peak exhibits a strong decrease in characteristic frequency $f_0$ around 100~K. This indicates that the glass transition temperature $T_g \approx 100$~K.
(iii) At $T < T_g$, aging behavior can be observed in the resistivity, as shown in Figure~\ref{fig:thetaglassaging}. It can be well described by the Kohlrausch-Williams-Watts-law based on stretched exponential function \cite{Ediger00,Ediger96}.
    Its relaxation time $\tau_{\rm aging}$ obeys the Arrhenius law reaching $100-1000$~s in the vicinity of $T_g$;
    again consistent with the conventional definition of glass transition.
 Most significantly, $\tau_{\rm noise}$ = 1/$f_0$ at $T > T_g$ follows the same behavior uncovering common dynamics of the equilibrium states at high temperatures and of the non-equilibrium states below $T_g$.
The observed Arrhenius behavior indicates strong-liquid nature meaning that the glassy dynamics on approaching $T_g$  can be described as a gradual slowing down probably reflecting an increasing number of dynamically correllated charge clusters.
(iv) Consistently, x-ray diffuse scattering data reveal that the spatial growth of the charge clusters is closely associated with the glassy charge dynamics.

Having in mind that these organic materials are rather clean systems with no significant amount of impurities,
and that \tetcz, in contrast to \tetrz, possess nearly equilateral triangular lattice,
Kanoda and collaborators suggested that the charge frustration works against long-range electronic crystallization and thus plays the primary role in the formation of charge glass in these two-dimensional strongly correlated systems \cite{Kagawa13,Sato17}.
In other words, the competition between the charge-ordered phase and
a charge glass may be governed by geometrical frustration.
Indeed, Dobrosavljevi{\'c} and collaborators demonstrated theoretically that the interplay of long-range interactions and geometric frustration in disorder-free Coulomb liquids consists
in lifting the ground state degeneracy produced by frustration and in generating a manifold of low-lying metastable states together with characteristic features of glassy dynamics, as observed experimentally.
Their results suggest that \tetrz\ and \tetcz\ constitute prominent examples of a self-generated Coulomb glass \cite{Mahmoudian15}. Still, we should recall that the lattice degrees of freedom are involved in the creation of the electronic-crystals states, both the charge glass as well as the charge order;
the extent to which they influence the crystallization mechanism remains to be clarified.

\subsubsection{Dirac electrons}
\label{sec:DiracElectrons}

\label{sec:tiltedcones}
For more than three decades the charge-transfer salt \aeti\ is one of the most fascinating members of the BEDT-TTF family \cite{Bender84a,Bender84b,Dressel94}.
Besides serving as the prime model for charge-order, the compound attracts enormous attention for the formation of a Dirac electronic state when pressure is applied (Figure~\ref{fig:alphapressure}).
Dirac materials are a novel class of solid state systems in which the low-energy electronic properties are not described in terms of a Schr\"{o}dinger wave equation, but by a relativistic Dirac equation resulting in a linear energy dispersion around the Fermi energy, where valence and conduction bands touch each other \cite{CastroNeto09, Wehling14}.
While three-dimensional Dirac and Weyl semimetals are subject to intense research nowadays \cite{Armitage18}, the underlying physics was first explored in the two-dimensional case of mono-layered graphite, {\it i.e.} graphene \cite{CastroNeto09}.
Whereas in graphene the Dirac cones are isotropic, less symmetric Dirac materials such as two-dimensional organic solid \aeti\ possess anisotropic cones with tilted axes and in presence of long-range and short-range Coulomb interaction provide novel exotic phenomena.

\begin{figure}
  \centering
  \includegraphics[width=0.9\columnwidth]{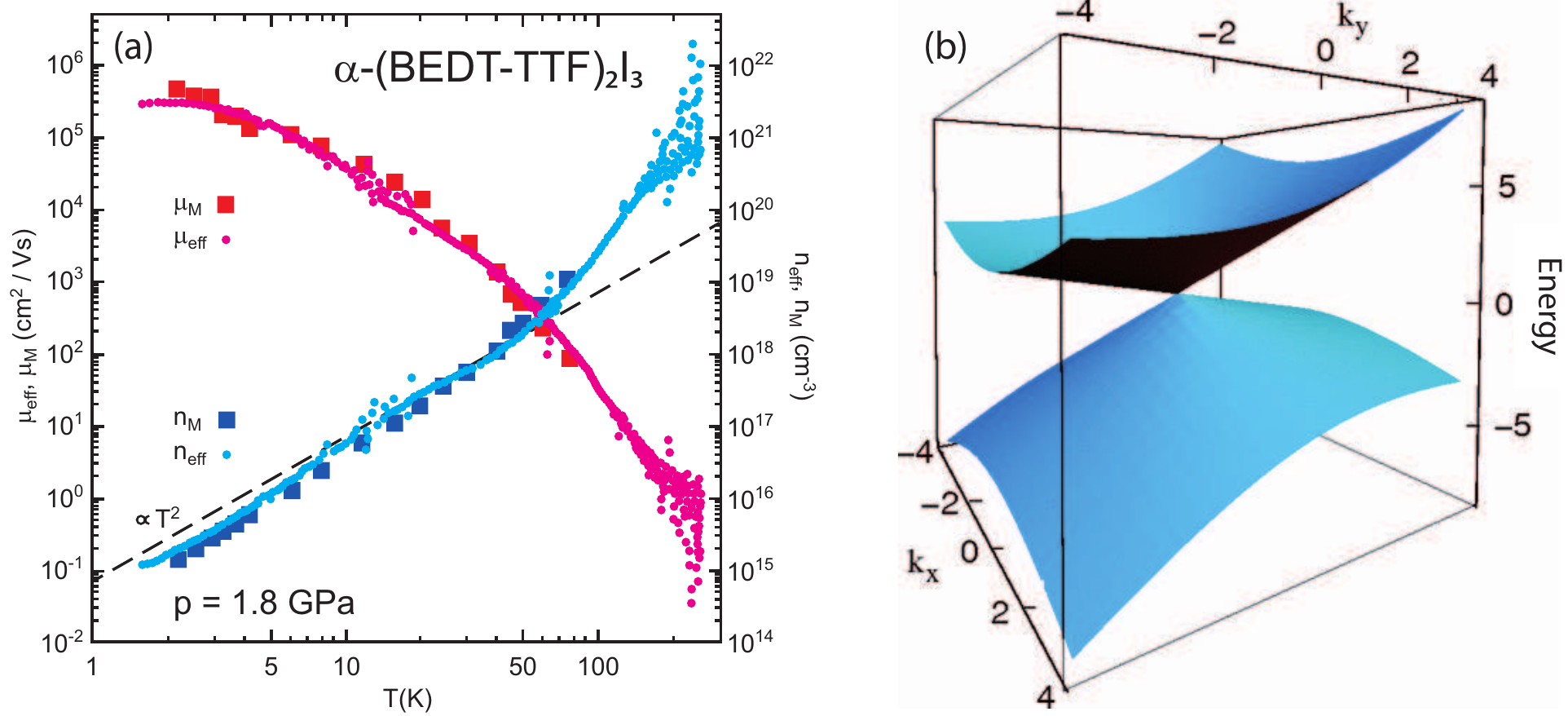}
    \caption{\label{fig:Dirac1}
(a)~The carrier density $n(T)$ and the mobility $\mu(T)$ under the pressure of $p=1.8$~GPa plotted as a function of temperature in double logarithmic manner. The cyan dots show the effective carrier density $n_{\rm eff}$ and the magenta dots the mobility $\mu_{\rm eff}$ estimated from the Hall coefficient ($R_H=1/ne$) and the conductivity ($\mu = \sigma_1/ne$).
The magnetoresistance mobility $\mu_M$ and the density $n_M$, on the other hand, is shown as red and blue squares, respectively. The carrier density obeys $n(T) \propto T^2$ from $T=10$~K to 50~K (indicated by the dashed line) (data from \cite{Tajima06}).
(b)~Band structure of \aeti\ near the Dirac point. The energy dispersion $E_{\lambda}(q) = w_0 + \lambda \sqrt{w_x^2q_x^2 + w_y^2 q_y^2}$ for the special choice of $w_x = w_y = 1$ and $w_0 = (0,0.6)$ is given in natural units.
The Dirac cone is tilted in the $y$-direction (reproduced from \cite{Goerbig08} with permission). }
\end{figure}

Experimentally this topic was pioneered by the magnetotransport studies of Tajima and collaborators
\cite{Tajima00,Tajima06,Tajima07,Tajima09,Tajima18,Kajita14}, quickly explained by bandstructure calculations
\cite{Katayama06,Kino06,Kobayashi07,Goerbig08,Kobayashi09,Kajita14}. Here the challenge is that
the atomic coordinates are not really known with high precision
in the required parameter range of high-pressure and low-temperature; and that the influence of the I$^-_3$ anions cannot be neglected \cite{Kondo05,Kondo09,Alemany12}.
Although the electrical conductivity is basically temperature independent, the carrier density drops quadratically with temperature while the mobility increases approximately by $\mu(T)\propto T^{-2}$, as shown in Figure~\ref{fig:Dirac1}(a).
Subsequent measurements of specific heat \cite{Konoike12}, nuclear magnetic resonance (NMR) \cite{Hirata16,Hirata17} and optical spectroscopy \cite{Beyer16,Uykur19} provide further evidence of massless Dirac Fermions.

Using an extended Hubbard model within the Hartree mean-field theory, Kobayashi {\it et al.} examined the band structure of the stripe charge-ordered state of \aeti\ under pressure and
found that with increasing pressure a topological transition occurs from a conventional insulator with a single-minimum in the dispersion relation at the M point in the Brillouin zone, toward a new phase, which exhibits a double minimum \cite{Montambaux09,Kobayashi11}. The transition is characterized by the appearance of a pair of Dirac electrons with a finite mass. Due to the topological nature of this transition, the Berry curvature vanishes in the conventional phase and has a double peak structure with opposite signs in the new phase \cite{Vanderbilt18}. Osada considered the possibility that \aeti\ becomes a Chern insulator
due to additional transfers, potentials and magnetic modulations \cite{Osada17,Osada19,Osada20}.

An interesting approach should be mentioned at this point; instead of applying hydrostatic pressure, two and four sulfur atoms in the BEDT-TTF molecule can be replaced by selenium in order to increase the orbital overlap and bandwidth \cite{Inokuchi95}. The resulting $\alpha$-(BEDT-STF)$_2$I$_3$ and $\alpha$-(BEDT-TSF)$_2$I$_3$ [better known as $\alpha$-(BETS)$_2$I$_3$] stay metallic down to $T_{\rm CO}=80$~K and 50~K, respectively. There have been suggestions that exotic Dirac cones can be achieved even under ambient pressure and temperature \cite{Morinari14,Naito20a},
unfortunately heavy disorder makes the experimental realization challenging  at present\cite{Naito97,Naito20c,Kitou20}.

\paragraph{Effect of correlations}
\label{sec:DiracElectrons_correlations}
In contrast to graphene, the Dirac cones in \aeti\ are strongly tilted \cite{Katayama06,Goerbig08}, as shown in Figure~\ref{fig:Dirac1}(b). The position of the bands can by tuned by pressure; also temperature is an important parameter. The real situation in \aeti, however, is not as clean as in the case of graphene, due to the existence of
additional electronic bands and the effect of the anion sheets \cite{Alemany12,Pouget18}.
Pressure-dependent optical studies \cite{Beyer16,Uykur19} reveal the presence of
additional charge excitations, also inferred from magnetotransport \cite{Monteverde13}.
The effect of interlayer magnetoresistance in the case of tilted Dirac cones
was subsequently discussed \cite{Tajima18,Mori19,Tani19} and suggested that angular-dependent measurements
could reveal the in-plane anisotropy of electronic structure.

\begin{figure}[h]
  \centering
  \includegraphics[width=0.8\columnwidth]{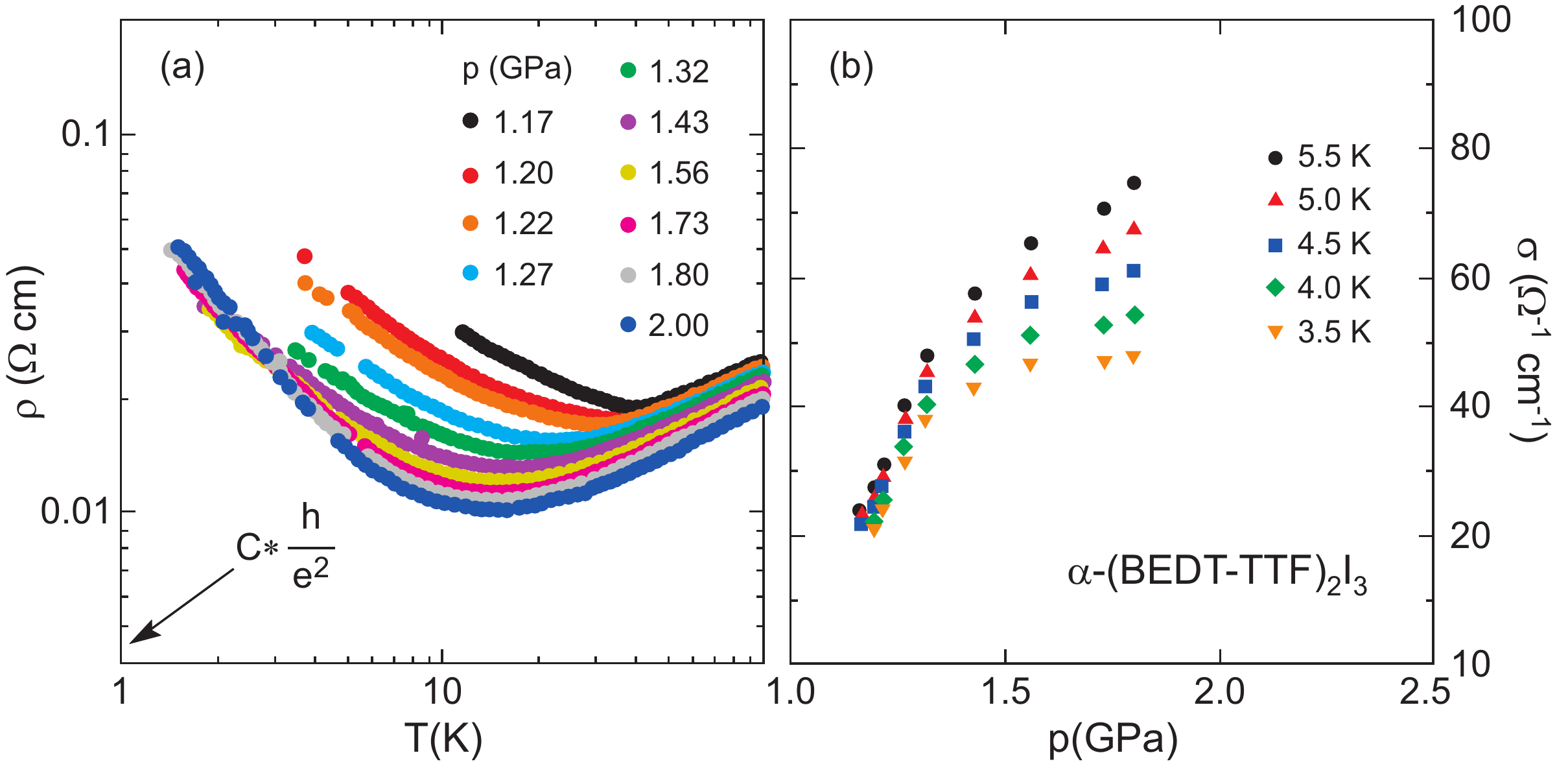}
    \caption{\label{fig:dcpressure}
(a) In the massless Dirac fermion state of \aeti\ above 1.1~GPa
the resistivity $\rho(T)$ increases at very low temperatures even when
the pressure is raised up to 4~GPa.
Also indicated is the bulk resistivity corresponding to the quantum sheet resistance,
$C \cdot (h/e^2) = 4.5 \times 10^{- 3}~\Omega{\rm cm}$, with the lattice constant along the $c$-axis under a pressure of around 2~GPa: $C=1.7$~nm. (b) Conductivity versus pressure for fixed temperatures above 1.1~GPa in the low-temperature region, where the resistivity upturn appears (after \cite{Liu16}). }
\end{figure}

Liu {\it et al.} {}\cite{Liu16} observe an increase of $\rho(T)$ at low temperatures \cite{Liu16} that cannot be completely suppressed and becomes stronger when approaching the charge-ordering transition (Figure~\ref{fig:dcpressure}).
They suggest that massless Dirac fermions  interact via short-range Coulomb repulsion \cite{Tanaka16}.
A pseudogap forms in the charge channel before opening a real charge gap in the charge-ordered phase;
this is in accord with pressure-dependent optical experiments, which infer that the closing of the charge-order gap with pressure comes to a halt around 1~GPa \cite{Beyer16}.
Recently is was pointed out \cite{Winter17}
that in organic materials spin-orbit coupling  might play an important role, causing spin-orbit gaps that render impossible the realization of a true zero-gap Dirac state in \aeti.

\begin{figure}
  \centering
  \includegraphics[width=0.7\columnwidth]{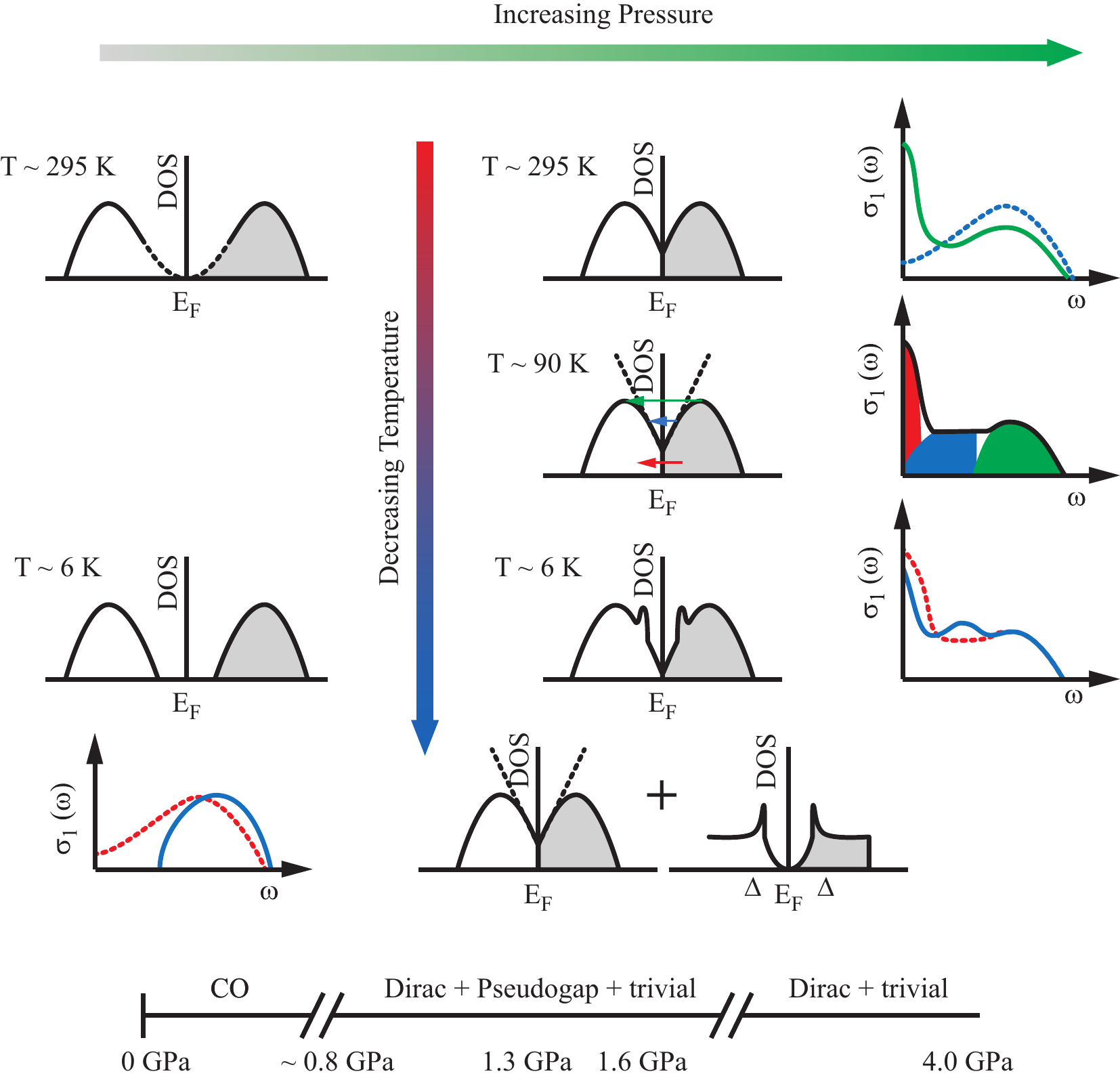}
    \caption{\label{fig:Dirac4}
Schematic diagrams for the electronic structure and corresponding optical conductivity of \aeti\ (reproduced from \cite{Uykur19}).
The diagrams are given for the low-pressure insulator state and the high-pressure Dirac state at various temperatures.
Here, $T_{PG}$ stands for the temperature, where the pseudogap starts to open.
At the bottom the pressure evolution of the various electronic phases of \aeti\ at low $T$ is summarized:
the charge-order insulating state at low pressure; a metallic states in the intermediate pressure regime consisting of massless Dirac electrons, next to carriers in correlation-split and trivial bands; and
above 4.0 GPa only the Dirac electronic state and carriers in trivial bands remain.}
\end{figure}

From extensive optical studies under high pressure up to 4.0~GPa, Uykur {\it et al.} concluded that at low pressure \aeti\ possesses a clear charge-order gap in the optical conductivity; but with  rising $p$ these bands approach each other and overlap, leading to a more-or-less narrow Drude contribution.
At low temperatures the edge of these two bands develop linear dispersions.
As illustrated in Figure~\ref{fig:Dirac4}, $\sigma_1(\omega)$ consists of three components in the range of high-pressure:
(i) a low-energy Drude response, (ii) a frequency-independent conductivity due
to the Dirac electrons and (iii) a mid-infrared band arising with the
incoherent transitions due to on-site and inter-site Coulomb repulsion.
Upon cooling, electronic correlations cave in a pseudogap with states piling up at the edges.
As a result, the Drude spectral weight is transferred to finite energies.
In other words, there are clear fingerprints of the electronic correlations among the Dirac
electrons; the interaction can be tuned by temperature and pressure.

\begin{figure}
  \centering
  \includegraphics[width=0.8\columnwidth]{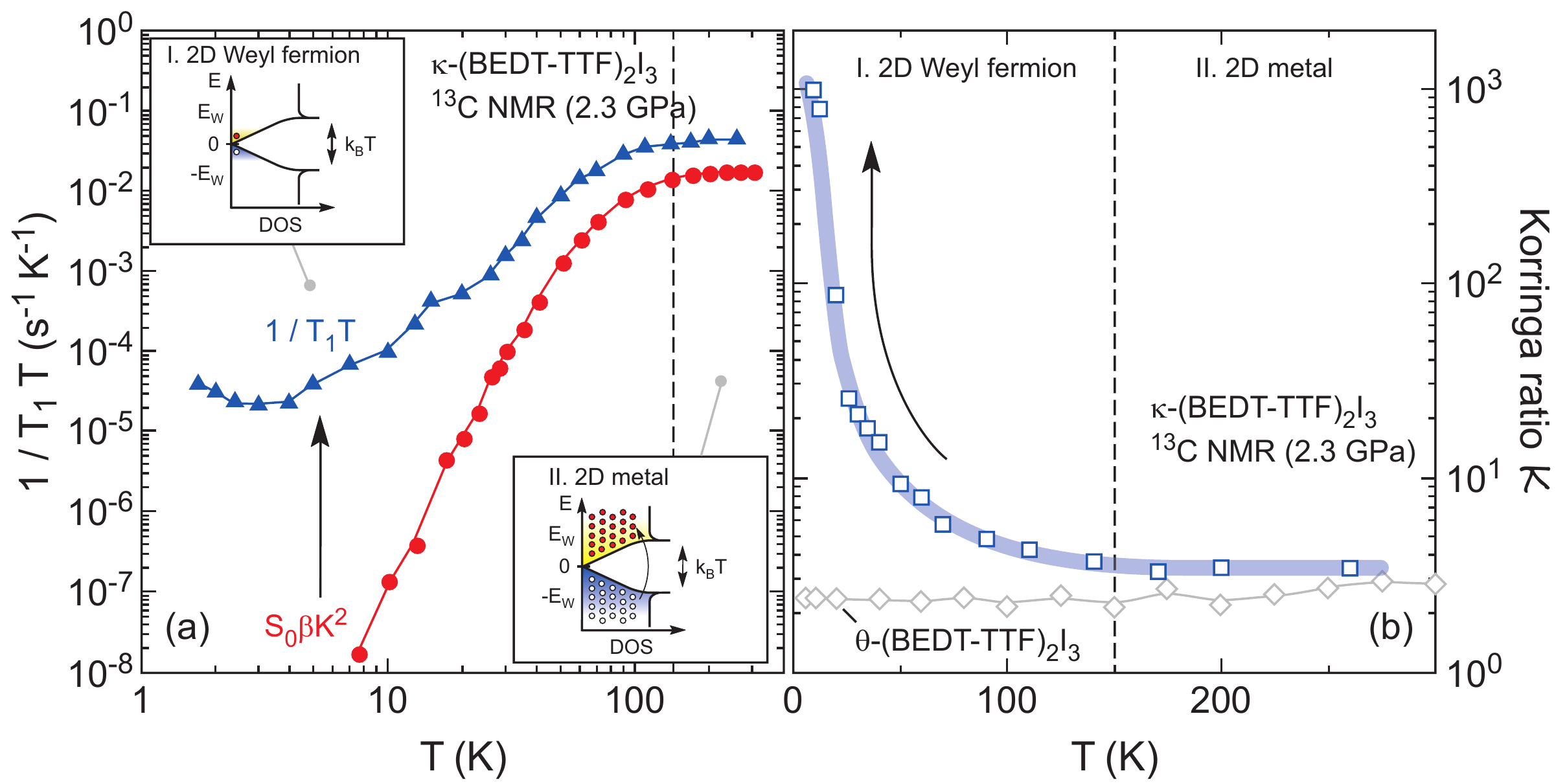}
    \caption{\label{fig:Hirata}
(a) Temperature dependence of the $^{13}$C-NMR spin-lattice relaxation rate $1/T_1T$
(triangles) and the squared Knight shift $K$ (circles) \cite{Hirata16} measured at a pressure of 2.3~GPa and a magnetic field of 6~T. The insets depict the density of electronic states near $E_F$ with thermally generated electron-hole pairs (circles) indicated for low (I) and high (II) temperatures, respectively.
The density of states is linear up to $|E_W – E_F|$ and levels off above it.
(b) The Korringa ratio $K$ (squares) dramatically increases as the temperature is lowered.
For comparison the results for $\theta$-(BEDT-TTF)$_2$I$_3$ (diamonds) are plotted \cite{Hirata12}
(after \cite{Hirata17}).
}
\end{figure}

Also NMR investigations lead to the conclusion that the Dirac fermions
in \aeti\ do interact \cite{Hirata16,Hirata17}.
Using  $^{13}$C-NMR spectroscopy an unusual temperature dependence of spin-related properties
was observed in \aeti, when pressure of 2.3~GPa is applied indicating strong correlations among the linearly dispersing electrons \cite{Hirata17}.
In regular metals, the electronic density of states is constant in energy,
resulting in a constant quantity $1/(T_1 T K^2)$ upon cooling, where $T_1$ is the spin-lattice relaxation rate, $T$ the temperature and $K$ is the Knight shift; this is known as the
Korringa law \cite{Moriya63,Narath68}. In contrast,
in systems with linearly dispersing bands and Dirac cones,
both $1/T_1 T$ and $K^2$ rapidly drop upon cooling, reflecting the vanishingly
small density of states (DOS) around the Fermi energy $E_F$.
Despite the fact that the charge-order transition is strongly suppressed at a pressure of 2.3~GPa as shown in Figure~\ref{fig:alphapressure},
a crossover from a Korringa-like metal to a gapless state occurs below $T\approx 150$~K, suggesting
a density of state profile depicted in the inset of Figure~\ref{fig:Hirata}(a).

The strength of short-range electronic correlations can be measured by a so-called Korringa ratio
\begin{equation}
\mathcal{K}=\frac{1}{T_1 T} \frac{1}{S_0 \beta K^2} \hspace*{5mm} \text{with} \hspace*{5mm}
S_0 = \frac{4\pi k_B}{\hbar} \frac{\gamma_n}{\gamma_e} \quad ,
\end{equation}
where $\gamma_n$ is the nuclear gyromagnetic ratio and $\gamma_e$ is the electron gyromagnetic ratio;
$k_B$ is the Boltzmann constant and $\hbar = 2\pi h$ the Planck constant;
$\beta$ is a form factor representing the anisotropy of the hyperfine interaction. In weakly correlated electron systems, $\mathcal{K}$ is constant in $T$ and of the order of unity, as demonstrated in Figure~\ref{fig:Hirata}(b) for the example of $\theta$-(BEDT-TTF)$_2$I$_3$.
The divergent increase of the Korringa ratio by a factor of 1000 upon cooling
evidences that \aeti\ at high pressure is far from behaving like a regular metal.
Combining these observations with model calculations, Hirata {\it et al.} suggest that this divergence stems from an interaction-driven velocity renormalization that almost exclusively suppresses zero-momentum spin fluctuations because the long-range part of the Coulomb interaction remains unscreened around the Dirac points. It addition the bandwidth  is reduced by short-range electron correlation.
The NMR results also indicate that preexisting excitonic fluctuations in close proximity to the charge order govern the electronic nature in the low-temperature region.
The excitonic instability is controlled by a small chemical-potential shift and an in-plane magnetic field.
NMR relaxation rate probes these excitonic-spin fluctuations \cite{Ohki19a}.

\paragraph{Domain Walls}
\label{sec:DiracElectrons_domainwalls}
Dielectric and optical studies of the anisotropic charge response in \aeti\ revealed
two low-frequency relaxation modes (Figure~\ref{fig:alphadielectricfreq}), as discussed in full detail in Sections~\ref{sec:DielectricResponse} and ~\ref{sec:dielectric_domainwalls}.
Both modes can be attributed to the motion of domain wall pairs between two types of domains which are created due to breaking the inversion symmetry. The first mode is attributed to the motion of charged-domain walls along the $a$-axis, while
the smaller second mode is associated with the motion of neutral $180^\circ$ domain-wall pairs along the $b$-axis (cf.\ Figure~\ref{fig:DomainWalls1}).

Ohki {\it et al.} examined the detailed temperature dependence of the electronic states of interacting two-dimensional Dirac electrons in pressurized \aeti\ by using a semi-infinite two-dimensional lattice model \cite{Ohki19b}. Above the charge-order gap, a peak structure emerges due to the two-dimensional Dirac cones, which drastically changes in the vicinity of $T_{\rm CO}$ when they  merge with the massive
Dirac electrons \cite{Ohki18c}; a behavior not uncommon in indirect semiconductors. They also address the problem of domain wall conductivity in a charge-ordered insulating phase and can explain the discrepancy of
a small energy gap extracted from resistivity data \cite{Liu16} compared to the optical gap \cite{Beyer16}
by  metallic conduction along a one-dimensional domain wall emerging at the border of two charge-ordered ferroelectric regions with opposite polarizations.
With increasing intersite interaction  $V$, a transition from the massless Dirac phase to the massive Dirac phase occurs simultaneously with the charge ordering \cite{Matsuno16,Omori17,Ohki18a}. Upon further increasing $V$, the system changes from the charge-ordered massive Dirac state to the charge-ordered state with no Dirac cones.

    \subsection{Electrodynamics of weakly dimerized ferroelectrics}
\label{sec:COFerroelectricity}

Electronic ferroelectricity describes an ordering phenomenon that involves mainly electrons \cite{VandenBrinkKhomskii08,Naka10,Yamauchi14,Ishihara10,Ishihara14,TomicDressel15}, in contrast to conventional ferroelectricity, for instance in BaTiO$_3$ or other titanates, where the ferroelectric properties are determined by the ions \cite{LinesGlassBook} (see the introductory part of this Chapter~\ref{sec:ChargeOrder} and references therein for more information). In the above Section \ref{sec:COweaklydimerized} we have discussed basic features characteristic for electronic ferroelectricity which have been observed in two-dimensional organic solids: the optical second harmonic generation, the development of charge disproportionation and the formation of ferroelectric domains. Here we continue by exploring dynamical features: hysteresis effect, specific dielectric response, and strong non-linear effects and ultrafast response.

\subsubsection{Polarization switching}
\label{sec:Switching}

\begin{figure}[h]
	\centering\includegraphics[clip,width=0.5\columnwidth]{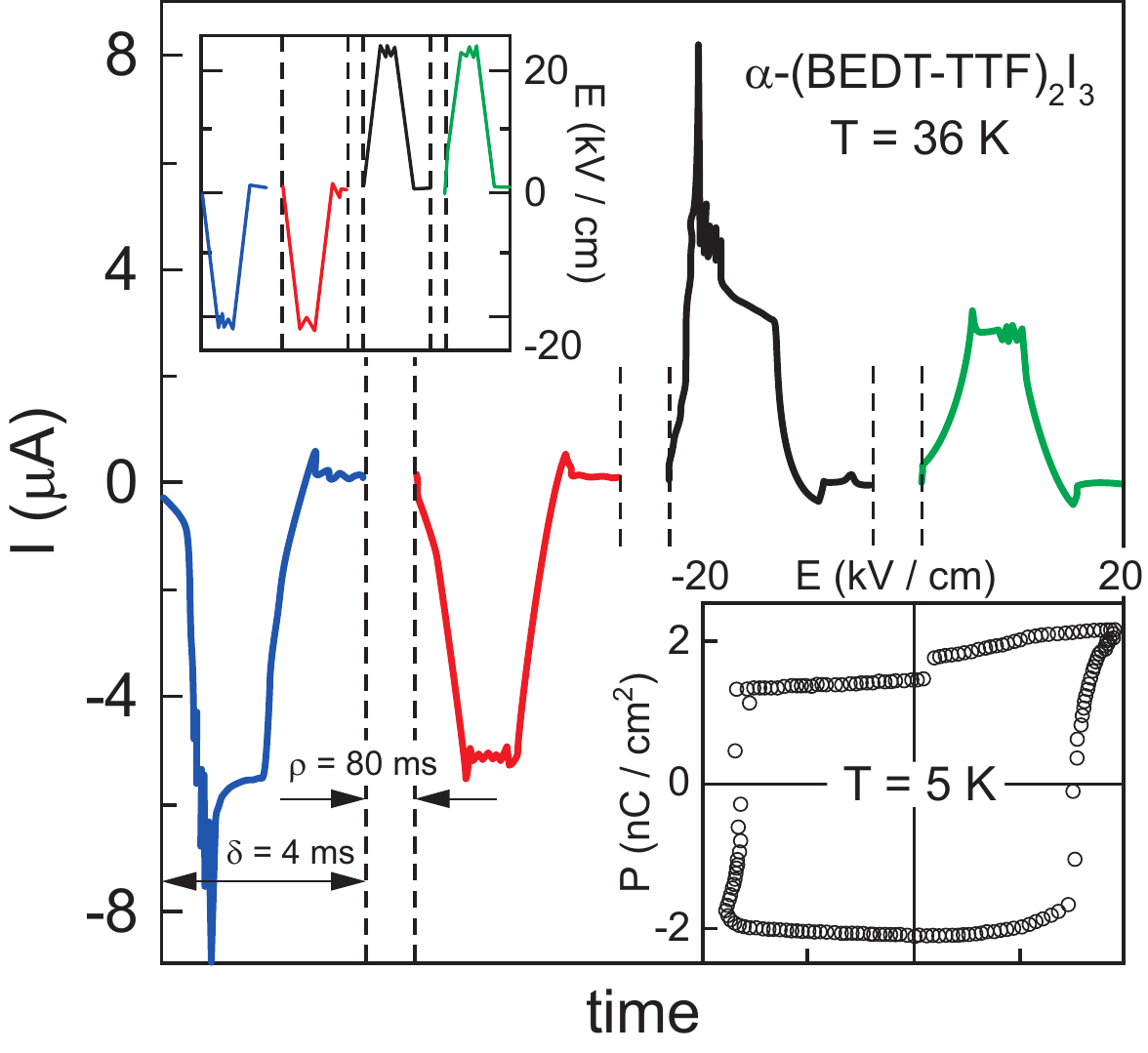}
	\caption{Time-dependent current of $\alpha$-(BEDT-TTF)$_2$I$_3$ at $T=36$~K (main frame) generated by the sequence of excitation signals plotted in the upper inset. The lower inset shows the polarization as a function of the electric field at $T=5$~K (after \cite{Lunkenheimer15}).
		\label{fig:alphahysteresis}}
\end{figure}

Polarization reversal or switching induced by an electric field is commonly considered as the most important property of ferroelectrics. The occurrence of a ferroelectric hysteresis loop is a direct consequence of the motion of domain walls
which is microscopically explained by the switching of polarization.
In the case of \aeti, Lunkenheimer {\it et al.} could proof ferroelectricity by conducting the respective polarization experiments \cite{Lunkenheimer15}.
In Figure~\ref{fig:alphahysteresis} the polarization-induced current response is plotted
as a function of time while the electric field is switched as shown in the upper inset.
The polarization shows a hysteresis loop that closes around 20~kV/cm and exhibits the form typically observed in true ferroelectrics (lower inset of Figure~\ref{fig:alphahysteresis}).
The latter result is obtained only at $T=5$~K, probably because the conductivity at elevated temperatures is still to high
and the remaining Ohmic losses prevent switchability \cite{Scott08}. Surprisingly, the saturation polarization of 2~nC/cm$^2$ is several orders of magnitude smaller than the values typically found in one-dimensional organic ferroelectrics such as TTF-CA \cite{TomicDressel15}. This suggests that at $T=5$~K only a tiny fraction of polar domains can be switched by the applied electric fields, indicating a high degree of cooperative freezing of the domain wall motion; for further discussion see Section~\ref{sec:DielectricResponse}.
Finally, we note that in \tetrz\ polarization switching  has not been reported by now, which may be due to either relatively high conductivity or due to a low breakdown field.

\subsubsection{Dielectric response}
\label{sec:DielectricResponse}

\begin{figure}[h]
	\centering\includegraphics[width=0.8\columnwidth]{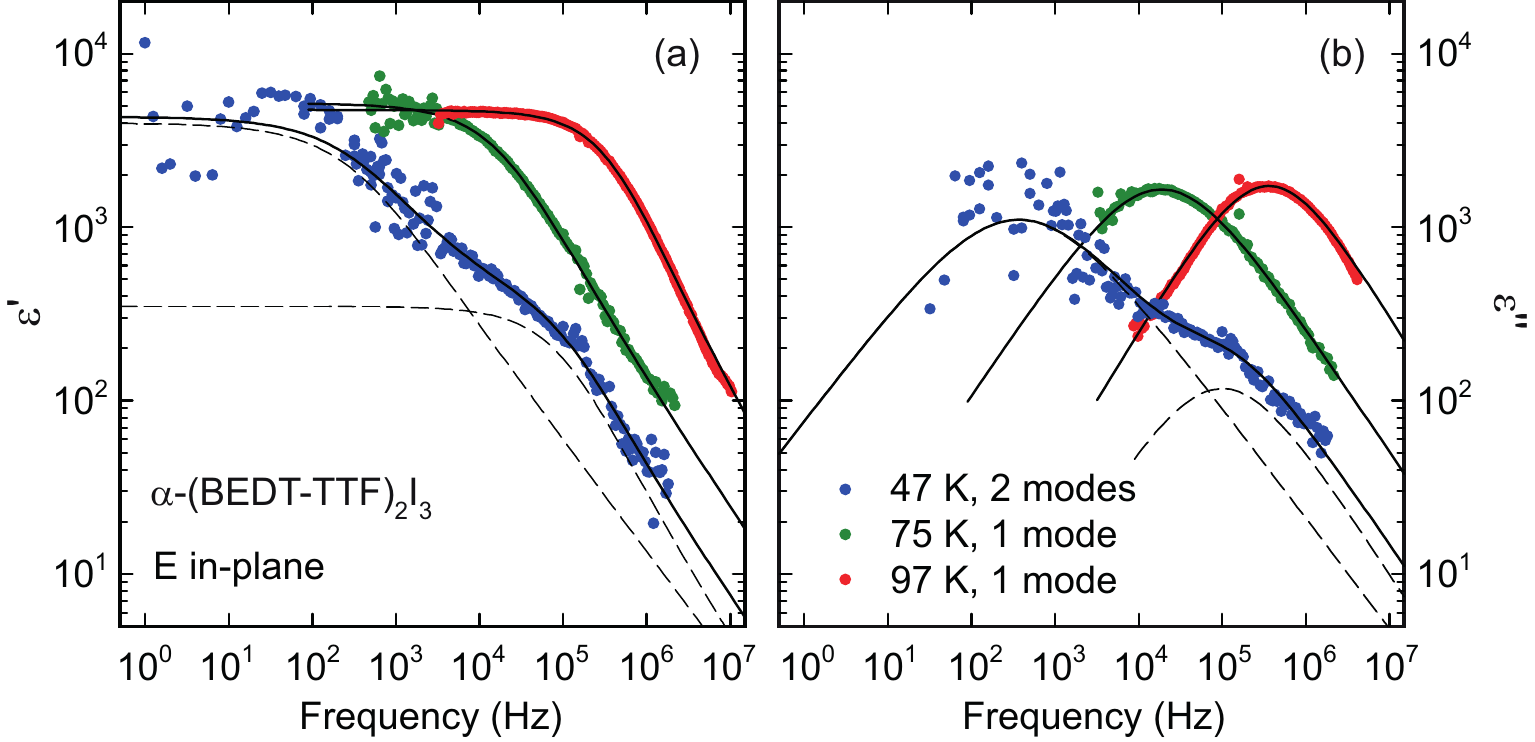}
	\caption{Double logarithmic plot of the real part  (a) and imaginary part (b) of the in-plane dielectric function of $\alpha$-(BEDT-TTF)$_2$I$_3$ for three representative temperatures. For $T=47$~K the full lines correspond to the fit to a sum of two generalized Debye functions;
the dashed lines represent contributions of the two single modes. Above $T=75$~K, only one mode is identified, and the full lines represent fits to single generalized Debye function (after \cite{IvekPRB11}).
		\label{fig:alphadielectricfreq}}
\end{figure}

For \aeti\ a rather complex and anisotropic dielectric response is observed in the Hz to MHz frequency range below $T_{\rm CO}$
\cite{IvekPRB11,Dressel94,IvekPRL10,Kodama12,Lunkenheimer15,IvekCulo17}.
To disentangle the different contributions, simultaneous fits of real and imaginary parts of the dielectric function
by a generalized Debye form, known as Cole-Cole function  \cite{Jonscher77,Jonscher99}, are performed as plotted in Figure~\ref{fig:alphadielectricfreq}.
The results reveal that within the molecular planes the dielectric spectra exhibit substantial dispersion with two discernible contributions:
\begin{equation}
\varepsilon(\omega)-\varepsilon_\infty
 = \frac{\Delta\varepsilon_\mathrm{LD}}{ 1 + \left(i \omega \tau_{0,\mathrm{LD}} \right)^{ 1-\alpha_\mathrm{LD} } }
 + \frac{\Delta\varepsilon_\mathrm{SD}}{1 + \left(i \omega \tau_{0,\mathrm{SD}} \right)^{ 1-\alpha_\mathrm{SD} } } \quad ,
\label{eqmodel}
\end{equation}
where $\varepsilon_\infty$ is the high-frequency dielectric constant,
$\Delta\varepsilon$ is the dielectric strength, $\tau_0$ the mean relaxation
time and $(1-\alpha)$ the symmetric broadening of the relaxation time distribution
function of the large (LD) and small (SD) dielectric modes, respectively. The broadening
parameter $(1-\alpha)$ of both modes is about $0.70 \pm 0.05$, and the dielectric strength does not
change significantly with temperature. Importantly, $\tau_{0,\mathrm{LD}}$ changes with temperature in a thermally activated manner, whereas $\tau_{0,\mathrm{SD}}$ is temperature-independent and reminiscent of a domain-wall-like behavior.

\begin{figure}
	\centering\includegraphics[width=0.7\columnwidth]{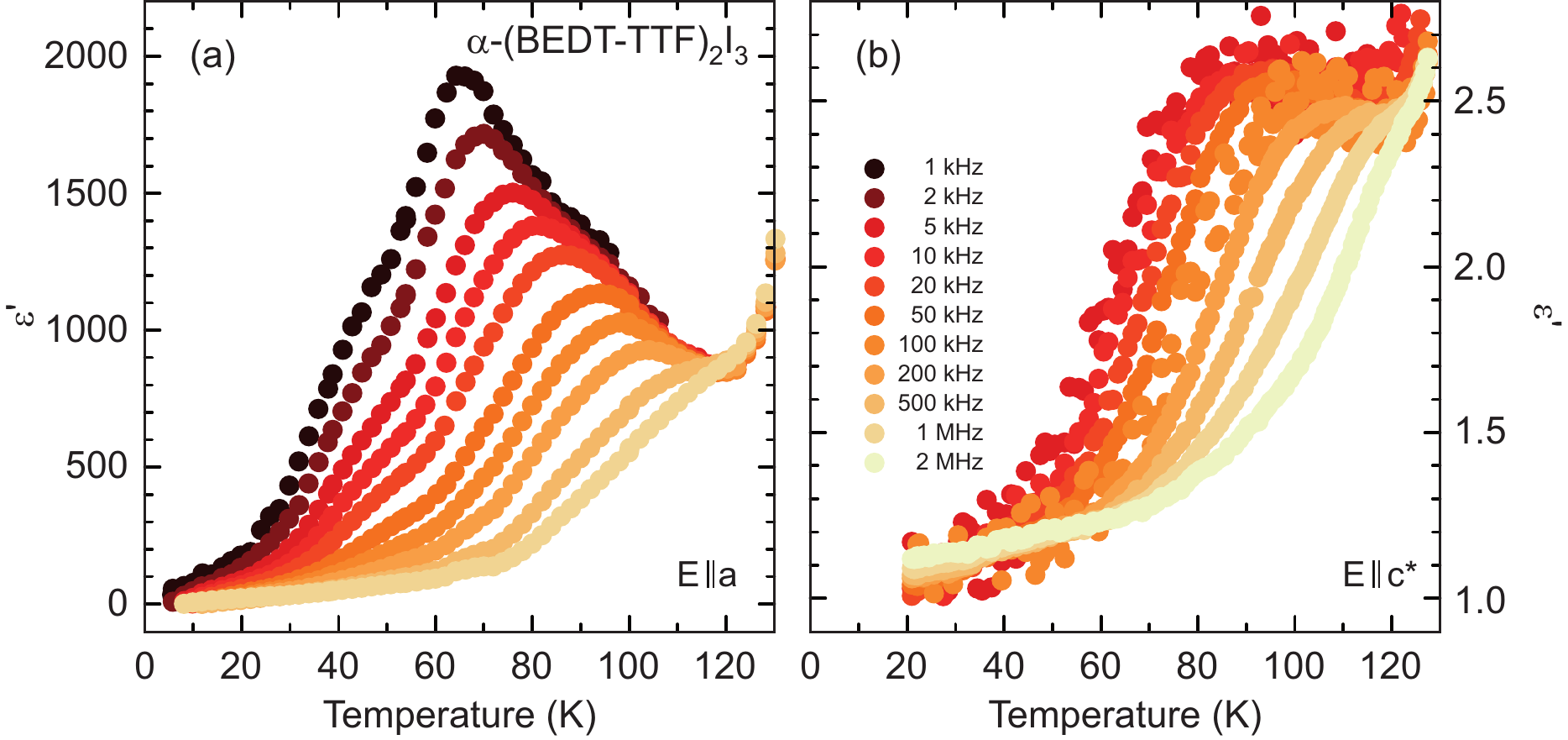}
	\caption{Temperature dependence  of the real part of dielectric function  of \aeti\ measured (a) in plane and (b) out of plane for various frequencies as indicated (after \cite{IvekCulo17}).
For the in-plane direction a second, smaller peak is resolved in the temperature sweep of $\varepsilon^{\prime}(T)$
that corresponds to the complex two-mode response detected in frequency space, plotted in Figure~\ref{fig:alphadielectricfreq}.
		\label{fig:alphadielectrictemp}}
\end{figure}

These features of the dielectric response are strikingly different from
the ones found in one-dimensional organic ferroelectrics, such as TMTTF$_2$$X$ and TTF-CA compounds (see \cite{TomicDressel15} and references therein) and also reported in \tetrz\ \cite{NadMonceautheta06}. In those ferroelectrics a clear Curie peak is observed at $T_\mathrm{CO}$, as expected for a regular ferroelectric phase transition.
In contrast, $\alpha$-(BEDT-TTF)$_2$I$_3$ shows no sign of a non-dispersive Curie-like peak
in $\varepsilon^{\prime}(T)$ at $T_{\rm CO}$, where a clear-cut charge order occurs, as discussed in Section~\ref{sec:COweaklydimerized}.
Instead, when recording the dielectric response as a function of temperature as diplayed
in Figure \ref{fig:alphadielectrictemp}, one finds a strongly dispersive peak in $\varepsilon^\prime(T)$ well below $T_\mathrm{CO}$. It is more pronounced within the plane compared to the perpendicular direction. A finite out-of-plane component of polarization is expected due to an asymmetric charge distribution along the molecular axis oriented almost parallel to the $c^*$-axis, the direction perpendicular to molecular $ab$-planes (cf.\ Figure~\ref{fig:structure_alpha}).
For the in-plane orientation a second, smaller peak can be resolved;
this particular detailed structure is due to the complex two-mode response as revealed by the data recorded in the frequency space. Distinct interpretations have been suggested to explain these observations
ranging from two-dimensional cooperative bond-charge-density wave with ferroelectric nature \cite{IvekPRL10, IvekPRB11, TomicDressel15}, over relaxor-based picture \cite{Lunkenheimer15, LunkenLoidl15} to disorder-induced scenario \cite{IvekCulo17, IvekPRB11}.

In the following we have to answer conclusively the fundamental question: what is the origin of the complex dielectric response in \aeti? The aim is to adequately describe both the large dispersive and small non-dispersive mode in the charge ordered state developed at long-range scale as testified by
optical second harmonic generation (Figure~\ref{fig:alphaSHG}) \cite{YamamotoJPSJ08}.

\subsubsection{Domain walls}
\label{sec:dielectric_domainwalls}

According to this interpretation, the long-range ferroelectric order remains in place. The failure in detection of the Curie peak is likely due to experimental problems related to the high conductivity at $T>T_\mathrm{CO}$ and/or restricted frequency range of the dielectric measurements. The dielectric response in the charge order state of \aeti{} is then naturally attributed to domain wall motion.
This motion commonly depends on the interaction with randomly located defects,
resulting in a distribution of relaxation times, just as observed. It indicates that the dielectric relaxation takes place between different metastable states,
which correspond to local changes of the charge distribution across the length scale of the  domain-wall thickness.
In \aeti{}, disorder originates in the anions \cite{Dressel94} and due to many short hydrogen bonds between anions and ethylene groups of BEDT-TTF molecules on A, A$^\prime$, and B sites \cite{Alemany12} directly influence the charge response in the BEDT-TTF conducting layers \cite{IvekCulo17, IvekPRB11}.

\begin{figure}
	\centering\includegraphics[clip,width=1\columnwidth]{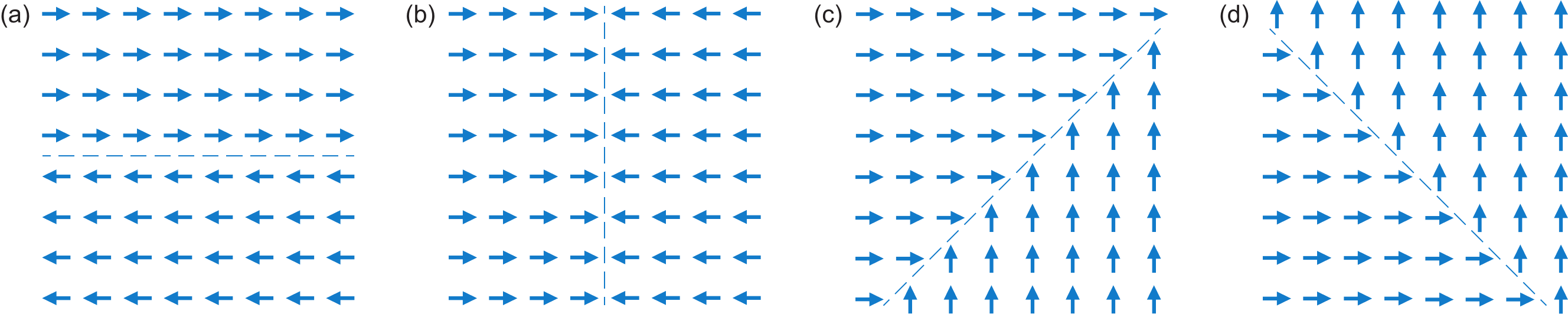}
	\caption{Different configurations of domain wall denoted by the dashed lines: (a)~Sketch of $180^\circ$ domain wall with dipoles oriented antiparallel next to each other. (b) Also a head-to-head arrangement is possible, leading to a charged domain wall. Panels (c) and (d) depict $90^\circ$ domain walls.
		\label{fig:DomainWalls1}}
\end{figure}

Domain walls in ferroelectric crystals  separate the symmetry-equivalent directions of the polarization;
their creation minimizes the electrostatic and elastic energies \cite{Catalan12}.
They are commonly classified in two types according to the relative angle between the domain-wall plane
and the polarization vector: electrically neutral domain walls and charged domain walls.
Figure~\ref{fig:DomainWalls1} illustrates that in two dimensions
the neutral domain wall orientation is restricted to $180^\circ$ and 90$^\circ$ walls;
the former one segregate domains with antiparallel polarization,
while the 90$^\circ$ walls separate regions with mutually perpendicular polarization.
On the other hand, domain configurations with a jump of the normal component of polarization can be separated only by domain walls that carry bound charge. The creation of this type of walls require almost a perfect compensation by free charge carriers \cite{Bednyakov15}.

\begin{figure}[h]
	\centering\includegraphics[clip,width=1\columnwidth]{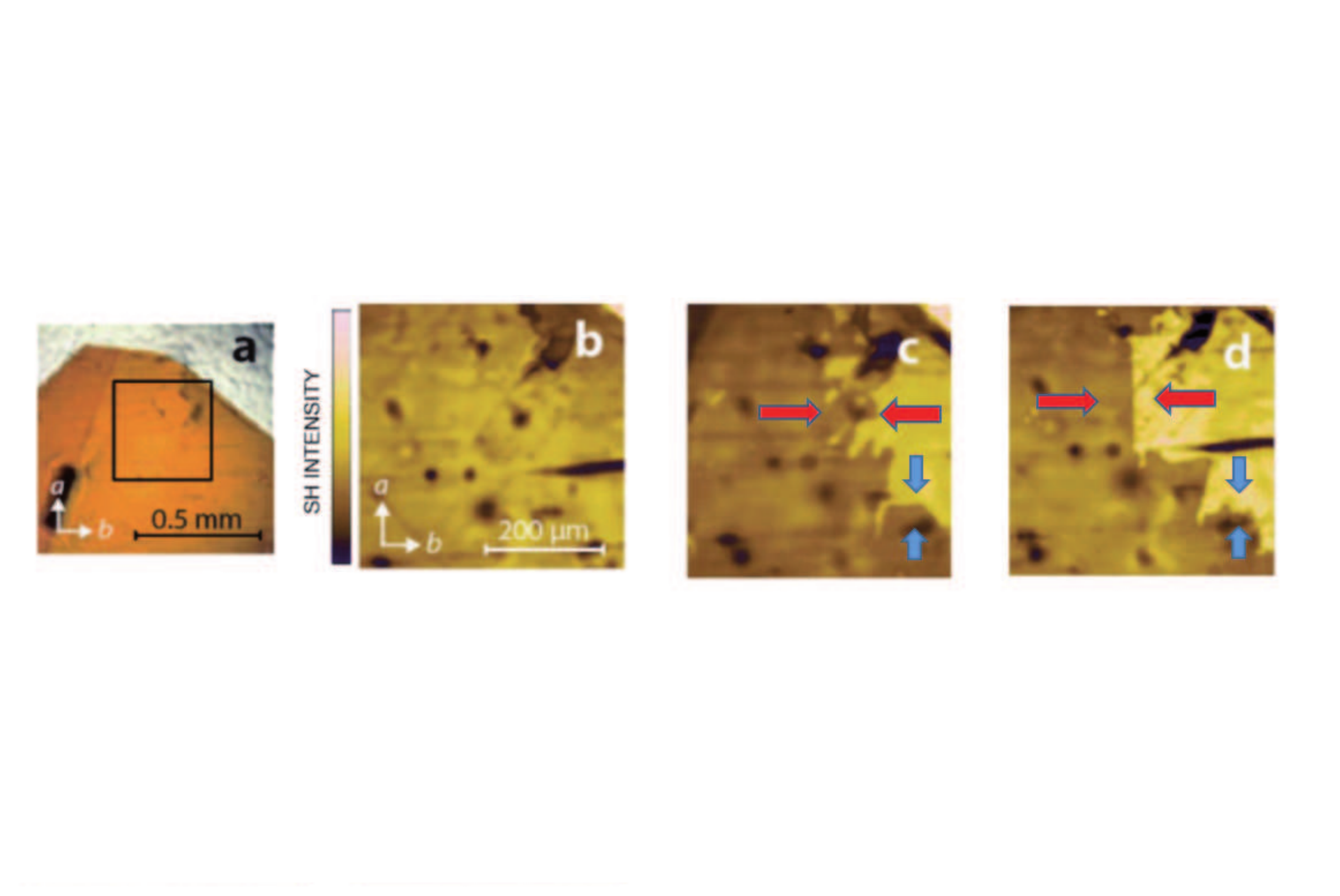}
	\caption{(a) Transmission image of a single crystal of $\alpha$-(BEDT-TTF)$_2$I$_3$. Square denotes the area of sample used in second harmonic generation (SHG) interferometry measurements; (b) SHG image taken at $T$=140\,K (above $T_\mathrm{CO}$) (c) SHG image taken at $T$=50\,K (below $T_\mathrm{CO}$) after rapid cooling. Bright and dark regions are split dominantly along the a-axis indicating 180$^\circ$ domain wall; (d) SHG image taken at 50\,K after high-temperature annealing; the image reveals that the domains were displaced from the original positions before annealing dominantly along the $b$-axis. Red and blue arrows in (c) and (d) denote neutral and charged domain walls, respectively (reproduced from Yamamoto {\it et al.} \cite{Yamamoto10} with the permission of AIP Publishing).
		\label{fig:alphadomainimagesA}}
\end{figure}

As mentioned in Section~\ref{sec:MITchargeorder}, x-ray diffraction and optical second-harmonic interferometry provide experimental evidence of domain walls in \aeti.
Yamamoto {\it et al.}  took images of large polar domains with opposite polarizations that are reproduced in Figure~\ref{fig:alphadomainimagesA}. They clearly resolved the formation of the boundary along the $a$-axis indicating 180$^\circ$ domains \cite{Yamamoto10}.
The mobility of these walls was determined by recording images of the same crystal area after rapid (panel c) and slow cooling (panel d) through the charge-ordering phase transition at $T_{\rm CO} = 135$~K.
The image of the domain structure in the annealed state demonstrates a lateral shift of
the neutral 180$^\circ$-wall and  growth of the bright domain.
Furthermore, the domain boundaries are rugged and a closer inspection reveals that they consist of both neutral and charged domain walls. Evidently, the latter are strongly pinned and thus much less mobile. A low degree of switchable polarization at $T=5$~K, shown in Figure~\ref{fig:alphahysteresis}, may be a result of mobility difference between neutral and charged domain walls: only the domain states that contain neutral 180$^\circ$-domain walls exhibit polarization switching.

On this basis, the complex dielectric response in \aeti\ can be understood. Ivek {\it et al.} \cite{IvekPRB11} 
suggested two possible types of domain-wall pairs with the constraint of charge neutrality implying that a change of stripes is equivalent to strictly replacing the unit cells of one twin type [(A,B)-rich unit cells] by another [(A$^\prime$,B)-rich unit cells], as depicted in Figure~\ref{fig:alphatwin}.
In Figure~\ref{fig:DomainWalls1}(a) we illustrate that the first type is a pair of neutral 180$^\circ$ domain walls,
{\it i.e.} a domain wall and the corresponding anti-domain wall between the charge-rich and charge-poor stripes along the $b$-axis. A second type of charged domain walls
that develops along the $a$-axis is shown in Figure~\ref{fig:DomainWalls1}(b):
here a jump of the normal component of the polarization is equal to the polarization itself.
Importantly, in contrast to neutral 180$^\circ$ domain walls,
the stability of charged domain walls depends on the compensation by free charge carriers.

The nearly temperature-independent mean relaxation time of the small dielectric relaxation
observed in \aeti\ evidences that resistive dissipation is not dominant; thus it is described appropriately by 180$^\circ$ domain walls.
Let's now return to the large dispersive dielectric mode observed in \aeti\ (Figure~\ref{fig:alphadielectrictemp}).
We propose that this dispersive mode is caused by the motion of charged-domain walls, whose formation should be promoted by screening
since the charge-ordered state is characterized by a relatively high conductivity.
They will remain stable as long as the free-charge-carrier screening effectively compensates their charges.
Since dispersion is determined by the free carrier screening, the mode should follow a thermally activated behavior similar to the dc resistivity, exactly as observed.

There still remain some open questions concerning the dielectric response in \aeti.
In order to reveal its microscopic origin, as well as to clarify the low degree of switchable polarization, further investigations of the topology of domain structure are highly desirable.
A first approach was recently reported utilizing scanning near-field infrared nanoscopy
in the vicinity of $T_{\rm CO}$ \cite{PustogowSciAdv18}.

\subsubsection{Non-linear effects and ultra-fast response}
\label{sec:nonlinear+ultafastresponse}
\begin{figure}
	\centering\includegraphics[clip,width=0.4\columnwidth]{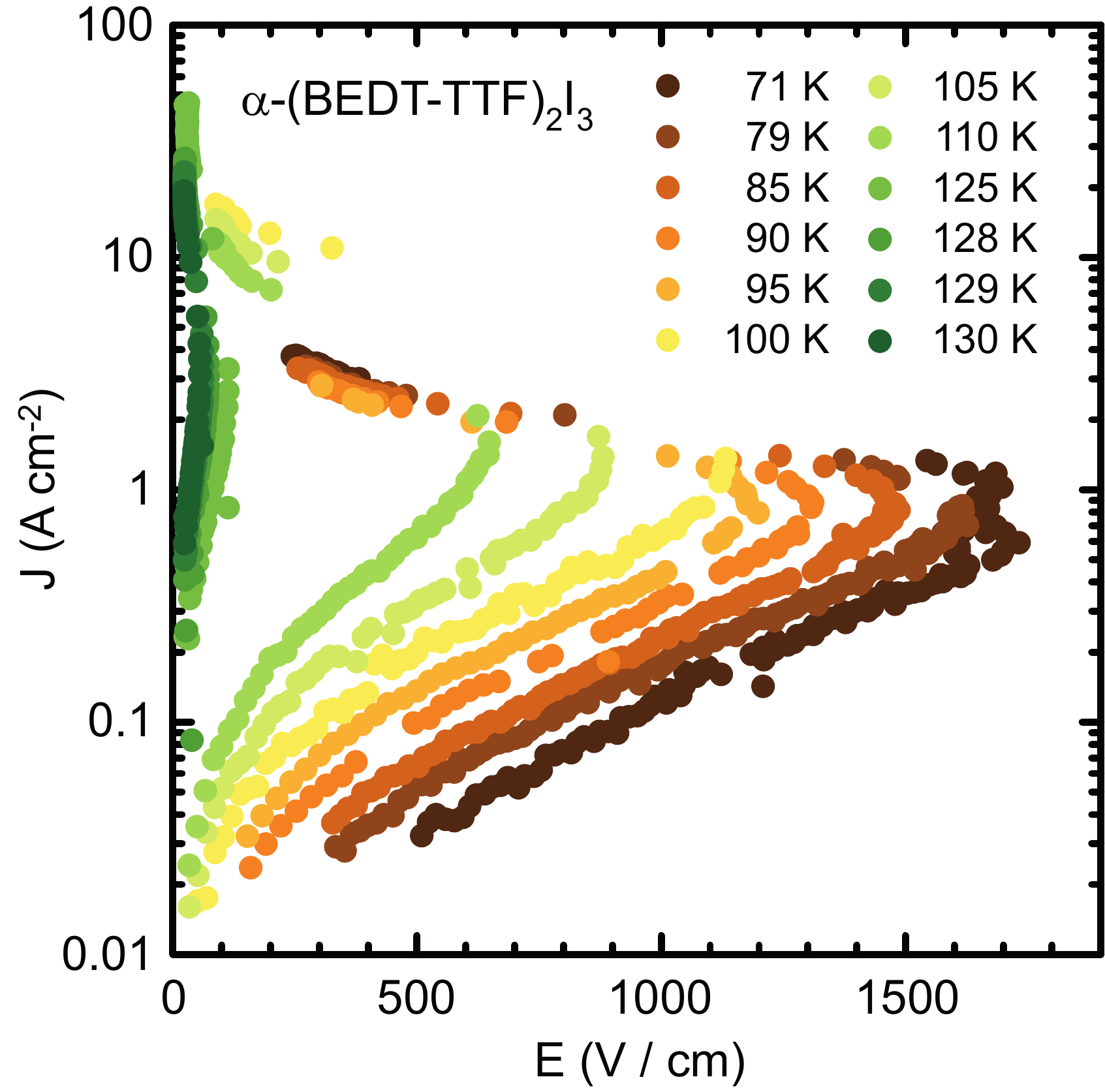}
	\caption{Current density $J$ as a function of the applied electric field $E$ measured
in \aeti\ at different temperatures below $T_\mathrm{CO}$.
A negative differential resistance behavior is observed above the threshold field,
which increases with decreasing temperature (after \cite{Peterseim16}).
		\label{fig:alphaNDRoptics}}
\end{figure}

Domain wall motion may also be responsible for the huge negative differential resistance above very high threshold fields and in reversible switching to transient high-conducting states, plotted in Figure~\ref{fig:alphaNDRoptics}.
Time- and field-dependent transport measurements on \aeti\ provide evidence \cite{Tamura10, IvekPRB12, Peterseim16} that the rate of domain wall formation strongly increases when the dc electric fields is high enough to overcome their formation due to thermal excitations.
Under these conditions, the motion of domain wall pairs becomes increasingly correlated and creates growing conduction regions until percolation promotes negative differential resistance. The time-dependent effects are
due to changes in the coupling between molecular and anion sublattices induced by the applied electric field,
otherwise involved in the charge-ordering phase transition \cite{Alemany12}.

\begin{figure}[b]
    \centering
       \includegraphics[width=0.8\columnwidth]{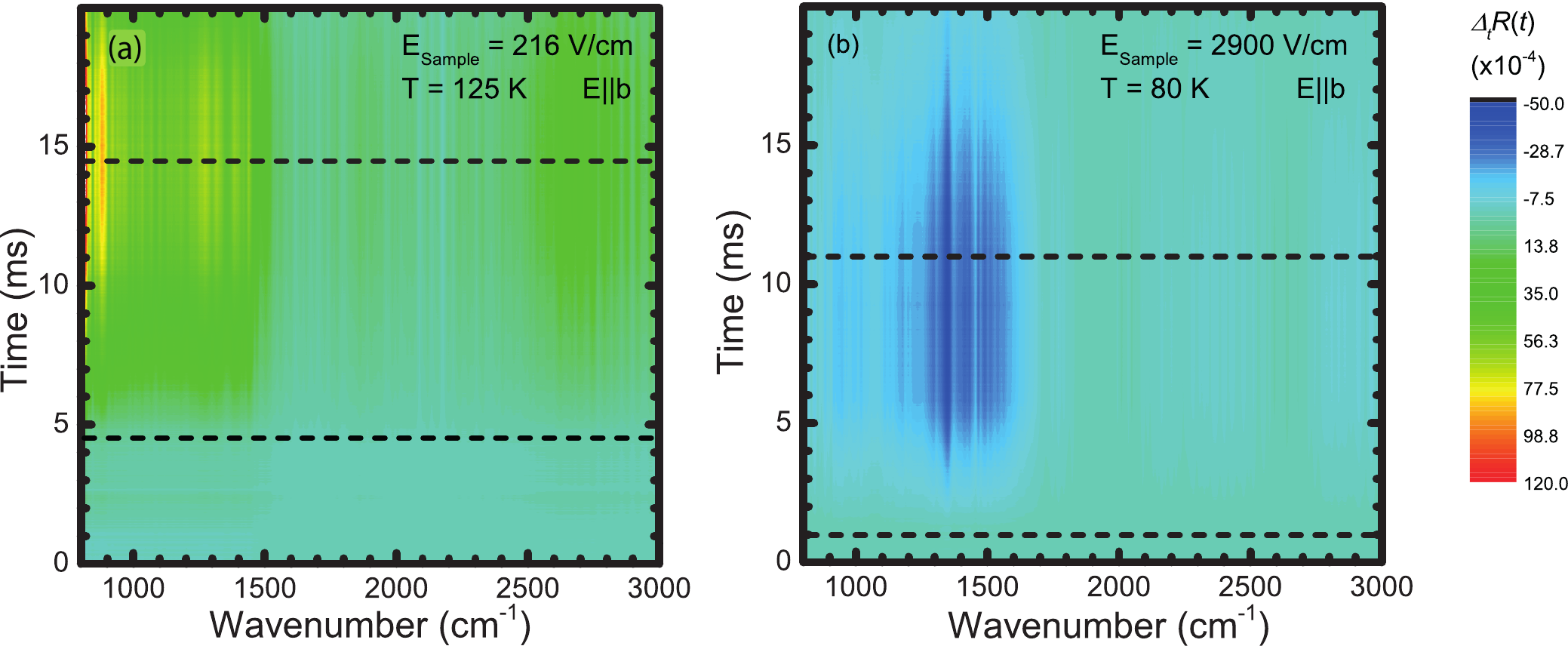}
    \caption{Contour plot of the reflectivity change $\Delta_t R(\nu,t)$ along the $b$-direction of \aeti\ at different temperatures: (a) At $T=125$~K an electric field of $E_{\rm sample}=216$~V/cm is applied along the $a$-axis lasting for 10~ms. (b) At lower temperatures, $T=80$~K,  $E_{\rm sample}=2900$~V/cm have to be applied. The dashed horizontal lines mark the onset and end of the voltage pulse. The vertical stripes in the spectrum are due to instabilities of the interferometer mirror during the step-scan run (after \cite{Peterseim16}).
    \label{fig:aeti_contour2}}
\end{figure}

Alternatively, the overall behavior can be explained
within a two-state model of non-equilibrium electrons by excitation of charge carriers with the high mobility \cite{Peterseim16}.
Especially optical studies show that transient optical properties of electric field-induced metallic state at about 125\,K, {\it i.e.} close to $T_\mathrm{CO}$, differ from the state induced deep in the insulating state at around 80\,K.
From Figure~\ref{fig:aeti_contour2}, we can see that the spectral signatures of the former are similar to the optical properties found for $T > T_\mathrm{CO}$, while the spectral response of the latter evidences a novel electronically induced metalic-like state. Peterseim {\it et al.} suggest that the novel state is characterized by excitations of charge carriers with an extremely high mobility, which resemble massless Dirac-like carriers with linear dispersion found in \aeti\ at high pressure \cite{Peterseim16} (cf.\ Section~\ref{sec:DiracElectrons}).

Another non-linear study performed at temperatures lower below $T=60$~K revealed a power-law behavior in the current-voltage characteristics that is explained by assuming an electric field dependent potential barrier of the thermally excited topological defects such as electron-hole pairs and/or domain walls from the charged-ordered state \cite{Uji13}.
On reducing the temperature, the power-law exponent increases from 1 to 3 around  $30~{\rm K} < T_\mathrm{KT} < 40$~K,
which is interpreted as a fingerprint of the Kosterlitz-Thouless type of transition.
This intrepretation implies that many topological defects exist above $T_\mathrm{KT}$, while there remain only few below $T_\mathrm{KT}$. Uji {\it et al.} \cite{Uji13} suggest that these very excitations, when polarized by an ac electric fields, also contribute to the in-plane dielectric function $\varepsilon^\prime$.
Finally, Okamoto and collaborators reported nonlinear electric transport measurements at 30\,K conducted simultaneously with a terahertz-radiation imaging method. The results showed that in the negative differential resistance state the ferroelectric order melts in the elongated region forming a nonlinear conducting path \cite{Sotome17}.

Electrical conductivity switching and negative differential resistance can also be induced by strong light pulses.
Usually, photoexcitated states undergo a rapid decay to the ground electronic state and most changes induced by photoexcitation occur transiently. Time-resolved measurements of the electrical conductivity are therefore required to detect a transient photo response. By irradiating the crystal with a pulsed laser, it is switched to a high conductivity state;
then the current response to an applied voltage pulses is measured.
Notably, the conductivity switching can be repeatedly retrieved by applied pulsed voltage without further irradiation, thus indicating a memory effect. The memory effect can be controlled by the temporal width and height of voltage pulses; it is understood in terms of a high current filament formation in the high conductivity state \cite{Iimori09, Iimori14}.

\begin{figure}
    \centering\includegraphics[width=\columnwidth]{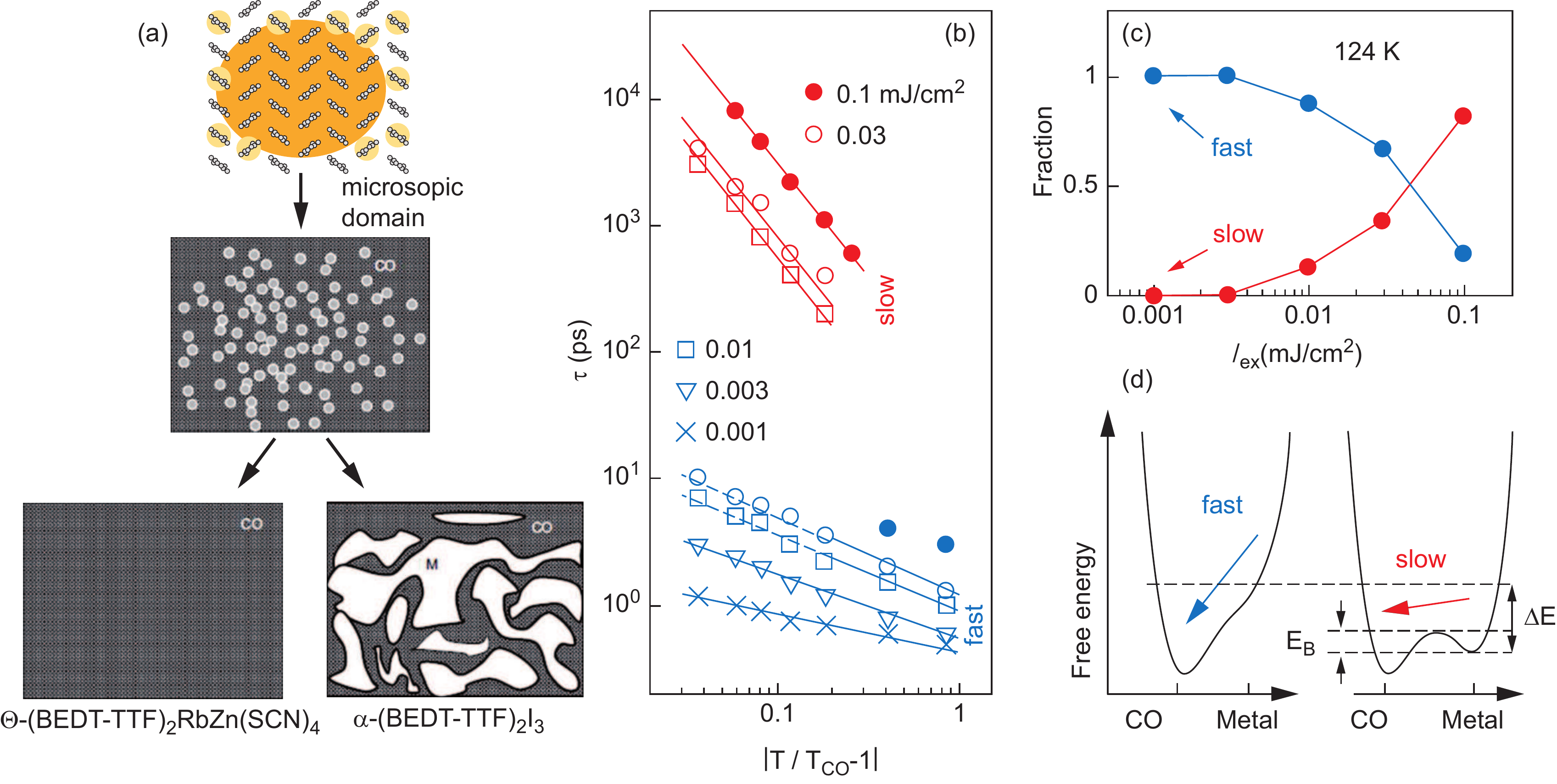}
   \caption{(a) Schematic representation of the primary processes of the photo-induced phase transition: intially the microscopic metallic domains are generated; they can quickly recover the charge ordered state as in $\theta$-(BEDT-TTF)$_2$RbZn(SCN)$_4$, or they can evolve in macroscopic metallic domains as in $\alpha$-(BEDT-TTF)$_2$I$_3$. (b) Relaxation times of the photoinduced metallic state in $\alpha$-(BEDT-TTF)$_2$I$_3$, $\tau_{\rm fast}$ and $\tau_{\rm slow}$ versus reduced temperature for several excitation intensities.
    (c) Fraction of the fast (blue dots: $\tau_{\rm fast} \approx 1$~ps) and slow (red dots: $\tau_{\rm slow} \approx 1$~ns) decay components as a function of $I_{ex}$ at $T=124$\,K.
    (d) Schematic illustrations of free energy surface for $T \ll T_{\rm CO}$ (fast relaxation) and $T\approx T_{\rm CO}$ (slow relaxation). The energy barrier $E_B$ is much smaller than the thermal energy $k_BT\approx 10$~meV   (after \cite{Iwai12}).}
	\label{fig:alphaPIPT2}
\end{figure}

As the last point in this Section, we briefly address the photo-induced phase transition phenomena observed by time-resolved femtosecond pump-probe spectroscopy on these charge-ordered systems \cite{Iimori07, Iwai07, YamamotoJPSJ08, Kawakami10, Miyashita10, Tanaka10,Iwai12}.
The findings strongly support a purely electronic mechanism of ferroelectricity,
while the electron-phonon interaction resulting in the molecular rearrangements contribute to the decay processes.
Iwai {\it et al.} found that charge order is destroyed by an initial femtosecond laser pulse; the charge-ordered state melts extremely fast on a sub-picosecond time scale,
indicating that the initial process is purely electronic and no structural instability is associated with it. Microscopic metallic domains form within 15~fs;
while in \aeti\ they condense further to a macroscopic metallic region on a timescale of about 200~fs,
in \tetrz\ a large potential barrier against molecular displacement prevents the evolution of macroscopic metallic islands;
this is sketched in Figure~\ref{fig:alphaPIPT2}(a).
Eventually, both systems relax back to the charge-ordered state within several picoseconds or nanoseconds.
This fast relaxation is clearly different from the one at the millisecond time scale observed  when a high dc electric field is applied, as shown in Figure~\ref{fig:aeti_contour2}.
The decay processes strongly depend on temperature and light intensity:  a significant slowing-down  is found when the temperature rises close to $T_{\rm CO}$ suggesting an inhomogeneous character of the photo-induced metallic state.
While for low intensities these microscopic metallic domains relax rapidly to the charge-ordered insulating state, for high laser intensities the decay time of metallic state gets markedly longer, enabling formation of macroscopic domains associated with some molecular rearrangement.
The decay process consists of two components; the two relaxation times of the metallic state follow a critical slowing down behavior. Slower recovery takes place at the temperatures close to $T_\mathrm{CO}$, while fast recovery occurs at the temperatures deep below $T_\mathrm{CO}$. The corresponding relaxation times, slow and fast, are associated with the relaxation of macroscopic and microscopic metallic islands, respectively [Figure~\ref{fig:alphaPIPT2}(b-d)].

Very recently an ultrafast response of the THz-wave generation was discovered in $\alpha$-(BEDT-TTF)$_2$I$_3$ upon photoexcitation \cite{Itoh14, Itoh18}. The THz-wave generation by means of optical rectification is a second-order nonlinear optical process -- the same way the generation of a second-harmonic signal (SHG) is. Remarkably, the electric field amplitude of THz-wave sets in at $T_\mathrm{CO}$, as SHG does (Figure~\ref{fig:alphaSHG}), indicating that it originates in charge ordering driven ferroelectric polarization.

The difference in the photo-induced dynamics found in \aeti\ and \tetrz\ are only quantitative.
Numerical studies using the extended Peierls-Hubbard model on an anisotropic triangular lattice show \cite{Miyashita10} that this difference arises from the distinct crystallographic symmetry and different degree of structural modification associated with the thermally-driven charge order phase transition: in the former low-symmetry system it is very small \cite{Kakiuchi07, Tanaka08}, while in the latter high-symmetry compound it is rather substantial \cite{HMoriPRB98,Mori99b,Watanabe04,Miyashita07}. A detailed account of this topic is given by Iwai in \cite{Iwai12}.

    \subsection{Ferroelectricity driven by charge order in dimerized solids}
\label{sec:COdimerized}
\label{sec:COHgCl}
Charge ordering phenomena are rare in dimerized two-dimensional BEDT-TTF compounds
because the organic layers consist of face-to-face pairs of {BEDT-TTF}$^{0.5+}$ molecules,
{\it i.e.} dimers, resulting in an effectively half-filled band.
In that case the onsite Coulomb repulsion $U$ is much more important compared to the nearest neighbor interaction $V$.
For a long time, there were only a few compounds reported to exhibit appreciable charge order,
such as $\kappa$-(BEDT\--TTF)$_4$\-PtCl$_6$$\cdot$C$_6$H$_5$CN,
the triclinic $\kappa$-(BEDT\--TTF)$_4$\-[$M$(CN)$_6$]\-[N(C$_2$H$_5$)$_4$]$\cdot$3H$_2$O
and the monoclinic $\kappa$-(BEDT\--TTF)$_4$[$M$(CN)$_6$]\-[N(C$_2$H$_5$)$_4$]\-$\cdot$2H$_2$O
(with $M$ =  Co$^{\rm III}$, Fe$^{\rm III}$, and Cr$^{\rm III}$)
salts.
However, all of them are challenging to grow and limited in crystal size; hence details of their physical properties and electronic states
are lacking at the moment \cite{Swietlik06, Ota07, Lapinski13}.

Due to the lack of appreciable charge disproportionation in most of the commonly studied Mott insulators and quantum spin liquid compounds, a family of dimerized BEDT-TTF compounds has drawn particular attention, the $\kappa$-{\rm (BEDT-TTF)$_2$Hg(SCN)$_2$}$X$ series,
where $X$ = Cl$^{-}$, Br$^-$, I$^-$, that was introduced almost twenty years ago by Lyubovskaya {\it et al.} \cite{Lyubovskaya91,Aldoshina93,Lyubovskii96,Zhilyaeva99,Lyubovskii02}.
In contrast to the $\kappa$-salts containing Cu-ion in the polymeric anion sheet,
the BEDT-TTF molecules of the Hg-compounds are slightly displaced with respect to each other
within a dimer, as shown in Figure~\ref{fig:structure_HgBrCl}(a). This results in a smaller intra-dimer transfer integral $t_d=129$~meV and correspondingly a weaker on-site Coulomb repulsion  $U\approx 250$~meV \cite{Drichko14}.
For the case of \hgcl, {\it ab initio} density functional theory calculations were performed
using the full potential local orbital basis,  generalized gradient approximation and
the room-temperature crystal structure \cite{Gati18a}.
With interdimer transfer integrals $t^{\prime} = 51.0$~meV and  $t= 40.4$~meV, as defined in Figure \ref{fig:structure_HgBrCl}(b),
the frustration is rather strong $t/ t^{\prime} \approx  0.8$,
although not as close to unity as the canonical spin liquid systems listed in Table~\ref{tab:1}, Section \ref{sec:propertiesQSL}.

\begin{figure}
	\centering\includegraphics[clip,width=1\columnwidth]{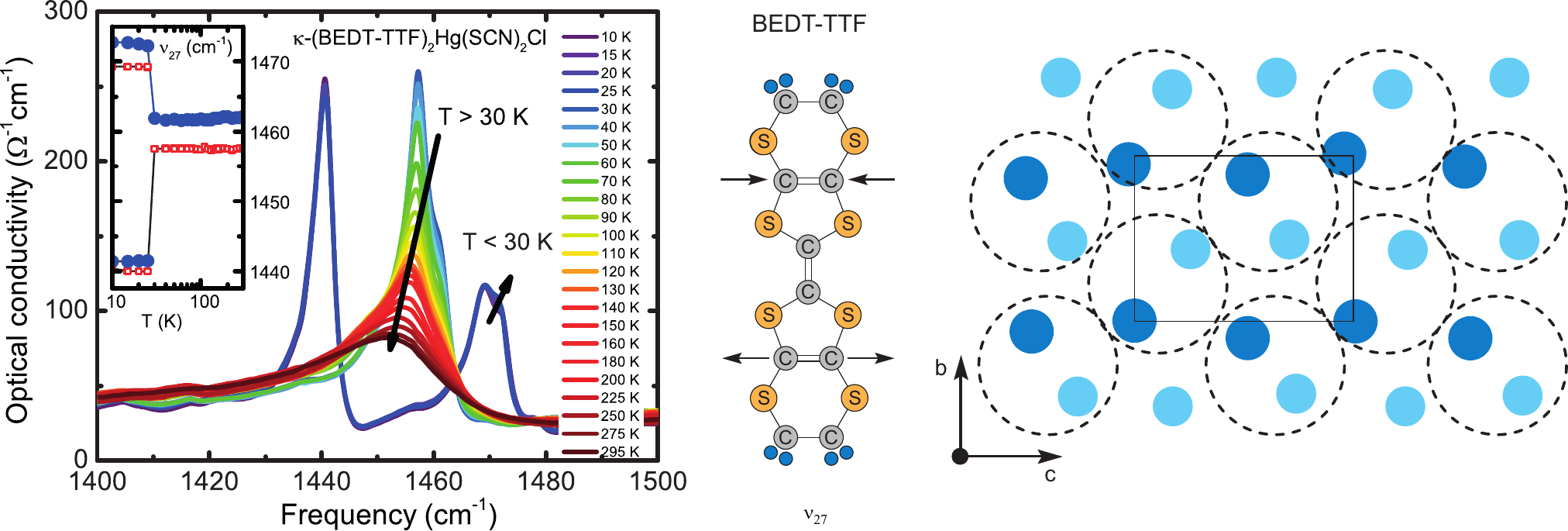}
	\caption{Molecular vibration $\nu_{27}({\rm b}_{1u}$) as a function of temperature of \hgcl\ showing clear splitting when the charge ordering is established at $T_{\rm CO}=30$~K. A shoulder in the mode above $T_{\rm CO}$ indicating two crystallographically different sites per unit cell; below $T_{\rm CO}$ this becomes even more pronounced in two doubled components. The results of fits with two, respectively four components are indicated by red and blue dots in the inset (after \cite{Ivek17}).
On the right we show a schematic representation of the horizontal charge stripe arrangement within the $bc$ dimer layer proposed for \hgcl\ \cite{Drichko14}.  The dark and light blue circles denote charge-rich and charge-poor molecules, respectively.}
	\label{fig:HgClnu27}\label{fig:HgClstripes}
\end{figure}

\hgcl\ is characterized by a moderate strength of dimerization, thus bridging the space between weakly dimerized, quarter-filled systems on one side (Chapter~\ref{sec:ChargeOrder}) and strongly dimerized, half filled dimer-Mott systems (Chapters~\ref{sec:MottTransition} and \ref{sec:Frustration}) on the other side (cf.\ Hotta's classification in Figure~\ref{fig:Hotta1}). At this point, it stands as a sole example of a dimerized solid that undergoes a pronounced metal-insulator phase transition due to charge ordering when cooled through $T_{\rm CO}=30$\,K \cite{Drichko14, Ivek17}.
In the metallic state, at temperatures above $T_{\rm CO}$, optical and dc resistivity properties are characteristic of a metal with half-filled band and strong electronic correlations. A broad band of the C=C vibration $\nu_{27}({\rm b}_{1u}$) peaks around 1455\,\cm{} at room temperature and narrows on cooling down. Below about 100~K, it becomes obvious that the band consists of two peaks, which are assigned to the different environment of the BEDT-TTF molecules in the unit cell. Charge ordering is evidenced by a clear splitting of the $\nu_{27}$ mode below $T_{\rm CO}$ into two well-separated bands at 1441 and 1470~\cm, plotted in Figure \ref{fig:HgClnu27}. The splitting takes place as abruptly as in \aeti\ and \tetrz\ that exhibit charge ordering and ferroelectricity (see Section~\ref{sec:COweaklydimerized}). Fitting the optical bands by two Fano resonances for temperatures above $T_{\rm CO}=30$\,K  and by four modes below $T_{\rm CO}$ reveals a charge imbalance of $2\delta_{\rho} \approx 0.2e$ between two different molecular sites, which are most likely located within the BEDT-TTF dimer.
This value is certainly smaller than the charge disproportionation $2\delta_{\rho} \approx 0.6e$ found in \aeti\ and \tetrz, but significantly larger than the charge disproportionation in other dimerized salts, such as \etcl\ or \etcn, that does not exceed the limit of $2\delta_{\rho} \approx \pm 0.01e$ (see Section~\ref{sec:quantumelectricdipoles}).

The charge ordering transition in \hgcl\ is of first order as testified by a strong change of the magnitude of low-frequency resistance fluctuations \cite{Thomas19}, and
by jump-like anomalies of the relative length change, implying a divergent thermal expansion coefficient, and by thermal hysteresis between cooling and warming cycles \cite{Gati18a}. The dominant lattice change takes place along the out-of-plane $a$-axis, indicating that the coupling between anions and molecular cations plays an important role in the formation of charge order. Lang and collaborators argue that the interplay between charge ordering and cation-anion coupling results in the charge stripes running along the $c$-axis (Figure~\ref{fig:HgClstripes}), as previously proposed by Drichko {\it et al.} based on the large anisotropy of the optical spectra \cite{Drichko14}.
Unfortunately, an experimental evidence for the charge pattern shown in Figure~\ref{fig:HgClstripes} is still missing: x-ray diffraction measurements performed down to $T=10$~K could not resolve any symmetry change \cite{Drichko14}.

\begin{figure}
	\centering\includegraphics[clip,width=0.6\columnwidth]{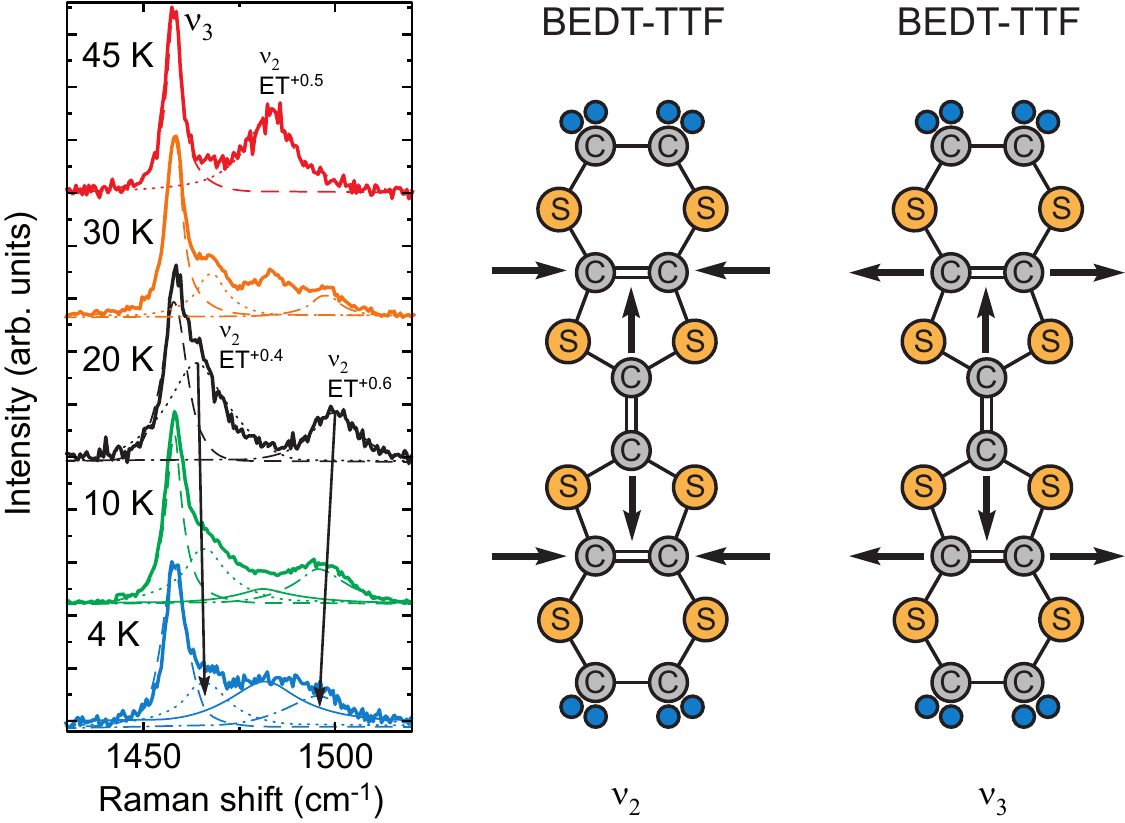}
	\caption{The fully symmetric molecular vibrations $\nu_{2}(a_g)$  and $\nu_{3}(a_g)$ are very sensitive to the local charge per molecule; the corresponding motion of the C=C bonds are indicated to the right.
The Raman shift measured for \hgcl\ at several representative temperatures above and below $T_{\rm CO}=30$~K evidences that the $\nu_{3}$ remains unaffected. Significant changes, however, are observed for the $\nu_{2}$ mode. {ET}$^{0.5+}$, {ET}$^{0.4+}$ and {ET}$^{0.6+}$ denote $\nu_{2}$ bands corresponding to $+0.5e$, $+0.4e$ and $+0.6e$ charge per BEDT-TTF molecule. In the metallic state at $T=45$~K {BEDT-TTF}$^{0.5+}$ band is observed, while in the charged ordered state at 20~K two bands {BEDT-TTF}$^{0.4+}$ and {BEDT-TTF}$^{0.6+}$ are clearly resolved. On further cooling at $T=10$~K and 4~K signatures of charge order melting are detected: the two bands {BEDT-TTF}$^{0.4+}$ and {BEDT-TTF}$^{0.6+}$ broaden and a {BEDT-TTF}$^{0.5+}$ band emerges (after \cite{HassanDrichko19}).}
	\label{fig:HgClnu2}
\end{figure}

The problem might be due to melting of charge order at low temperatures.
Results from a recent Raman scattering study suggest that the charge-ordered state
formed at $T_{\rm CO}=30$~K is not the ground state; rather it persists only down to $T=15$~K
and gradually melts below \cite{HassanDrichko19}.
In Figure~\ref{fig:HgClnu2} the temperature evolution of the Raman spectra is plotted in the frequency range of the charge-sensitive $\nu_{2} (a_{g})$ stretching vibration that involves the central C=C bond of the BEDT-TTF molecules.
The  single $\nu_{2}$ mode at 1490\,\cm{} corresponding to {BEDT-TTF}$^{0.5+}$ observed above $T_{\rm CO}$,
splits into two bands at 1475~\cm\ and 1507~\cm{} at $T=20$~K,
corresponding to {BEDT-TTF}$^{0.4+}$ and {BEDT-TTF}$^{0.6+}$ below $T_{\rm CO}$.
On cooling below 15~K, these bands gradually broaden, move closer in frequency
and lose spectral weight concomitantly as the band corresponding to {BEDT-TTF}$^{0.5+}$ gains it.
At $T = 2$~K the distribution of differently charged BEDT-TTF molecules $+0.4e : +0.5e : +0.6e$ is approximately equal to 0.3. The question remains whether this charge distribution is the final state or
whether it changes on further cooling.

The charge-order phase transition in \hgcl\ is extremely sensitive to external pressure: once the charge order is suppressed the compound remains metallic without indications of superconductivity \cite{Lohle17,Lohle18,Gati18a}.
This behavior is distinct from the linear pressure dependence of charge imbalance observed in \aeti\ \cite{Beyer16}.
It is very surprising that such small pressure variations lead to so substantial changes in the electronic properties; it is
only comparable to \etcl, which is located next to the Mott insulator-metal transition, discussed in Section~\ref{sec:afmMott}.
In the case of the charge-ordered \hgcl\ one does not expect that reducing the effective Coulomb repulsion $V/W$ or $U/W$, where $V$, $U$ and $W$ are inter-site and on-site Coulomb interaction, and bandwidth, respectively,
is responsible for the pronounced effect. L{\"o}hle {\it et al.} suggested that the lattice plays an important role in the phase transition \cite{Lohle17}. However, from Raman spectroscopy no major changes of lattice phonons are observed at $T_{\rm CO}$ implying that the coupling of the charge order to the lattice is weak.


Finally, indications for ferroelectricity driven by charge-order in \hgcl, were provided by dielectric spectroscopy,
albeit no polarization switching could be recorded so far,
probably due its rather high in-plane conductivity \cite{Gati18a}.
For that reason, the dielectric measurements were performed with the ac electric field applied along the out-of-plane $a$-axis only.
The dielectric response observed in the MHz frequency range
exhibits some features expected for conventional ferroelectrics:
Curie-like peak occurs in the real part of dielectric function $\varepsilon^{\prime}(T) \approx 400$ right at $T_\mathrm{CO}=30$~K with negligible dispersion. Also, the two branches of $1/\varepsilon^{\prime}(T)$ above and below $T_\mathrm{CO}$ are close to linear;
the slope in the ordered state is much larger than the one above $T_\mathrm{CO}$.
It was suggested that the ferroelectric order formed below the metal-insulator transition
is of order-disorder type \cite{Gati18a}. However, a pronounced frequency dependence of $\varepsilon^{\prime}(T)$ is expected in this case \cite{Krohns19}, just in contrast to what is observed.
Order-disorder type implies that disordered electric dipoles exist already in the metallic phase above $T_{\rm CO}$, but compelling experimental evidence for this proposal is lacking.
Another important issue in this regard is the anionic contribution to the measured dielectric response.
As pointed out \cite{Gati18a}, the anions are mobile in some way and shift towards charge-rich molecular sites in the vicinity of $T_\mathrm{CO}$ in order to minimize overall Coulomb energy; this arrangement stabilizes the long-range charge order.
This implies that the dielectric measurements along the $a$-axis probe not only the electronic but also
the anionic contribution to the dielectric response, because this is exactly the direction the cationic BEDT-TTF molecules and the anion layers alternate.
Recently is was shown \cite{deSouza18} that similar effects happen in the quasi-one-dimensional organic ferroelectrics (TMTTF)$_2$$X$.

We close by noting that the subject of ferrolectricity driven by charge order in dimerized solids deserves more attention in future. Specifically, efforts should be invested to unravel conclusive experimental evidences for the charge stripe pattern in the ground state, the role of electron-phonon coupling and the intrinsic dielectric  response.

\section{Mott Metal-Insulator Phase Transition}
    \label{sec:MottTransition}
    The properties of materials are only understood when their ground states and phase transitions can be described and predicted. In the case of correlated electron systems, the Mott metal-insulator transition (MIT) serves as the litmus test: no structural modification  drives the phase transition but the increasing influence of electronic repulsion \cite{MottBook,EdwardsBook,GebhardBook,DobrosavljevicBook}.
Tremendous efforts have been devoted to this fundamental problem, nevertheless
there is still no general agreement on all detail how the Mott transition actually takes place.
The validation of theoretical advances by meaningful experiments on suitable model systems turns out to be the real challenge.
Research was boosted on the experimental and theoretical frontier by the discovery of high-temperature superconductivity in cuprates, which are doped Mott insulators \cite{Imada98}. But one should keep in mind that raising the carrier concentration yields a rather different scenario than varying correlations at constant half band filling. In this regard the bandwidth-tuned metal-insulator transition in vanadium oxides has been intensely studied for decades \cite{Morin59,Jayaraman70,Limelette03b,Lupi10},
however, there is still no unanimous conclusion to what extend it is affected by corresponding structural transitions or solely  driven by strong electron–electron correlations \cite{Hansmann13,Najera17,Najera18,Shao18}.
Unfortunately, the majority of transition-metal oxides undergoes antiferromagnetic (afm) order in the insulating state at low temperatures preventing the exploration of the genuine Mott state and metal-insulator transition \cite{Imada98}. At the phase transition not only the electronic properties change but also the magnetic degrees of freedom.

Having said that, organic charge-transfer salts are realized by now as superior model systems for studying Mott physics:
\begin{itemize}
\item The energy scales are much smaller compared to transition metal compounds,
allowing the coverage of a significantly wider range in effective temperature $T/W$ and correlations $U/W$.
The investigations are also carried out at lower temperatures -- but still accessible for regular helium cryostats -- where thermal fluctuations are of minor importance and quantum effects may become decisive.
\item Since the compounds are rather soft, external pressure is a powerful tool to substantially increase the orbital overlap $t$ and hence the bandwidth $W$. Alternatively, isovalent substitution of anions with different size acts as chemical pressure. Furthermore, the bandwidth can be readily varied by chemical substitution in the organic donor molecules, for instance, when replacing S by Se.
\item For comparison with theory it is advantageous that most organic systems under study, contain only a single band of relevance.
\item In analogy to transition metal oxides, the simple or extended Hubbard model can be applied to these strongly correlated molecular compounds. The Mott transition results in splitting the conduction band into lower and upper Hubbard bands with a gap at the Fermi energy of typically 0.2 to 1 eV. No charge-transfer band disturbs this ideal scenario.
\end{itemize}

\begin{figure}
  \centering
    \includegraphics[width=1\columnwidth]{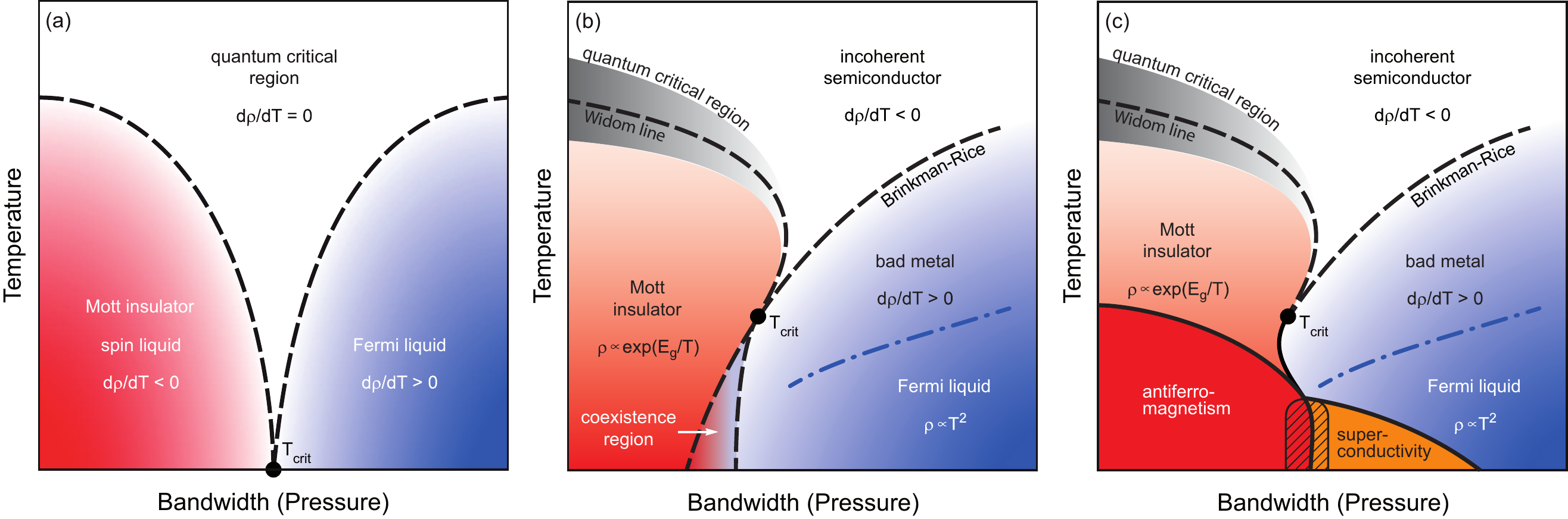}
    \caption{\label{fig:schematicPhaseDiagrams}
Upon tuning the bandwidth by chemical or physical pressure, for instance, a paramagnetic Mott insulator transforms to a correlated metal. Different theoretical scenarios of the Mott insulator-metal transition are considered.
(a)~Based on mean-field theory, a quasi-continuous transition from a Mott insulator to the Fermi liquid through a
quantum critical point at $T_{\rm crit} = 0$ is suggested  for quantum spin liquids with spinon quasiparticles on the Mott insulating side \cite{Florens04,Senthil08}.
(b)~For a fully frustrated system with no magnetic order, dynamical mean-field theory predicts a first-order transition with phase coexistence up to the critical endpoint $T_{\rm crit}$~\cite{Georges96}, and a fan-shaped quantum-critical regime associated with the quantum Widom line at higher temperatures~\cite{Terletska11,Vucicevic13}, in excellent agreement with experiment~\cite{Furukawa15b,Pustogow18a,Rosslhuber19,Rosslhuber19}. The metallic state is confined by the Brinkman-Rice temperature~\cite{Radonjic12,Deng13}, identified by a maximum in $\rho(T)$; the coherent Fermi-liquid regime occurs at lower temperatures. When thermal fluctuations exceed the bandwidth $W$ and interaction strength $U$, semiconducting behavior prevails; neither a gap nor a quasiparticle peak are stabilized.
(c)~Strong magnetic interactions commonly leads to antiferromagnetic order of the Mott insulator at low temperatures with significant impact on the phase boundary.
A spatial coexistence of antiferromagnetic order and superconductivity is observed in the vicinity of the first-order phase transition.
}
\end{figure}

In a first approach, we assume that the Mott gap gradually closes by reducing the electron-electron interaction $U$ or increasing the bandwidth $W$, implying  that the Mott transition is of second order \cite{Kadanoff67,Florens04,Senthil08}.
For the particular case of two-dimensional systems with no long-range magnetic order,
one expects a direct and continuous transition between a paramagnetic metal and a paramagnetic Mott insulator
as sketched in Figure~\ref{fig:schematicPhaseDiagrams}(a).
This would be a particular example of
a quantum critical point at $T=0$ unrelated to any mechanism of spontaneous symmetry breaking.
As correlations become more effective, the weakly-correlated Fermi liquid with Landau quasiparticles vanishes at the Mott critical point. However, there remain gapless spin excitations (spinons) on the insulating side, {\it i.e.} a Mott insulating spin liquid with a spinon Fermi surface.
Both the gap in the spectral function and the quasiparticle weight vanish at the critical point. At finite temperatures there is a crossover to a marginal spinon liquid and a marginal Fermi liquid, respectively. Above the Mott critical point a quantum critical regime develops with non-Fermi-liquid properties \cite{Senthil08}.

In contrast to this scenario, already Mott predicted that the transition between the insulating and metallic states should be discontinuous, implying a regime of coexistence, as depicted in Figure~\ref{fig:schematicPhaseDiagrams}(b).
Calculations based on dynamical mean field theory (DMFT) \cite{Georges93,Rozenberg94,Georges96,Bulla01,Kotliar04,Vollhardt12} confirm the picture of a first-order phase transition with a coexistence region on both sides limited by spinodal lines that end at the critical endpoint $T_{\rm crit}$, which now has moved to finite temperatures.
In the high-temperature crossover region ($T>T_{\rm crit}$) paramagnetic DMFT calculations on a single-band Hubbard model at half filling yield quantum critical transport \cite{Terletska11,Vucicevic13,Vucicevic15} that spreads out following the quantum Widom line. For experimental verification of this behavior, it is decisive that the critical endpoint is only around 20 to 40~K for $\kappa$-(BEDT-TTF)$_2$$X$ salts, while $T_{\rm crit}\approx 450$~K is an order of magnitude larger for (V$_{1-x}$Cr$_x$)$_2$O$_3$ \cite{Limelette03b}.

The electronic phase diagram resembles the thermodynamic phases of classical gases and liquids. The phase boundary in the $p$-$T$-diagram is well defined only up to the critical endpoint $T_{\rm crit}$; in the supercritical regime above, there is no way to strictly distinguish gas from liquid \cite{Kadanoff67,StanleyBook}. The Widom line is the crossover  between liquid-like and gas-like behavior \cite{Fisher69,Simeoni10} and can be identified by changes in the derivative, as illustrated in Figure~\ref{fig:MIT3} for the case of electronic analogues of interest here.

Except for a few quantum spin liquid candidates, organic Mott insulators are also prone to antiferromagnetic order at low-enough temperatures \cite{Powell11}, which gives an additional twist to the insulator-metal boundary [Figure~\ref{fig:schematicPhaseDiagrams}(c)] rendering the common negative slope ${\rm d}T/{\rm d}p <0$ due to the magnetic contribution to the entropy \cite{Dang15}.
The common border of antiferromagnetism and superconductivity implies important contributions of spin fluctuations to the superconducting glue, similar to transition metal oxides. Let us consider this case first before we discuss the fully-frustrated
paramagnetic insulators.

  \subsection{Antiferromagnetic Mott insulator}
\label{sec:afmMott}

\begin{figure}
  \centering
  \includegraphics[width=0.9\columnwidth]{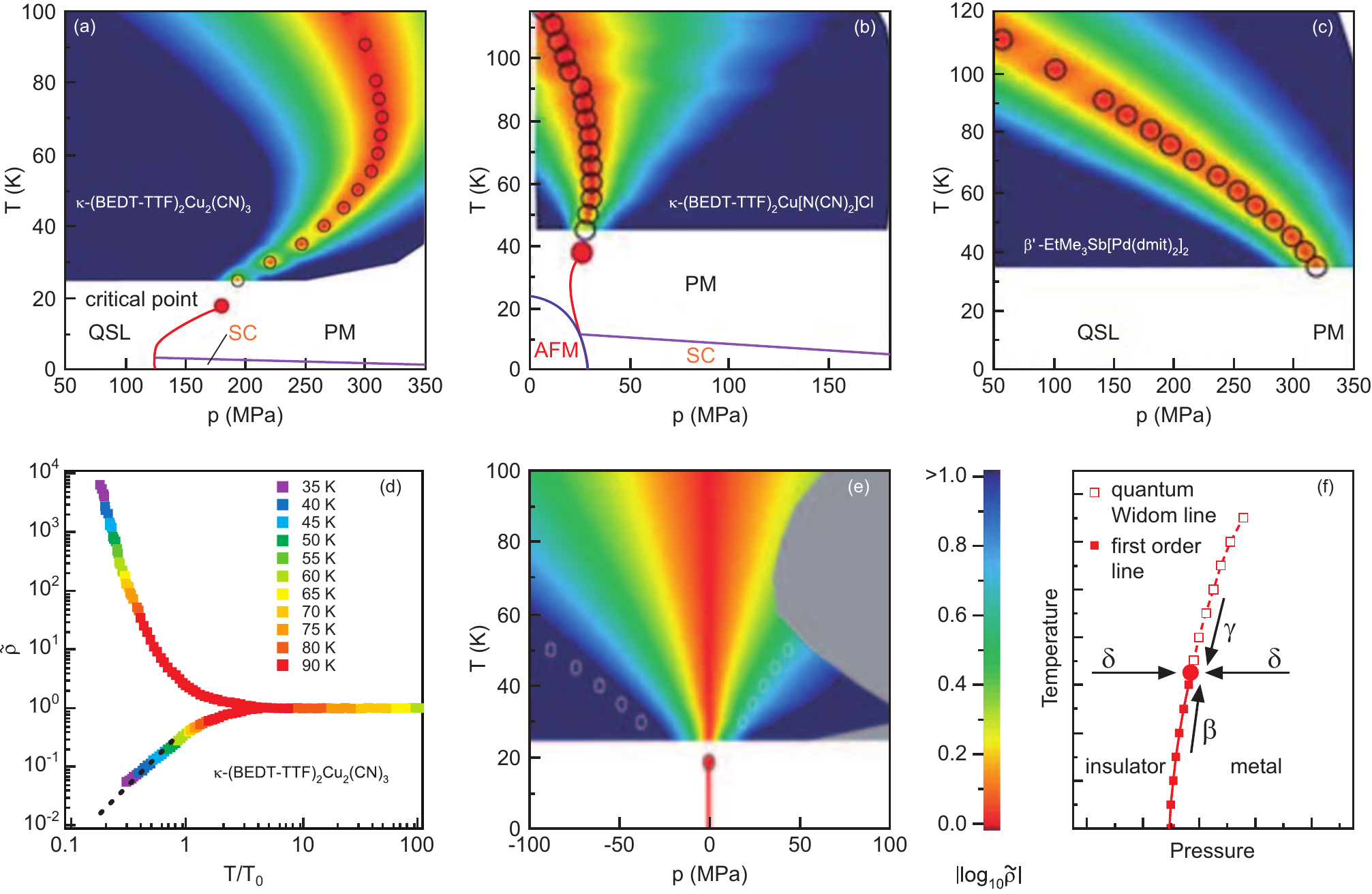}
    \caption{\label{fig:MottCriticality}
Pressure–temperature phase diagram of (a) \etcn, (b) \etcl\ and (c) \dmit\ obtained from resistivity measurements in a gas-pressure cell. QSL stands for quantum spin liquid, SC for supercondcutor, AFM for antiferromagnet and PM for paramagnetic metal.
The red line represents the first-order Mott transition line terminating at a critical end point. The open circles indicate the metal–insulator crossover pressure $p_c(T)$.
The color represents the magnitude of $\left|\log_{10} \tilde{\rho}\right|$, where $\tilde{\rho}$ is the normalized resistivity.
(d)~The scaling plot of the normalized resistivity $\tilde{\rho}(\delta p, T)$ {\it versus} $T/T_0 = T/|c\, \delta p|^{z\nu}$ with the present values, $z\nu = 0.62$ and $c = 20.9$ for \etcn.
(e)~A symmetric quantum critical region can be obtained by renormalizing the axes.
A range of colour represents the  magnitude  of  $|\log_{10}\tilde{\rho}|$ for \etcn,
where $\tilde{\rho}$ is the normalized  resistivity. The grey-colored  region  is  experimentally  inaccessible.
The bold red line represents the first-order Mott transition line  terminating  at  a  critical  endpoint.
The insulating and metallic states form at $\delta p < 0$  and $\delta p >  0$, respectively.
The open circles indicate the characteristic temperatures for the entrance to the low-temperature regimes of the gapped Mott insulator or the Fermi liquid, $T^*$, defined by the value of $T/T_0$ at which $\tilde{\rho}(\delta p,T)$ starts to deviate from the  critical behavior $\tilde{\rho} =  \exp\left\{\pm (T/T_0)^{-1/z\nu}\right\}$ ($z\nu = 0.62$). Below these circles, the system  departs from the quantum critical regime toward the Mott insulator ($\delta p < 0$) or Fermi liquid ($\delta p > 0$)
(data from \cite{Furukawa15a}). (f)~Scheme how the different critical exponents $\beta$, $\gamma$ and $\delta$ are determined from scans in the phase diagram as a function of temperature and pressure (suggested in \cite{Gati16}).}
\end{figure}

Here the antiferromagnetic Mott insulator \etcl\ certainly is of superior importance
because a tiny pressure of 30~MPa suffices for entering the metallic or superconducting phase with the record high $T_{c}\approx 13$~K.
At elevated temperatures, the overall charge transport is incoherent; but upon cooling the behavior bifurcates.
At ambient pressure, the system becomes a pronounced insulator: a maximum of the transport gap derived from the logarithmic derivative of $\rho(T)$ indicates the so-called quantum Widom line \cite{Pustogow18a,Pinteric15}.
This crossover from a more insulating to a more metallic regime is best seen in electrical transport studies using pressure sweeps at fixed temperatures \cite{Furukawa15a}.
Furukawa {\it et al.} investigated several organic Mott insulators
and could extract extended crossover regimes as plotted in Figure~\ref{fig:MottCriticality}.
The observed behavior collapses into a genuine phase diagram for all compounds, displayed in
Figure~\ref{fig:MIT3}, if the temperature $T$ and the electronic correlations $U$ are normalized by the respective bandwidth $W$ \cite{Pustogow18a}.
Using advanced temperature-dependent scaling, a mirror-symmetric behavior on the insulating and metallic  sides can be obtained, leading to a fan-shaped  quantum  critical region in the pressure-temperature phase diagram.
A material-independent quantum-critical scaling relation was identified with a clear distinction into a Fermi liquid or a Mott insulator, irrespective of the ground states of the organic compounds \cite{Furukawa15a}.
The observed Mott quantum criticality confirms the predictions of Dobrosavljevi{\'c} and collaborators \cite{Terletska11}, who calculated the incoherent charge transport in the high-temperature crossover region.

\subsubsection{Mott quantum criticality}
\label{sec:Mottquantumcriticality}
\begin{figure}
  \centering
  \includegraphics[width=\columnwidth]{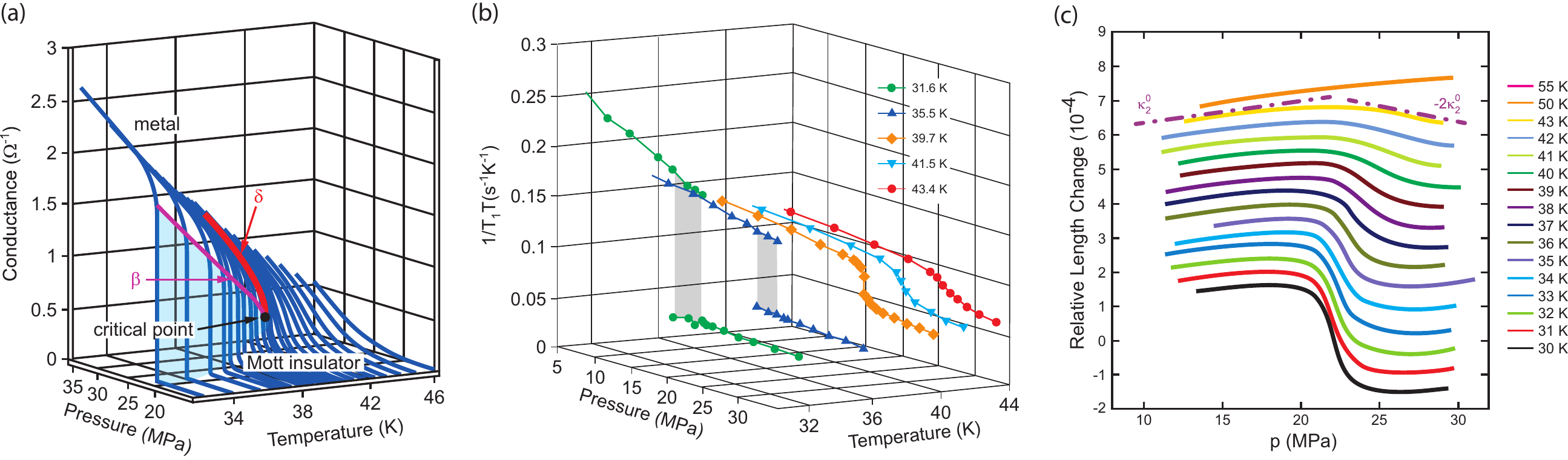}
    \caption{\label{fig:CriticalPoint}
    The critical behavior is observed in the electrical, magnetic and lattice degrees of freedom.
(a)~Pressure-dependent conductance $G_T(p)$ of \etcl\
measured at various temperatures around the critical endpoint (filled circle).
The shaded area indicates the conductance jump. The red and green curves represent the critical behaviour at
$T = T_{\rm crit} \approx 39.7$~K and $T < T_{\rm crit}$, which give the critical exponents $\delta$ and $\beta$, respectively.
The hysteresis of the conductance jump (for example, $\sim 0.2$~MPa at $\sim 32$~K) is not appreciable at this scale. (reproduced from \cite{Kagawa05} with permission).
(b)~$^{13}$C-NMR relaxation rate around the Mott critical endpoint $T_{\rm crit}$. The pressure dependence of $1/T_1 T$ at various fixed temperatures; note the reversed pressure scale, compared to panel (a). The gray shaded areas represent the coexistence of insulating and bad metallic phases (taken from \cite{Kagawa09}).
(c)~Relative length changes, $\Delta L/L$, as a function of applied pressure at constant temperatures between 30 and 55~K measured on \etcl\ along $a$-axis. The data have been offset for clarity. The broken lines close to the data at $T=43$~K, that is, distinctly above $T_{\rm crit} \approx 36.5$~K of the critical endpoint, are guides to the eyes and serve to estimate the pressure-induced changes in compressibility. The strong non-linearities, which are observed here, reflecting nonlinear strain-stress relations, highlight a breakdown of Hooke’s law of elasticity (data from \cite{Gati16}).}
\end{figure}

The critical point $T_{\rm crit} \approx 40$~K has drawn particular interest as it ends the first-order phase boundary, and might act as the starting point for quantum critical behavior.
In the seminal work on Cr-substituted V$_2$O$_3$, Limelette {\it et al.} analyzed the critical exponents and scaling functions close to the critical endpoint of the Mott metal-insulator transition [Figure~\ref{fig:MottCriticality}(f)]
\begin{equation}
\sigma-\sigma_c \propto(T_{\rm crit} - T)^{\beta} ~,~~~~~
\sigma-\sigma_c \propto (p - p_{\rm crit})^{1/\delta} ~,~~~~
\left({\rm d}\sigma/{\rm d}p\right)|_{p_c(T)} \propto (T - T_{\rm crit})^{\gamma} \quad ,
\label{eq:criticalexponents}
\end{equation}
and found the universal properties of a liquid-gas transition \cite{Limelette03b}; {\it i.e.}  $\beta = 0.5$,  $\delta \approx 3$, and $\gamma = 1$ as predicted by mean-field theory \cite{Castellani79,Kotliar00}. These critical exponents are believed to be universal, {\it i.e.} independent on the microscopic details of the system \cite{Kadanoff67,StanleyBook}.
Along these lines, Kagawa {\it et al.} investigated \etcl\ using a variable gas pressure cell down to $T=33$~K \cite{Kagawa05,Kagawa04b}; the  critical exponents extracted ($\beta = 1$, $\delta = 2$, and $\gamma = 1$) did not correspond to any known universality class [Figure~\ref{fig:CriticalPoint}(a)]. The analysis extended up to $|T-T_{\rm crit}|/T_{\rm crit}<0.2$ with significant deviations close to the critical endpoint.

Although the Mott transition takes place in the charge sector, it is interesting to  consider also the spin degrees of freedom in these strongly correlated electron systems. The spin-lattice relaxation rate $1/T_1$
measured by $^{13}$C-NMR is a probe of spin fluctuations. A clear jump in $1/T_1 T$ is observed at the Mott transition at low temperatures when the system is isothermally tuned across the transition by He-gas pressure \cite{Kagawa09}, as depicted in Figure~\ref{fig:CriticalPoint}(b).
The jump vanishes at the critical point and only a smooth cross-over remains at higher temperatures.
As seen in Figure~\ref{fig:CriticalPoint}(a) and (b), the conductance is enhanced when the Mott transition is crossed to a metal with increasing pressure; the paramagnetic spin fluctuations are strongly suppressed.
$\left[1/T_1(p)T-1/T_1(p_c)T\right]$ follows a power law $|p-p_c|^{1/\delta}$ with $\delta=2$, again. Any deviation from the average value of one electron per site obviously makes charge carriers contribute to the conductivity, but in addition spin fluctuations are suppressed \cite{Sentef11}. The subject remains open for future investigations.

Commonly neglected by theory, the electronic system couples to the underlying compressible crystal lattice.
Lang and collaborators thoroughly measured the temperature and pressure dependence of the thermal expansion
in order to study the lattice response near the Mott transition.
Fully deuterated $\kappa$-(d8-BE\-DT\--TTF)$_2$\-Cu\-[N\-(CN)$_{2}$]Br falls extremely close to the critical point already at ambient pressure \cite{deSouza07}; it was found that near the critical endpoint the Gr{\"u}neisen scaling breaks down
\cite{Nakazawa96,Papanikolaou08,Bartosch10,Zacharias12,deSouza15}.
In a next step \etcl\ could be deliberately tuned across the Mott transition using a continuously controlled helium-gas pressure.
The relative length change $\Delta L/L$ with pressure exhibits a strong nonlinear variation around $T_{\rm crit}$;
here Hooke’s law of elasticity breaks down \cite{Gati16}.
Figure~\ref{fig:CriticalPoint}(c) displays $\Delta L/L$ as a function of pressure measured within the plane at various temperatures; similar results are recorded perpendicular to the $ac$-plane. At $T=30$~K an abrupt jump is observed, reflecting the first-order character of the phase transition. As the temperature rises, the discontinuity gradually decreases until it becomes a continuous crossover for $T>T_{\rm crit}$. Slightly above the critical endpoint the relative length change is rather nonlinear, which is explained by a critical elasticity as a result of the coupling of the critical electronic background to the lattice; in other words Hooke's law does not hold in the temperature-pressure regime close the critical endpoint. The critical exponents [Eq.~(\ref{eq:criticalexponents})] extracted from the experimental data, $\beta= 0.52$, $\delta = 3.2$, and $\gamma = 1.0$, are in good consistence with the values for the mean-field universality class.
There are valid arguments that this behavior at the Mott transition holds for all systems that are amenable to pressure tuning \cite{Gati16}.

While \etbr\ is well on the metallic side of the phase boundary, the successive substitution of deuterated (d8-BE\-DT\--TTF) molecules effectively increases correlations and eventually shifts the alloy across the Mott transition. Magnetotransport measurements by Sasaki {\it et al.} \cite{Sasaki08b} found that a suppression of $T_{c}$ in magnetic fields;
for $H > H_{c2}\approx 12$~T the superconducting phase has completely vanished.
The critical endpoint $T_{\rm crit}$, however, seems not to be affected.
An alternative approach utilizes x-ray irradiation to systematically influence the critical behavior in \etcl\ \cite{Urai19}.
The introduced disorder reduces the superconducting transition temperature $T_c$ only slightly but strongly affects the resistivity around the critical endpoint. Pressure-sweeps at fixed temperatures down to 15~K yield a broadening of the phase transition and a corresponding shift of $T_{\rm crit}$. Analyzing their data according to Figure~\ref{fig:MottCriticality}, Urai {\it et al.}  found that the exponent $z\nu \approx 0.46$ basically does not change upon irradiation up to 70~h. The drastic suppression of the critical end point $T_{\rm crit}^*$ is interpreted as disorder-enhanced critical fluctuations of the metal-insulator transition. They speculate that the Mott quantum critical fluctuations are hidden behind the first-order transition and can be revealed by disorder. However, this implies that the system is spatially homogeneous for  $T > T_{\rm crit}^* \rightarrow 0$ due to disorder, which is in contrast to the coexistence regime.
As discussed in more detail in Sec.~\ref{sec:randomness}, extended irradiation blurs any clear signatures of a discontinuous metal-insulator transition \cite{Gati18a}; {\it i.e.} above a certain disorder level, the Mott transition becomes a smeared first-order transition with some residual hysteresis.

\subsubsection{Coexistence regime}
\label{sec:coexistenceregime}
\begin{figure}[h]
  \centering
  \includegraphics[width=0.8\columnwidth]{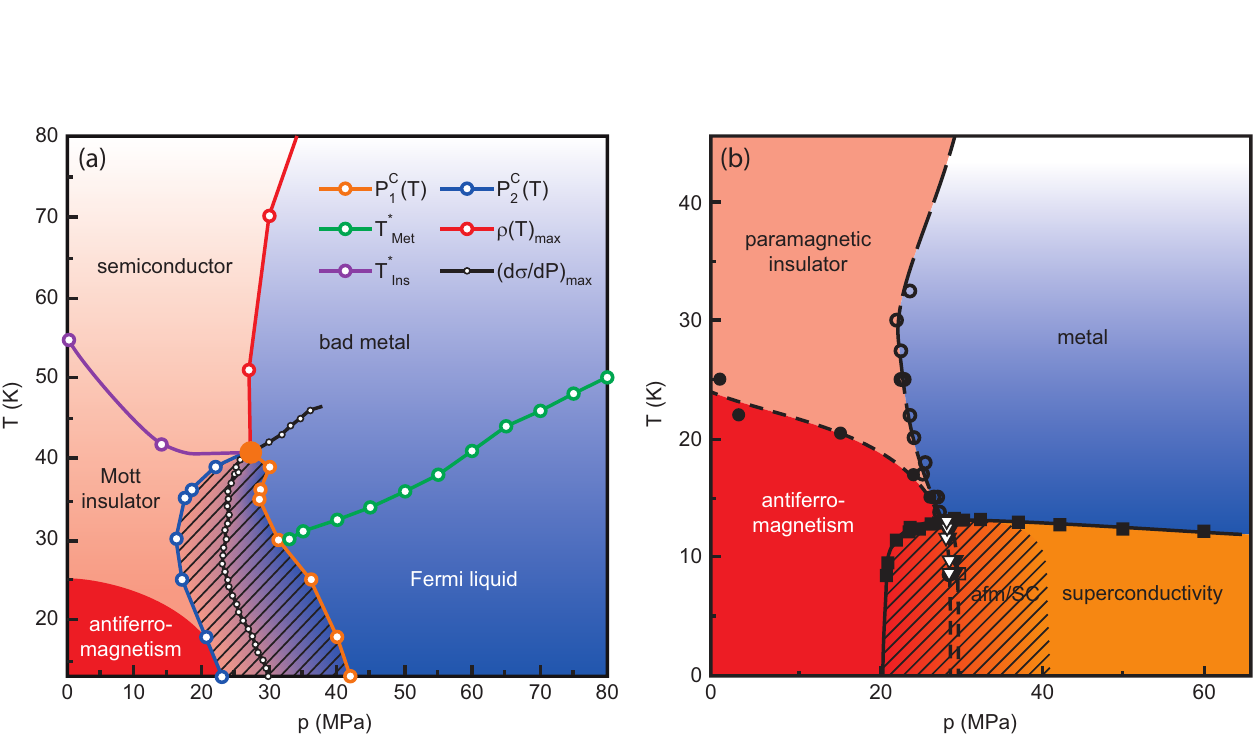}
    \caption{\label{fig:PhaseDiagram_k-Cl}
Pressure-temperature phase diagram of the \etcl\ salt. (a)~Four regions can be identified from transport measurements
\cite{Limelette03a}: bad metal, Fermi liquid, semiconductor and insulator, which orders antiferromagnetically below $T_N\approx 25$~K \cite{Lefebvre00,Yasin11}.
The spinodal lines define the region of coexistence of insulating and metallic phases, which is indicated by the hatched area.
The first-order transition line is identified by the maximum slope d$\sigma$/d$P$ and terminates at the critical end point.
(b)~The antiferromagnetic critical line $T_N(P)$ (dark circles) was determined from NMR relaxation rate while $T_c(P)$ for unconventional superconductivity and the metal-insulator $T_{\rm MI}(P)$ (open circles) lines were obtained from
the ac susceptibility. The afm/SC boundary (double-dashed line) is determined from the inflection point of
$\chi^{\prime}(P)$ and, for 8.5 K, from sublattice magnetization. This boundary line separates
two regions of inhomogeneous phase coexistence (shaded area) (adopted from \cite{Limelette03a,Lefebvre00}).}
\end{figure}

The Mott transition is supposed to be of first order below the critical endpoint $T_{\rm crit}$, implying
a pressure region in which correlated metal and insulator coexist \cite{Georges96,Vollhardt12,Vollhardt20}.
Limelette {\it et al.} measured the in-plane electrical transport of \etcl\ at fixed temperatures below $T_{\rm crit} \approx 40$~K in a gas pressure cell \cite{Limelette03a} and  observed a marked hysteresis around the metal-insulator transition that was  attributed to spatial inhomogeneities.
As depicted in Figure~\ref{fig:PhaseDiagram_k-Cl}(a), the coexistence regime extends over a pressure range of approximately 20~MPa, as determined from the inflection in the conductivity data $\sigma(p)$ upon increasing and decreasing pressure.

Below $T_N\approx 25$~K, antiferromagnetic ordering occurs that was investigated by magnetic resonance spectroscopy and other methods \cite{Yasin11,Lefebvre00}.
Extended ac susceptibility measurements as a function of pressure for selected temperatures as well as by taking cooling and heating curves at fixed pressure values were used by Lefebvre {\it et al.} to map the coexistence regime between the antiferromagnetic Mott insulator and superconducting states \cite{Lefebvre00} as plotted in Figure~\ref{fig:PhaseDiagram_k-Cl}(b).
Below a characteristic temperature $T^* \approx 20$~K
the NMR lines split into two groups corresponding to a metallic (superconducting) and an insulating (antiferromagnetic) phase that spatially coexist.
These remarkable findings evidence percolative superconductivity \cite{Kagawa04a,Muller09,Muller17}. The results indicate the absence of 
itinerant antiferromagnetism in \etcl, confirming previous suggestions \cite{Kanoda97c}; the interacting spins are localized on the dimers.
This scenario is in contrast to (TMTSF)$_2$PF$_6$, where superconductivity coexists with the itinerant antiferromagnetism of the spin-density-wave phase \cite{Vuletic02}. In the present case superconductivity can be directly stabilized from the antiferromagnetic Mott insulator.

These findings were confirmed by ultrasonic velocity and attenuation measurements on \etbr\ \cite{Fournier07} where the coexistence zone of the antiferromagnetic and superconducting phases was observed deep in the metallic part of the pressure-temperature phase diagram. The system was tuned by varying the cooling cycle, {\it i.e.} fast, slow cooling, low-temperature annealing. The two phases are found to compete, whereas superconducting fluctuations begin to contribute to the attenuation at 15~K, namely,
at the onset of magnetic order that is well above the superconducting transition temperature $T_{c}=11.9$~K.

\begin{figure}[h]
  \centering
  \includegraphics[width=0.6\columnwidth]{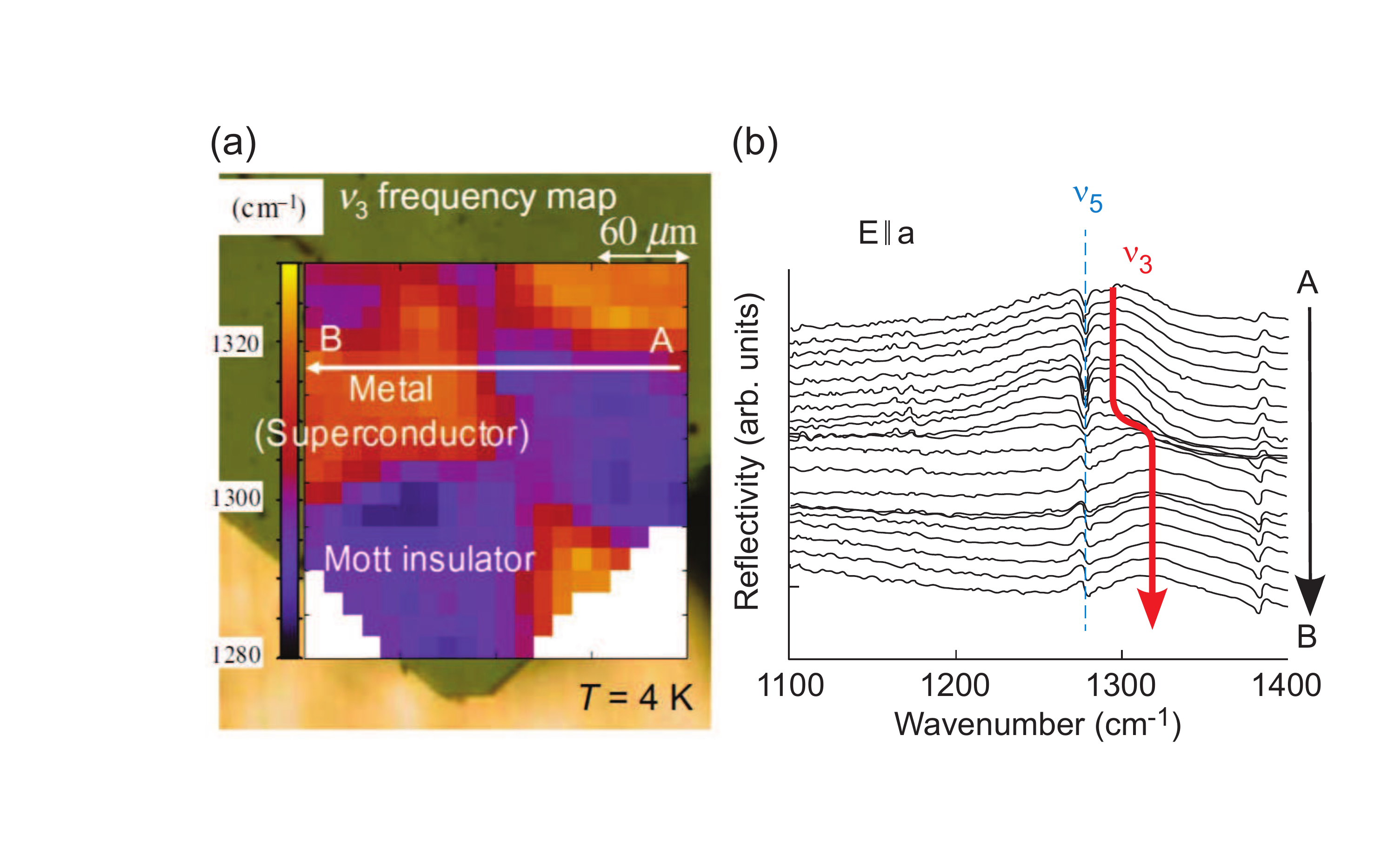}
    \caption{\label{fig:Scan}
 (a) Peak frequency contour map of the $\nu_3$ mode ($E\parallel a$-axis) of $\kappa$-[(h8-BEDT-TTF)$_{0.4}$(d8-BEDT-TTF)$_{0.6}$]$_{2}$Cu[N(CN)$_2$]Br at $T=4$~K. Bright orange colors (higher frequency) indicate a metallic nature and dark violet colors (lower frequency) indicate an insulating nature. The metal–insulator phase separation can be observed as a spatial image. (b)~Reflectivity spectra along the arrow A–B in panel (a) with a $12~\mu$m steps (reproduced from \cite{Sasaki09}).}
\end{figure}
In order to demonstrate the spatial phase separation Kimura and collaborators \cite{Nishi05,Nishi07} studied the properties of partially deuterated \etbr\ using spatially resolved magneto-optical spectroscopy.
While they can conclude a coexistence of the metallic or superconducting phases with the insulating phase,
the present resolution of approximately 10~$\mu$m constitutes only an upper limit. Interestingly, with increasing magnetic field, the insulator-metal phase boundary shifts towards smaller $U/W$, in accord with transport experiments \cite{Sasaki08b};
above $H_{\rm c2}$, however, they suggest an enlargement of the metallic regime. Concomitantly, Sasaki {\it et al.} focussed on the vibrational features of \etbr\ to obtain the crucial contrast between metallic and insulating regions when tuning through the phase transition by different amount of deuteration \cite{Sasaki04a,Sasaki05,Sasaki09}. The fully symmetric $\nu_3(a_g)$ mode around 1300~\cm\ is strongly electron-molecular vibrational (emv) coupled \cite{Maksimuk01,Dressel04a,Girlando11a} and provides a local probe for detecting changes in the electronic state.
Figure~\ref{fig:Scan} displays a contour map of metallic and insulating regions having micrometer sizes and irregular shapes. The boundary between the insulating and metallic regions is within the instrumental resolution, and hence no intermediate state appears at the frontier.
This observation indicates a macroscopic phase separation between the metal/superconductor and Mott insulator. Since the system falls right at the characteristic S-shape phase boundary, the sample crosses the first-order transition when cooled from room temperature. The phase separation occurs after intersection with the first-order transition line.

M{\"u}ller {\it et al.} ~\cite{Muller11,Muller12} established fluctuation spectroscopy as a powerful method for investigating the dynamics of correlated charge carriers in the vicinity of the Mott transition in the quasi-two-dimensional charge-transfer salts, looking in particular at $\kappa$-(d8-BEDT-TTF)$_2$Cu[N(CN)$_2$]Br. The observed $1/f$-type fluctuations are quantitatively very well described by a phenomenological model based on the concept of non-exponential kinetics. The main result is a correlation-induced enhancement of the fluctuations accompanied by a substantial shift of spectral weight to low frequencies in the vicinity of the Mott critical end point. This sudden slowing down of the electron dynamics is considered as a universal feature of metal-insulator transitions. The findings support the idea of electronic phase separation in the critical region of the phase diagram \cite{Muller11,Muller12}.

\subsection{Frustrated Mott insulator}
\label{sec:QSLMott}

In order to target investigations of the genuine Mott transitions without being affected by magnetic order, quantum spin liquids are the systems of choice. As illustrated in Figure~\ref{fig:schematicPhaseDiagrams}(b), the phase boundary is solely determined by the electronic degrees of freedom; hence magnetic contributions to the entropy are absent. In this case, Clausius-Clapeyron relation implies that the phase boundary acquires a positive slope ${\rm d}T_{\rm IM}/{\rm d}p >0$: the thermodynamic ground state is metallic.
In other words, the mobile electrons in the Fermi liquid state have less entropy than those localized in the Mott state.

Pustogow {\it et al.} succeeded
to compose a genuine phase diagram of Mott insulators by combining optical and transport experiments on three representative spin liquid compounds \dmit, \agcn\ and \etcn\ \cite{Pustogow18a}.
\begin{figure}[h]
  \centering
  \includegraphics[width=0.9\columnwidth]{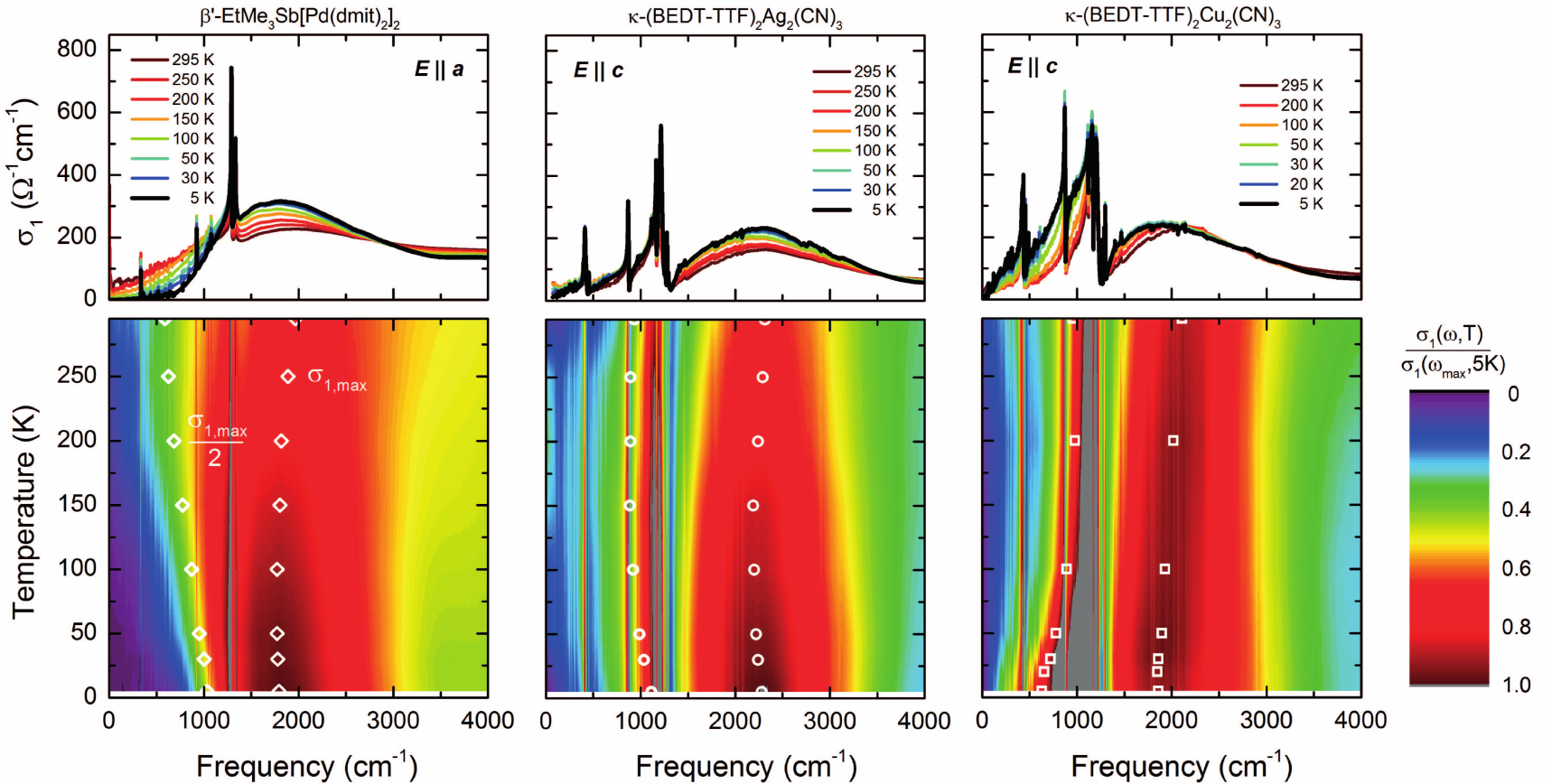}
    \caption{\label{fig:Mott-optics}
Temperature evolution of the optical conductivity of the three quantum-spin-liquid compounds \dmit, \agcn\ and \etcn. The dominant feature that contains all information of the intrinsic Mott physics is the Mott-Hubbard band centered around 2000~\cm. At low frequencies narrow phonon modes appear on top. The contour plots illustrate the tem\-per\-ature-dependent changes of the Mott-Hubbard band, where the open white symbols denote the maximum and half-maximum positions (cf. Figure~\ref{fig:MIT3} upper right). Note that the band shape and the low-frequency conductivity show distinct behavior for each compound which is related to the respective position in the phase diagram (following \cite{Pustogow18a}). }
\end{figure}
Figure~\ref{fig:Mott-optics} presents the temperature changes of the frequency-dependent conductivity measured along one representative orientation. While in \dmit\ the low-frequency Mott-gap opens upon cooling, and spectral weight shifts toward the Mott-Hubbard band, for \agcn\ only little variation is observed below 1000~\cm. Most surprising, however, is the increase of the in-gap absorption
in \etcn\ \cite{Kezsmarki06,Elsasser12} that could be assigned to metallic fluctuations.
One can imagine that upon approaching the metal-insulator-phase boundary, the number and effect of metallic puddles increases, reflecting the phase coexistance close to the first-order phase boundary, as depicted in Figure~\ref{fig:schematicPhaseDiagrams}(b).
When normalizing the temperature $T$ and the correlation strength $U$ by the respective bandwidth $W$ as determined from optical experiments, the materials behavior collapses on a quantitative level  as displayed in Figure~\ref{fig:MIT3}. The quantum Widom line is clearly seen as well as the back-bending towards the critical endpoint. Below $T_{\rm crit}$, it becomes a first-order phase boundary with a range of coexisting phases, which have been subject of subsequent studies.

Tuning the bandwidth enables us to follow the electron system from the strongly correlated Mott insulator through the range of phase coexistence into the metallic regime even at lowest temperatures.
\begin{figure}
  \centering
  \includegraphics[width=0.7\columnwidth]{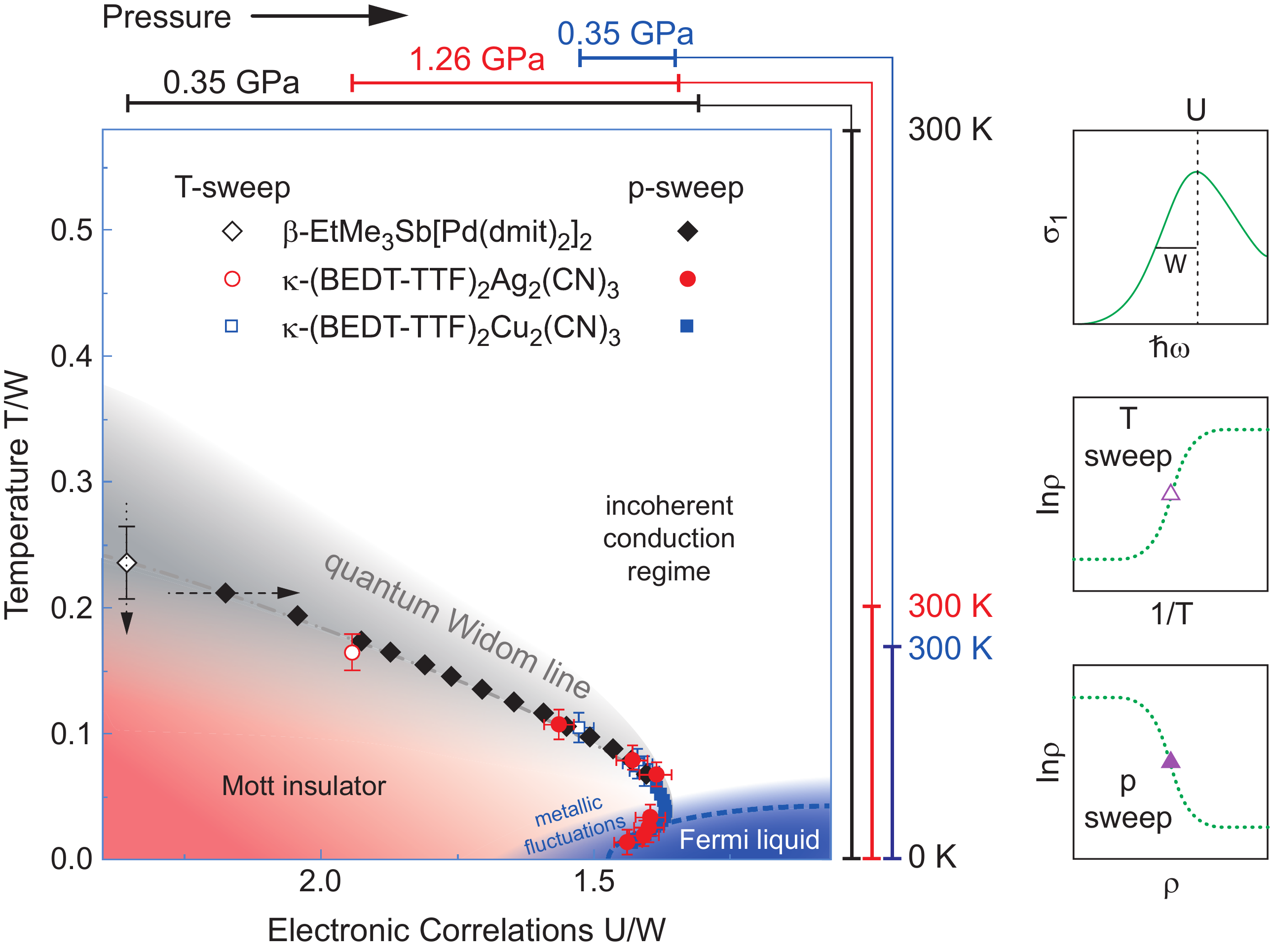}
    \caption{\label{fig:MIT3}
Quantitative phase diagram of pristine Mott insulators. The temperature $k_BT$ and Coulomb repulsion $U$ are normalized to the bandwidth $W$ extracted from optical spectroscopy; note that the direction of the bottom axis is reversed in order to mimic
pressure. Since in these quantum spin liquids magnetic order is suppressed, the large residual entropy causes a pronounced back-bending of the quantum Widom line at low temperatures leading to metallic fluctuations (semi-transparent blue area) in the Mott state close to the phase boundary. As the effective correlations decrease further, a metallic phase forms (blue area) with Fermi liquid properties \cite{Yasin11}. The universal phase diagram guided by the quantum Widom line is constructed on basis of optical and transport data \cite{Pustogow18a} as well as the pressure-dependent transport
studies \cite{Furukawa15a,Shimizu16} on \dmit\ (black diamonds), \agcn\ (red circles), and \etcn\ (blue squares).
Upon rescaling temperature (right bars) and pressure (top bars), the quantum Widom line is found to be universal for all compounds. The curvature, as well as the $T=W$ and $U=W$ values match well with theoretical calculations \cite{Vucicevic13}. On the right we illustrate how $U$ and $W$ are determined from the optical spectra and the quantum Widom line from the electrical
resistivity $\rho(T;p)$ measured as a function of temperature (open symbols) and pressure (solid symbols), indicated by arrows in the main graph.}
\end{figure}
Unfortunately, for the paradigmatic quantum spin liquids, like \etcn, $T_{\rm crit}$ is around 20~K, {\it i.e.} below the temperature range accessible for continuously-tunable helium gas pressure cells due to solidification of the pressure medium \cite{Jeftic97}; preventing a continuous pressure sweep at fixed temperatures that could be applied in the case of V$_2$O$_3$ and \etcl. Nevertheless, temperature-dependent experiments at a large number of pressure values or chemical substitutions make it possible to  densely map the region around the phase transition.

\begin{figure}
  \centering
  \includegraphics[width=0.7\columnwidth]{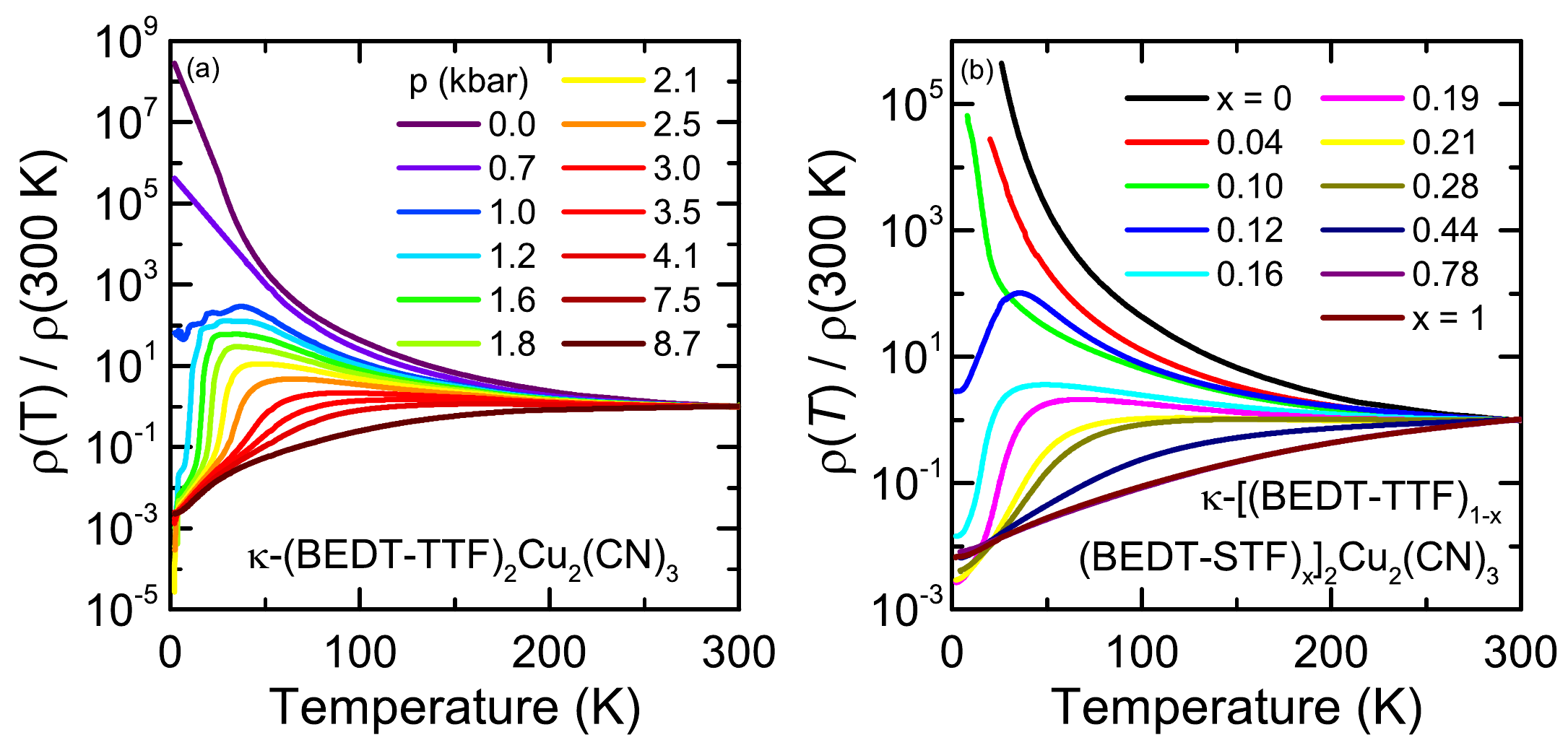}
    \caption{\label{fig:dcPressureAlloy}
(a)~Temperature dependence of the dc resistivity of \etcn\ measured within the highly conducting $bc$-plane when applying different values of hydrostatic pressure as indicated \cite{Lohle18}. (b)~Temperature dependent resistivity of
\stfcn\ where in the inner BEDT-TTF rings sulfur was partially substituted by selenium. The data are normalized to room temperature \cite{Saito20}. }
\end{figure}

Figure~\ref{fig:dcPressureAlloy} demonstrates the evolution of temperature-dependent dc transport of the highly-frustrated Mott insulator \etcn\ as the bandwidth is enlarged in two alternative ways: hydrostatic pressure and chemical substitution. The strongly insulating behavior turns into a metallic, first at low temperatures, eventually in the entire temperature range. The data provide a wealth of information on the insulating as well as on the conducting regime that is discussed in Section \ref{sec:randomnessQSL}. Concerning the insulator-metal transition, however, the temperature sweeps do not really cross
the first-order phase boundary because the critical pressure changes only weakly with temperature \cite{Geiser91,Kurosaki05,Furukawa18}; hence no hysteresis is observed.

The hallmark of a first-order phase transition is the involvement of latent heat, often seen as a hysteric behavior in various physical quantities. Maybe more compelling is the real-space phase coexistence in a sizeable tuning range. In Sec.~\ref{sec:afmMott} we discussed how optical methods allow a mapping of the phase segregation at low temperatures,
but common infrared spectroscopy is limited to a resolution of approximately $10~\mu$m, as illustrated in Figure~\ref{fig:Scan}(a).
Utilizing scanning near-field microscopy in the infrared spectral range \cite{Knoll99,Hillenbrand00,Keilmann04}, spatial inhomogeneities on the nanometer scale can be visualized, as first demonstrated by Basov and collaborators on vanadium oxides \cite{Qazilbash07,Qazilbash09,Qazilbash11,Huffman18,Stinson18}. In Figure~\ref{fig:alphanearfield} the example of \aeti\ is presented where cryogenic near-field nanoscopy was first applied to the charge-order phase transition at $T_{\rm CO}=135$~K \cite{PustogowSciAdv18}.
At present, however, temperatures below 20~K are not routinely accessible for this method, and only first attempts have been made so far to apply it to the problem under discussion here \cite{Iakutkina20}.

Nevertheless, indirect methods have been employed in order to prove that at this first-order phase transition, metallic regions coexist in the insulating matrix. As a matter of fact, the transformation from an insulator to a metal resembles classical
percolation phenomena, where a conduction threshold is hit when the metallic filling fraction rises. The statistical problem of percolation has been subject to numerous theoretical and numerical treatments \cite{Kirkpatrick73,StaufferBook,BollobasBook}; the behavior depends on the dimension and particular type of percolation (bond, site). The electrodynamic properties of mixed media can be well described by effective-medium approximations suggested by Garnett, Bruggeman and others \cite{TorquatoBook,ChoyBook}. A characteristic of the percolation regime is the pronounced divergence of the dielectric constant \cite{Efros76,Bergman78,Hovel10,Hovel11} shortly before the percolation threshold is reached.
In order to look at the coexistence regime of the first-order Mott transition from this perspective,
comprehensive dielectric measurements have been conducted on \etcn\ as a function of temperature and frequency, when the bandwidth is increased via hydrostatic pressure as well as chemical substitution of the organic molecules  \cite{Pustogow19,Rosslhuber19,Saito20} (cf.\ Figure~\ref{fig:dcPressureAlloy}).

\begin{figure}
  \centering
  \includegraphics[width=0.8\columnwidth]{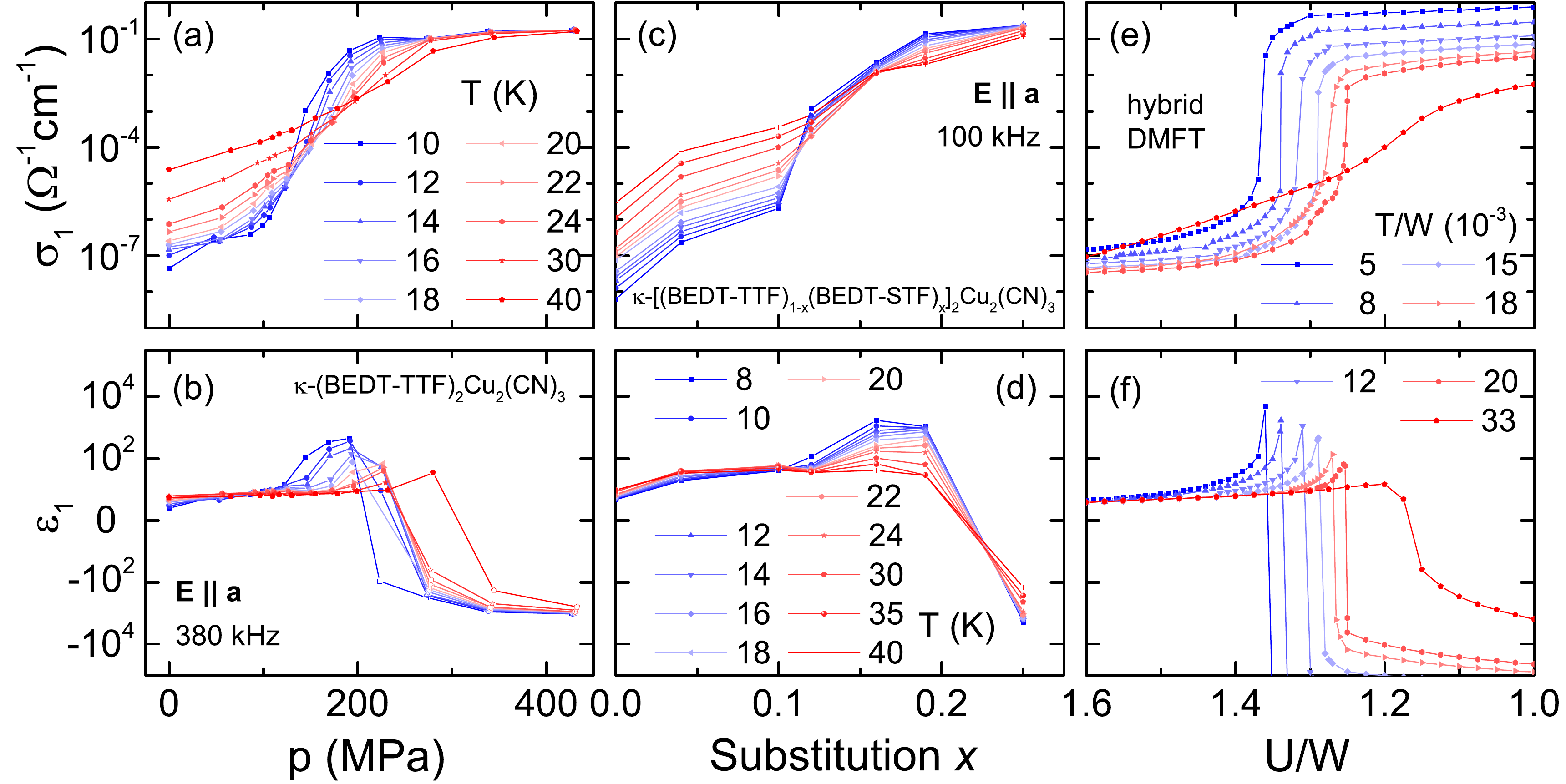}
  \caption{\label{fig:percolation1}
  (a) The Mott insulator-to-metal transition of \etcn\ appears as a rapid increase of $\sigma_1(p)$ that smoothens at higher temperatures; above $T_{\rm crit}$ a gradual crossover remains. (b) $\epsilon_1(p)$ exhibits a sharp peak below $T_{\rm crit}$. The results are taken at $f=380$~kHz and plotted on a logarithmic scale. (c,d) Similar behavior is observed for chemical BEDT-STF substitution. (e,f) Fixed-temperature line cuts of hybrid DMFT simulations (see text) as a function of correlation strength $U/W$ and $T/W$~\cite{Vucicevic13,Pustogow18a} resemble the experimental situation in minute detail, including the shift of the transition with temperature. The lack of saturation of $\sigma_1$ as temperature is lowered, which is seen in DMFT modeling, simply reflects the neglect of elastic (impurity) scattering within the metallic phase (outside the phase coexistence region) (from \cite{Pustogow19}).}
\end{figure}
Figure~\ref{fig:percolation1}(a) demonstrates the temperature evolution of the conductivity step as the pressure increases across the insulator-metal transition. The rapid enhancement of six orders of magnitude within a narrow range of less than 100~MPa at $T=10$~K becomes smoother and more gradual as the temperature rises towards and across $T_{\rm crit}$. A rather similar behavior is seen in Figure~\ref{fig:percolation1}(c) where the substitutional-dependence of $\sigma_1(T,x)$ is plotted for various $T$. Although the density of $p$- and $x$-dependent data points do not allow to extract critical exponents, the overall behavior resembles well the case of the antiferromagnetic Mott insulator \etcl\ discussed in Sec.~\ref{sec:afmMott}, in particular Figure~\ref{fig:CriticalPoint}(a).
In the corresponding lower panels of Figure~\ref{fig:percolation1}, the permittivity $\epsilon_1$ is plotted for fixed temperatures as a function of pressure and BEDT-STF substitution, respectively.
Around the critical pressure $p\approx 150$~MPa and critical concentration of $x\approx 0.15$, $\epsilon_1$ reaches a maximum that shifts to higher $p$- and $x$-values as $T$ increases, corresponding to the positive slope
predicted in Figure~\ref{fig:schematicPhaseDiagrams}(b). Well above the critical temperature $T_{\rm crit}$, the increase in $\epsilon_1$ is minuscule. The experimental observations are well reproduced by DMFT calculations for a single-band Hubbard model [Figure~\ref{fig:percolation1}(e,f)], which are supplemented by an appropriate electrical-network model representing such spatial inhomogeneity utilizing the standard effective-medium approximation \cite{Pustogow19,Rosslhuber19}.

\begin{figure}
  \centering
  \includegraphics[width=0.9\columnwidth]{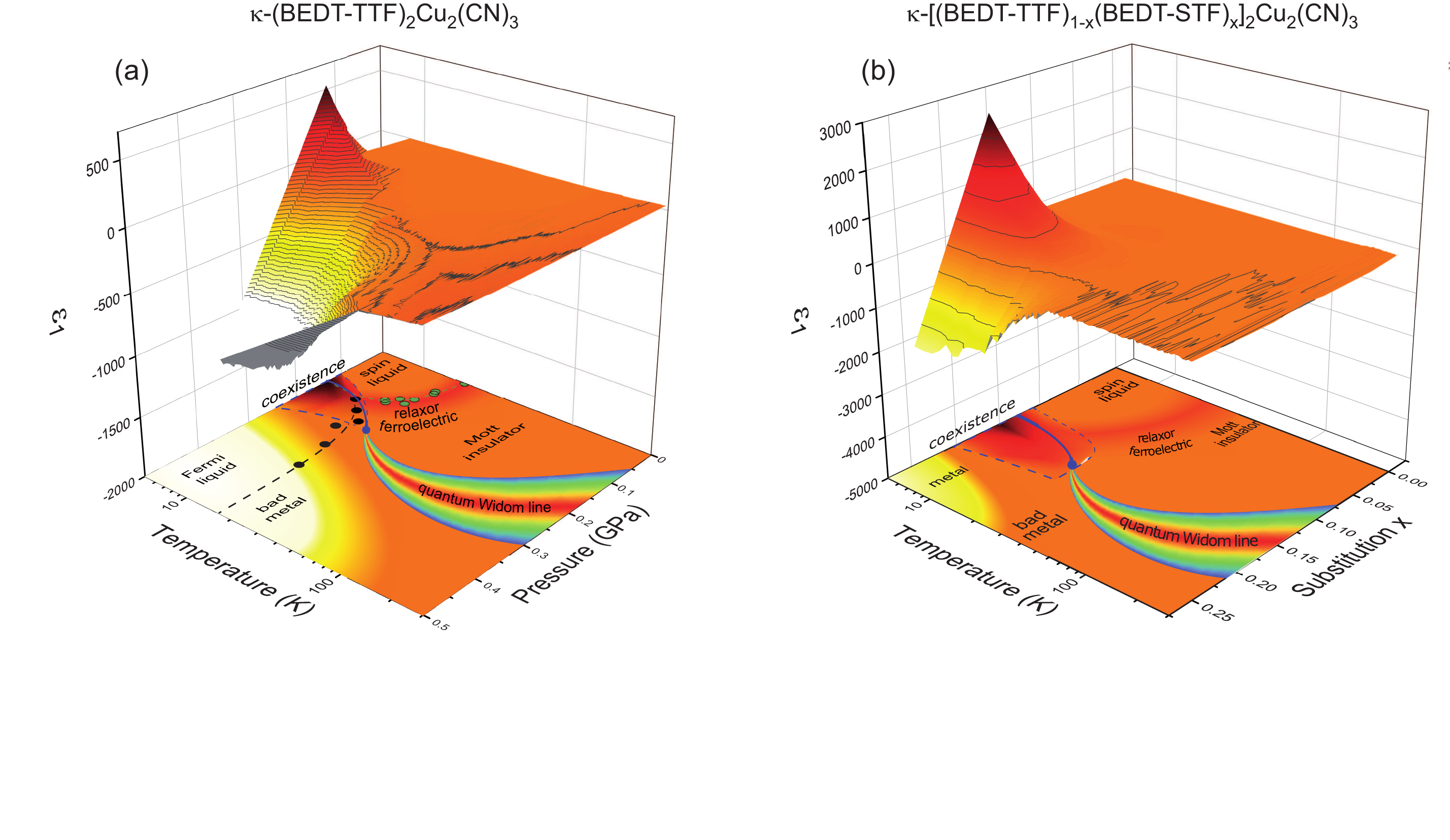}
  \caption{\label{fig:3D_eps1}
(a)~Temperature-pressure contour plot of the dielectric permittivity of \etcn. The permittivity of $\epsilon_{1}(T,p)$ probed at $f=380$~kHz increases up to 600 centered around $p=180$~MPa and below $T=20$~K, close to the first-order Mott transition.
(b)~The temperature-substitution plot of the permittivity for \stfcn\ exhibits its maximum of $\epsilon_{1}\approx 2500$ around $x=0.15$.
This is ascribed to a range of spatially separated metallic and insulating regions. Projected phase diagram also includes $\rho(T,p)$ data:
at $T^{\ast}$, the resistivity deviates from the Fermi-liquid behavior (after \cite{Rosslhuber19,Saito20}). }
\end{figure}
R{\"o}sslhuber {\it et al.} mapped the dielectric catastrophe as a function of $T$ and effective correlations \cite{Rosslhuber19,Saito20,Pustogow19}, displayed in Figure~\ref{fig:3D_eps1}. The two alternative methods of tuning the bandwidth in the highly frustrated spin-liquid compound \etcn\ yield rather similar features as far as the overall behavior is concerned as well as most of the details. This provides compelling evidence that in fact the intrinsic physics is probed consisting of correlation effects and spatial percolation.

Note, the dielectric catastrophe at the Mott transition was first observed and discussed in the context of critically doped semiconductors, such as Si:P \cite{MottBook,Castner75,Rosenbaum83,Hering07,Lohneysen90,Lohneysen00,Lohneysen11}, which is a particular blend of percolation, correlations and disorder.
It was suggested \cite{Terletska11} that the transition from a quantum spin liquid as the fully frustrated Mott insulator to a Fermi-liquid metal is a much cleaner and well defined situation. Although disorder might be an issue for the substitutional series, the agreement between both approaches infer that the overall behavior is not severely affected. It would be of interest to apply this dielectric method to intentionally disordered systems using progressive irradiation.

\subsection{Coherent transport}
\label{sec:coherenttransport}

In a seminal paper, Merino and McKenzie \cite{Merino00} demonstrated that
the transport properties of $\kappa$-BEDT-TTF salts can be described even on a quantitative level
within dynamical mean-field theory (DMFT) \cite{Rozenberg94,Rozenberg95,Limelette03a,Merino08}.
At low temperatures, \etbr, \etscn\ and  similar organic conductors constitute prime examples of Fermi-liquids:
The resistance exhibits a quadratic temperature dependence
\begin{equation}
\rho(T) = \rho_0 + A T^2 \quad ,
\label{eq:FL1}
\end{equation}
obeying  the Kadowaki-Woods rule \cite{Jacko09}
\begin{equation}
A \propto \gamma_e^2 \quad ,
\label{eq:KadowakiWoods}
\end{equation}
with $\gamma_e$ the Sommerfeld coefficient, characterizing the electronic contribution to the specific heat, {\it i.e.} the electronic density of states at the Fermi level $E_F$.
This coherent state extends up to the Fermi-liquid temperature $T_{\rm FL}$,
where the parabolic increase in $\rho(T)$ is lost.
Metallic transport, defined by ${\rm d}\rho(T)/{\rm d}T > 0$, prevails up to
the so-called Brinkman-Rice temperature $T_{\rm BR}$,  well above the Ioffe-Regel-Mott limit where metallic transport is supposed to break down \cite{Pustogow19}, see Figure~\ref{fig:schematicPhaseDiagrams}(b).
Above this maximum in $\rho(T)$ around 80 to 100~K, the systems exhibit some incoherent behavior resulting in a semiconducting
temperature dependence.
The smooth crossover from coherent Fermi-liquid to more incoherent excitations leads to a non-monotonic $T$-dependence not only in the electrical resistance, but also in thermopower and Hall coefficient.
While the overall behavior is quite generic, details are sample dependent
and due to intrinsic remnant disorder caused by an incomplete ordering of the ethylene-end-groups of BEDT-TTF molecule at about 75\,K ~\cite{Pinteric02,Strack05}; cf.\ Figure~\ref{fig:BEDT-TTF} in Chapter~\ref{sec:structure}.

Among the molecular quantum materials, the $\kappa$-phase BEDT-TTF family serves as the primary testfield for the correlation dependence of their electronic properties. The starting point is either the antiferromagnetic Mott insulator \etcl\ or the highly-frustrated sister compound \etcn. The systems can be tuned into the metallic or even superconducting Fermi-liquid state by increasing the bandwidth via external pressure or chemical modification, as depicted in Figure~\ref{fig:schematicPhaseDiagrams}(b) and (c).

Limelette {\it et al.} investigated \etcl\ under pressure $p>30$~MPa in order to explore the emerging Fermi-liquid regime \cite{Limelette03a}: they could characterize the dc resistivity
according to Eq.~(\ref{eq:FL1}) up to $T_{\rm FL}$, which is of the order of 35~K at 50~MPa.
The prefactor $A(p)$ strongly increases as the pressure is reduced and diverges
with a critical pressure of approximately 20~MPa (see\ Figure~\ref{fig:PhaseDiagram_k-Cl}).

\begin{figure}
\centering
\includegraphics[width=0.9\columnwidth]{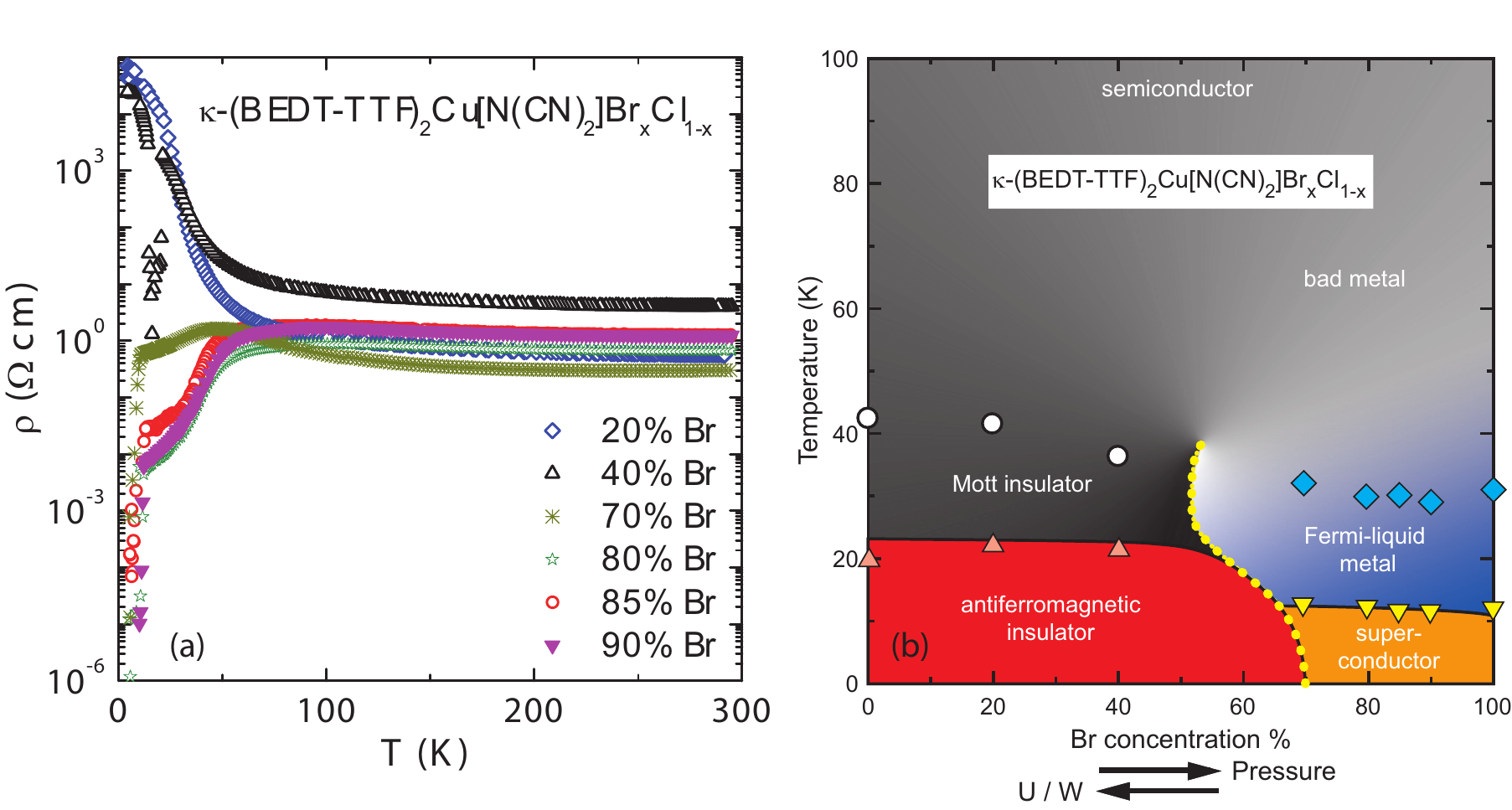}
\caption{
(a) Temperature dependence of the in-plane dc resistivity $\rho(T)$ of
$\kappa$-(BEDT-TTF)$_2$\-Cu[N(CN)$_2$]Br$_{x}$Cl$_{1-x}$ with
different Br substitution  as indicated. For $x=40\%$
metallic fluctuations indicate the closeness to the metallic phase.
\label{fig:kappa-BrCl_dc}
(b)~Schematic phase diagram of
$\kappa$-(BEDT-TTF)$_2X$. The on-site Coulomb repulsion with
respect to the bandwidth $U/W$ can be tuned either by external
pressure or modifying the anions $X$. The bandwidth controlled
phase transition between the insulator and the Fermi
liquid/superconductor can be explored by gradually replacing Cl by
Br in $\kappa$-(BEDT-TTF)$_2$Cu[N(CN)$_2$]Br$_{x}$Cl$_{1-x}$. The first-order insulator-to-metal transition line is sketched after the generic diagram for $\kappa$-(BEDT-TTF)$_2X$ \cite{Ito96,Kanoda97a,Limelette03a}. The
data points are obtained from transport measurements (a) and magnetic susceptibility (from \cite{Yasin11}).
\label{fig:kappa_phasediagram}
}
\end{figure}
Replacing Cl by Br in \etbrcl\ provides an alternative route to shift the compound across the metal-insulator transition.
Despite the larger ions, the orbital overlap increases
and the effective Coulomb repulsion $U/W$ decreases, as depicted in the phase diagram of Figure~\ref{fig:kappa_phasediagram}(b).
Comprehensive optical, transport and magnetic investigations on \etbrcl\ yield the charge and spin dynamics as the metallic state evolves, when moving across the phase boundary or reducing the temperature \cite{Yasin11,Faltermeier07,Merino08,Dumm09,Dressel09}. As shown in Figure~\ref{fig:kappa-BrCl_dc}(a),
the pristine and weakly substituted compounds behave strongly isulating; but for $x=0.4$ metallic fluctuations become obvious at low temperatures. Above $x=0.7$, metallic and superconducting properties are present. It is important to note, however, that in these cases the coherent charge carrier response develops only below approximately $50$~K, in accord with theory \cite{Merino00}. In Figure~\ref{fig:kappa-scattering}(d)-(f), $\rho(T)$ is presented on a quadratic temperature scale in order to better visualize the Fermi liquid behavior. The $[\rho(T)-\rho_0] \propto T^2$ dependence holds up to the Fermi-liquid temperature $T_{\rm FL} \approx 30$~K basically independent on $x$. Pressure-dependent studies \cite{Limelette03a} shown in Figure~\ref{fig:PhaseDiagram_k-Cl}(a) reveal a somewhat stronger increase with $p$.

\subsubsection{Fermi-liquid behavior}
\label{sec:FermiLiquid}
The obtained prefactor $A(x)$ in Eq.~(\ref{eq:FL1}) increases, when going from the pristine Br compound to $x=0.7$.
Within Landau's Fermi-liquid theory, the factor $A$ characterizes the strength of the electron-electron
interaction and is related to the effective carrier mass $m^*$ or the effective Fermi temperature $T_F^*$,
like
\begin{equation}
A\propto (m^*)^2 \propto (T_F^*)^{-2} \quad ,
\label{eq:KadowakiWoods2}
\end{equation}
in accord with the Kadowaki-Woods
relation (\ref{eq:KadowakiWoods}), that is well established for heavy
fermions or transition-metal compounds, and was also seen in  organics \cite{Kadowaki86,Ito96,Miyake89,Maeno97,Jacko09}.
The product $A\times (T_{\rm FL})^2$ was suggested to remain constant
when going through the phase diagram either by chemical or physical pressure \cite{Limelette03a,Yasin11}; more detailed studies are required to confirm this point.
We should note that results on $\kappa$-(BEDT-TTF)$_2$Cu[N(CN)$_2$]Br have been reported, which deviate from this behavior \cite{Strack05}, likely caused by a remnant disorder \cite{Pinteric02}.

In addition to the temperature dependence, optical spectroscopy provides the
possibility to extract the frequency dependence of the scattering rate and of the effective mass, when performing an extended Drude analysis of the conductivity spectra \cite{DresselGruner02,BasovRMP}. Eventually we can consider a $\omega$-$T$ scaling of the scattering and interaction processes with respect to these two energies.
For \etbrcl\  single crystals with $x = 0.9$ and 0.85 and 0.73 one obtains a gradual increase in the effective mass  from 2  to $m^*/m_b\approx 6$ \cite{Dumm09,Dressel11}.
This agrees with the general prediction by Brinkman-Rice theory \cite{Imada98} and DMFT
calculations \cite{Georges96} that the electronic correlations become stronger on approaching the Mott transition
from the metallic side.
More recent calculations by resonating-valence-bond theory of the Hub\-bard-Heisenberg model also predict a gradual increase in $m^*$ for values of effective repulsion $U/W$ not too close to the
first-order Mott transition and a strong increase very close to
the transition \cite{Powell05}.
Kagawa  {\it et al.} \cite{Kagawa09} noted that $m^*$ is well defined only in the Fermi-liquid regime at low temperatures; in the `bad metallic' regime of the phase diagram [Figure~\ref{fig:kappa_phasediagram}(b)], however,
$m^*$ is probably not an adequate physical parameter to feature the Mott criticality at the critical end point.

\begin{figure}
\centering
\includegraphics[width=0.6\columnwidth]{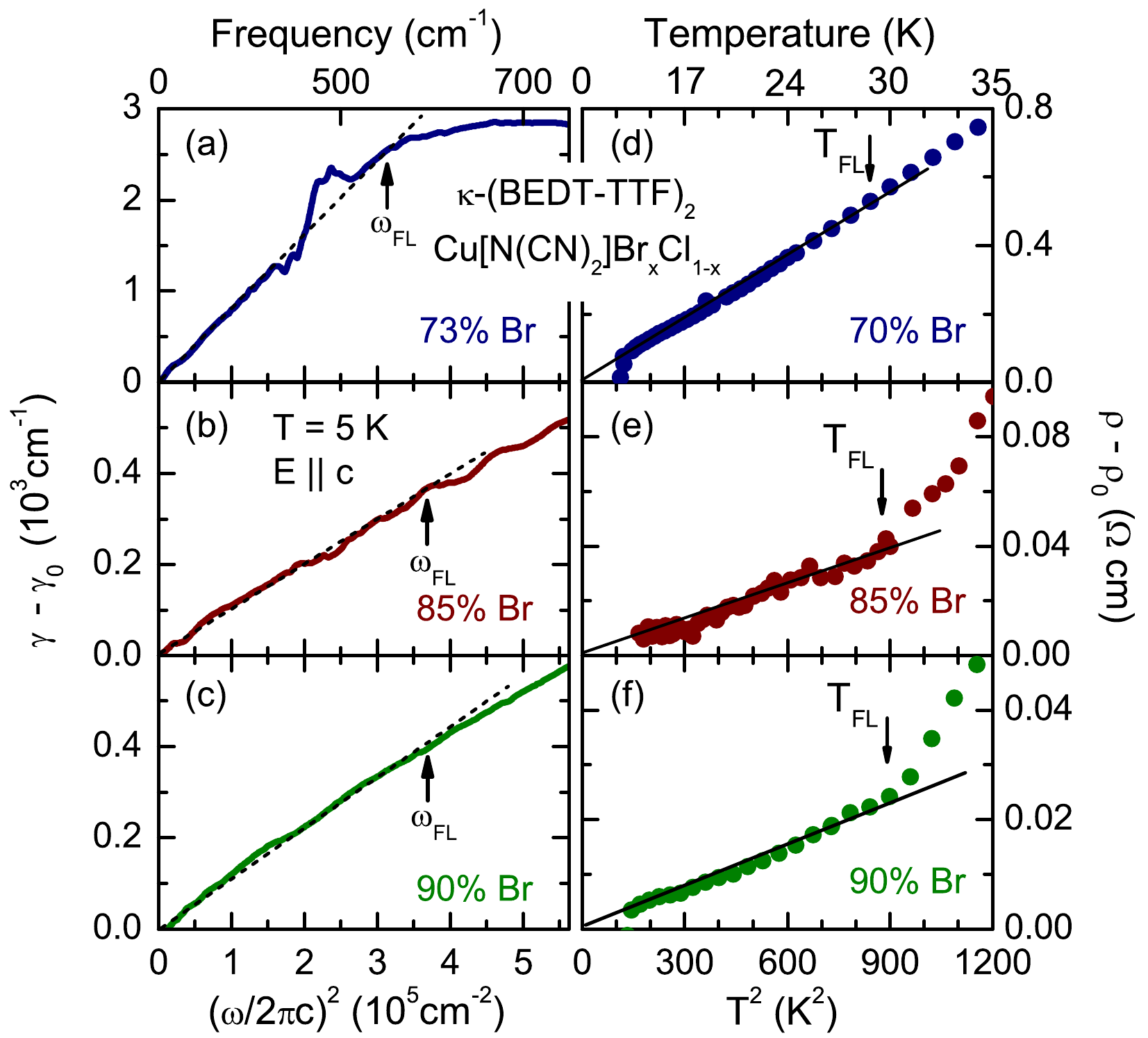}
\caption{\label{fig:kappa-scattering} Temperature and frequency dependence of the scattering rate for \etbrcl\ with different substitution values $x$.
(a) - (c) The frequency-dependent part of
the scattering rate $\gamma(\omega) - \gamma_0$ at $T=5$~K of
$\kappa$-(BEDT-TTF)$_2$Cu[N(CN)$_2$]Br$_{x}$Cl$_{1-x}$ as function
of the squared frequency. We determined for the
frequency-independent dc limit of the scattering rate
$\gamma_0$ the following values: 315~cm$^{-1}$ ($x=0.73$),
48~cm$^{-1}$ ($x=0.85$), and 280~cm$^{-1}$ ($x=0.9$). Note the
different vertical scales for the frames.
(d) - (f) The temperature-dependent scattering rate $\tau^{-1}(T)=\gamma(T)$ is obtained from the in-plane dc resistivity ($\rho(T)-\rho_{0})\propto 1/\tau(T)$. The plots as a function of $T^2$ in the low-temperature region yield a quadratic behavior up to $T_{0}$. Below $T_c\approx 12$~K the systems
become superconducting (data from \cite{Dumm09,Yasin11}).
}
\end{figure}
Fermi-liquid theory makes predictions about
the temperature dependence of the scattering rate as well as about its frequency dependence
\cite{AshcroftBook,AbrikosovBook,PinesBook,Maslov17}:
\begin{equation}
\gamma(T,\omega) = \tau^{-1} = \gamma_0 + B\left[(p \pi k_B T/\hbar)^2 +
\omega^2\right] \quad , \label{eq:FL2}
\end{equation}
where the constant $\gamma_0$ describes the residual scattering processes at zero energy due to impurities, surface etc. corresponding to the residual resistivity $\rho_0$ in Eq.~(\ref{eq:FL1}).
The parameter $B$ is related to the density of electronic states at the Fermi level $E_F$.
For purely inelastic scattering, $p=2$ is predicted \cite{Landau57,Gurzhi59,Rosch06,Berthod13} but experimental verification is scarce \cite{Yasin11,Nagel12,Mirzaei13,Stricker14,Tytarenko15}.
As seen from Figure~\ref{fig:kappa-scattering}(a)-(c), a quadratic frequency dependence of the low-temperature scattering rate was obtained in the case of \etbrcl\ up to approximately 600~cm$^{-1}$ where unconventional non-Fermi-liquid behavior
prevails even at $T=5$~K \cite{Dumm09,Dressel11}.
The corresponding temperature-dependent dc resistivity is plotted in panels (d) to (f).
Consistently, the slope rises as the Mott transition is approached: the effective correlations $U/W$
increase due to the reduction of bandwidth.
From the optical data taken at different temperatures, $p=2.3$ is extracted close to the expectation.
As seen in Figure~\ref{fig:kappa-scattering}, the upper limit of the Fermi-liquid regime is determined by deviations from the $T^2$ and $\omega^2$ line. It is interesting to note
that in all cases the rise in temperature has a much stronger effect than the increase in frequency.

An indirect method of probing the metallic phase boundary was suggested by Sasaki {\it et al.}  \cite{Sasaki04b} similar to the spatially resolved study presented in Figure~\ref{fig:Scan}:
by carefully following the $\nu_3(a_g)$ mode of several $\kappa$-(BEDT-TTF)$_2X$ salts
when the temperature is reduced, they can identify the border $T^*$ between the bad metallic behavior at elevated temperature and the Fermi-liquid behavior of a correlated good metal at low temperatures. Variations of the electronic properties affect the emv coupling of the molecular vibrations. They also found a shift in the $\nu_3(a_g)$ mode when the boundary to the Mott insulator is crossed by cooling \etcl.

\subsubsection{$\omega$-$T$ scaling}
\label{sec:wTscaling}
\begin{figure}[h]
\centering
\includegraphics[width=0.8\columnwidth]{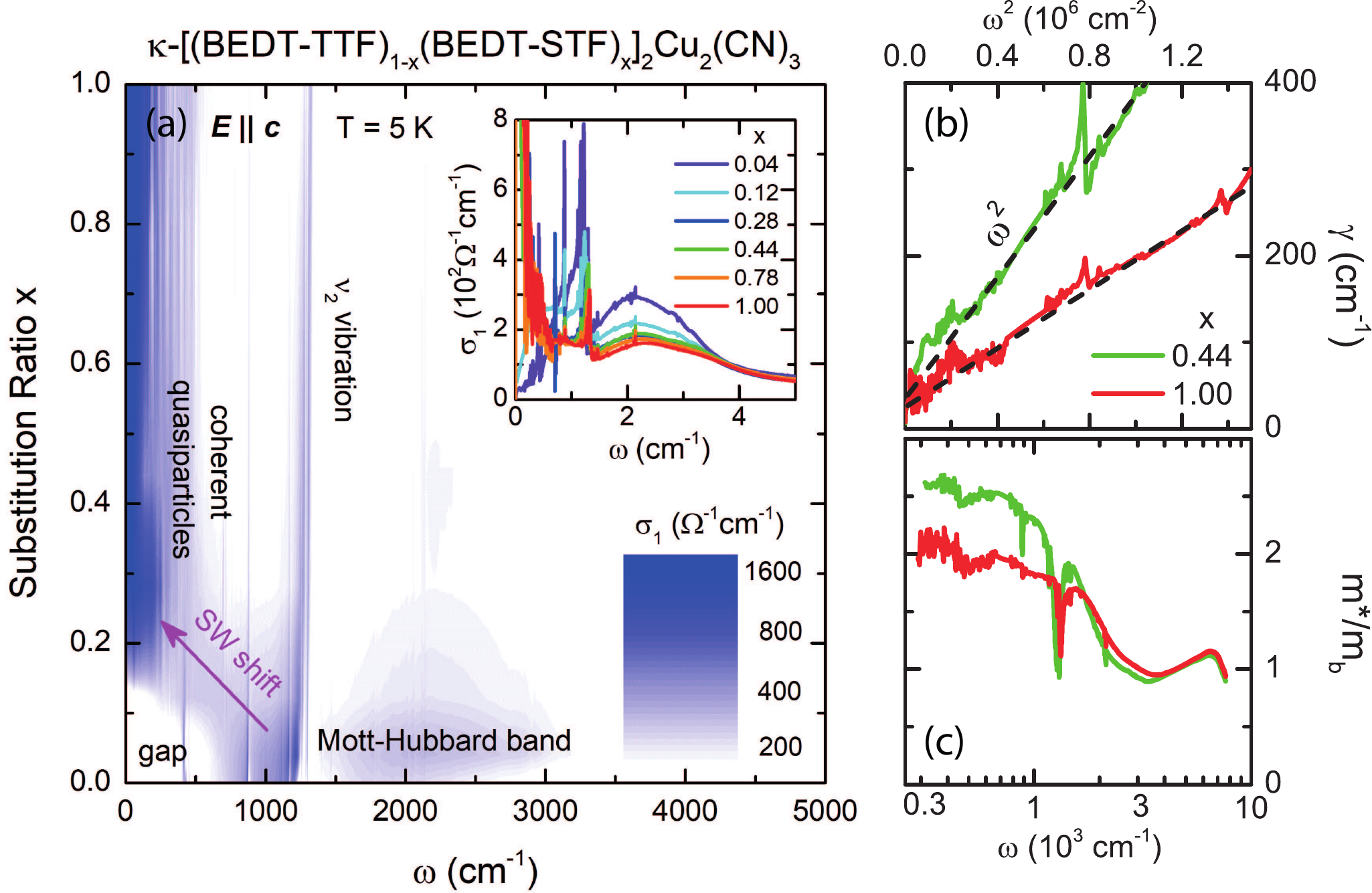}
\caption{
Optical properties of the series \stfcn. The inset in panel (a) displays the optical conductivity
for several crystals with different substition $x$ as indicated. The data are recorded at $T=5$~K
for the polarization $E\parallel c$. The most prominent feature is the Mott-Hubbard band around 2000~\cm.
Upon increasing $x$, this peak shifts towards higher energies.
At the same time, a strong Drude-type zero-frequency conductivity emerges
indicating the transition from a gapped Mott insulator to a strongly correlated
metallic state with renormalized quasiparticles. The main frame contains the contour plot of the conductivity spectra
at different substitutional values $x$ ($\sigma_1$ is plotted on a logarithmic blue-white scale)
illustrating the spectral-weight shift from high to low energies as the BEDT-STF content increases. As correlations become less pronounced, the Mott gap closes and
coherent quasiparticles are stabilized at low frequencies.
The strong but substitution-independent vibrational feature at 1250~\cm\ corresponds to the emv coupled $\nu_2(a_g)$ mode.
(b) From the extended Drude analysis, the energy-dependent scattering rate $\gamma$ and effective mass $m^*$ is obtained. For the metallic compounds $x\geq 0.28$, $\gamma(\omega)$ increases quadratically with frequency, providing strong evidence for a Fermi liquid
response. With increasing $x$, the slope $B$ is reduced and the
$\gamma(\omega) \propto \omega^2$ range expands to higher frequency, shown for the two examples $x=0.44$ and 1.00.
(c) Effective mass as a function of frequency for different BEDT-STF substitutions. The mass enhancement becomes more pronounced as the Mott transition is approached; for $x=0.44$ one finds $m^*/m_b=2.6$ (taken from \cite{Pustogow20,Pustogow21}).
\label{fig:STF-optics}}
\end{figure}

A comprehensive investigation of the Fermi-liquid regime in the two-dimensional organic compounds
was conducted by Pustogow {\it et al.} \cite{Pustogow20}, who compared the
transport and optical properties of the substitutional series \stfcn\ with $0 \leq x \leq 1$. Due to the extended selenium orbitals, the bandwidth $W$ is enlarged as the  BEDT-STF content increases; the Mott transition takes place around $x = 0.12$ as shown in Figure~\ref{fig:dcPressureAlloy}(a).
The optical conductivity displayed in the inset of Figure~\ref{fig:STF-optics}(b) resembles an
insulator [$\sigma_1(\omega\rightarrow 0) = 0$] for low $x$ with a pronounced mid-infrared absorption band ($U \approx 2000$~\cm) that corresponds to excitations across the Mott-Hubbard gap. With increasing $x$, spectral weight shifts to lower energies and eventually a Drude-like term appears as the metallic phase is entered \cite{Saito20}. The contour plot illustrates the evolution of the conductivity as $x$ rises. Fueled by the vanishing Mott-Hubbard excitations, the coherent quasiparticle response appears above $x \approx 0.2$. In that range,
an extended Drude analysis of the optical conductivity \cite{DresselGruner02} yields the energy dependence of the scattering rate and effective mass, presented in Figure~\ref{fig:STF-optics}(b) and (c). When plotting $\gamma(\omega)$ as a function of $\omega^2$, a straight-line behavior can be observed up to $\omega_{\rm FL}$, which corresponds to the upper limit of the Fermi liquid regime. With increasing 
$x$ the boundary considerably shifts to higher energies, in accord with the observations made on \etbrcl\ (Figure~\ref{fig:kappa-scattering}).
It is worth to note that the Fermi-liquid behavior is observed over an extremely large temperature and frequency range because of the narrow bands compared to transition metal oxides, for instance.
The slope determined by the pre\-fac\-tor $B$ decreases with rising $x$, {\it i.e.} as we move away from the insulator-metal transition.
Correspondingly, from the frequency dependence of the renormalized mass, a strong increase of $m^*$ is observed as $x$ is reduced towards the Mott transition at 0.12.
The low-frequency limit is taken as a measure of the correlation strength
that is related as $B \propto (m^*/m_{\rm b})^2$ in analogy to Eq.~(\ref{eq:KadowakiWoods}).

The crucial point is the $\omega$-$T$ scaling expressed in Eq.~(\ref{eq:FL2}): The quadratic dependence of the scattering rate on frequency, plotted in Figure~\ref{fig:STF-optics}(b),
is observed at different temperatures; they can be unified by scaling the curves with a temperature independent factor $p$. This is taken as compelling evidence for the universality of Landau's Fermi-liquid concept upon varying the correlation strength and do not leave much space for theories of a quasi-continuous Mott transition that involve a divergent Kadowaki-Woods ratio \cite{Senthil08}.
Interestingly, the proportionality between the $T^2$  and $\omega^2$ dependence of $\gamma(T,\omega)$ exceeds the inelastic limit $p = 2$ and exhibits a pronounced enhancement towards the Mott
metal-insulator transition \cite{Pustogow20,Pustogow21}.
This behavior is in contrast to the observations in various strongly correlated electron systems \cite{Maslov17}, and can be explained by the fact that in \stfcn\ the narrow quasiparticle peak centred at $\omega = 0$ coexists and considerably overlaps with the broad, non-metallic Hubbard bands at $\pm U/2$ (with a width $W$) as we are close to the Mott transition.  In other words, the strongly correlated charge carriers are not as freely moving as expected for a good metal; in part they are localized due strong Coulomb repulsion. Further theoretical studies are required in order to elucidate these issues in more depth.

\subsubsection{Bad-metal regime and Ioffe-Regel-Mott limit}

The Fermi-liquid transport at low temperatures ($T<50$~K or so) is followed by a vaguely-defined `bad metal' regime  \cite{Emery95}. Although quasiparticles are still present, their lifetime is significantly limited \cite{Deng13}. This regime --~often identified with a linear-in-$T$ resistivity~-- is not well understood, although continuously subject to theoretical treatises \cite{Hartnoll15}.
At even high temperatures, when transport becomes incoherent and dominated by the large scattering rate,
metallic conduction breaks down. The so-called Ioffe-Regel-Mott limit is the maximal resistivity that can be reached in a metal according to the Boltzmann semiclassical theory \cite{Gunnarsson03,Hussey04}.
As first realized by Ioffe and Regel \cite{Ioffe60}, the metallic conductivity is restricted as the mean free path $\ell$ cannot get smaller than the lattice spacing $d$ \cite{Mott72}; in other words, the linear-in-temperature increase of the resistivity should saturate at one point. There are numerous examples of strongly correlated systems, however, for which $\rho(T)$ does not show a sign of saturation up to fairly elevated temperatures.

For that reason, it is rather illuminative to look at the optical response, where the low-frequency spectral weight becomes reduced and the simple Drude-like behavior modified well before $\rho(T)$ saturates. We can define `bad metals' by the loss of coherence.
Dynamical mean field calculations \cite{Rozenberg95,Merino00,Limelette03a,Deng13} predict the modification of the Drude peak that extends well above the Ioffe-Regel-Mott limit, in full agreement with experiments on several $\kappa$-(BEDT-TTF)$_2X$ salts \cite{Merino08,Dumm09,Li19,Saito20,Pustogow20}, as shown in Figure~\ref{fig:STF-optics}(a), but also $\theta$-(BEDT-TTF)$_2$I$_3$.

\begin{figure}
 \centering
\includegraphics[width=1\columnwidth]{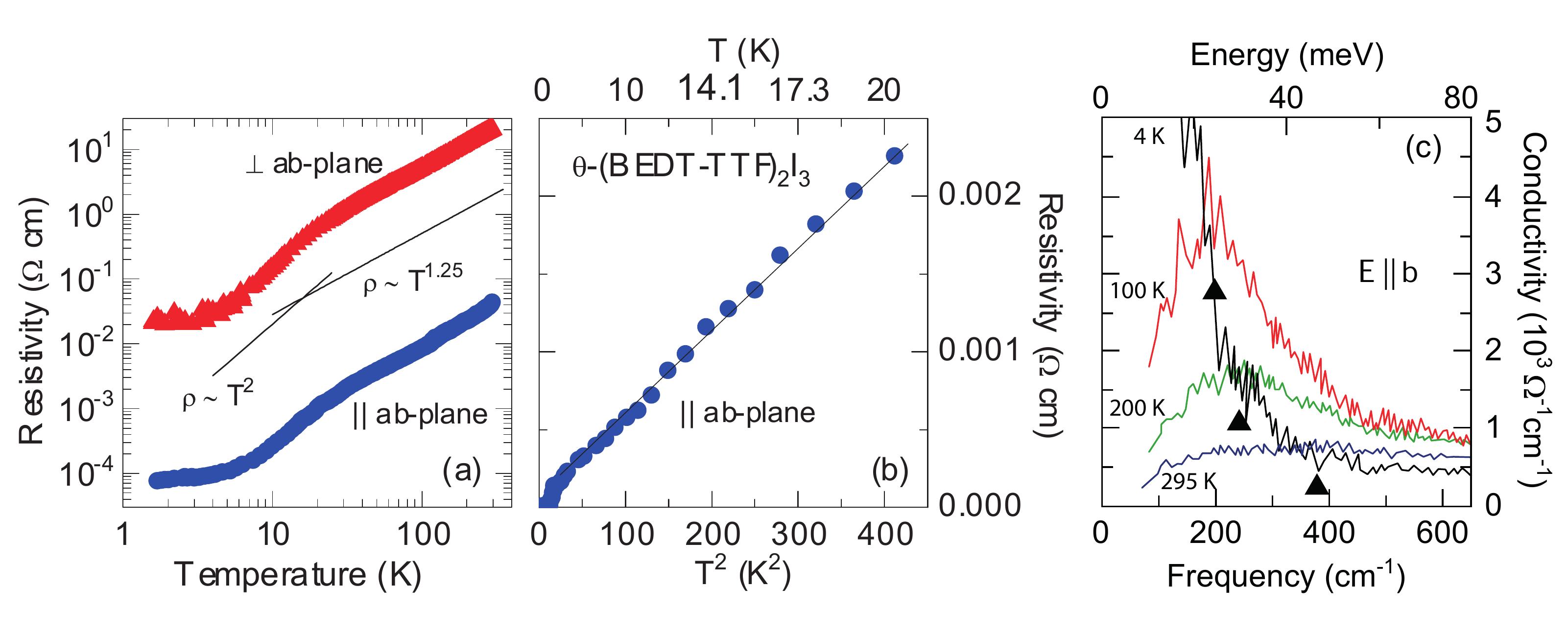}
 \caption{\label{fig:theta-I3_dc} Temperature-dependent electronic properties  of  $\theta$-(BEDT-TTF)$_2$I$_3$.
 (a) Parallel and perpendicular to the $ab$-planes, the resistivity exhibits a $\rho(T)\propto T^{1.25}$ power-law behavior at elevated temperatures. (b) Below $T_0\approx 20$~K a quadratic dependence reveals a Fermi-liquid   (after \cite{Dressel11}).
(c) In-plane conductivity spectra of $\theta$-(BEDT-TTF)$_2$I$_3$ at different temperatures. The Drude component at low temperatures diminishes and shifts to a finite energy peak (indicated by black triangles) with increasing $T$  (after \cite{Takenaka05}).
  \label{fig:theta-I3_optics} }
\end{figure}
The two-dimensional organic conductor $\theta$-(BEDT-TTF)$_2$I$_3$
draws attention for numerous reasons. In contrast to other $\theta$-phase salts
that undergo
charge-ordering transitions into an insulating ground state \cite{Mori99b,Mori06,Alemany15},
the dc resistivity of $\theta$-(BEDT-TTF)$_2$I$_3$ remains metallic down to $T_c=3.6$~K.
The transition temperature can be raised above 5~K, when the specimens are tempered at 70\,$^{\circ}$C
\cite{Salameh07}. Quantum oscillations
prove the presence of a Fermi surface with two-dimensional orbits and one-dimensional trajectories
\cite{Salameh07,Terashima94,Nothardt04,Nothardt06}.
The presence of a Fermi-liquid state is also seen from a quadratic temperature increase in the resistivity
observed at low temperatures. As  displayed in Figure~\ref{fig:theta-I3_dc}(a,b),
it turns into a $T^{1.25}$ dependence above $T_{\rm FL}\approx 20$~K or so. A closer inspection reveals a slight kink around 120~K where also the anisotropy $\rho_{\parallel}/\rho_{\perp}$ changes \cite{Salameh07}.

As pointed out by Takenaka {\it et al.} \cite{Takenaka05}, above $T\approx 50$~K the in-plane
resistivity exceeds the Ioffe-Regel-Mott limit $\rho_{\rm IRM}=(h/e^2)d=4.4\times 10^3~\Omega$\,cm for an interlayer
distance of $d=17$~\AA\ \cite{Tamura88}, implying incoherent transport above.
The transition from a coherent quasiparticle state to an incoherent state was addressed by temperature-dependent optical spectroscopy on $\theta$-(BEDT-TTF)$_2$I$_3$. Despite the metallic behavior of $\rho(T)$, a well defined Drude peak is present only at low-temperatures. Upon warming to $T=100$~K, the peak moves to finite frequencies in the far infrared spectral range. The origin of the finite energy peak seen in Figure~\ref{fig:theta-I3_optics}(c) is not clear.
For both directions ($E\parallel a$ and $b$) the maximum moves to approximately $250 - 400$~\cm\ when going up to room
temperature. It is discussed whether this spectral behavior represents a `dynamical localization' theoretically explored for other organic semiconductors in more detail \cite{Fratini14},
or can be treated like fluctuating density waves \cite{Delacretaz17}.

The disappearance of the Drude peak is accompanied by a shift of spectral weight to energies above 1~eV.
The loss of coherence and the redistribution of a substantial amount of spectral weight is explained by strong electronic correlations, as has been observed in a large variety of correlated bad metals \cite{Basov11}.
The disappearance of the Drude peak indicates a strong suppression of kinetic energy with increasing $T$: the charge carriers transport becomes incoherent. The spectral weight shifts to energies higher than the bandwidth.
Gunnarsson and collaborators pointed out \cite{Gunnarsson07} that the reason for this violation
of the Ioffe-Regel-Mott limit
is not simply due to interactions, like in some transition metal oxides, or due to the small bandwidth compared to the temperatures.
\begin{equation}
\sigma_1(0) = \frac{\varphi}{W} \int_0^{\infty}\sigma_1(\omega){\rm d}\omega \approx \frac{\left| E_k \right| }{W} \quad ,
\end{equation}
where $\varphi = 1$-2 and $W$ is the bandwidth of the order of 500~meV. In the case of $\kappa$-(BEDT-TTF)$_2X$ salts
\cite{Faltermeier07,Dumm09}, $\sigma_1(\omega)/\rho_{\rm IRM}$ is suppressed due to correlations reducing $|E_k|$ and expanding the energy scale \cite{Gunnarsson07}.

It is interesting to note that the behavior of the optical conductivity $\sigma_1(\omega)$ in the bad metallic regime at high temperatures is rather similar to that seen in disordered materials at low $T$ \cite{Hussey04}.
Radonji{\'c} {\it et al.} \cite{Radonjic10} showed that with increasing disorder the bandwidth $W$ increases, but disorder does not lead to qualitative  differences. The large bandwidth makes the compounds more metallic, in accord with experiments \cite{Sasaki12c} as discussed in Sec.~\ref{sec:randomness}. With proper scaling, the optical conductivity remains even quantitatively very similar. No question,
the `bad metallic' regime
remains a desideratum for future research.

Using numerical methods Wessel and collaborators considered the Mott metal-insulator transition while varying the degree of frustration \cite{Dang15}. Most interesting, the slope of the phase boundary changes sign, from positive slope of ${\rm d}T/{\rm d}p$ at $t^{\prime} = t$ to negative slope for the unfrustrated case $t^{\prime}=0$ This basically reflects the difference in entropy of the insulating ground state. Similar results have been received earlier by Tremblay and collaborators. They also provided a detailed thermodynamic description of the correlation-driven Mott transition \cite{Walsh19a,Walsh19b,Sordi19}, where they can trace the quantum Widom line above $T_{\rm crit}$ and the spinodal lines $U_{c1}$ and $U_{c2}$ below. Particular attention was paid to the local entropy.

\subsection{Effects of disorder on the Mott transition}
\label{sec:randomness}
The electronic properties of real materials are strongly affected by electronic interaction and randomness \cite{Lee85,Belitz94,DobrosavljevicBook}.
Localization of particles can be caused by Coulomb repulsion and
disorder. While electronic correlations are the driving force behind the Mott transition, the Anderson transition is due to coherent backscattering of non-interacting particles from randomly distributed \begin{figure}[h]
  \centering
  \includegraphics[width=0.4\columnwidth]{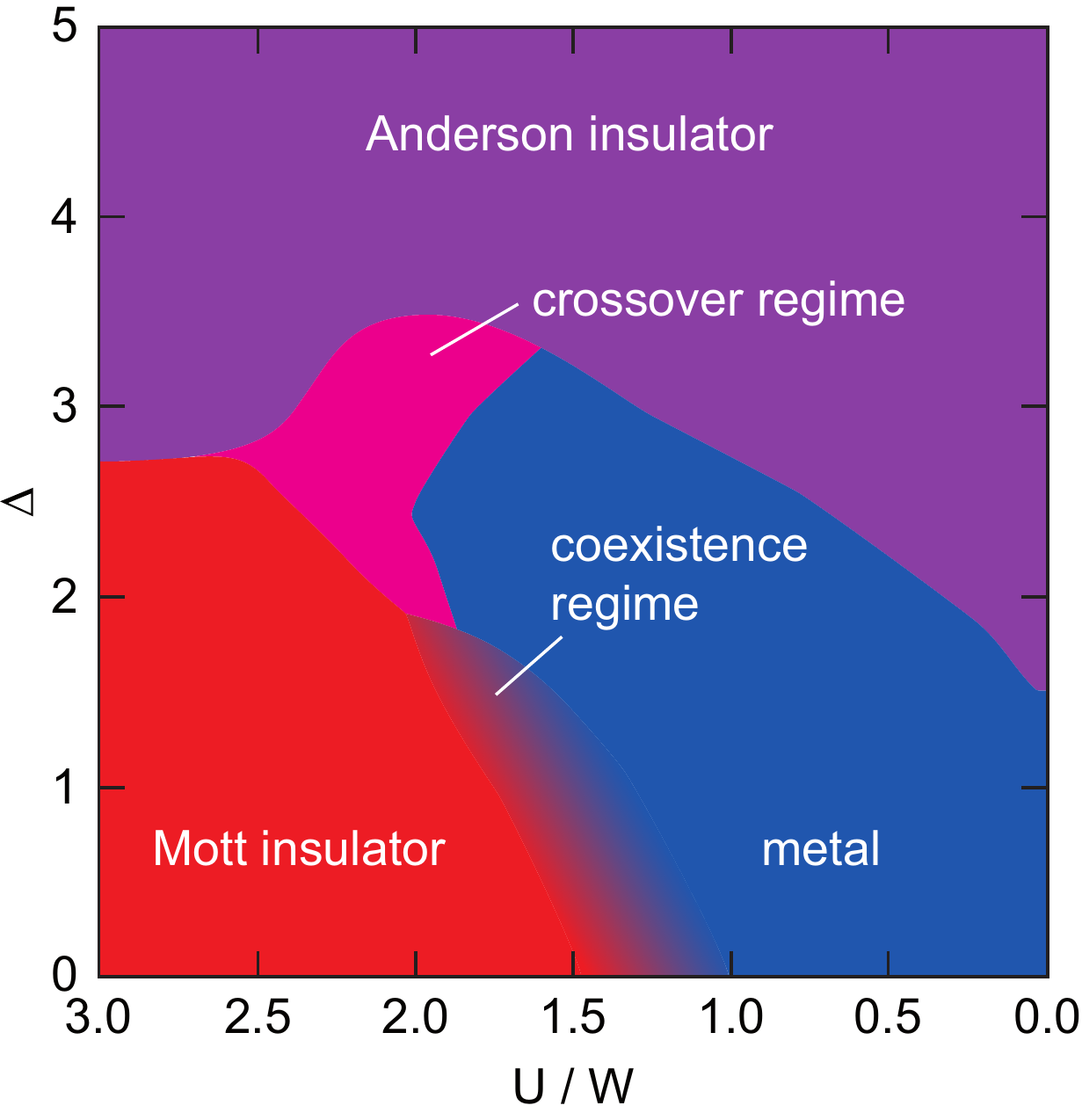}
    \caption{\label{fig:Anderson1}
    Phase diagram of the non-magnetic Anderson-Hubbard model as calculated by dynamical mean field theory with the typical local density of states
(based on \cite{Byczuk05}). Note the reversed scale for $U/W$.}
\end{figure}
impurities. The challenge for a theoretical understanding is that disorder and interaction effects are known to compete in subtle ways \cite{Finkelshtein83, Castellani84, Tusch93, Lohneysen00,Kravchenko04, Byczuk05,AbrahamsBook}.

In Figure~\ref{fig:Anderson1} the ground state phase diagram of the Anderson-Hubbard model is displayed, based on dynamical mean field theory (DMFT) calculations of the disordered Hubbard model at half filling with the typical local density of states \cite{Byczuk05}. The interaction strength is given by $U/W$, the disorder strength by $\Delta$.
Two different phase transitions are found: a Mott insulator-metal transition for weak disorder $\Delta$ and
an Anderson transition for weak interaction $U$. Two insulating phases surround the correlated, disordered metallic phase.

\begin{figure}[h]
  \centering
  \includegraphics[width=0.8\columnwidth]{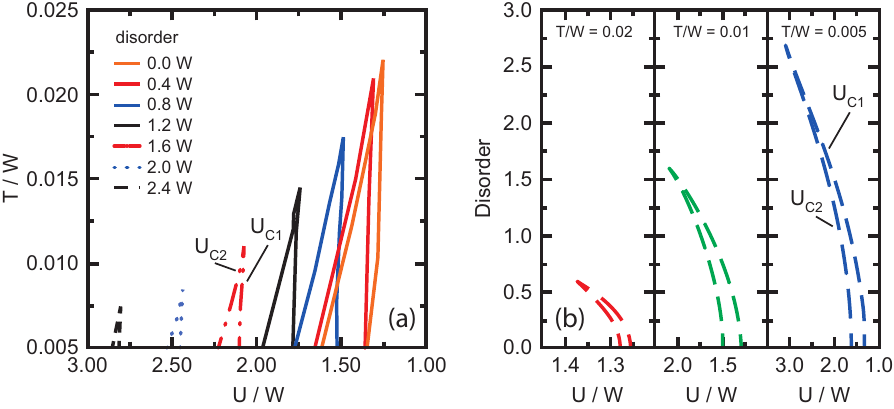}
    \caption{\label{fig:disorder}
Phase diagram for the disordered Hubbard model at nonzero temperature (a) for different degrees of disorder and (b) for
different temperatures $T$ given in units of the bandwidth $W$ (after \cite{Aguiar05}); note the axis $U/W$ goes from right to left in order to mimic the pressure dependence. The spinodal lines $U_{c1}$ and $U_{c2}$ indicate the boundaries to the insulating and metallic solutions.
}
\end{figure}

Dobrosavljevi{\'c} and collaborators also considered the temperature evolution when investigating the effect of disorder on the Mott transition by DMFT combined with typical medium theory \cite{Dobrosavljevic97,Dobrosavljevic03,Aguiar05,Braganca15}; in particular they looked at the coexistence regime of insulating and metallic solutions.
While for weak disorder the coexistence region is found to be similar to that in the clean case, with increasing disorder Anderson localization effects are responsible for shrinking the coexistence  region, as depicted in Figure~\ref{fig:disorder}. Both the $U_{c1}$ and $U_{c2}$ lines move toward larger interaction potential and become closer to each other when disorder increases.
As disorder becomes even stronger and exceeds twice the bandwidth, the region drastically narrows and the critical temperature $T_{\rm crit}$ abruptly goes to zero. Here the transition occurs at $U \approx W$ and Anderson and Mott routes to localization become equally important; in other words the effects of interaction and disorder are of comparable relevance for charge localization. The observation of a metal-insulator transition without a coexistence region suggests that the nature of the transition has changed from first to second order as disorder increases. This is in accord with the idea of a quantum-critical regime for  $T > T_{\rm crit}$, that was suggested by theory \cite{Terletska11} and experiments \cite{Furukawa15a} on various dimerized organic Mott insulators, as shown in Figure~\ref{fig:MottCriticality} and discussed in Sec.~\ref{sec:Mottquantumcriticality}.

Single crystals from organic charge transfer salts are renowned for their superior quality, mainly because they are grown by
electrochemical methods \cite{WilliamsBook,Montgomery94}. Occasionally some effect of the starting molecules,
solvent, or atmospheric environment has been reported \cite{Strack05,Lang06,Pinteric14}, but
the crystal structure and overall behavior is barely affected.
An intrinsic and reversible way of introducing disorder in BEDT-TTF salts is the adjustment of the cooling rate \cite{Su98b,Stalcup99}, which affects the freezing of the terminal ethylene-group disorder (Figure~\ref{fig:BEDT-TTF}) at around 70 to 80~K and thereby the resistivity profiles \cite{Tanatar99,Saito99,Muller02,Wolter07} and the pairing symmetry of the superconducting ground state \cite{Kanoda90,Kanoda93,Achkir93,Le92,Harshman90,Lang92,Dressel94a,Lang96,Pinteric00,Pinteric02}.

\begin{figure}
  \centering
  \includegraphics[width=0.8\columnwidth]{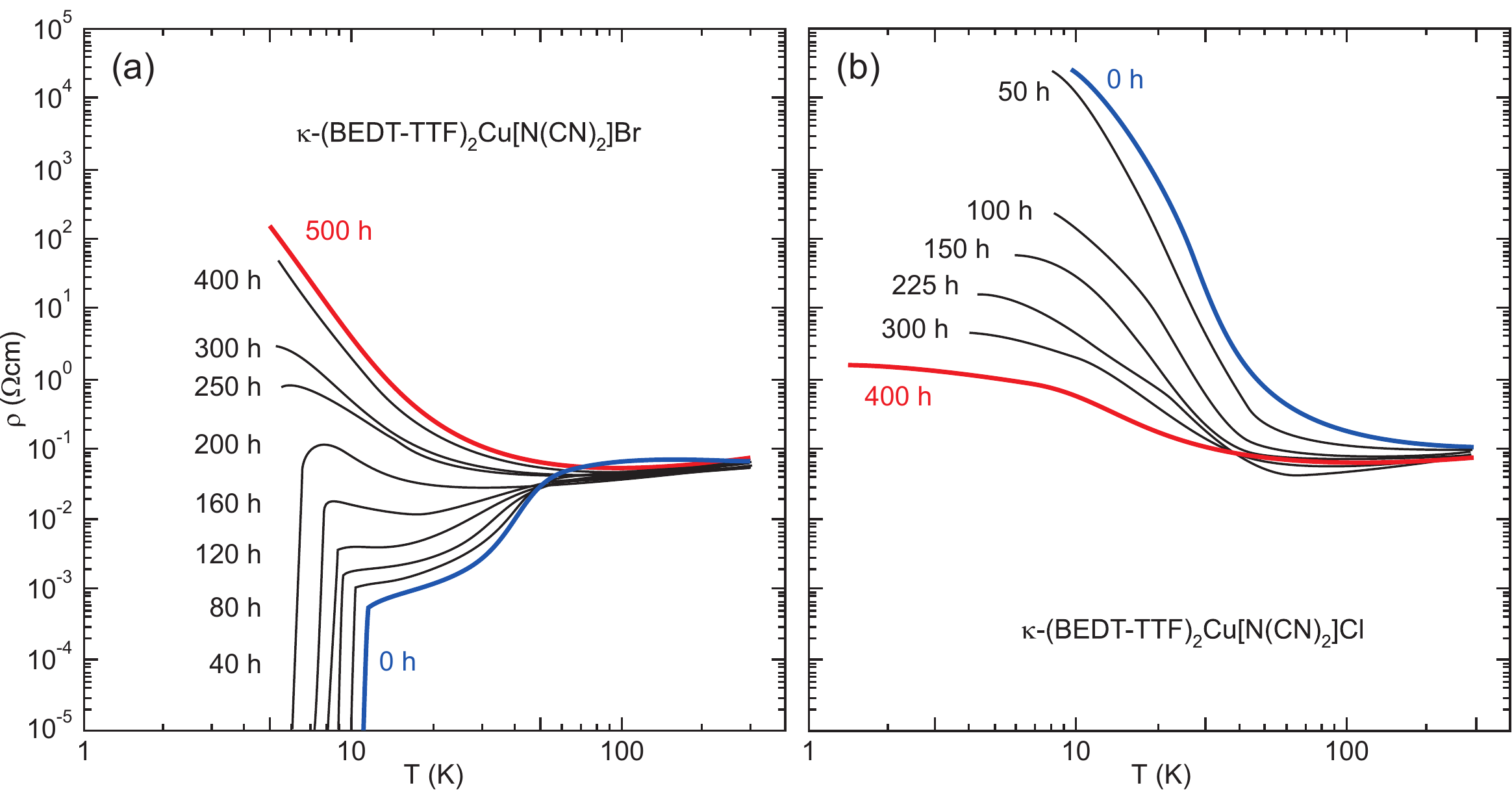}
    \caption{\label{fig:irradiation1}
Temperature dependence of the dc resistivity of (a) \etbr\ und (b) \etcl\ irradiated by x-ray.
The time indicated is the total x-ray exposure time at room temperature (data from \cite{Sasaki07,Sano10,Sasaki12c}).}
\end{figure}

This provides the possibility
to study the disorder dependence of physical properties by targeted artificial defects.
X-ray irradiation is commonly utilized to create disorder in molecular solids in a controlled way.
Sasaki and collaborators could show \cite{Sasaki11,Yoneyama10,Sasaki12,Sasaki12c} that the defects are introduced mainly in the anion layers separating the organic molecules leading to a modulation of the potential rather than a creation of electronic charges \cite{Sasaki07,Matsumoto12}. In general, the room-temperature resistivity substantially decreases with irradiation dose.
Also the characteristic hump around $T=100$~K quickly disappears \cite{Analytis06,Sano10,Sasaki12c}, in good agreement with DMFT calculations \cite{Radonjic10}.
The low-temperature properties are even more sensitive to irradiation as revealed by extensive studies of the normal and the superconducting state of \etscn\ and \etbr, as well as of $\beta$-(BEDT\--TTF)$_2$AuI$_2$. Already small doses increase the residual resistance $\rho_0$ and reduce the transition temperature $T_{c}$ \cite{Dolanski92,Analytis06};
however it takes rather extensive irradiation to completely kill the superconducting state \cite{Sano10,Sasaki11}.
The metallic compound \etbr, displayed in Figure~\ref{fig:irradiation1}(a) as an example, turns insulating when the electrons become localized due to the randomness of the potential: the charge transport takes place via hopping between sites \cite{Sano10}.

\begin{figure}[h]
  \centering
  \includegraphics[width=0.8\columnwidth]{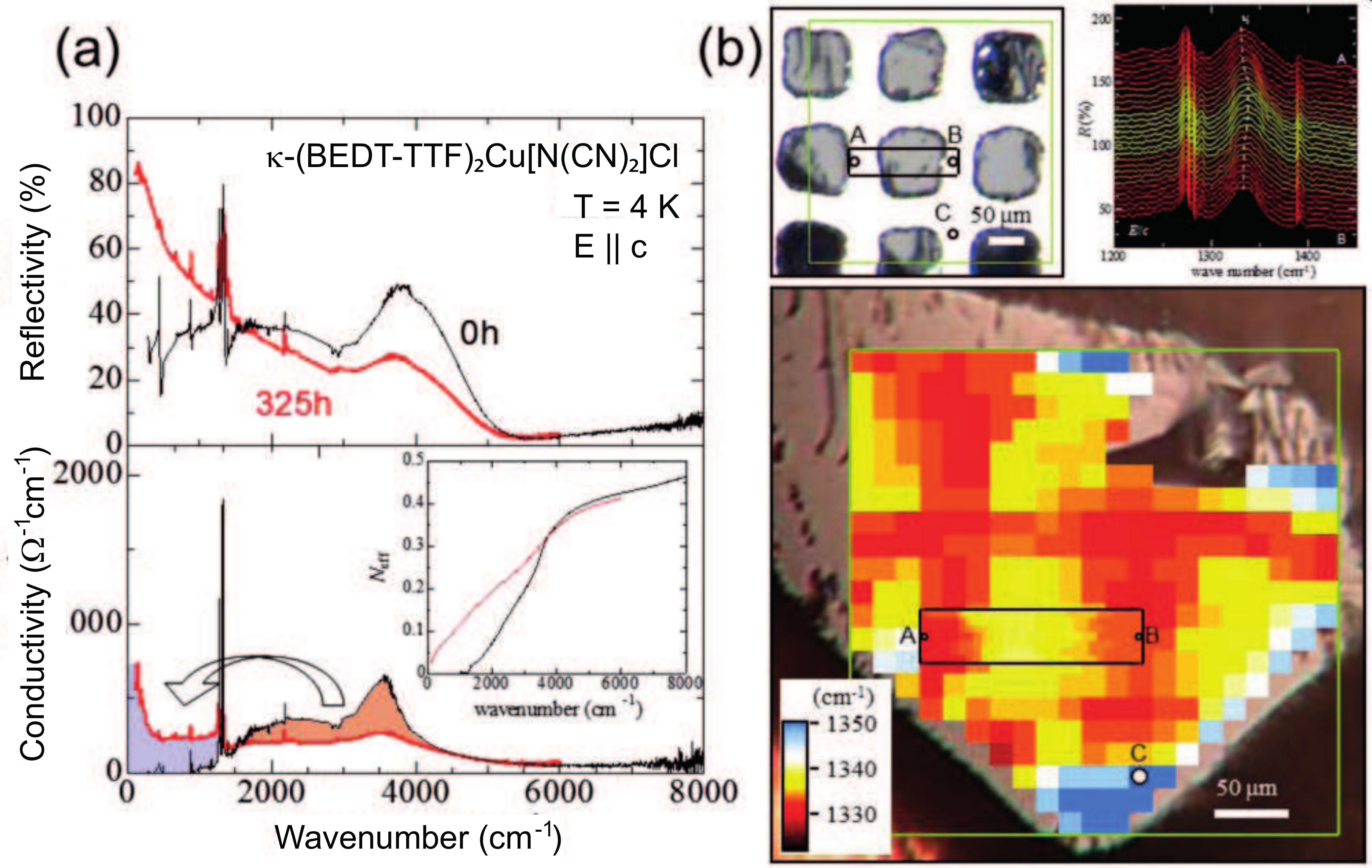}
    \caption{\label{fig:irradiation2}
(a) Infrared optical reflectivity and conductivity of \etcl\ before and after x-ray irradiation.
The inset of the lower panel indicates the effective number of carriers;
(b)~Scanning micro-region infrared reflectance spectroscopy scanning map of the partly x-ray irradiated \etcl. The sample is irradiated through the molybdenum mesh mask (reproduced from \cite{Sasaki12c}).}
\end{figure}
The Mott insulators, on the other hand, such as \etcl\ or \etcn, loose their high-resistivity state upon irradiation: the resistivity decreases in the whole temperature range. For \etcl\ a metallic-like temperature dependence down to about 50\,K is reached at a rather low dose \cite{Sasaki12c,Urai19}, as shown in Figure~\ref{fig:irradiation1}(b).
With increasing defect concentration, the Mott gap is filled, resulting in a shift of spectral weight from the interband transition to low energies: optical measurements  \cite{Sasaki08a} show a Drude-like behavior at low temperatures, shown in Figure~\ref{fig:irradiation2}(a).
DMFT calculations by Radonji{\'c} {\it et al.} explain this behavior by an increase of the bandwidth \cite{Radonjic10}.
By using an x-ray microbeam, the spatial dependence can be investigated. To that end a \etcl\
crystal is irradiated through a molybdenum mesh ($90~\mu{\rm m} \times 90~\mu$m) \cite{Sasaki04a,Sasaki05}
and the fabricated pattern studied by scanning micro-region infrared reflectance spectroscopy with a spatial resolution of 5 to $15~\mu$m as already introduced in Sections~\ref{sec:coexistenceregime} and \ref{sec:FermiLiquid}. Mapping the emv coupled molecular vibration $\nu_3(a_g)$ provides local information on the electronic states (Figure~\ref{fig:irradiation2})
because the frequency of the mode reflects the electronic state via electron molecular vibration coupling \cite{Sasaki04b,Sasaki12c}.

\begin{figure}[h]
  \centering
  \includegraphics[width=0.55\columnwidth]{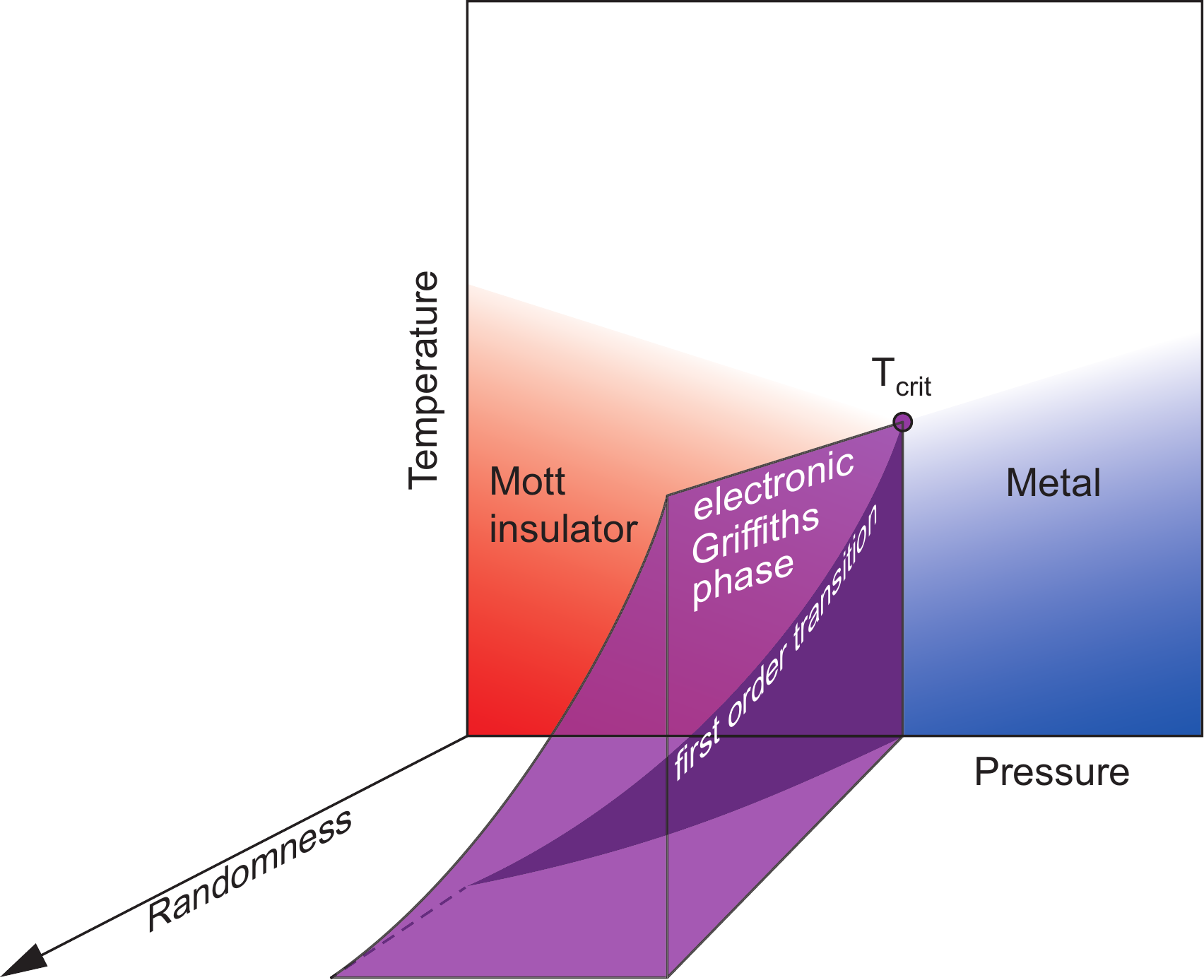}
    \caption{\label{fig:Griffiths}
    Schematic phase diagram how the Mott transition develops with increasing randomness, as suggested from NMR experiments on \etcl\ subject to x-ray irradiation \cite{Yamamoto20}. The coexistence regime at the first-order phase transition gradually diminishes as shown in Figure~\ref{fig:disorder}. Above that spatial inhomgeneities give rise to an electronic Griffiths pase.}
\end{figure}

Spatial inhomogeneities in the two-dimensional electron and magnetic system may give rise to a Griffiths phase at the metal-insulator transition \cite{Dagotto05,Vojta06,Krivoruchko14}. Following theoretical studies \cite{Tanaskovic04,Andrade09} recent $^{13}$C-NMR investigations on strongly irradiated (500~h x-ray exposure) \etcl\ found a slow dynamics that was interpreted in this regard \cite{Yamamoto20}. The electronic system exhibits long-length self-organization without long-range order. A temperature-pressure-randomness phase diagram with an electronic Griffiths phase is proposed as displayed in Figure~\ref{fig:Griffiths}.

The quantum-spin-liquid Mott insulator \etcn\  seems to be less susceptible to x-ray irradiation.
The resistivity varies only little \cite{Sasaki07,Sasaki12c,Sasaki15} even for large doses and also the magnetic properties remain unchanged by chemical substitution \cite{Saito18}. The Mott gap does not collapse due to irradiation, but hopping conduction, present even in the pristine crystals \cite{Sedlmeier12,Elsasser12,Sasaki15}, becomes dominant.

\v{C}ulo {\it et al.} measured the dc resistivity and Hall effect of several quantum-spin disordered Mott insulators with rather similar chemical compositions and crystal structures: \etcn, \agcn, and $\kappa$-(BEDT-TTF)$_2$B(CN)$_4$ \cite{Pinteric14,Culo15,Culo19}.
While around room temperature the transport properties are mainly determined by the strength of the electron correlations,
upon reducing $T$, hopping transport takes over due to inherent disorder. The most disordered
compound \etcn\ exhibits the lowest dc resistivity and the highest charge carrier density,
{\it i.e.}, in the phase diagram it is located closest to the metal-insulator transition. The least disordered compound
$\kappa$-(BEDT-TTF)$_2$B(CN)$_4$ shows the highest resistivity and the lowest carrier density, {\it i.e.}, lies farthest from the phase transition. The observations are explained within the theory of Mott-Anderson localization
as a consequence of disorder-induced localized states within the correlation gap \cite{Pinteric14,Culo19}.

\section[Quantum Spin Liquid versus Magnetic Order]{Frustration in Two Dimensions: Quantum Spin Liquid versus Magnetic Order}
    \label{sec:Frustration}
Stimulated by the pioneering work of L. Pauling on geometrical frustration of electric dipoles
\cite{Pauling35}, P.W. Anderson considered spin models possessing an extensive degeneracy of states that prevent long-range magnetic order even at $T=0$~K \cite{Anderson56,Anderson73,Fazekas74}.
The idea is that Heisenberg spins form pairs, {\it i.e.} quantum mechanical singlets, called valence bonds.
A resonating valence-bond state can be obtained by superposition of different valence-bond solids, where the singlet pairs are not necessarily restricted to nearest neighbors.
These systems do not develop long-range magnetic order  despite the strong interactions,
hence they are called quantum spin liquids.
In contrast to spin glasses, which eventually become static \cite{FischerBook,ContucciBook,DiepBook,Mydosh15},  
spin liquids keep some dynamics down to zero temperature.
Although the singlet bonds can be randomly oriented on the triangular lattice, the valence-bond solid involves some symmetry breaking that becomes even more tangible in real materials where the valence-bond solid states are intertwined with lattice distortions that stabilize singlet pairs on bonds with stronger interaction \cite{Powell11,Balents10,Savary16,Zhou17,Kimchi18,Knolle19,Li20,Broholm20}.
In contrast to bosonic ($S=1$) magnon excitations in long-range-ordered antiferromagnets,
these quantum spin liquids exhibit fractional, fermionic $(S=\frac{1}{2}$) magnetic excitations, called spinons,
that can propagate independently without involving charge. The absence of an energy gap in the excitation spectrum implies the existence of a spinon Fermi surface.

After searching numerous compounds for experimental realization, molecule-based systems were finally identified as the best suited models for exploring the physics of quantum spin liquids on a triangular lattice \cite{Powell11,Kanoda11,Savary16,Zhou17,Dressel18,Wen19},
since alternative systems, such as YbMgGaO$_4$ or 1T-TaS$_2$, have much smaller energy scales  \cite{Li15,Ribak17,Chamorro20}.
Despite enormous theoretical efforts, the origin of the spin-liquid phase in two dimensions
is not completely solved because pure geometrical frustration should not suffice to stabilize the quantum spin-liquid state \cite{Huse88,Bernu92,Capriotti99,Kaneko14} and disorder might play an important role, too \cite{Watanabe14,Shimokawa15,Wu19,Kawamura19}. As a matter of fact, numerous questions remain unanswered concerning the ground state and magnetic excitations spectrum of these materials.

Owing to several reviews on the organic spin liquid compounds \etcn\ and \dmit\ \cite{Powell11,Kanoda11,Savary16,Zhou17,Dressel18}, in Section \ref{sec:propertiesQSL} we
give only a brief summary
of their properties, with the focus on \cat\ as the latest example. In Section~\ref{sec:gaplessQSL} we start with the question, whether the quantum spin liquid state exhibits a gapped excitation spectrum \cite{Vasilev05}.
In the following sections, open issues and hot topics of current research will be discussed, such as the 6~K-anomaly and charge-spin coupling in \etcn\ (Section~\ref{sec:6Kanomaly}),
or the influence of magnetic field and disorder on the spin liquid state as well as the formation of a valence-bond solid (Sections~\ref{sec:magneticfieldQSL} and \ref{sec:randomnessQSL}).

\subsection{Properties}
\label{sec:propertiesQSL}
\begin{figure}
 \centering
 \includegraphics[width=0.3\columnwidth]{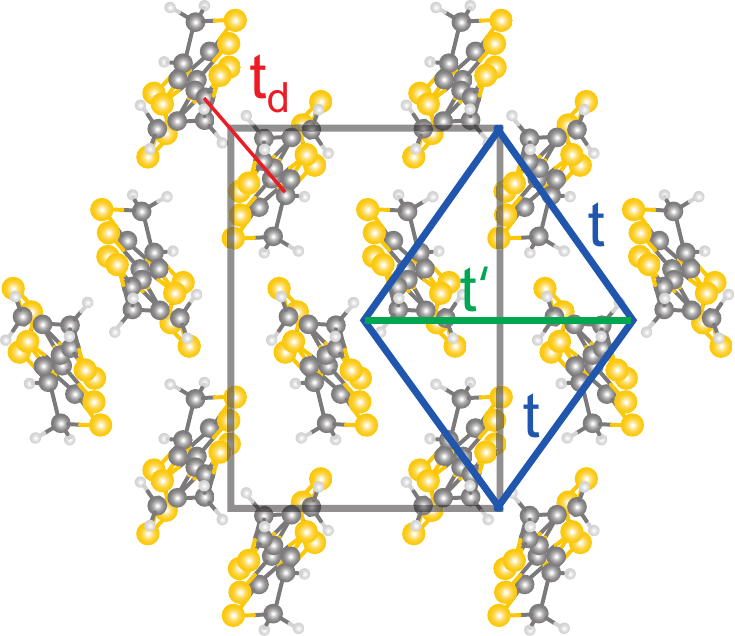}
\caption{For $\kappa$-(BEDT-TTF)$_2X$ the molecules are arranged in dimers that constitute an anisotropic triangular
lattice within the conduction layer. The nearest-neighbor and the second-nearest-neighbor inter-dimer transfer integrals are labeled  by $t$ and $t^{\prime}$, respectively; the inter-dimer transfer integral is marked by $t_d$. }
\label{fig:k-structure}
\end{figure}
For the most prominent examples
\etcn\ \cite{Shimizu03,Kurosaki05},  \agcn\  \cite{Shimizu16,Pinteric16}, \dmit\ \cite{Itou08,Itou10}
and \cat\ \cite{Isono13,Isono14,Shimozawa17} the starting point is rather similar: here
molecular dimers each hosting $S=\frac{1}{2}$ form a highly frustrated triangular lattice  \cite{Kandpal09,Nakamura09,Nakamura12}. Defining the frustration as the ratio of transfer integrals $t^{\prime}$ and $t$ by the sketches in Figures~\ref{fig:k-structure} and \ref{fig:CAT-structure}, the compounds are close to a perfect triangular arrangement as summarized in Table~\ref{tab:1}.
The intra-dimer transfer integral of \etcn\ and \agcn\ is $t_d\approx 200$~meV and 264~meV, respectively.
When estimating the onsite Coulomb repulsion $U\approx 2t_d$ \cite{McKenzie98}, at ambient
conditions one obtains $U/t=7.3$ and 10.5 with the ratio of the two inter-dimer
transfer integrals $t^{\prime}/t \approx 0.83$  and 0.90 very close to unity.
For comparison, the related compounds \etbr\ and \etcl\ possess a frustration
\begin{table}[h]
\caption{Four different molecular-based quantum spin-liquids compounds on a triangular lattice with antiferromagnetic coupling $J$; the degree of frustration $t^{\prime}/t$ is calculated by tight-binding approximation based on extended H{\"u}ckel (EH) studies of molecular orbitals or on {\it ab-initio} density-functional-theory (DFT) calculations \cite{Komatsu96,Nakamura09,Kandpal09,Jeschke12,Jacko13a,Shimizu16,Hiramatsu17}.
The effective correlations defined as the ratio of inter-dimer Coulomb repulsion $U$ and bandwidth $W$ are listed as extracted from the optical conductivity; the electronic contribution to the specific heat is also shown $\gamma_e$ \cite{Oshima88,Komatsu96,Shimizu16,Hiramatsu15,Itou08,Kato12a,Kato12c,Pustogow18a,Isono14,Shimozawa17,
Yamashita11,Yamashita17}.
\label{tab:1}}
\hspace*{2mm}
\begin{center}
{
\begin{tabular}{l|c c c c}
Compound                                       &$J$ &$t^{\prime}/t$ & $U/W$ & $\gamma_e$ \\
                                                &(meV; K) & (DFT; EH)  &   &(mJ\,K$^{-2}$mol$^{-1}$)\\
\hline
$\kappa$-(BE\-DT\--TTF)$_2$\-Cu$_2$\-(CN)$_{3}$ & 19; 220 & 0.83; 1.1~~& 1.52 & 12-15 \\
$\kappa$-(BE\-DT\--TTF)$_2$\-Ag$_2$(CN)$_{3}$   & 19; 220 & ~~--~~; 0.90 & 1.96  & 10 \\
$\beta^{\prime}$-EtMe$_3$\-Sb\-[Pd(dmit)$_2$]$_2$ & 22; 250 & 0.77; 0.90 & 2.35 & 19.9 \\
$\kappa$-H$_3$(Cat-EDT-TTF)$_2$ & ~~7-8; 80-90 &1.25; 1.48 & & 58.8
\end{tabular}}
\end{center}
\end{table}
$t^{\prime}/t \approx 0.42$ and 0.44, respectively \cite{Kandpal09,Koretsune14}. Theoretical considerations suggest a frustration $t^{\prime}/t$ slightly off perfect triangle \cite{Tocchio09,Dang15,Kyung06,Laubach15,Yamada14}

\begin{figure}
 \centering
 \includegraphics[width=0.5\columnwidth]{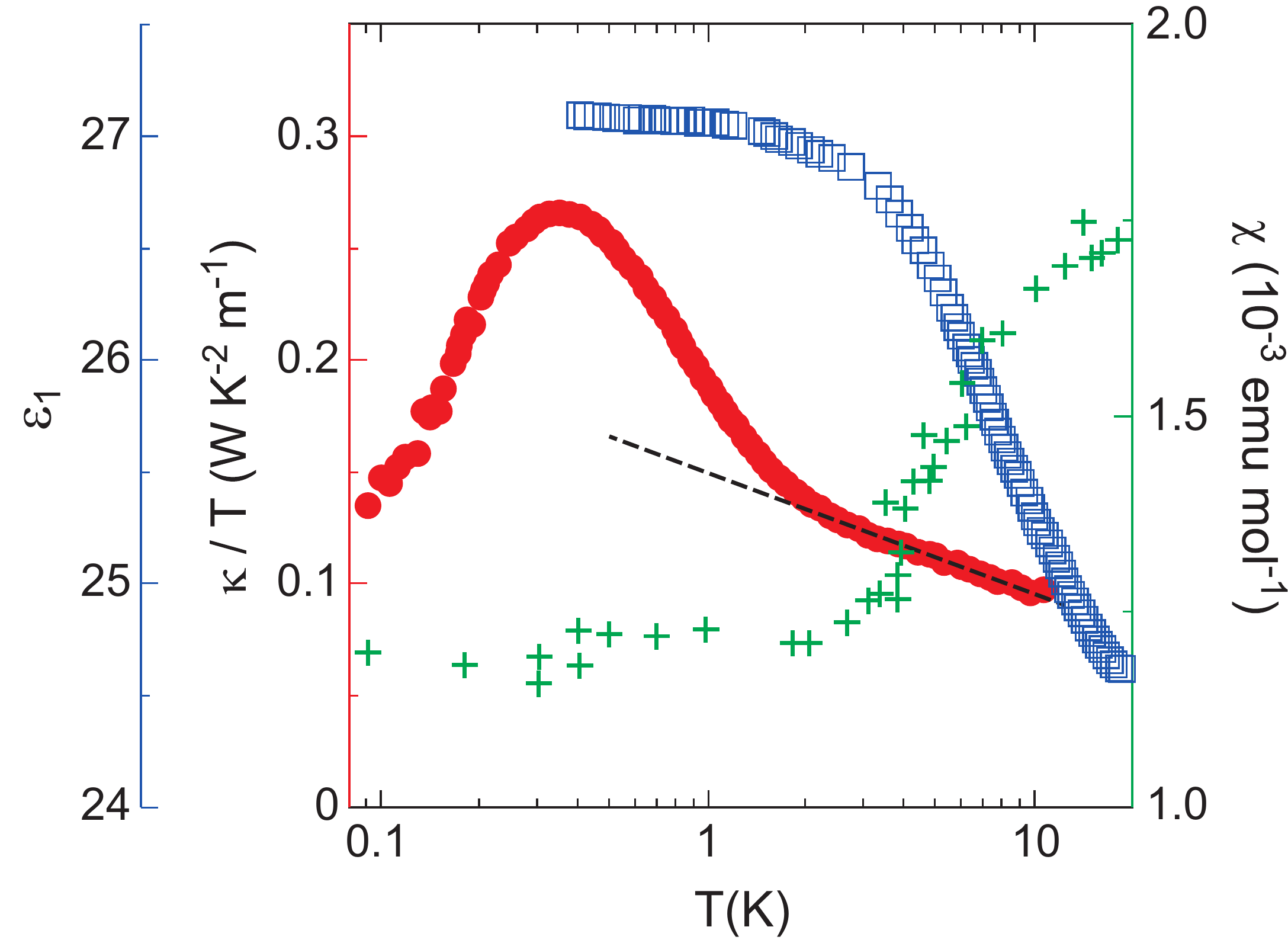}
\caption{Low-temperature physical properties of \cat\ illustrating the spin-liquid and paraelectric states. The temperature dependence of the dielectric constant $\epsilon_1(T)$ corresponds to the blue open squares and left axis, the thermal conductivity divided by temperature $\kappa/T$ is shown by the red dots on the red left axis, and the magnetic susceptibility $\chi(T)$ refers to the green symbols and right axis. The dashed line is an eye guide. The concurrent occurrence of quantum paraelectric and quantum-spin-liquid phases below $T\approx 2$~K is inferred from the saturation of $\epsilon_1(T)$ and $\chi(T)$  (after \cite{Isono14,Shimozawa17}).}
\label{fig:cat1}\label{fig:CatBarrett}
\end{figure}

At ambient pressure, no indication of N{\'e}el order is observed for temperatures as low as 20~mK, despite the considerable antiferromagnetic exchange of $J\approx 220-250$~K for the dimerized
$\kappa$- and $\beta^{\prime}$-salts \cite{Shimizu03,Shimizu16,Itou08}
and slightly less in the case of the hydrogen bonded \cat.
As an example, the low-temperature physical properties of \cat\ are displayed in Figure~\ref{fig:cat1}. Below a maximum
around 20~K, the spin susceptibility $\chi(T)$ drops upon cooling, but no magnetic transition can be identified down to $T=50$~mK using SQUID and torque magnetometry  \cite{Isono14}. The behavior is reasonably well fitted by the $S=\frac{1}{2}$ antiferromagnetic Heisenberg model on an isotropic triangular lattice. The thermal conductivity shows a maximum around $T=0.4$~K and then decreases in a linear fashion with a finite intercept as $T\rightarrow 0$.
The dielectric permittivity rises as the temperatures is lowered but $\epsilon_1(T)$ saturates below 2~K; this behavior is associated with quantum fluctuations and described by the Barrett formula (\ref{eq:Barrettformula}) developed for perovskites \cite{Barrett52} (see Section \ref{sec:cat}).

\subsection{Gapless quantum spin liquid?}
\label{sec:gaplessQSL}
Thermodynamic measurements on \etcn\ reveal a linear term in the heat capacity
\begin{equation}
C_p(T) = \gamma_e T + \beta T^3 \label{eq:specificheat}
\end{equation}
that provides evidence for gapless spinon excitations \cite{Shimizu06,Yamashita08,Isono18}; the second term contains the typical phonon contributions. This is in stark contrast to thermal transport data, which exhibit a vanishing $\kappa/T$ at $T=0$~K; the result implies the presence of a small gap of $\Delta \approx 0.5$~K {\it i.e.} the absence of gapless fermionic excitations \cite{Yamashita09}.
The presence of a spinon Fermi surface is of crucial importance for the low-energy excitations in these Mott insulators.
While in inorganic compounds the dispersion of the excitation spectrum is directly mapped by inelastic neutron scattering, molecular quantum magnets evade these investigations by the abundance of hydrogen and their limited crystal size. Since most methods applied are indirect, the dispute is not resolved easily one should carefully  consider what they are sensitive to. The results of complementary approaches have to be included eventually forming a consistent story; although the peculiarities of certain compounds might stain the unified picture.

\begin{figure}
  \centering
  \includegraphics[width=0.8\columnwidth]{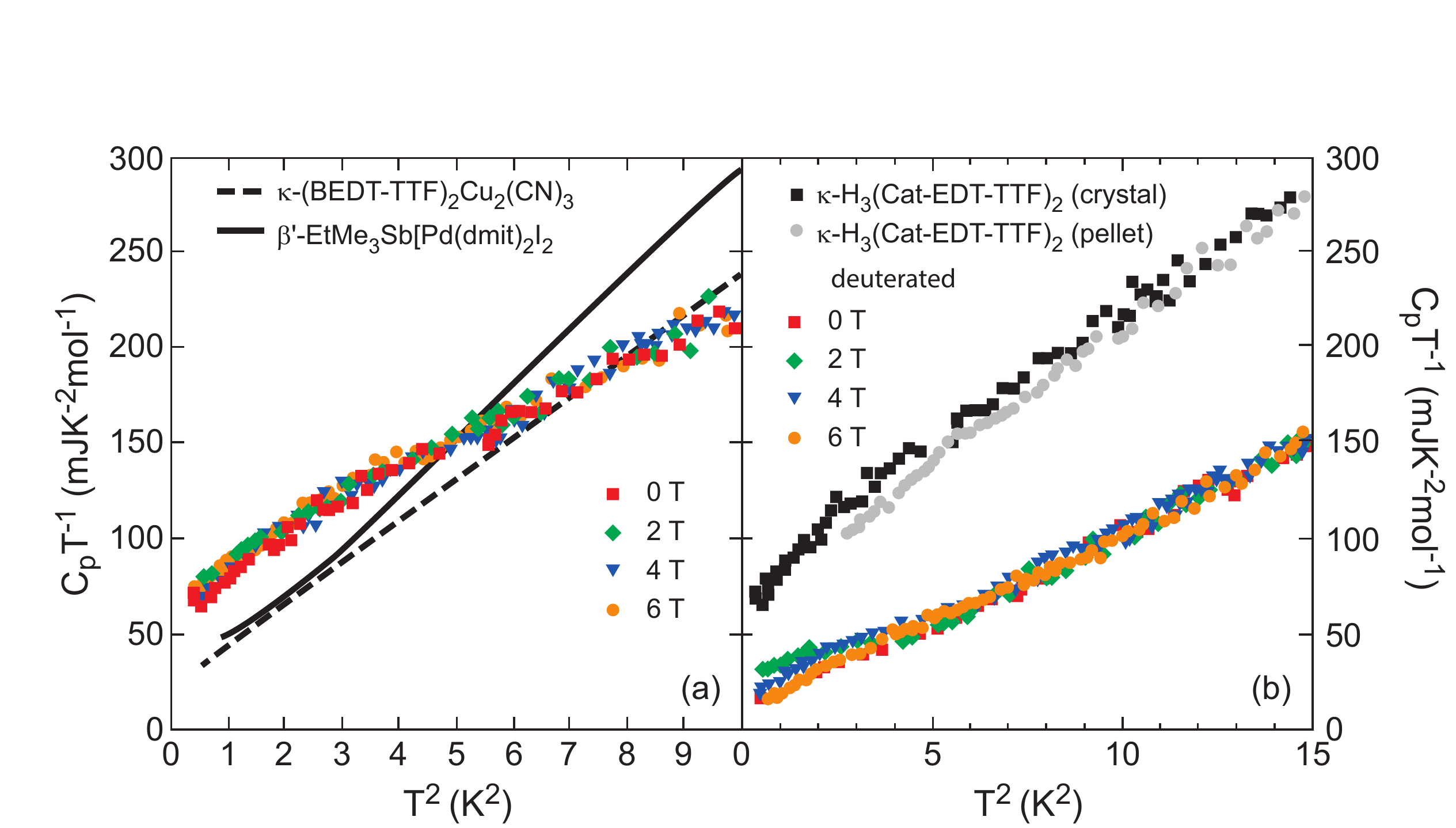}
    \caption{\label{fig:cat3}
(a)~The $C_p T^{-1}$ {\it versus} $T^2$ plot of the specific heat capacity of \cat\
in the low temperature region that allows the separation of the contributions according to
$C_p(T) = \gamma_e T + \beta T^3$.
No dependence on magnetic field  is observed up to 6~T.
The dashed and solid lines represents the data of other two spin-liquid compounds,
\etcn\ and \dmit.
(b)~Comparison of $C_p T^{-1}$ {\it vs.} $T^2$ plots of several single crystals of \cat\
with a pressed pellet.
The data of $\kappa$-D$_3$(Cat-EDT-TTF)$_2$ obtained under fields up to 6~T are also plotted
(after \cite{Yamashita17}).}
\end{figure}

In Figure~\ref{fig:cat3}(a) the temperature dependence of $C_p(T)$ is plotted for \cat\ together with the results of other spin-liquid compounds. The extrapolation to $T=0$ yields a rather large offset of $\gamma_e = 58.8$~mJ\,K$^{-2}$mol$^{-1}$, three to four times bigger than observed for \etcn\ and \dmit,
in accord with $J = 80$~K compared to 220 - 250~K of the other materials listed in Table~\ref{tab:1}. The large electronic density of states has been related to spinon excitations in the gapless spin liquid.
The thermodynamic observations are confirmed by susceptibility measurements ($\chi_0 = 1.2\times 10^{-3}$~emu/mol). It is instructive to consider the Wilson ratio
\begin{equation}
R_W = \frac{4\pi^2 k_B^2 \chi_0}{3(g\mu_B)^2\gamma_e} \quad ,
\label{eq:Wilson}
\end{equation}
where $g\approx 2.002$ is the $g$-value of non-interacting electrons, $\mu_B$ is the Bohr magneton and $k_B$ the Boltzmann constant.
The Wilson ratio is also seen as an empirical indicator for the importance of spin-orbit coupling,
which is commonly taken as a minor issue in organics; albeit this aspect has come under scrutiny in some systems recently \cite{Winter17,Osada18,Riedl19}.
Most theories on  quantum spin liquids predict that $R_W\ll 1$, in contrast to measured ratios \cite{Balents10,Prelovsek19}.
For all organic compounds, we find a slightly enhanced Wilson ratio compared to a free electron gas: $R_W = 1.4 - 1.6$ \cite{Yamashita17},
but significantly less than the anomalously large Wilson ratio ($R_W= 70$) reported for the three-dimensional quantum spin liquid Na$_4$Ir$_3$O$_8$ \cite{Okamoto07n,Chen13}.

According to Ng and Lee \cite{Ng07}, optical studies may provide important information; if spinon excitations exhibit a Fermi surface, they should show up not only in the thermal conductivity but also contribute to the optical conductivity \cite{Ioffe89,Ng07,Potter13}. A power-law behavior $\sigma_1(\omega)\propto \omega^{2}$ is expected for low-temperatures and frequencies that becomes $\sigma_1(\omega)\propto \omega^{3.3}$ above the exchange coupling $J$ if impurity scattering is negligible.
In the case of \etcn\ the challenge is that close to the Mott transition, inadvertent effects of metallic fluctuations and regions dominate at finite temperatures, hampering the extraction of spinon contributions \cite{Kezsmarki06,Elsasser12,Pustogow18a}.
Hence, only strongly correlated systems deep in the Mott insulating phase are suitable candidates.
\begin{figure}
\centering
\includegraphics[width=0.7\columnwidth]{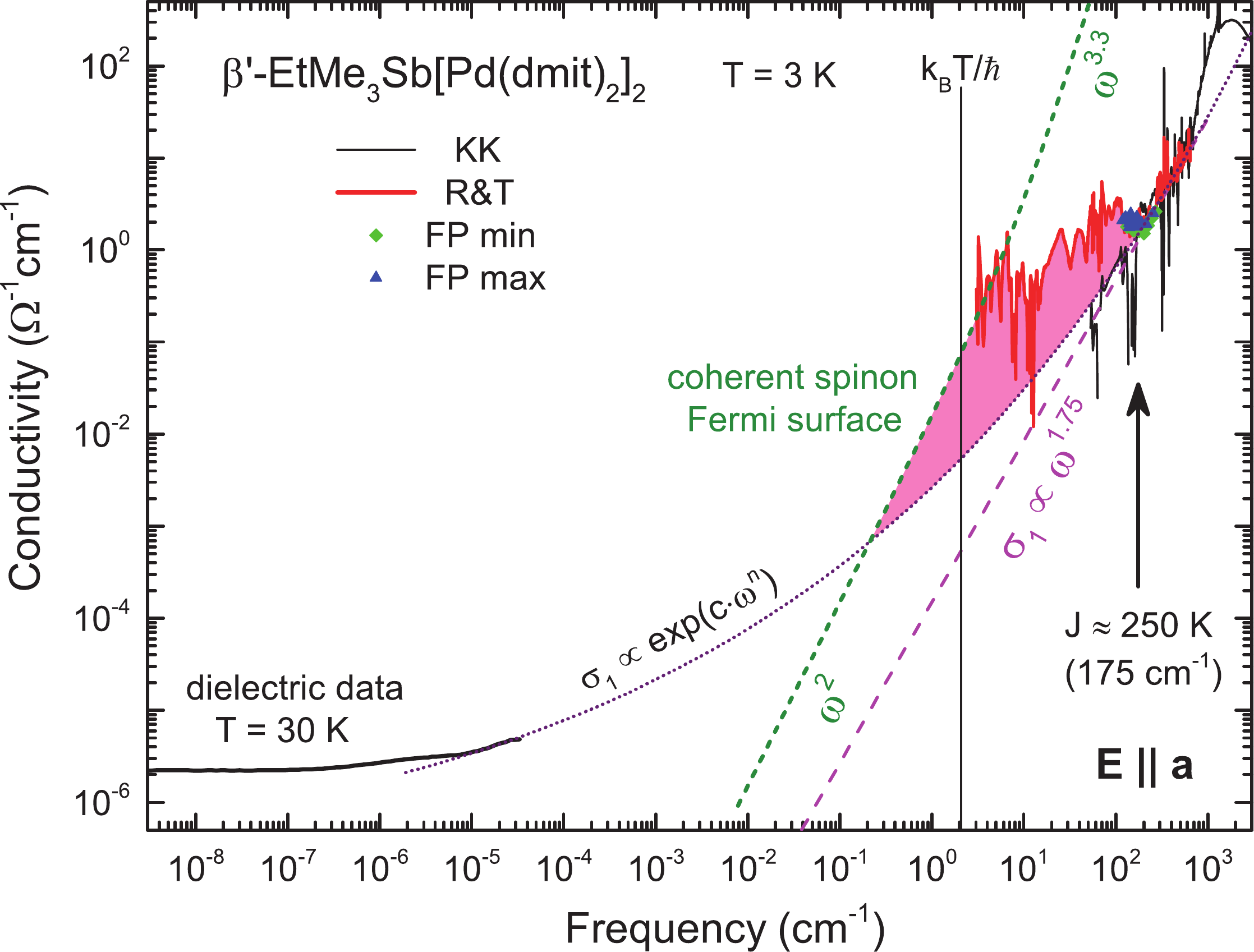}
\caption{
Delectric and optical conductivity of \dmit\ measured along the $a$-direction at low temperatures ($T=3$~K)
in a broad range from radio frequencies up to the infrared. Infrared reflectivity measurements yield $\sigma_1(\omega)$ via the Kramers-Kronig analysis (KK), while in the THz range reflectivity and transmission data (R \&\ T) allow a direct evaluation of $\sigma_1(\omega)$. Minima and maxima of Fabry-Perot oscillations (FP) are also used in this regard.
The optical data in the THz and far-infrared ranges ($2~{\rm cm}^{-1} < (\omega/2\pi c) <
200$~\cm) is much larger than the background, which was approximated by
a power law $\sigma_1\propto \omega^{1.75}$ from the optical region down to lower frequencies; and
by a stretched exponential $\exp\{\omega^{0.1}\}$ that
smoothly connects the dielectric range with the optical data.
The dashed green line represents the theoretically predicted frequency-dependence of a
coherent spinon Fermi surface coupled to the optical conductivity via an emergent gauge field \cite{Ng07}.
The effective range of spinons (shaded in red) extends from the crossing of the $\omega^2$ line with the electronic
background conductivity up to the antiferromagnetic exchange coupling energy
$\hbar\omega/ k_B \approx J \approx 250$~K
(data taken from \cite{Pustogow18b,Dressel18}).
\label{fig:spinons}
}
\end{figure}
By preparing several extremely thin single crystals of the strongly correlated spin-liquid compound \dmit\ the optical transmission could be measured down to the THz range of frequency. Figure~\ref{fig:spinons} shows that clear indications of spinon contributions to the
optical conductivity can be identified. They become pronounced only at rather low temperatures ($T=3$~K) and low frequencies $\hbar\omega < J = 22$~meV, corresponding to 250~K. For $\hbar\omega < k_BT$ one expects a drop according to $\sigma_1(\omega) \propto \omega^2$ \cite{Ng07} until hopping transport dominates at lowest frequencies \cite{Pinteric14,Dressel16,Dressel18}. This is a confirmation of gapless excitations and the presence of a spin Fermi surface \cite{Pustogow18b}. Unfortunately, comparable experiments on \cat\ have not been performed yet.

Measurements of the magnetocaloric effect of \etcn\ down to mK temperatures reveal a decoupling of the electron spins from the lattice in the quantum-spin-liquid state, which is seen as indication for gapless spinon excitations \cite{Isono18}. Isono {\it et al.} then apply a magnetic field in order to move the system away from a quantum critical point, and they find the number of spin states that interact with the lattice vibrations strongly reduced; in other words, there exists only a
weak spin–lattice coupling. The picture of a quantum critical behavior in the quantum-spin-liquid compound \etcn\ was previously suggested on the basis of magnetic torque measurements \cite{Isono16}, where a universal critical scaling was observed. Winter and collaborators \cite{Riedl19}, however, challenged this interpretation and proposed that disorder-induced spin defects (Figure~\ref{fig:RVB}) better explain the low-temperature properties. These spins are attributed to valence-bond defects that emerge spontaneously as the quantum spin liquid enters a valence-bond glass phase.
This suggestion is strongly supported by recent ESR experiments \cite{Miksch20}, presented in Section~\ref{sec:magneticfieldQSL}.
Disorder effects may also resolve the dispute on the existence or non-existence of a spin gap,
since they make the spinons localized and thus not conduct heat. Indeed,
two independent recent studies of \dmit\ --~in contrast to \cite{Yamashita10} but
similar to previous studies of \etcn{} \cite{Yamashita09}~--
do not find a finite residual term in the thermal conductivity suggesting an absence of mobile gapless fermionic excitations. They conclude that the low-energy excitations responsible for the sizeable $\gamma_e$ are localized; the heat transport is entirely due to phonons scattered by low-energy spinon excitations of the spin liquid state \cite{Bourgeois-Hope19,Ni19}.
We will come back to this topic in Section~\ref{sec:randomnessQSL} shortly.

\subsection{6~K-anomaly}
\label{sec:6Kanomaly}
Although the spin-liquid state is supposed to be homogeneous down to lowest temperatures, for \etcn\ a remarkable anomaly was observed around $T^*=6$~K in a large number of different physical quantities:
it is seen in NMR \cite{Shimizu06}, ESR \cite{Komatsu96,Padmalekha15,Miksch20}, and magnetic susceptibility measurements \cite{Manna10}, in heat capacity and thermal conductivity \cite{Shimizu03,Yamashita08,Yamashita09}, by anisotropic lattice effects \cite{Manna10}, and as an anomaly in the out-of-plane phonon velocity and ultrasound attenuation \cite{Poirier14}.
There is no doubt that the 6~K-anomaly in \etcn\ is omnipresent and robust;
a broadly accepted explanation, however, remains elusive to date.
It was suggested that the second-order phase transition reflects some instability of the quantum-spin-liquid phase. This may be related to a change in the spin chirality with local $Z_2$ formation \cite{Baskaran89,Kawamura84,Moessner01,Kimchi18}, or Amperean pairing of spinons \cite{Lee07,Galitski07,Grover10,Li10} or the formation of an excitonic condensate by charge neutral pairs of charge $+e$ and charge $ -e $ fermions \cite{Qi08}.

\begin{figure}[h]
  \centering
  \includegraphics[width=0.7\columnwidth]{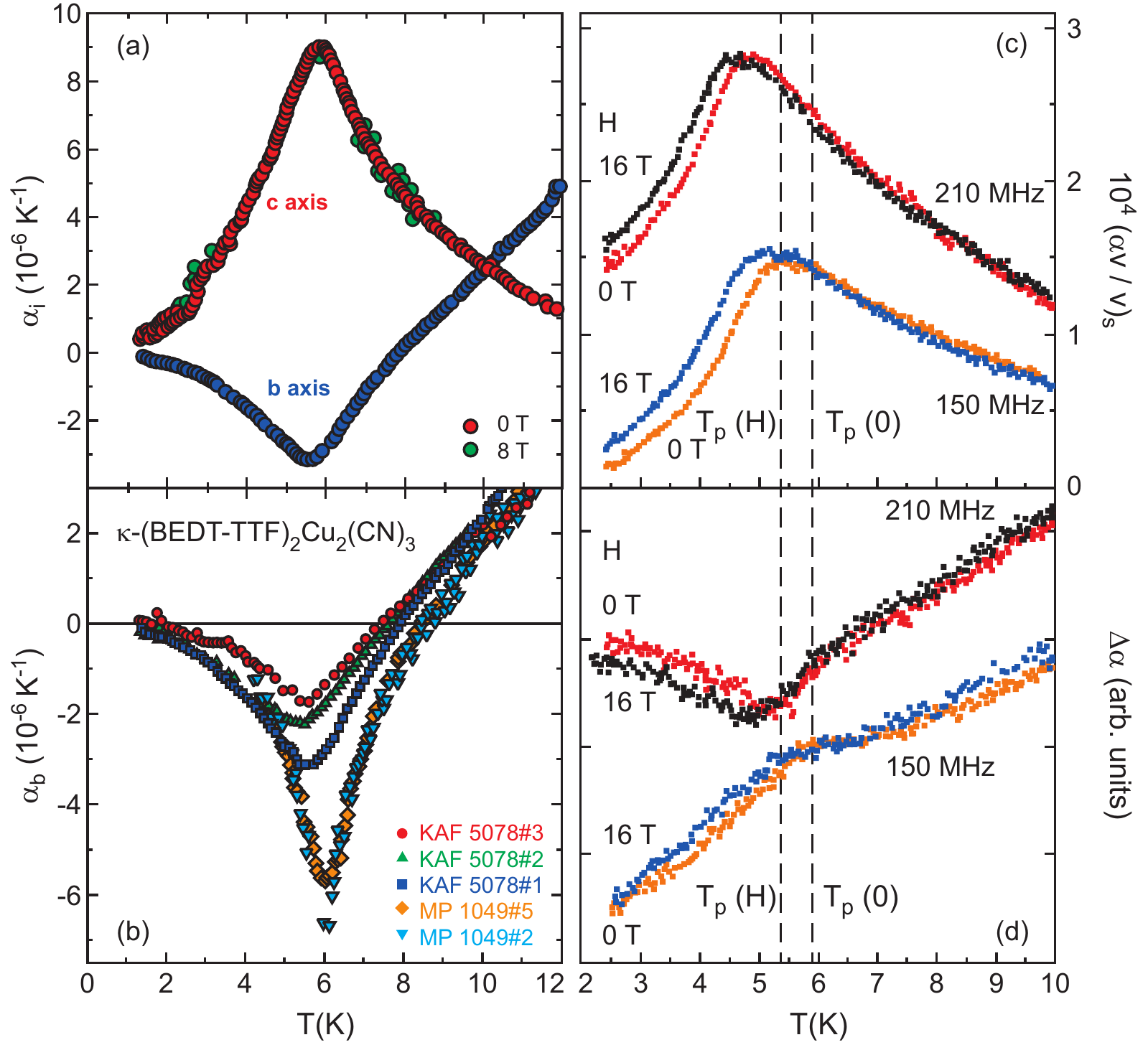}
    \caption{\label{fig:CuCNanomaly}
(a)~Thermal expansion coefficients $\alpha_i(T)$ for \etcn\ measured along the in-plane $b$- and $c$-axes around the 6~K phase transition at $H=0$ and 8~T (green dots) (after \cite{Manna10,Manna12}). (b)~When results of different crystals are compared, a large sample-to-sample variation becomes obvious as far as the size of the transition is concerned while the temperature shifts less than 0.5~K (after \cite{Manna18}).
(c)~Softening peaks $(\Delta V / V)_S$ and (d) variation of the attenuation $\Delta \alpha$ for \etcn\ as a function of temperature below 10~K: $H=0$~T, 150~MHz (red dots) and 210~MHz (blue dots); $H=16$~T (black symbols). The vertical dashed lines indicate the transition temperatures $T_p(0)$ and $T_p(H)$ (data from  \cite{Poirier14}). }
\end{figure}

Several results evidence that the lattice is involved via spin-phonon coupling. In Figure~\ref{fig:CuCNanomaly}(a)
the thermal expansion coefficient $\alpha_i = l_i^{-1} \partial l_i / \partial T$ of \etcn\ is plotted for different crystal direction $i$. The distinct $\alpha_b$ versus $\alpha_c$ anisotropy implies in-plane lattice distortions. They are temperature dependent and most pronounced for $T < 50$~K, where upon cooling the $b$-axis lattice parameter strongly contracts (large positive $\alpha_b$) while the $c$-axis lattice constant expands ($\alpha_c<0$).
Since the hopping amplitudes $t^{\prime}$ and $t$ among the dimers depend sensitively on the lattice parameters (cf. Figure~\ref{fig:k-structure}),
the degree of frustration $t^{\prime}/t$ increases in this temperature range.
Around 10~K the behavior reverses sign.
The distinct peak around $T^*=6$~K clearly indicates a phase transition. While the overall feature is robust, there is a sizeable variation among crystals from the same and different batches, as illustrated in Figure~\ref{fig:CuCNanomaly}(b).
Surprisingly, no dependence on magnetic field is observed up to 8~T.
The absence of any hysteresis is consistent with a second order phase transition.
The entropy release found experimentally \cite{Manna10} exceeds the residual spin entropy considerably, providing strong arguments that spin degrees of freedom alone are insufficient to account for the phase transition.
Charge degrees of freedom are a possible candidate to account for this mismatch. In fact, according to optical conductivity measurements \cite{Kezsmarki06,Elsasser12}, the charge gap in \etcn\  is strongly suppressed because the material is next to the insulator-metal transition where metallic quantum fluctuations are present \cite{Pustogow18a,Dressel18} due the vicinity of the coexistence regime.

Ultrasound measurement of \etcn\ by Poirier {\it et al.} \cite{Poirier14} reveal a softening of the longitudinal velocity
along the direction perpendicular  to  the  $bc$-plane for temperatures below 20~K. As shown in Figure~\ref{fig:CuCNanomaly}(c), the  maximum value is found around 5\,K and it slightly depends on a frequency. The behavior below the peak is affected by an applied magnetic field.
Although more susceptible to extrinsic effects, the corresponding attenuation displayed in
Figure~\ref{fig:CuCNanomaly}(d) is in accord with the sound velocity.
Neither a charge-lattice coupling nor a classical magnetoelastic coupling appear  appropriate  to  render  an  account  of  these  elastic anomalies. On the one hand, vibrational spectroscopy could prove the absence of charge order and any variation around $T^* \approx 6$~K \cite{Sedlmeier12}. On the other hand, classical model of the magnetoelastic coupling cannot  predict a  softening  peak  located  near  6~K,  when  the exchange interaction $J$ is 40 times larger at 250 K. Relaxation effects  were  also  discarded  because  of  inconsistencies  with the frequency dependence. There are attempts to explain the frequency-dependent velocity softening observed in ultrasound at low temperatures to a  spinon-phonon coupling; this effect is reduced below $T^*$ due to a pairing instability transition of the spinons \cite{Poirier14}.

Below $T\approx 100$~K, numerous physical properties in both the spin and charge
sectors evidence some anomalous behavior that infers an exotic
charge-spin coupling and the possibility that it may play a
pertinent role in the formation of the quantum spin liquid at low temperatures.
Microwave investigations reveal an anomaly of the dielectric function
around $T^*=6$~K that depends on frequency and power as well as on external magnetic field \cite{Poirier12a}. Even more surprising is the Curie-Weiss behavior observed in the audio- and radio-frequency dielectric constant below 60\,K;  $\epsilon_1(T) \propto~ $C$/(T-T_C)^{-1}$, where $C$ is the Curie constant and $T_C$ was fixed to 6\,K in line with previous observations \cite{Abdel10}; see Figure~\ref{fig:kappaQSLdieltemp} for illustration.
In order to verify this assumption, Pinteric {\it et al.} worked with both $C$ and $T_C$ as free fit parameters and found that the Curie-Weiss parameters differ for the in-plane and out-of-plane crystallographic directions and for single crystals of different syntheses \cite{Pinteric14}. This stresses the involvement of the lattice with anionic-cationic layers and the effect of disorder.

Abdel-Jawad {\it et al.} claimed that the permittivity of \etcn\ exhibits a tiny anomaly around $T_C\approx  6 K$ that is almost independent on frequency; the absence of any remnant polarization was interpreted as an indication of antiferroelectric ordering of electric dipoles on dimers.
Recent dielectric measurements by R{\"o}sslhuber {\it et al.}, however, cast doubt on this interpretation, as
with increasing hydrostatic pressure as well as by chemical substitution, a low-temperature feature evolves and grows rapidly that could be related to percolative effects due to spatial inhomogeneities when the first order phase transition is approached
\cite{Pustogow19,Rosslhuber19,Saito20}, as discussed in Sec.~\ref{sec:QSLMott}. Here no anomaly is found around 6~K. We should also note that
the relaxor dielectric anomaly observed in all quantum-spin liquid candidates at temperatures below about 50~K cannot simply be assigned to electric dipoles; the absence of sizeable static electric dipoles has been demonstrated by overall vibrational, dielectric and structural studies thereby eliminating also any connection to the spin-liquid state. However, this conclusion may  not discard fluctuating dipoles and the charge-spin coupling in the quantum-spin-liquid formation since fast charge oscillations are observed. We come back to this topic in Section~\ref{sec:randomnessQSL} and discuss it at length in Sections \ref{sec:quantumelectricdipoles} and \ref{sec:QSL+afm}.

\subsection{Valence-bond solid}
\label{sec:magneticfieldQSL}
\label{sec:VBS}
Since magnetic field is expected to affect the spin-arrangement in a quantum spin liquid \cite{Zhou17},
the application of an external $H$-field was considered for most experiments on \etcn .
Hence it is surprising that the influence reported for common laboratory scale magnetic fields up to 8 - 14~T is negligible, e.g.\ Figure~\ref{fig:CuCNanomaly}.
Low-temperature NMR measurements reveal an anomalous field-dependent spectral broadening
of the $^{13}$C line for fields along the $a^\star$-axis ({\it i.e.} perpendicular to the planes) that is attributable to spatially nonuniform staggered magnetization induced in the spin liquid under magnetic fields. Nonmagnetic impurities also cause a broadening of the NMR-line \cite{Gregor09}.
The thermal conductivity $\kappa$ in the low-temperature regime is slightly enhanced for fields above 4~T in the $a$-direction, which Yamashita {\it et. al.} interpreted as the closing of a small spin gap in the quantum-spin-liquid state \cite{Yamashita09,Yamashita12}. A negligible field effect in \dmit\ quantum spin liquid was recently reported by two independent groups \cite{Bourgeois-Hope19,Li19}, in contrast to previous measurements \cite{Yamashita10}. The absence of reproducible experimental data on the thermal conductivity $\kappa(T)$ represents a challenge to understanding quantum-spin-liquid nature in organic solids.

Based on  muon spin relaxation ($\mu$SR) measurements, a macroscopic phase separation was claimed in the quantum spin liquid \etcn\ below $T=0.3$~K \cite{Nakajima12}. Already at zero field two phases are identified by different spin dynamics; but with increasing magnetic field the difference stabilizes and observed as static inhomogeneity.
Similar investigations have been conducted by Pratt {\it et al.} down to mK~temperatures \cite{Pratt11}.
They report that applying a small magnetic field of 5~mT to \etcn\ at very low temperatures produces a quantum phase transition between the spin-liquid phase and an antiferromagnetic phase with a strongly suppressed moment.
This is described as Bose–Einstein condensation of spinon excitations with an extremely small spin gap of 3.5~mK.
A weak-moment  antiferromagnetic phase dominates the low-temperature phase diagram, with several subphases depending on magnetic field and temperature.
At higher fields, a second transition is found that suggests a threshold for deconfinement of the spinon excitations. Several of these sub-phases also show up in temperature and field-dependent ESR experiments \cite{Miksch20}.

Electron spin resonance (ESR) has the advantage to most other spectroscopic methods that it actually uses the electron spins to directly probe the magnetic properties of the quantum-spin-liquid state.
In line with first indications observed previously \cite{Komatsu96,Padmalekha15},
comprehensive ESR experiments conducted in a large frequency and temperature range recently \cite{Miksch20}
\begin{figure}[h]
  \centering
  \includegraphics[width=1\columnwidth]{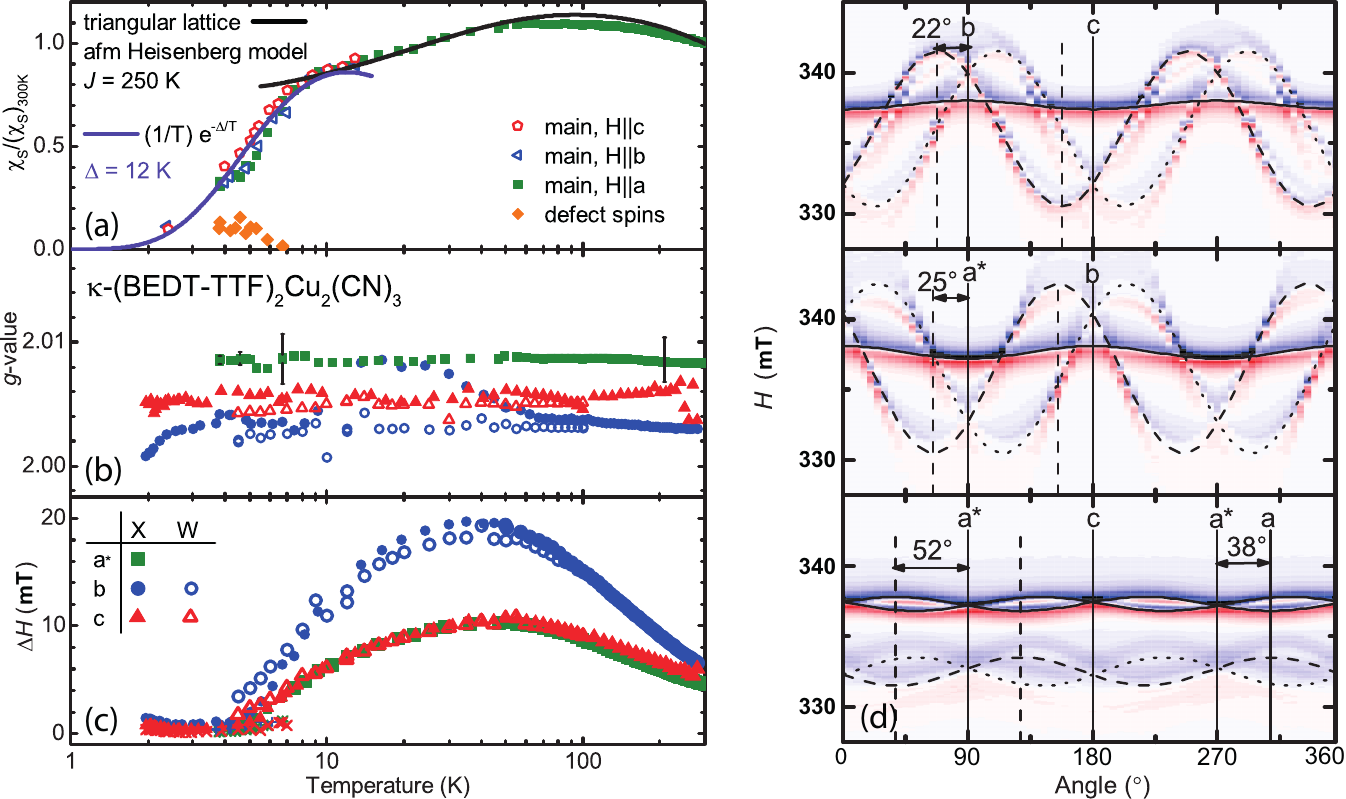}
      \caption{\label{fig:ESR}
Temperature dependence of the (a)~spin susceptibility $\chi_S$, (b)~$g$-value, and (c)~line width $\Delta H$ obtained from fits to the ESR spectra measured in the three directions ${H}\parallel a^*, b, c$ for \etcn\ single crystals.  The experiments have been performed in the X-band (9.5~GHz; solid symbols) and the W-band (95~GHz; open symbols).  The spin susceptibility is described by an antiferromagnetic Heisenberg model on a triangular lattice at elevated  temperatures. Below $T^*=6$~K an exponential decay of the main signal evidences the opening of a spin gap $\Delta = 12$~K. Below $T^*$ we notice the simultaneous appearance of the lone-spin component: The orange diamonds corresponds to the defect spins; the green $\star$, blue $+$, and red $\times$ indicate the respective line width.
The $g$-value does not exhibit any change with temperature within the experimental uncertainty estimated as 5\% of the line width.
The line width $\Delta H(T)$ is basically independent of the measurement frequency and field; it is largest along the $b$-direction. The minimum of $\Delta H(T)$ around $T \approx 3$~K coincides with features that can be identified in specific heat data at the same temperature \cite{Yamashita08,Manna10}.  Note the logarithmic temperature scale.
(d)~Anisotropy of the X-band ESR resonance field of \etcn\ measured at $T=2$~K.
Besides the main signal (solid line) from the BEDT-TTF radical, additional lines (dashed lines) appear at low-temperatures. The main line has its maximum along the crystallographic $b$-direction whereas the maxima of the defect signals are shifted by an angle of $\pm 22^\circ$ in the $bc$-plane. Similar shifts
can be identified for the other directions: $\pm 25^\circ$ in the $a^*b$-plane $\pm 38^\circ$ in the $a^*c$-plane. The mirror-image doubling of the signal corresponds to the two distinct orientations of the molecules that occur due to stacking faults.  (data from  \cite{Miksch20}).
}
\end{figure}
give compelling evidence that the spin-liquid state of \etcn\ is modified or even vanishes at $T^*\approx 6$~K.
In the high-temperature region, the magnetic properties can be described by an antiferromagnetic Heisenberg model on a triangular lattice, yielding a coupling of $J=250$~K in accord with previous estimates \cite{Shimizu03} listed in Table~\ref{tab:1}. Figure~\ref{fig:ESR}(b) shows that the $g$-values do not depend on temperature and frequency.
Below $T^*$ the signal becomes very narrow, and an additional component appears. Most important, the main signal completely vanishes upon further cooling, as plotted in Figure~\ref{fig:ESR}(a). In other words, the spin susceptibility $\chi_S(T)$ drops in an exponential way due to  opening of a spin gap $\Delta \approx 12$~K in the excitation spectrum.
This is explained by the formation of valence bonds (sketched in Figure~\ref{fig:RVB}) causing a singlet state; by spin-phonon coupling the lattice is affected like in spin-Peierls systems \cite{Dumm00}, comprising the observations by lattice expansion and ultrasound, depicted in Figure~\ref{fig:CuCNanomaly}.
Similar conclusions can be drawn from the NMR spin-lattice relaxation rate $(T_1T)^{-1}$, which is proportional to the susceptibility: a drop in the spin susceptibility is seen below $T^*$ before
it strongly increases due to defects \cite{Shimizu03,Shimizu06}.

The minor ESR signal shifts to lower fields and can be clearly identified as a separate maximum below $T=2.5$~K.  In Figure~\ref{fig:ESR}(d) the angular dependence of the ESR signal is displayed, revealing two low-temperature features with a large anisotropy of about 10~mT, which are symmetrically shifted by $\pm 22^{\circ}$ with respect to the main signal. A similar behavior has been reported for \etcl, due to crystallographically inequivalent adjacent layers \cite{Antal09,Antal12c,Antal15}, depicted in Figure~\ref{fig:structure_kappa}. For the monoclinic symmetry of \etcn, this explanation does not hold; hence stacking faults in the crystal are suggested to cause the doubling of the line. As the temperature is reduced further to the mK~range, this ESR line separates even more. From its pronounced dependence on the  external magnetic field and the particular anisotropy shown in Figure~\ref{fig:ESR}(d), the minor ESR feature is assigned to defect spins, which can interact with local magnetic moments of the Cu$^{2+}$ sites in the otherwise non-magnetic Cu$^{+}$ anion layers \cite{Komatsu96,Padmalekha15} by dipolar coupling eventually forming some weakly coupled antiferromagnetic state \cite{Miksch20}. Its strength
might vary from sample to sample, but the effect in general is robust.
\begin{figure}[h]
  \centering
      \includegraphics[width=0.8\columnwidth]{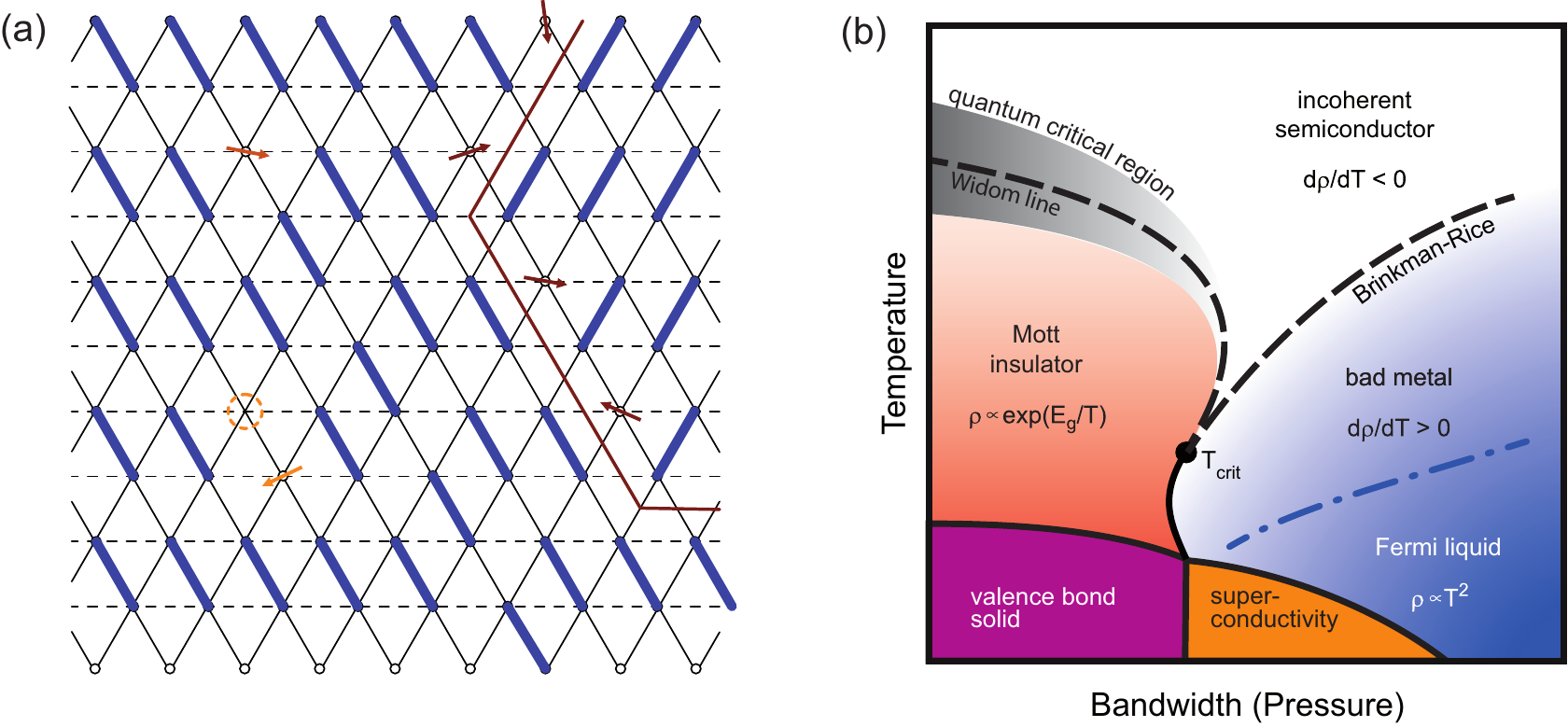}
    \caption{\label{fig:RVB}
(a)~Quasi-static valence-bond solid on an anisotropic triangular lattice, as suggested for \etcn\ \cite{Riedl19,Kawamura19,Miksch20}. The spin singlets indicated in blue preferably form along the ($b\pm c$)-directions; domain walls between the valence-bond patterns indicated in red.
Local valence-bond defects can occur at various kinds of grain boundaries; a local spin $\frac{1}{2}$
can also be caused by the breaking of a singlet bond probably due to an anion layer vacancy, emphasized by the red circle.
(b)~The valence bond solid state (violet) will affect the low-temperature phase boundaries sketched in Figure~\ref{fig:schematicPhaseDiagrams}. In the case of \etcn\ the insulating phase is next to the superconducting state.
\label{fig:VBS+phasediagram}}
\end{figure}
The findings are in agreement with the $\mu$SR experiments by Pratt {\it et al.} \cite{Pratt11} and the suggestion of Riedl {\it et al.} analyzing the low-temperature magnetic torque measurements \cite{Isono16,Riedl19}.

Figure~\ref{fig:RVB}(a) summarizes the present understanding of the ground state in \etcn\ \cite{Riedl19,Kawamura19,Miksch20}: below $T^*$, spin singlets form a valence-bond solid accompanied by a lattice distortion, similar to a spin-Peierls coupling.
There remain numerous orphan spins, {\it i.e.} local magnetic moments that might interact by dipolar coupling with each other, but also with magnetic moments in the anions layers.
Although there exists no long-range magnetic order, the valence-bond solid does influence the low-temperature phase diagram compared to the spin-less case sketched in Figure~\ref{fig:schematicPhaseDiagrams}(b).
In Figure~\ref{fig:VBS+phasediagram}(b) we suggest a revised phase diagram resembling the
quantum spin liquid candidate \etcn\ where the transition to the superconducting state is assumed vertical since no experimental
data are available due to low temperatures \cite{Kurosaki05,Lohle18,Furukawa18}.

It is crucial to extend this sort of low-temperature ESR investigations to other spin-liquid compounds in order to conclude whether the case of \etcn\ is particular. The effect of Cu$^{2+}$ ions, for instance
should not be an issue in the \agcn\ and \dmit\ salts. However, the formation of a disordered valence-bond solid implying the opening of a spin gap could be a general scenario \cite{Shimizu07,Tamura09,Kimchi18,Clay19}. In this regard, the recently synthesized compound
$\kappa$-(BEDT-TTF)$_2$Cu[Au(CN)$_2$]Cl is of interest as it possesses no disorder in the anions \cite{Tomeno20}.

\subsection{Randomness}
\label{sec:randomnessQSL}
This brings us to the long debate whether the quantum-spin-liquid properties in the organic compounds originate solely from geometrical frustration or whether some sort of randomness or disorder might be crucial \cite{Watanabe14}. The interplay between mutual interaction and quenched disorder is a fundamental issue not only in condensed matter research but also in the physics of cold atoms. Recently this was further investigated theoretically in three and two dimensions  \cite{Watanabe14,Shimokawa15,Liu18,Wu19,Uematsu19,Kawamura19}.

Within randomness-induced quantum-spin-liquid models,
disorder supports the cooperative action of quantum fluctuations and triangular frustration in the stabilization of
a quantum-spin-liquid state \cite{Watanabe14, Kawamura19}.
This is important since the frustrated triangular lattices in two dimensions are unable to destroy the long-range magnetic order on their own \cite{Huse88,Bernu92,Capriotti99}.
Randomness helps constituting the quantum spin liquid in two-dimensional organics; its nature is two-fold: 
(i) The intrinsic randomness originates in the charge sector and acts via charge-spin coupling;
Hotta \cite{Hotta10} suggests that quantum electric dipoles are formed on the dimers, which interact with each other and thus modify the exchange coupling $J$ between the spins on the dimers, crucial for the formation of the spin-liquid state. Experimentally, only inhomogeneous charge fluctuations are observed \cite{Sedlmeier12, Yakushi15, Nakamura17}, which indicate fluctuating electric dimers at the rate of about 0.1\,THz. We discuss this topic in Section \ref{sec:quantumelectricdipoles}.
(ii) The extrinsic quenched randomness comes into play as a result of disorder in the anion layers [Figures~\ref{fig:structure_kappa}(h) and (i)], and
was revealed to be inherent to \etcn\ and \agcn\ \cite{Dressel16,Pinteric16}; this aspect is crucial for understanding the dc transport and electrodynamic properties of these materials in general and is discussed in Sections \ref{sec:randomness} and \ref{sec:QSL+afm}.
Hence, the quenched randomness originating in the anions might provide a spatially random effective interaction to the spin degrees of freedom.

\etcl\ is an antiferromagnetically ordered Mott insulator at low temperatures with a coupling $t^{\prime}/t$ far from frustration. Extended x-ray irradiation of 500~h (0.5 MGy/h) introduces sufficiently strong disorder  to suppress the antiferromagnetic state in \etcl.
It was suggested  \cite{Furukawa15b,Urai20} that the system evolves towards a quantum spin liquid, because
the antiferromagnetic ordering observed in the pristine crystal disappears; no
spin freezing, spin gap, nor critical slowing down are observed by $^1$H-NMR experiments, instead
gapless spinon excitations emerge.

\begin{figure}
  \centering
  \includegraphics[width=0.4\columnwidth]{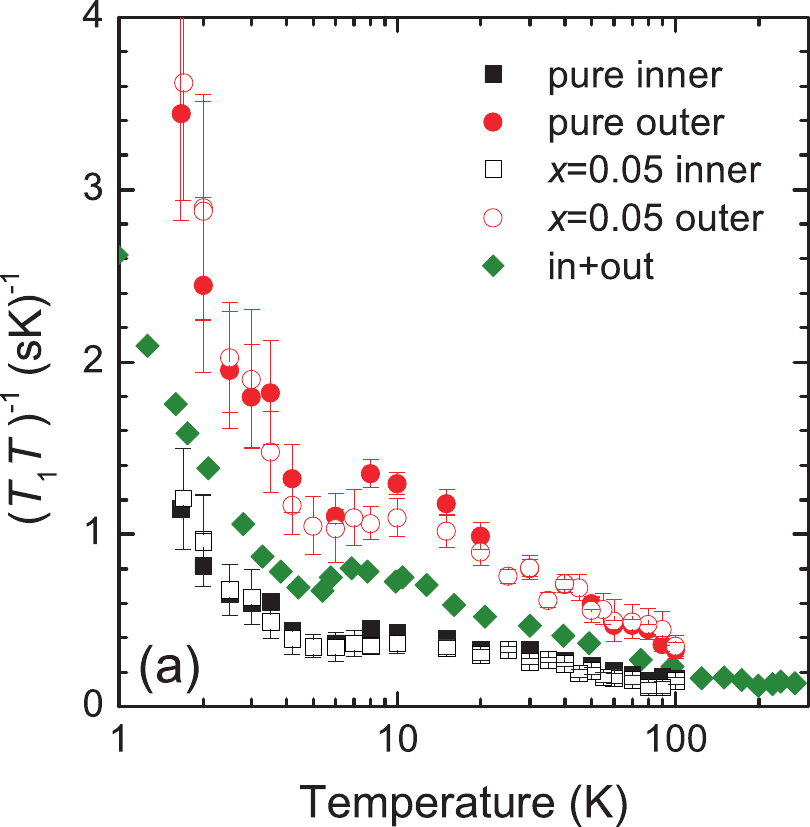}~~~~
  \includegraphics[width=0.47\columnwidth]{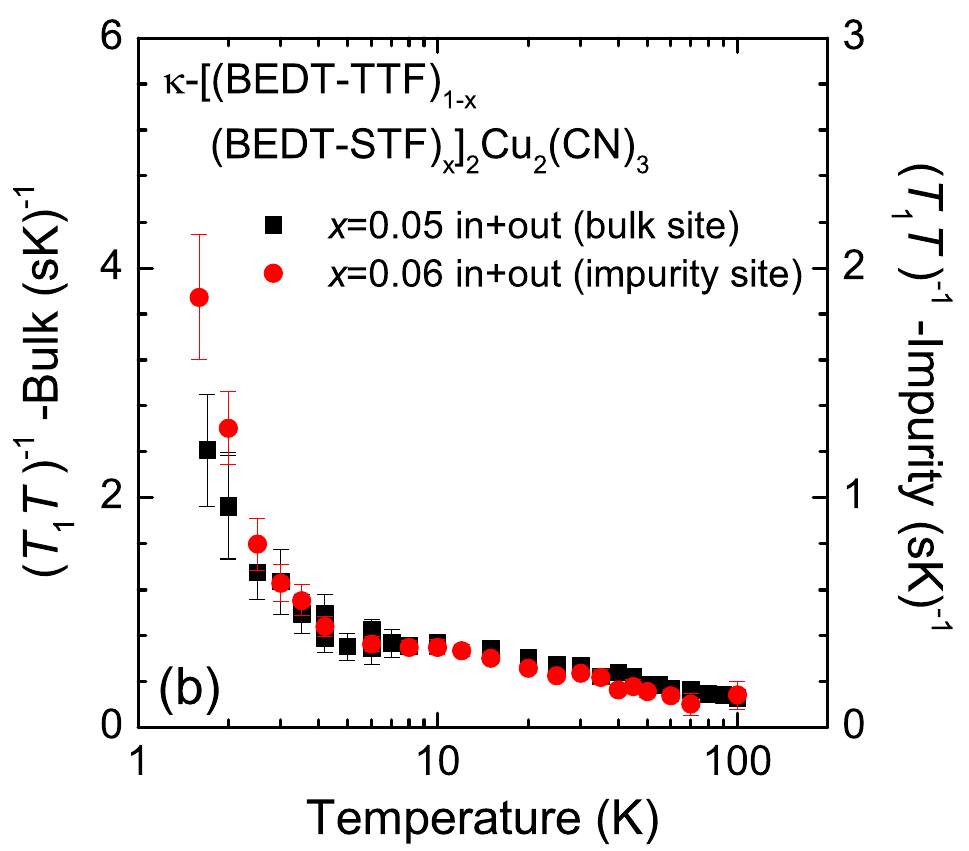}
    \caption{\label{fig:random3}\label{fig:NMR1}\label{fig:NMR2}
(a) Temperature dependence of $(T_1 T)^{-1}$ for the pure and $x=0.05$ crystals of \stfcn.
The green diamonds indicate data of a pure samples without distinguishing inner and outer sites \cite{Saito18,Shimizu06}.
(b) Mean value of $(T_1 T)^{-1}$ of the associated inner and outer sites for bulk and impurity sites.
The results of the bulk site refer to the left axis, and the results of the impurity site are plotted with the right axis \cite{Saito18}.
}
\end{figure}

Saito {\it et al.} went the other way by slightly modifying the BEDT-TTF molecules with partially replacing sulfur by selenium in the inner rings \cite{Saito18}.
As discussed in Section \ref{sec:QSLMott}, BEDT-STF substitution enlarges the bandwidth and drives the system across the Mott transition [Figure~\ref{fig:dcPressureAlloy}(b)]. However, the magnetic characteristics of \stfcn\ with $x=0.05$ are quantitatively similar to those of the pristine crystals (Figure \ref{fig:random3}). Moreover, magnetically the substituted sites are also the same as in the bulk.
NMR spectra from the impurity site suggest a decrease in local spin susceptibility and that no staggered moments are induced. Thus, the results indicate that the static and dynamic susceptibilities do not change, even at very low temperatures. This led the authors to the conclusion that disorder might already play a role in the pure compound
and that the observed magnetic quantum-spin-liquid properties are not solely caused by geometrical frustration.

This idea was extended to $\lambda$-(BEDT-STF)$_2$GaCl$_4$ \cite{Saito19}, where also site-selective NMR was utilized to investigate the non-magnetic insulating phase of the stripe lattice system consisting of triangular and square tilings.
As the temperature decreases antiferromagnetic spin fluctuations develop. The spin-lattice relaxation rate $T_1^{-1}$
is strongly enhanced, but saturates below 3.5 K with no indications of long-rang magnetic order.
$\lambda$-(BEDT-STF)$_2$GaCl$_4$  is a disordered electronic system where a novel quantum disordered state is realized.  The non-magnetic ground state is discussed in terms of geometrical frustration,  disorder, and  the quantum critical point between  the  antiferromagnetic  phase  and  the  spin  gap phase.

In a new class of hybrid organic crystals [EDT-TTF-CONH$_2$]$_2$[BABCO] Szirmai {\it et al.} \cite{Szirmai20} suggested that a quantum-spin-liquid state is introduced not by frustration -- which is only $t : t^{\prime} : t^{\prime\prime} = 1.00 : 0.75 : 060$ -- but mainly by disorder that originates in  the molecular rotor BABCO$^-$ \cite{Lemouchi12,Lemouchi13}. Despite the rather strong coupling of $J \approx 314$~K, the compound does not show indications of a magnetic phase transition down to 20~mK.
From a variety of advanced magnetic probes (multi-frequency ESR, $^1$H-NMR, zero-field $\mu$SR), the authors conclude that intrinsic randomness due to the configuration of the frozen BCO Brownian rotors causes subtle disorder potential that suppresses magnetic order. The $^1$H-NMR relaxation rate $T_1^{-1}$ exhibits a weak increase below 10~K following an unusual power law, similar to \etcn\ and other quantum-spin-liquid candidates, cf. Figures~\ref{fig:NMR2} and \ref{fig:HgCl_NMR1}. The electron spin relaxation rate of $\mu$SR spectra exhibits only a slow change with temperatures without a critical slowing down in the temperature range from 0.5~K down to 20~mK, which is a universal characteristic of several quantum-spin-liquid compounds \cite{Kermarrec11,Pratt11,Fak12,Clark13}.

Dynamic magnetic inhomogeneities are indicated by the strong field dependence of $T_1^{-1}$ observed at around 1~K in \etcn\ and \agcn\ \cite{Shimizu03,Shimizu06,Shimizu16}. Disorder is also identified in magnetically coupled defect spins linked to the frustrated BEDT-TTF organic lattice discussed in previous Section \ref{sec:magneticfieldQSL}.
Defect spins coupled to kagome and triangular lattices are also observed in inorganic quantum spin liquids herbertsmithite ZnCu$_3$(OH)$_6$Cl$_2${} \cite{Zorko17,Khuntia20} and 1T-TaS$_2${} \cite{Klanjsek17}. The observation of disorder outside frustrated lattices might present a common feature of quantum-spin-liquid candidates; further studies are needed to clarify the interplay between the defect spins and the inherent frustrated lattices, as well as whether the presence of this disorder is related to the establishment of quantum-spin-liquid states.

\section{Coupling of Quantum Electric and Magnetic Dipoles}
    \label{sec:quantumdisorder}
The hallmark of the spin-liquid state is the absence of long-range magnetic order and the presence of persistent spin dynamics down to $T=0$.
From a quantum mechanical point of view there is no unique ground state, but a superposition of numerous equi\-valent solutions.
The random variation in time leads to tiny fluctuations.
This implies that the electronic distribution is neither solely determined nor static.
In Chapter~\ref{sec:Frustration}, we discussed the example of quantum spin liquids, where strongly correlated spins residing on the spin-frustrated lattice fluctuate
and thus prevent an antiferromagnetic ordering even at absolute zero temperature.
Likewise, a quantum electric dipole liquid are expected to emerge in frustrated dielectrics with fluctuating electric dipoles residing on a triangular lattice in two dimensions and preventing long-range order \cite{Shen16,Powell18}.
In this regime these aspects are not isolated but a certain coupling can be expected. In other words, the development of some magnetic order
might affect the dielectric properties and vice versa; in general this also acts on the crystal lattice via magnetoelastic and piezoelectric coupling. While multiferroicity in inorganic compounds has drawn significant attention for the last twenty years \cite{Spaldin05,VandenBrinkKhomskii08,Dong15,Spaldin20}, similar observations in organic charge-transfer compounds remain elusive 
\cite{Lunkenheimer12,Lang14,Pinteric18}.

In the subsequent Sections, we first give an overview of relevant theoretical studies concerning the interplay between spin and charges and of the experimental efforts to disclose electric dipoles in quantum spin liquids \etcn\ and \agcn\, and in antiferromagnet \etcl\ (Sections \ref{sec:quantumelectricdipoles} and \ref{sec:QSL+afm}). We continue with a unique example of a quantum liquid composed of both magnetic and electric dipoles in \cat\ (Section \ref{sec:cat}), followed by a presentation of a quantum electric dipole liquid with glassy signatures and an exotic coupling with magnetic degrees of freedom in \hgbr{}.


\subsection{Quantum electric dipoles}
\label{sec:quantumelectricdipoles}

\begin{figure}
  \centering
  \includegraphics[width=0.9\columnwidth]{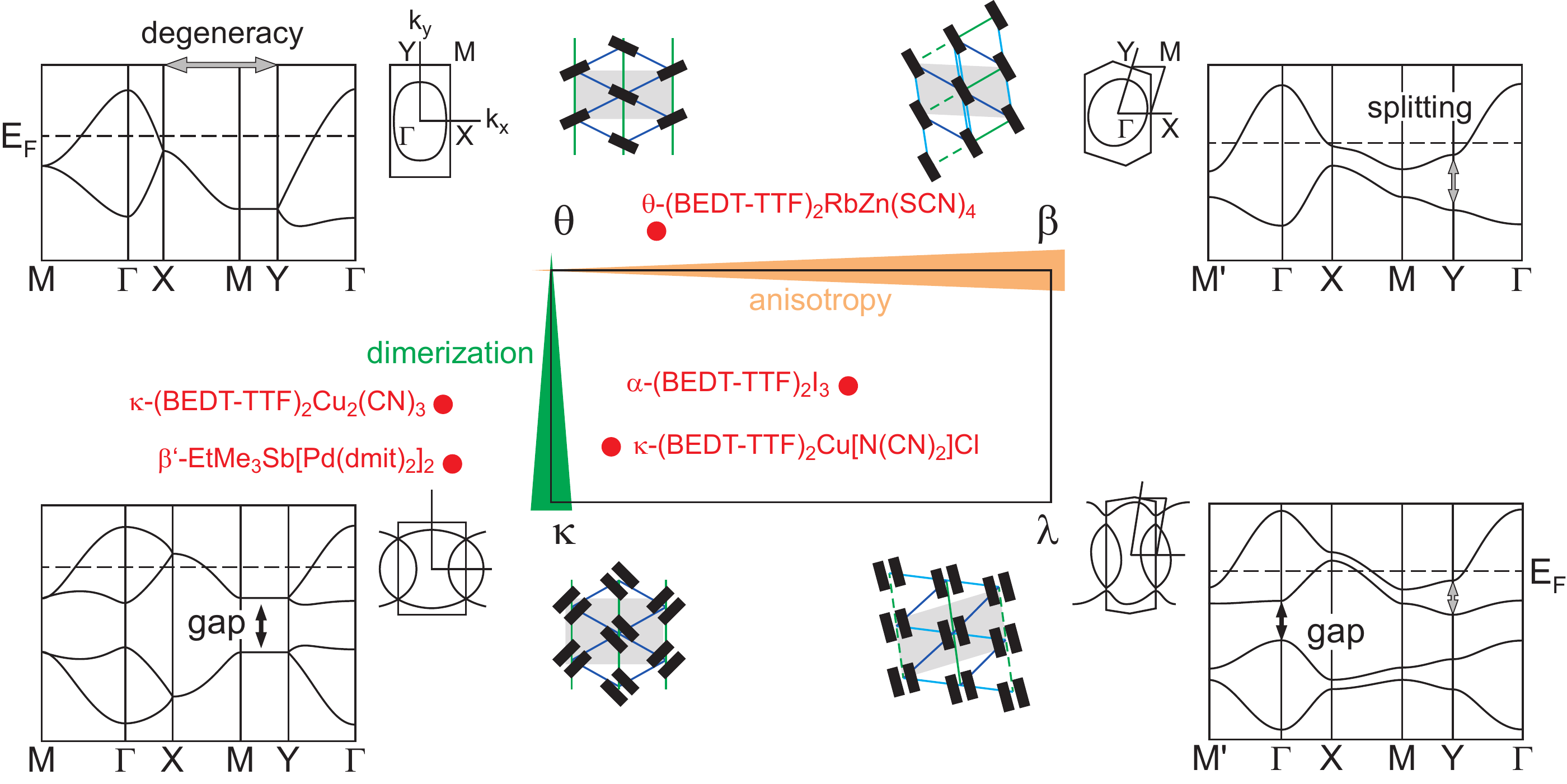}
    \caption{\label{fig:Hotta1}
The classification of quarter-filled electronic system in the weak-coupling
regime. The $\theta$- ($\alpha$-) and $\beta$-type with weak or no dimerization form a triangular lattice in unit
of molecule, whereas those with strong dimerization, $\kappa$- and $\lambda$-type do so in unit of dimer. The distortion of the ideal triangular lattice in terms of unequal bonds is described by the anisotropy (after \cite{Hotta03,Hotta12}). }
\end{figure}

Due to the $A_2B$ stoichiometry of most BEDT-TTF salts with monovalent anions $B^-$, the systems are supposed to possess a quarter-filled conduction band; dimerized structures, such as the $\kappa$- or $\lambda$-phases (Figure~\ref{fig:structure_general}), however, lead to half-filled bands. In the latter case, on-site Coulomb repulsion $U$ dominates the electronic interactions, and the compounds serve as prime models for Mott physics.
subject of Chapter~\ref{sec:MottTransition}. In the former case, however, inter-site Coulomb interaction $V$ cannot be neglected; the $\alpha$-, $\beta$- and $\theta$- compounds (Figure~\ref{fig:structure_general}) are the most important and heavily studied ones.
A general  classification of these quasi-two dimensional organic conductors is given
by Hotta's minimal model based on an anisotropic triangular lattice \cite{Hotta03,Seo04,Hotta12}.
The main factors that determine the band structure are dimerization and anisotropy, {\it i.e.} geometrical frustration, as illustrated in Figure~\ref{fig:Hotta1}. The inter-molecular Coulomb interaction $V$ governs the degree of geometrical frustration in the limit of strong dimerization; the dimerization is given by strength of interaction $t_d$ within the molecular dimer.

\begin{figure}
	\centering\includegraphics[clip,width=0.35\columnwidth]{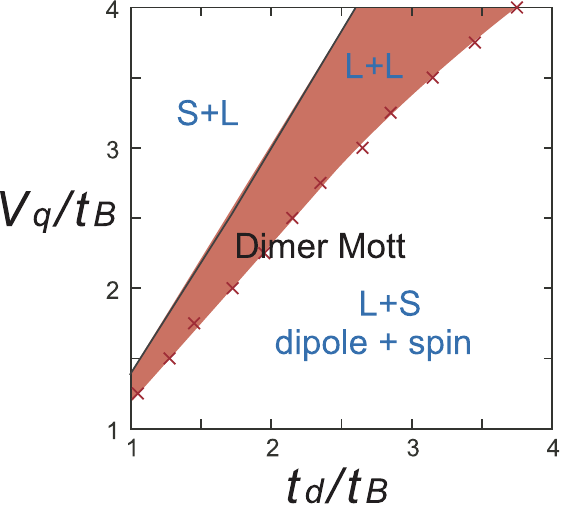}
	\caption{Phase diagram calculated within an effective dipolar-spin model by Hotta \cite{Hotta10}. $t_B$ and $t_d$ are the interdimer and intradimer hopping integrals, and $V_q$ is the intersite Coulomb interaction. S+L, L+S and L+L denote a charge ordered spin liquid, a spin-ordered charge liquid, and a charge-spin liquid phase, respectively  (after \cite{Hotta10}).}
	\label{fig:DSliquid}
\end{figure}

When dimerization is negligibly small but finite, the charge degrees of freedom are governed by $V$ and the transfer integral  $t$ along the
nearest neighbor bonds. In this limit, the system tends to a charge-ordered insulator, as discussed in Chapter~\ref{sec:ChargeOrder}. Frustration weakens the inter-site repulsion and finally destroys charge order; an exotic metallic state exists with limited charge mobility. The compounds are paramagnetic with no tendency towards magnetic order.
As the dimerization $t_d$ develops, the electron becomes confined within the dimer (dimer Mott insulator, cf.\ Chapter~\ref{sec:MottTransition})
but still delocalized, providing the possibility that dielectric degrees of freedom form a so-called `quantum electric dipole', which is subject to fluctuations and correlations depending on $t_d$ and $V$. The electric moments start to compete as frustration weakens $V$, leading to some unusual dielectric properties. Exchange coupling of the spins associated with those dielectric moments may give rise to some nontrivial magnetic behavior, as depicted in Figure~\ref{fig:DSliquid}.
Increasing $t_d$ suppresses the dielectric moments, while the spin degrees of freedom become more important. By weakening the interaction strength and approaching the insulator-metal transition (for instance, by reducing the effect of on-site repulsion $U/t$), spatial fluctuations of electrons develop and nontrivial long-range spin exchange interactions when the geometrical frustation becomes large. The long-range antiferromagnetic order may be destroyed and a quantum spin liquid state appears, as discussed in Chapter~\ref{sec:Frustration}.

\begin{figure}[h]
  \centering
  \includegraphics[width=0.7\columnwidth]{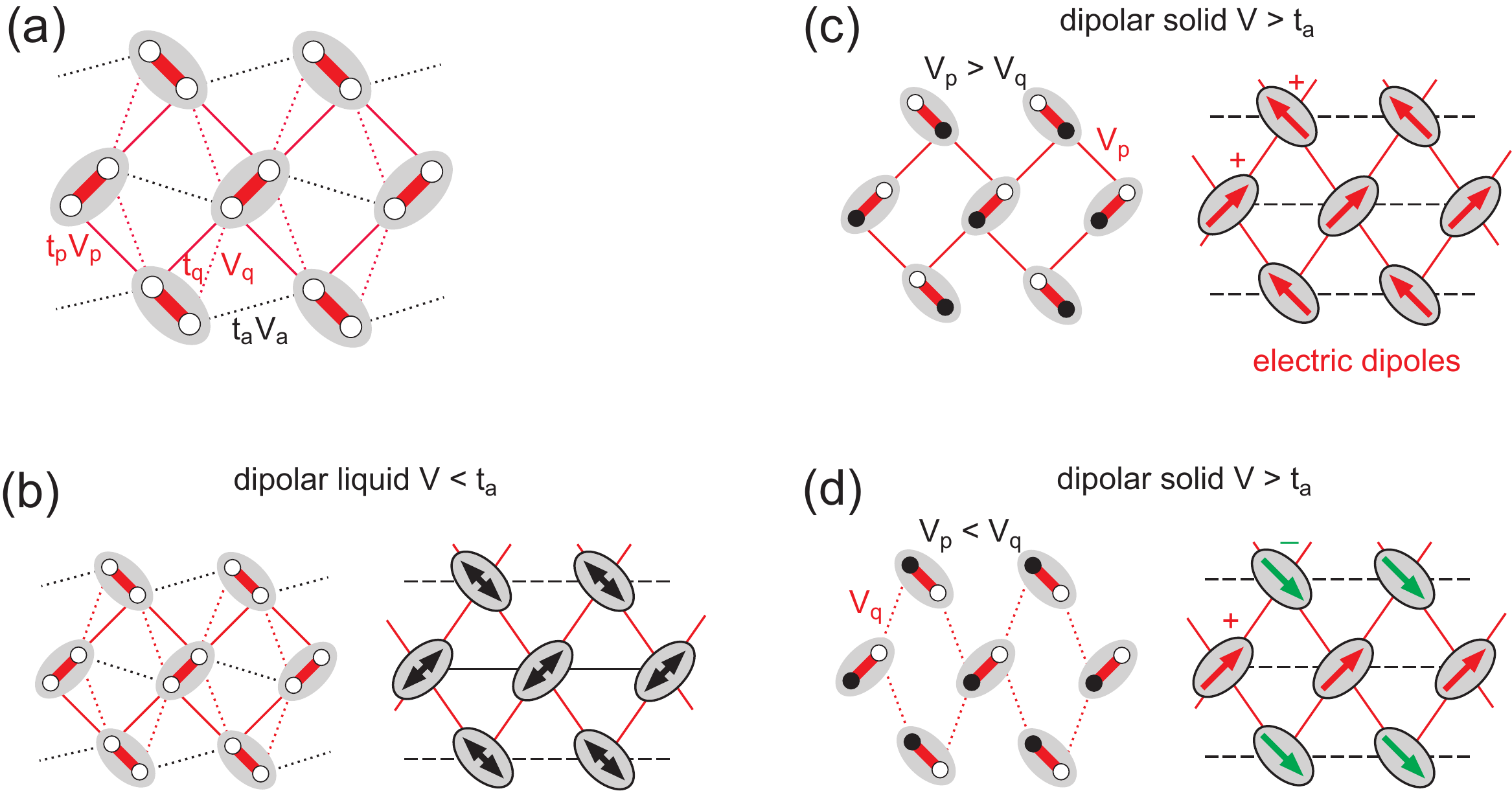}
    \caption{
(a)~In this $\kappa$-type lattice structure the molecules are presented by circles,
the shaded area and bold red line represent the dimers ($t_d$, $V_d$).
The interactions between the dimers are indicated by the overlap $t_q$ and $t_p$ and by
the Coulomb repulsion $V_q$ and $V_p$.
(b)~Unpolarized and (c,d) polarized configurations of quantum dipoles, depending on the
hierarchy of interactions.
Charges avoid neighboring alignment along the bond with strong interaction
(according to \cite{Hotta10}). }
\label{fig:Hotta3}
\end{figure}

Based on the extended Hubbard model with on-site and nearest-neighbor interaction $V$ that includes intra- as well as inter-dimer terms on a model lattice of $\kappa$ compounds, Hotta \cite{Hotta10} considered quantum electric dipoles as depicted in Figure~\ref{fig:Hotta3}.
In order to describe the observations on \etcn{}, she suggests that at low temperatures \etcn\ falls in the dipolar-spin liquid phase, where both spins and charges remain short-range ordered.
The dielectric anomaly that develops below 50~K in most of these charge transfer salts \cite{Pinteric99,Abdel10,Lunkenheimer12,Tomic13,Iguchi13,Pinteric14,Pinteric16} could be explained
under the condition that electric dipoles exist when the charge is unbalanced within the dimers.
A similar conclusion of ferroelectric charge order (dipolar order) is obtained by mean field and classical Monte Carlo methods \cite{Naka10}.

Very recently Powell and collaborators performed a comprehensive study on several $\kappa$-(BEDT-TTF)$_2$$X$ salts combing first-principle density functional calculations with empirical relationships for the Coulomb interactions \cite{Jacko20} in order to evaluate the model of Hotta \cite{Hotta10,Hotta12}, Naka and Ishihara \cite{Naka10} developed for the coupled dipolar and spin degrees of freedom.
The transverse-field in the Ising model is governed by the intradimer coupling $t_d$, which is
significantly smaller in \hgcl\ and $\kappa$-(d8-BE\-DT\--TTF)$_2$\-Cu\-[N\-(CN)$_{2}$]Br, compared to \etcl\ or \agcn. More important, however, is the more one-dimensional arrangement in the mercury-containing compounds compared to the others, which are in the quasi-two-dimensional limit.
We should note, however, that all these models do not include the underlying ionic lattice and coupling to phonons.

\begin{figure}
  \centering
  \includegraphics[width=0.7\columnwidth]{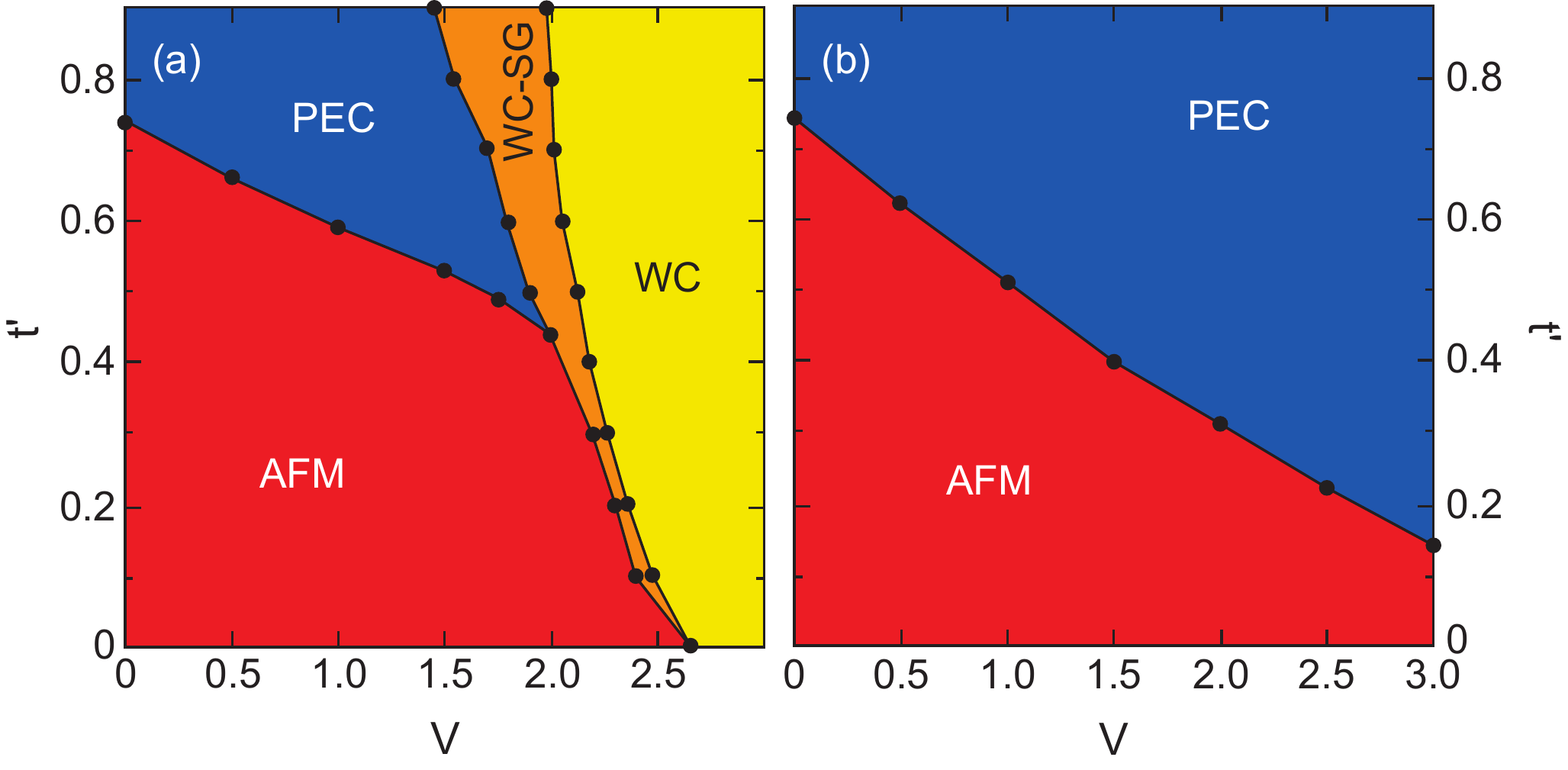}
    \caption{\label{fig:Clay2}
(a)~Ground state phase diagrams from self-consistent calculations by Dayal {\it et al.}
on periodic $4 \times 4$ lattice as a function of $t^{\prime}$ and $V=V_x=V_y$, $V^{\prime}=0$, for $U=6$, $\alpha =1.1$, $\beta=0.1$, and $K_{\alpha} = K_{\beta}=2.0$.
(b) Same as panel (a), but with $V=V_x=V_y=V^{\prime}$.
Here antiferromagnetic (AFM) and paired-electron crystal (PEC) phases are shown together with Wigner crystal (WC) and spin gap phase. Lines are guides to the eye.
(from \cite{Dayal11,Clay19}).  }
\end{figure}
An interplay of spins with charges and bonds
is pursued in the exact diagonalization study by Clay, Mazumdar and collaborators \cite{Li10,Dayal11,Clay12,Clay19} on the extended Hubbard model with both Holstein- and Peierls-type of electron-lattice couplings on the anisotropic triangular lattice.
The freezing of bonds and charges simultaneously with a spin-Peierls singlet formation
results in a so-called `paired-electron crystal', which exhibits moderate degrees of charge
disproportionation  and lattice displacement compared to those of the charge ordered state at larger $V$. The geometrical frustration completely replaces the charge ordered phases with the
novel paired-electron crystal. As shown above, geometrical frustration in $V$
destabilizes charge order against the dimer Mott insulator; however, compared to the pure dimer Mott case, the pair-electron crystal is more significantly stabilized, due to the spin degrees of freedom or to the spontaneous lattice modulation instead of the intrinsic dimerization. Figure~\ref{fig:Clay2} illustrates how electron-electron interaction affect the `paired electron crystal'. As $U$ increases moderately, the coupling $t^{\prime}$ also increases. The phase diagram depends critically on the form of
nearest neighbor interaction $V$. With $V_x = V_y= V$ but the diagonal term $V^{\prime}=0$, the Wigner crystal is found for sufficiently strong $V$, along with a narrow region, where it coexists with a spin gapped phase. Since most two-dimensional charge transfer salts have a tendency towards triangular lattices, the assumption $V^{\prime} = 0$ is not realistic. For $V_x =V_y = V^{\prime}$ the Wigner crystal is completely replaced by the paired electron crystal.

Fukuyama {\it et al.} considered the crossover from a quarter-filled system with charge-ordered ground state to the dimer Mott insulator due to strong dimerization \cite{Fukuyama17}. At high energy (in the range
of eV, {\it i.e.} optical frequencies) the dimer Mott insulator is stable, whereas at very low energy ($10^{-10}~{\rm eV} \approx  10$~kHz) charge order becomes dominant leading to extended domains
of different charge polarities. As a consequence, domain walls form in the system, giving rise to
the dielectric anomaly observed. This aspect is presented in full detail in Section~\ref{sec:COFerroelectricity}.

Despite considerable experimental efforts, no spectroscopic evidence has been obtained
for sizeable charge disproportionation in any of these $\kappa$-phase quantum spin liquid candidates and antiferromagnets \cite{Sedlmeier12,Pinteric16}.
There is a broad consensus that static charge order is less than 1\%. This is supported by the most recent x-ray diffraction measurements of \etcn\ and \agcn{} by Foury-Leylekian {\it et al.} \cite{Foury18,Foury20}. At the same time, the new data provide evidence for the symmetry breaking of the non-polar mean P2$_1/c$ structure thereby proving the presence of non-equivalent crystallographic sites and allowing the formation of electric dipoles. Interestingly, despite the fact that the structural signatures of symmetry breaking are not weak, the resulting static charge disproportionation is negligibly small and in the range of the error bar.
However, fluctuating electric dipoles might be concluded from dynamic charge fluctuations inferred from the charge-sensitive molecular-vibrational modes observed in Raman and infrared spectroscopy \cite{Yakushi15,Nakamura17}, which are broader than those of typical BEDT-TTF compounds; similar conclusion are drawn from NMR experiments. These fluctuating dipoles may show up in the microwave \cite{Poirier12a} and terahertz response \cite{Itoh13}, but the latter has been demonstrated to involve coupled anion-dimer vibrations instead \cite{Dressel16}. Lastly, these fast charge fluctuations cannot account for the dielectric response in the audio- and radio-frequency range. We discuss this topic in subsequent Section \ref{sec:QSL+afm}.

\subsection[Ferroelectric signatures in quantum spin liquid and afm states]{Ferroelectric-like signatures in quantum spin liquid and antiferromagnetic states}
\label{sec:QSL+afm}

\begin{figure}[h]
	\centering\includegraphics[width=0.6\columnwidth]{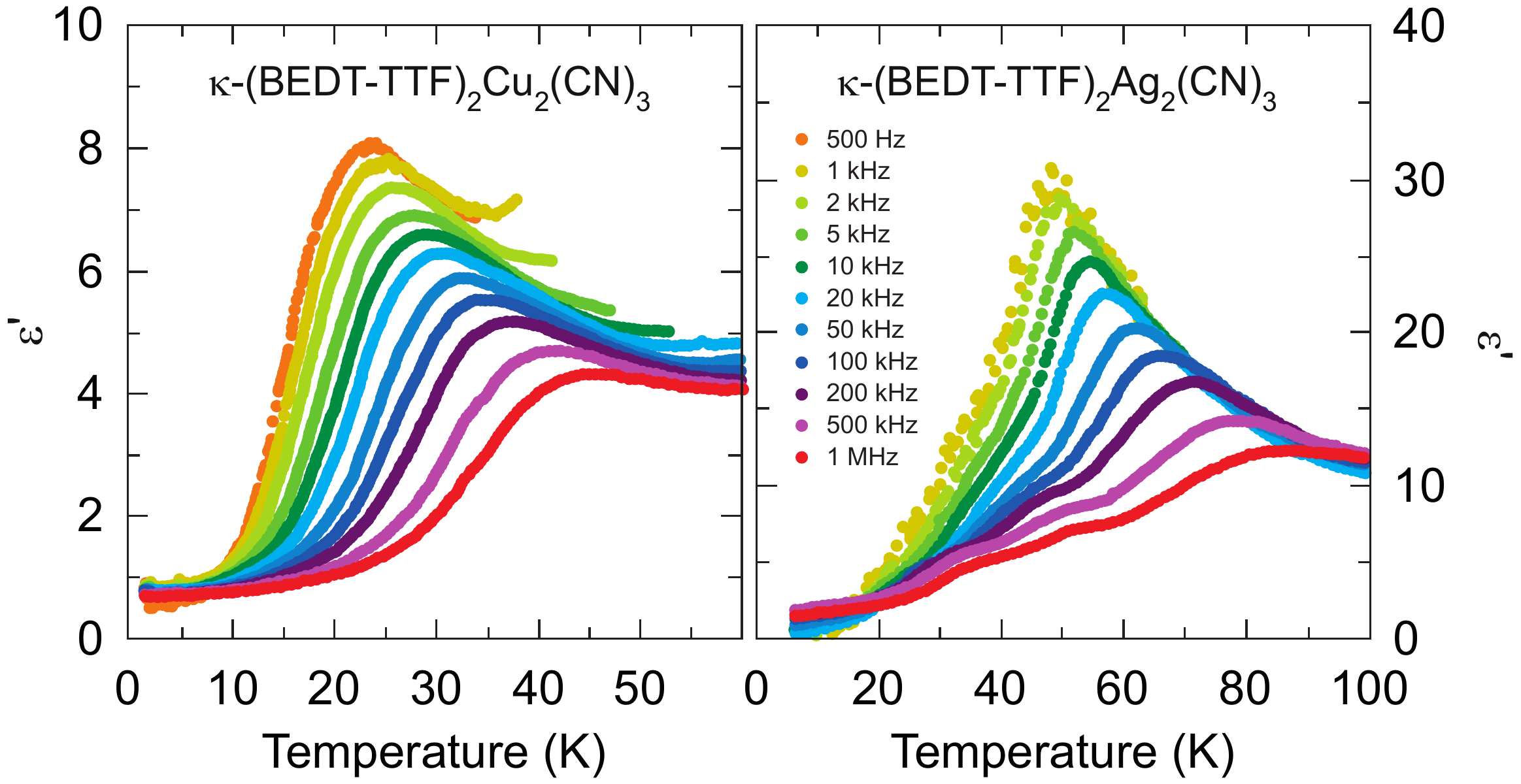}
	\caption{Temperature dependence of the real part of dielectric function $\varepsilon^{\prime}$ measured with the ac electric field applied ${E}\parallel a^\ast$ {\it i.e.} along the direction perpendicular to the molecular layers (a) of \etcn\ and (b) of \agcn\ (after \cite{Pinteric18}).
		\label{fig:kappaQSLdieltemp}}
\end{figure}

\begin{figure}
	\centering\includegraphics[width=0.9\columnwidth]{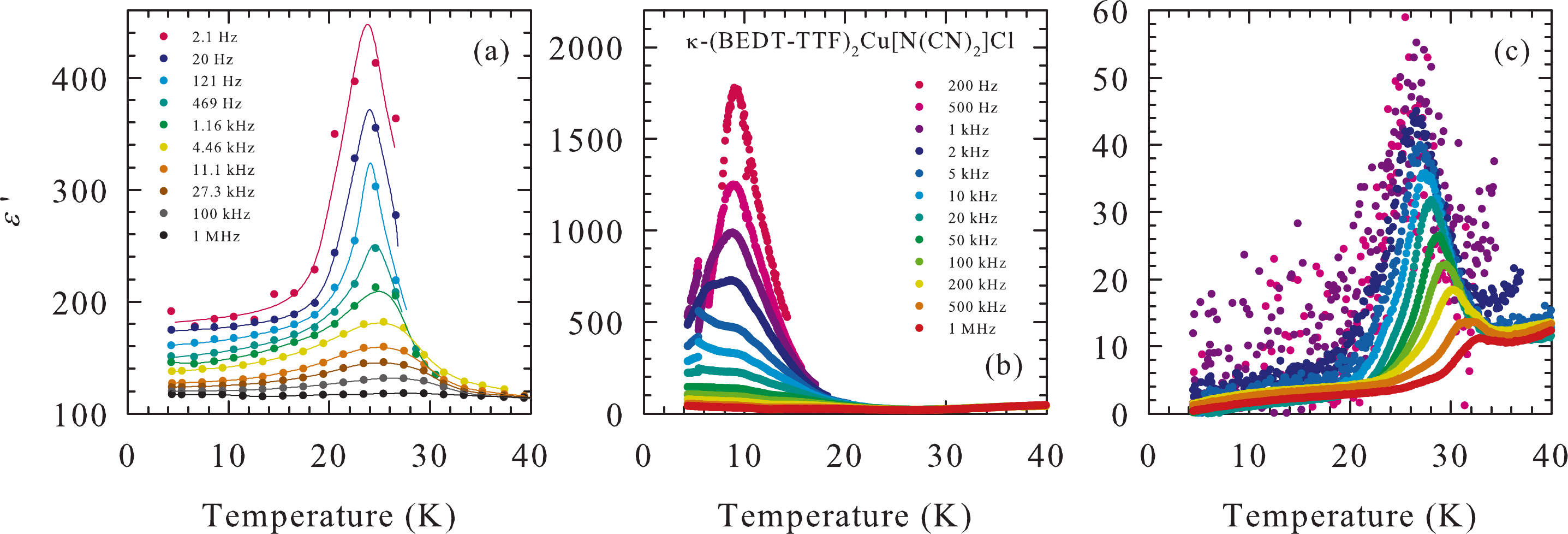}
	\caption{Temperature dependence of the real part of dielectric function $\varepsilon^{\prime}$ measured with the ac electric field applied ${E}\parallel b$, {\it i.e.} along the direction perpendicular to the molecular layers of three different single crystals of \etcl. Different behaviors are observed ranging from  (a) Curie-like at the magnetic transition $T_{N} \approx 25$~K suggesting multiferrocity, over (b) Curie-like at 8\,K to (c) relaxor-like  (after \cite{Pinteric18,Lunkenheimer12}).
		\label{fig:kappaCldieltemp}}
\end{figure}

An anomalous dielectric peak shows up as an ubiquitous property independent of the nature of the ground state in the spin sector \cite{Abdel10, Pinteric14, Pinteric16, Lazic18, Pinteric18, Lunkenheimer12}: in spin liquid compounds \etcn\ and \agcn\ (Figure~\ref{fig:kappaQSLdieltemp}), as well as in \etcl\ with antiferromagnetic ground state [Figure~\ref{fig:kappaCldieltemp}(c)].

The relaxor-type dielectric response suggests
a short-range ferroelectric-like order \cite{Abdel10, Abdel13}
and the existence of domains of low symmetry equivalent with the average high symmetry, in accord with density functional (DFT) calculations and structural refinements \cite{Dressel16,Pinteric16,Lazic18,Foury18,Foury20}.
However, both spectroscopic \cite{Sedlmeier12, Pinteric16, Pinteric18,Lazic18} and structural measurements \cite{Foury18, Foury20} consistently demonstrate the equally distributed charge in quantum spin liquid compounds.
Moreover, the persistence of strong quantum fluctuations due to charge or dipolar fluctuations in the kHz-MHz frequency range is unlikely because no Barrett-like behavior of the dielectric constant [cf. Equation~(\ref{eq:Barrettformula})] has been detected.

Since no appreciable static electric dipoles have been found, the dielectric anomaly in quantum spin liquid compounds has been attributed to the cooperative motion of charged domain walls within random domain structure \cite{Pinteric14,TomicDressel15,Pinteric18}.

In the BEDT-TTF-based quantum spin liquids, this structure originates from the cyanide groups located at the inversion centers bridging the polymeric CuCN/AgCN chains in the anionic layer.
In the \dmit\ with quantum spin liquid ground state, disorder occurs due to two equally probable orientations of the Et groups (static disorder) and
to internal rotational degrees of freedom of the Me groups (dynamic disorder);
both Et and Me groups reside in the non-conducting cations \cite{Lazic18}.
Interestingly, a quantum-spin-liquid state is not established in $\beta^{\prime}$-Me$_4$P[Pd(dmit)$_2$]$_2$  and $\beta^{\prime}$-Me$_4$Sb[Pd(dmit)$_2$]$_2$:
here static disorder is absent, but dynamic disorder is still present and causes a relaxor peak in $\varepsilon^{\prime}(T)$ \cite{Abdel13}.
In all cases, the disorder-induced domain structure is mapped onto the organic dimers and results in relaxor dielectric response with glassy signatures.
The charged domain walls increasingly contribute to the dielectric constant, and the response time gets longer as the temperature is lowered because screening becomes weak when the number of charge carriers is reduced \cite{Pinteric14,Pinteric16}. The domain structure changes significantly under x-ray irradiation: the relaxor peak
shifts to lower temperatures with extended irradiation  before a second feature emerges and becomes dominant, which is attributed to the direct response of anion defects \cite{Sasaki15}.

The broadened charge-sensitive molecular vibrational modes observed in Raman and infrared spectroscopy \cite{Sedlmeier12, Yakushi15, Nakamura17} also infer inhomogeneous charge fluctuations; however, these fast fluctuations cannot be invoked for explaining the dielectric response in kHz-MHz frequency range.
Similar conclusions on the presence of an inhomogeneous charge distribution
are drawn from the dc conductivity behavior showing variable range hoping,
as well as from the broadening of NMR linewidth \cite{Kawamoto02, Pinteric14, Culo19}.

In order to shed light on the microscopic understanding of the observed dielectric anomalies in kHz-MHz range, Fukuyama {\it et al.} proposed a one-dimensional tight binding model with on-site and inter-site Coulomb repulsion \cite{Fukuyama17}. The approach reveals that charge fluctuations are possible in the boundary region between Mott and charge order insulator. At low frequencies, oscillations occur in the double-well potential corresponding to a small charge disproportionation of opposite charge polarity; importantly, spatially extended domain walls connecting two respective domains are also present; they give rise to the observed dielectric response. The structural results of Foury-Leylekian {\it et al.} support the relevancy of this theoretical consideration as they evidence a symmetry breaking;
however the tiny inter-dimer charge imbalance is hardly above the resolution limit \cite{Foury18,Foury20}.
The extension to a larger number of randomly distributed domains, which lift off the inversion symmetry in the anionic layers, is straightforward, as identified by density-functional theory calculations of Lazi{\'c} {\it et al.}; the calculations find the ground state quasi-degenerate in energy and with reduced symmetry \cite{Dressel16,Pinteric16,Lazic18}.

On the other hand, a long-range dipolar order has been proposed to occur in $\kappa$-(BE\-DT\--TTF)$_2$Cu[N(CN)$_2$]Cl acting as a driving force for the antiferromagnetic ground state \cite{Lunkenheimer12}.
The inferred multiferroicity is based on the Curie-like dielectric peak [Figure~\ref{fig:kappaCldieltemp}(a)], as well as on the observed hysteresis and
time-dependent phenomena in the vicinity of the antiferromagnetic transition at $T_{N} \approx 25$~K.
Intriguing results obtained on another single crystal of \etcl\ suggest
that ferroelectricity is proximate to the magnetic and superconducting phases  [Figure~\ref{fig:kappaCldieltemp}(b)] \cite{Pinteric18}.
However, the relaxor-like dielectric peak as depicted in Figure~\ref{fig:kappaCldieltemp}(c)
has been observed in the most number of single crystals studied by different groups; most importantly, no symmetry breaking has been detected until now \cite{Matsuura19}.
Since there is no disorder in the anionic layers, the tendency to phase segregation has been invoked, in order to support the relevancy of the charged domain wall motion scenario \cite{Pinteric18}. Thus, further experimental efforts are vital in search for structural inversion symmetry breaking in order to clarify the ferroelectric-like signatures in this antiferromagnetic system.

\subsection{Quantum electric dipoles in a quantum spin liquid}
\label{sec:cat}
\begin{figure}[h]
	\centering
	\includegraphics[width=0.4\columnwidth]{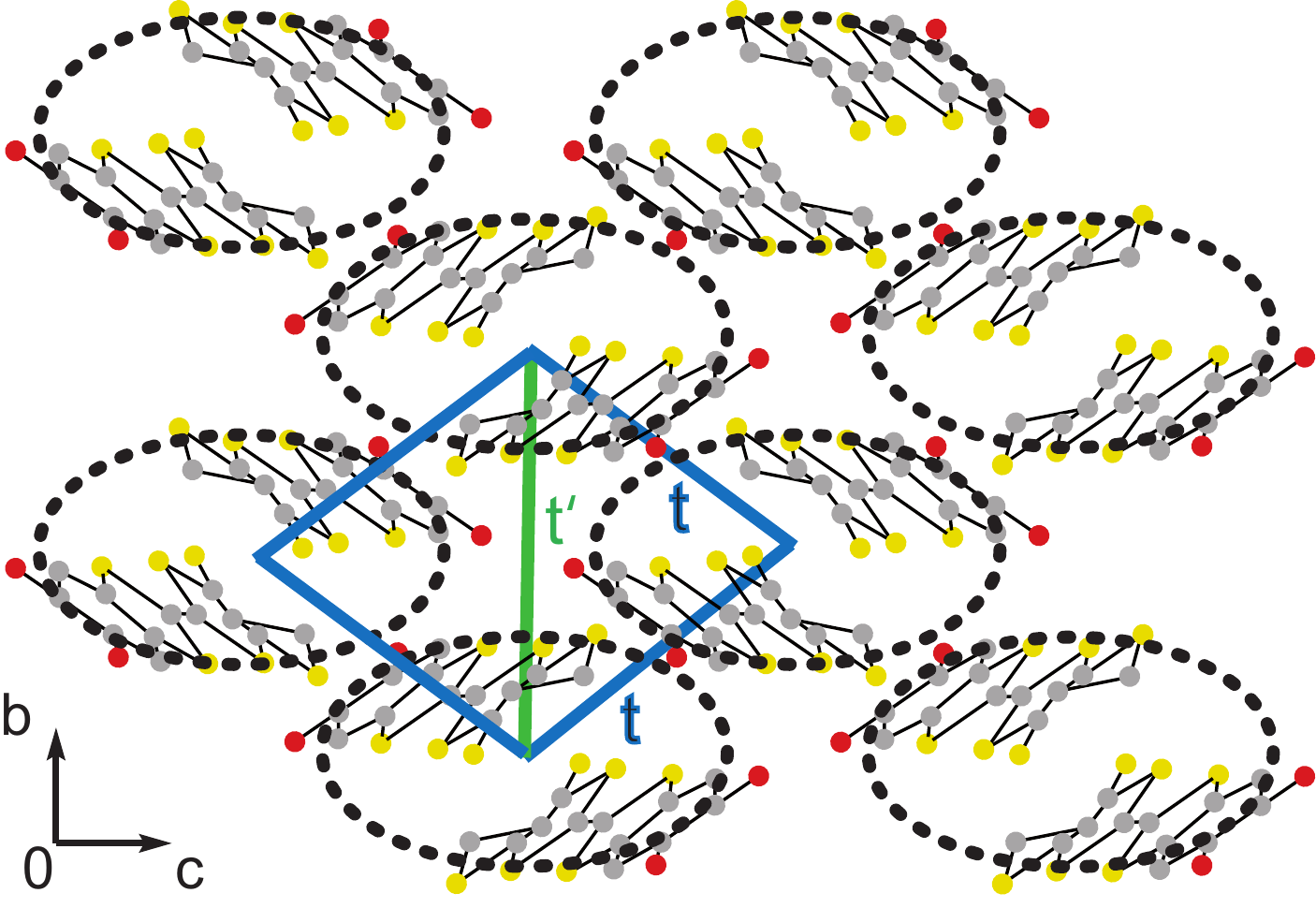}
	\caption{For \cat\ the molecules are arranged in dimers that constitute an anisotropic triangular
		lattice within the conduction layer. The inter-dimer transfer integrals $t$ and $t^{\prime}$  are defined along the sides of rhomboids and along one diagonal, respectively (after \cite{Isono14}).	
		}
	\label{fig:CAT-structure}
\end{figure}

A quantum liquid consisting of both electric and magnetic dipoles
might be realized in the hydrogen-bonded single-molecular Mott insulator $\kappa$-H$_3$(Cat-EDT-TTF)$_2$,
in which Cat-EDT-TTF spin-$\frac{1}{2}$ dimers are arranged on a two-dimensional triangular lattice as depicted in Figure~\ref{fig:CAT-structure} \cite{Isono14,Shimozawa17}. The moderately one-dimensional frustration $t^{\prime} / t \approx$ 1.25 is distinict from 0.83 of
quantum spin liquid candidates \etcn\ and \agcn\ (see Section \ref{sec:propertiesQSL}), but close to the $\kappa$-(BE\-DT\--TTF)$_2$B(CN)$_4$ with $t^{\prime}/t \approx$ 1.44 \cite{Yoshida15}. This material, similar to \cat{} compound, maintains a magnetically disordered Mott insulating state with enhanced quantum fluctuations over a wide temperature range, but in contrast to \cat\ undergoes a phase transition below 5~K into a spin-gapped phase.

The discovery of the purely organic metal \cat\ by Mori and collaborators \cite{Kamo12,Isono13,Isono14,Ueda14} is of particular interest in so far,
as the organic molecules are linked by hydrogen bonds as illustrated in Figure~\ref{fig:structure_CAT}.
Figure~\ref{fig:cat1} shows the temperature dependence of the dielectric constant $\varepsilon^{\prime}(T)$,
which exhibits a  quantum paraelectric behavior as described by the Barrett formula \cite{Barrett52}:
\begin{equation}
\varepsilon^{\prime}(T) = A + \frac{C}{ (T_1/2)\coth(T_1/2T) - T_C} \quad ,
\label{eq:Barrettformula}
\end{equation}
where $C = n\mu^2/k_{B}$ is the Curie constant, $n$ the density of dipoles, $\mu$ the dipole moment, and $k_{B}$ the Boltzmann constant.
$T_C=-6.4$~K is the Curie-Weiss temperature of ferroelectric order in the absence of strong quantum fluctuations;
while $T_1$ is the characteristic crossover temperature from the classical Curie-Weiss regime to the quantum paraelectric regime. From Figure~\ref{fig:cat1} we see that quantum fluctuations become important below $T_1 \approx 8$~K.
The development of strong quantum fluctuations may be associated with enhanced proton fluctuations arising from the zero-point motion of hydrogen atoms, which persist down to low temperatures. Cooperative action between proton and electron degrees of freedom induce intra- or inter-dimer charge fluctuations, and thereby a disordered state of quantum electric dipoles develops at low $T$.
Interestingly, the quantum spin liquid state emerges simultaneously with the quantum electric dipolar liquid.
Shimozawa {\it et al.} suggested that the localized spins couple with the zero-point motion of the protons. First principle density functional theory (DFT) calculations confirm this idea \cite{Tsumuraya15}.
The potential energy surface for the H atom is very shallow near the minimum points;
hence, there is a certain probability that the proton is delocalized between the two oxygen atoms.
Overall results thereby suggest that the quantum proton fluctuations give rise to a combined quantum liquid consisting of electric and magnetic dipoles. Interestingly, the Barrett-like behavior of the dielectric constant $\varepsilon^{\prime}(T)$ is absent in the other organic quantum-spin-liquids candidates
consisting of $\pi$-electron molecular layers that are separated by the anion layers.

Deuterated crystals, however, undergo an ordering transition at $T_{\rm CO} =185$~K leading to a charge disproportionation ($+0.94e : +0.06e$) associated with deuterium localization within the Cat-EDT-TTF layers.
The resistivity abruptly jumps at that temperature with a hysteresis of 4~K \cite{Ueda14}.
The $D_3$-compound exhibits a small and constant permittivity and a non-magnetic ground state below the phase transition at 185~K.
In other words, the quantum-disorder due to the fluctuating proton bond is crucial for the low-temperature properties, in the spin sector as well as for the electric dipoles.
And {\it vice versa}, the arrangement of the charge order and formation of spin singlets in the charge-rich dimers lead to the non-magnetic ground state of \dcat. As displayed in Fig.~\ref{fig:cat3}(b), the heat capacity of the deuterated compound remains at smaller values for the entire temperature range due to the lack of spin contributions. Except for very low $T$,
$C_p T^{-1}$ increases linearly with  $T^2$ but the Sommerfeld $\gamma_e$ term is much smaller. An additional Schottky term infers paramagnetic spins due to disorder.

\begin{figure}
  \centering
  \includegraphics[width=0.7\columnwidth]{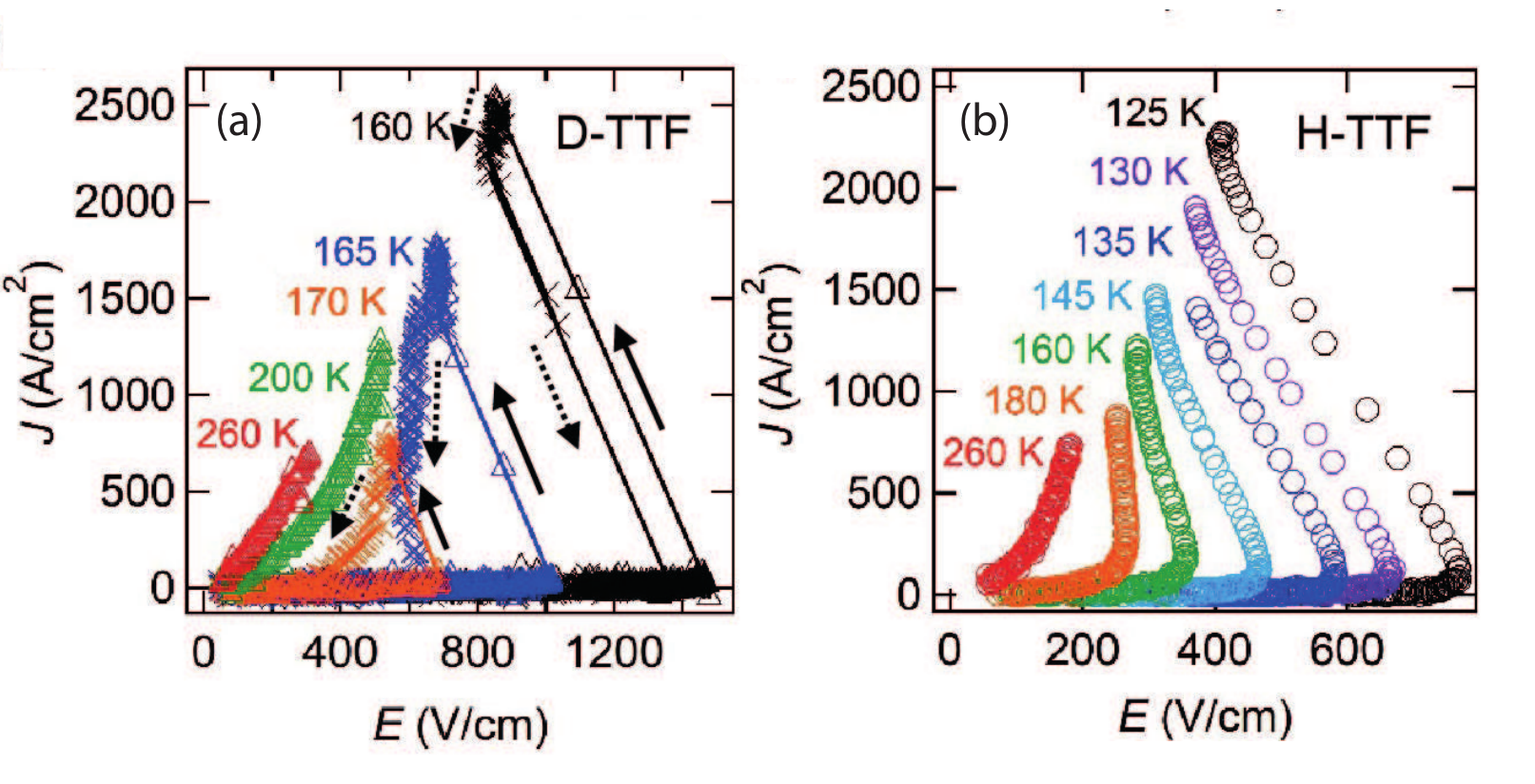}
    \caption{\label{fig:cat5}
$J$ {\it versus} $E$ curves of (a) \cat\ and (b) the deuterated analogue at several temperatures as indicated. The solid and dashed arrows represent the voltage-increasing and voltage-decreasing processes, respectively. The hysteresis appears only in the deuterated compound but not in the hydrogenated one (after \cite{Ueda19}). }
\end{figure}

Measuring the current density–electric field characteristics of \dcat{} above and below the charge-ordering transition
Ueda {\it et al.} found a negative differential resistance and hysteresis, which is considered to be induced by the deuterium dynamics  \cite{Ueda19}.
Applying a pulsed voltage, the initial charge-ordered state changes to a metastable state through a high-conducting (excited) state, which results in the appearance of the hysteresis, as illustrated in Figure~\ref{fig:cat5}(a).
Raman spectroscopy suggests that this metastable state is a non-charge-ordered dimer-Mott state.
The results can be understood by considering the temperature-dependent dynamics of hydrogen-bonded deuterium ({\it i.e.}, localization/fluctuations) coupled to the $\pi$-electrons in the conducting layers.
Figure~\ref{fig:cat5}(b) shows that on the contrary the hydrogen analogue \cat, which is a dimer-Mott insulator without charge-ordering phase transition, does not exhibit such hysteretic behavior, but it does display a similar negative differential resistance as well, indicating that some degree of proton localization is present. This idea is confirmed by DFT calculations
\cite{Tsumuraya15}, which find another H-localized state with reduced symmetry which lies only 2\,meV above the optimized state with minimum energy. The quasi-degenerate electronic state implies that random domains are present; domain wall motion may be responsible for the nonlinear effects.

It is interesting to recall the cooperative charge dynamics observed in the charge-ordered state of \aeti,
where also a negative differential resistance is observed (Figure~22) with a significant change in shape of the measured resistivity in time \cite{IvekPRB12,Peterseim16}. In Section~\ref{sec:dielectric_domainwalls} we discussed how the findings of negative differential resistance and switching to transient states are explained by cooperative domain-wall dynamics inherent to the ferroelectric state driven by charge-ordering.

 \subsection{Quantum electric dipoles with glassy signatures}
\label{sec:QEL}
\label{sec:DipoleLiquid}

\subsubsection{Quantum electric dipole lattice}
\label{sec:hgbr}

\begin{figure}
	\centering\includegraphics[clip,width=0.65\columnwidth]{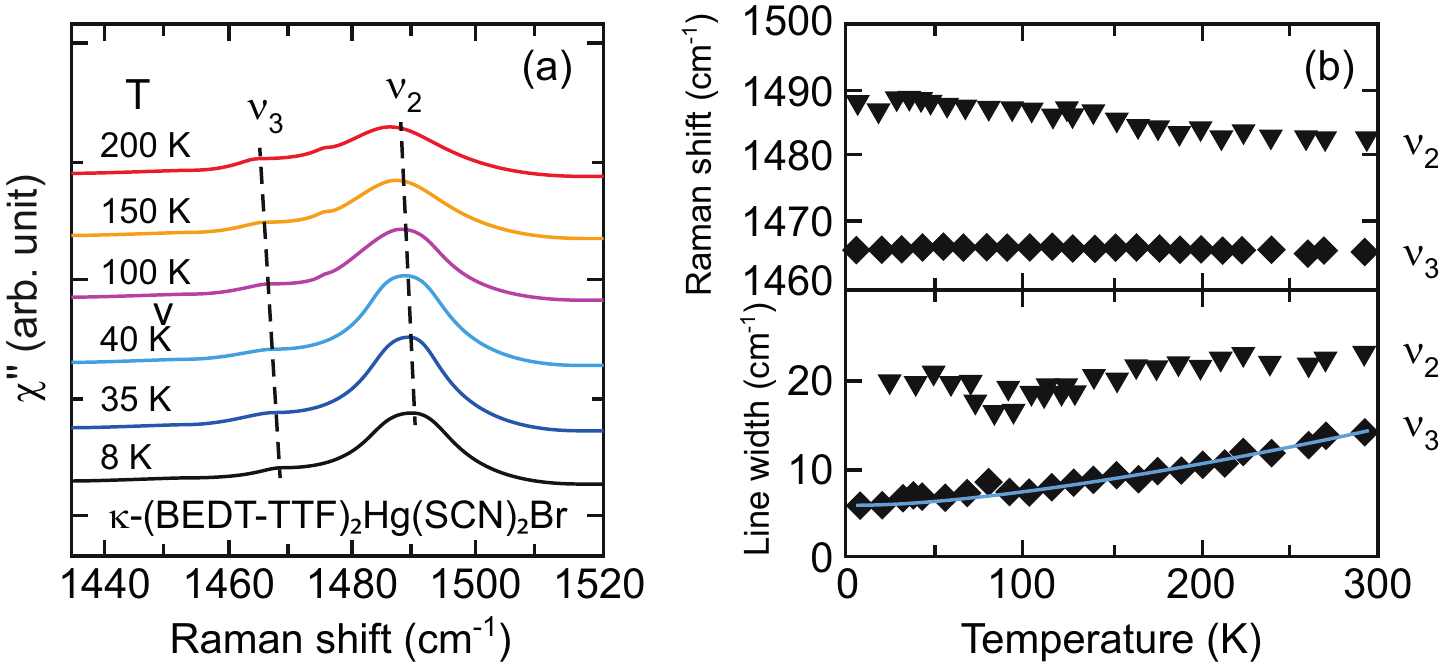}
	\caption{(a) Raman spectra of \hgbr\ in the region of the molecular vibrations $\nu_{2}$ and $\nu_{3}$
at temperatures between 200 and 8~K. (b) No splitting due to charge ordering is observed.
For the $\nu_{2}$ mode, the linewidth increases below 80\,K.
For comparison, the behavior of the charge-insensitive $\nu_{3}$ band is shown;
its linewidth decreases continuously from 300~K down to low temperatures (after \cite{Hassan18}).
	\label{fig:HgBrnu2}}
\end{figure}

Recently, Drichko and collaborators suggested that \hgbr\ may serve as an example of a quantum dipole liquid,
based on Raman scattering investigations \cite{Hassan18}. The system is {\it a priori} a good candidate because electric dipoles are arranged on a triangular lattice with moderate frustration
(Figure~\ref{fig:structure_HgBrCl}) \cite{Hotta10}. Raman and infrared vibrational spectroscopy shows that static charge order is absent \cite{Ivek17,Hassan18} despite a pronounced metal-insulator transition at $T_{\rm MIT}=80$~K, the origin of which is still elusive.
Charge fluctuations are invoked to explain the behavior observed in the $\nu_{2}(a_g)$ Raman band: Figure~\ref{fig:HgBrnu2} demonstrates that the band broadens on cooling and that the linewidth goes through
a slight minimum at around $T_{\rm MIT}$ before it increases again below.
The explanation starts from the two types of BEDT-TTF molecules with charge $+0.6e$ and $+0.4e$, respectively,
that are present in the charge-ordered sister system \hgcl\ (Section~\ref{sec:COdimerized});
it is assumed {\rm ad hoc} that these charges are also present in \hgbr, but
fluctuate between two molecules in a dimer; the assumption is in line with our estimate of charge fluctuations indicated by the broad $\nu_{27}({\rm b}_{1u}$) mode as discussed in Section \ref{sec:COHgCl}.
Applying Kubo’s two-states-jump model \cite{Yakushi12,Girlando14}, the shape of the band can be mimicked
assuming charge fluctuations with an exchange frequency $\omega_{\rm ex} \approx 30-40$\cm.

Most important, the Raman spectra in the $A_{1g}$ channel reveal a broad feature around 40~\cm\ of non-phonon origin that grows strongly below $T_{\rm MIT}=80$~K and attains maximum intensity below around 40~K, shown in Figure~\ref{fig:HgCl_Raman1}(a).
The energy of mode is lower than expected for magnetic excitations;
instead it is associated with the exchange frequency of charge fluctuations $\omega_{ex} \approx 30-40$~\cm.
The results are interpreted as fluctuating dipoles forming a quantum electric dipole liquid state.

\subsubsection{Glassy behavior}

Along this line one expects a quantum paraelectric behavior described by the Barrett formula \cite{Barrett52},
Eq.~(\ref{eq:Barrettformula}), as it is observed in other compounds with a triangular lattice, such as  BaFe$_{12}$O$_{19}$ and $\kappa$-H$_3$(Cat-EDT-TTF)$_2$ \cite{Shen16,Shimozawa17} (Figure~\ref{fig:CatBarrett}).
Instead, the dielectric response of \hgbr\ exhibits only a weak relaxor-like behavior in the audio and radio-frequency range,
which is screened by the conduction electrons \cite{Ivek17}.
Simultaneous fits of $\varepsilon^{\prime}(\omega)$ and $\varepsilon^{\prime\prime}(\omega)$ to the generalized Debye formula (see Section~\ref{sec:DielectricResponse}) disclose a strongly diminishing dielectric constant
and a broadening of the relaxation time distribution when the temperature decreases;
this behavior might indicate glassy freezing \cite{Staresinic02}.
The mean relaxation time $\tau_0$ follows the Arrhenius type of gradual slowing down.
When we take the temperature where $\tau_0$ extrapolates to the value of 100~s as the glass transition,
we obtain $T_g \approx$ 5\,K.
In the glassy phenomenology, the Arrhenius behavior is characteristic of strong slow-cooled glass formers
and implies that glassy dynamics involves only local rearrangements of charge configurations.
From this point of view, the low-frequency Raman mode is interpreted as a Boson peak arising from charge fluctuations between two molecular dimer sites.
In analogy to the charge-density wave phenomenology, we speculate that the low-frequency Raman mode (Boson peak) and the dielectric mode in kHz-MHz range represent fingerprints of the same phenomenon \cite{Biljakovic12, Remenyi15}:
the formation of dipole liquid state occurs on the local scale and bears glassy signatures.
This interpretation is in accord with previous observations: systems with strong Boson peaks
above $T_g$ tend to be those with strong liquid character \cite{Ediger96,Schroeder04,Nakayama02}.

\begin{figure}
  \centering
  \includegraphics[width=1\columnwidth]{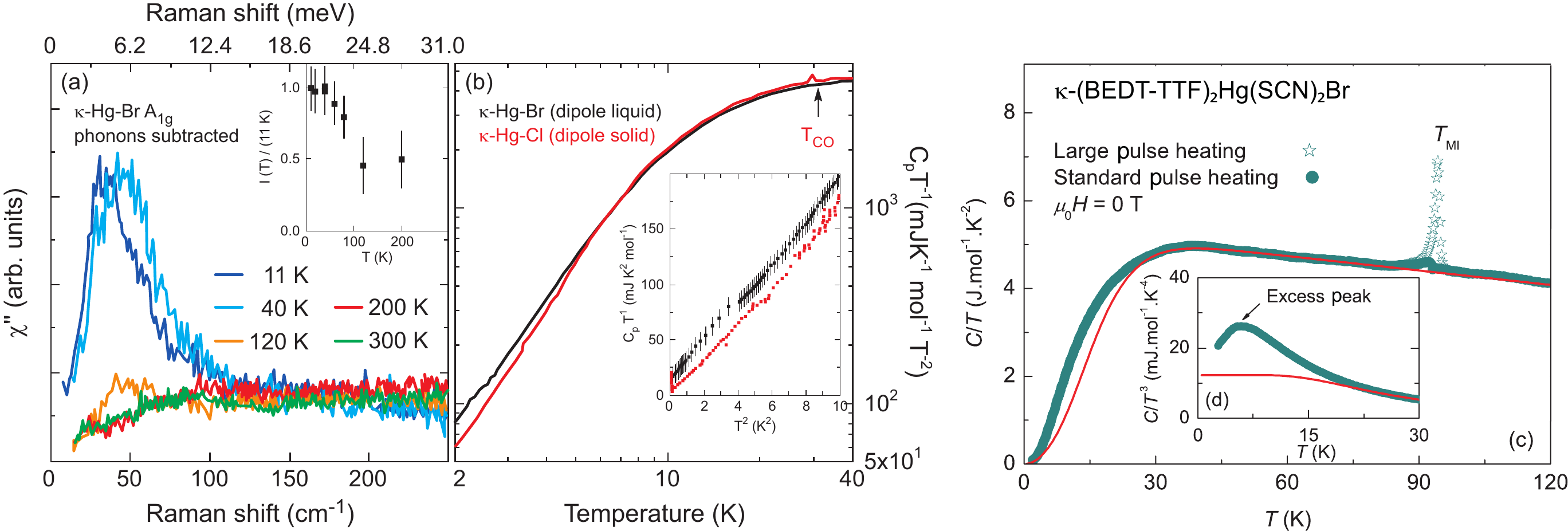}
    \caption{\label{fig:HgCl_Raman1}
(a) Temperature dependence of the collective mode in the A$_{1g}$ scattering channel for \hgbr, determined by subtracting phonons from the full Raman spectrum. Inset shows the intensity of the mode as a function of temperature. (b) Temperature dependence of the heat capacity $C_p$ for \hgcl\  (red line) and \hgbr\ (black line) below $T=40$~K. The two curves deviate from each other below approximately 6 K. The inset shows the low-temperature data with linear behavior of heat capacity for \hgbr.
(c) Temperature dependence of specific heat of \hgbr. The $C/T^3$ {\it vs.} $T$ plot demonstrates
the excess contribution from the Boson peak located at around $T=5$~K (after \cite{Hassan18,Hemmida18}). }
	\label{fig:HgBrBoson}
\end{figure}
Also specific heat measurements reveal the presence of a Boson peak below $T=20$~K, as displayed in Figure~\ref{fig:HgBrBoson}.
The significant non-Debye behavior indicating an excess of low-energy vibrational states \cite{Hemmida18}
is often taken as a universal signature of heterogeneity and glassy properties of the liquid state \cite{Ando18}. Importantly, the entropy associated with  the excess peak is significantly larger than what is expected
for the pure magnetic entropy of a spin-$\frac{1}{2}$ system,
indicating that the glassy state should be mainly attributed to heterogeneity in the charge sector.
Since no estimates of the Boson peak frequency, either on the basis of specific heat or low-frequency Raman data exist, we cannot finally conclude on their common origin. This issue is certainly worth clarifying in future studies.

\subsubsection{Coupling to magnetic degrees of freedom}

While in the regular dimer Mott insulator the magnetic coupling between the dimers in \hgbr\ is given by $J\approx 20$~meV (Table~\ref{tab:1}), unequal charge distribution on the dimers causes a renormalization \cite{Hotta10,Hotta12} to $J_{\rm DS}\approx 6-7$~meV, as consistently estimated from ESR and Raman experiments \cite{Hemmida18,Hassan18}; still too high to assign  the 40~\cm-peak  observed in Figure~\ref{fig:HgCl_Raman1}(a) to purely magnetic excitations.
From other metallic BEDT-TTF salts close to the charge-order transition, it is known \cite{Merino03,Dressel03b,Dressel10,Kaiser10} that charge fluctuations can cause collective modes present in optical spectra.
The coupling of these electric dipole fluctuations to $S=\frac{1}{2}$ spins on a triangular lattice might serve as mechanism for spin-liquid behavior \cite{Hotta10,Naka16}.
Starting with a Kugel-Khomskii type Hamiltonian, Naka and Ishihara show that the spin-charge interaction promotes an instability of the long-range magnetic ordered state around a parameter region where two spin-spiral phases  merge \cite{Naka16}. As a matter of fact, the fluctuating dipole liquid is rather similar to an orbital liquid \cite{Balents10,Savary16}. Specific heat measurements down to low temperatures are also consistent with the idea of a spin-liquid behavior in \hgbr, since a linear term is present only in the Br-compound but not in the Cl-analogue, where the electric dipoles are well ordered [Figure \ref{fig:HgBrBoson}(b)].
It is interesting to recall the proposal of Mazumdar and Clay \cite{Li10,Dayal11,Clay12,Clay19} about a paired electron crystal where magnetic interaction acts as a driving force for the charge order in a frustrated dimer lattice and a singlet ground state. In the case of \hgcl, where an abrupt charge-ordered phase occurs at $T_{\rm CO} = 30$~K (Figure~\ref{fig:HgClnu27}), it is likely that the long-range magnetic order must be excluded below $T_{\rm CO} = 30$~K, as discussed below.

At the first sight, evidence of magnetic properties in the charge-ordered state of \hgcl{} is pretty vague. In the region of $T_{\rm CO}$, one finds a kink in dc resistivity together with a decrease of the susceptibility;
this might be seen as tendency to a spin-gapped state. On the other hand, early ESR data suggested antiferromagnetic state formation slightly below $T_{\rm CO}$ \cite{Yasin12b}, while more recent ESR and specific heat measurements on samples from different sources  failed to detect any signature of magnetic transition \cite{Gati18a}. Although there is general agreement of both results, the details seem to be affected by impurities.

\begin{figure}
	\centering
	\includegraphics[width=0.7\columnwidth]{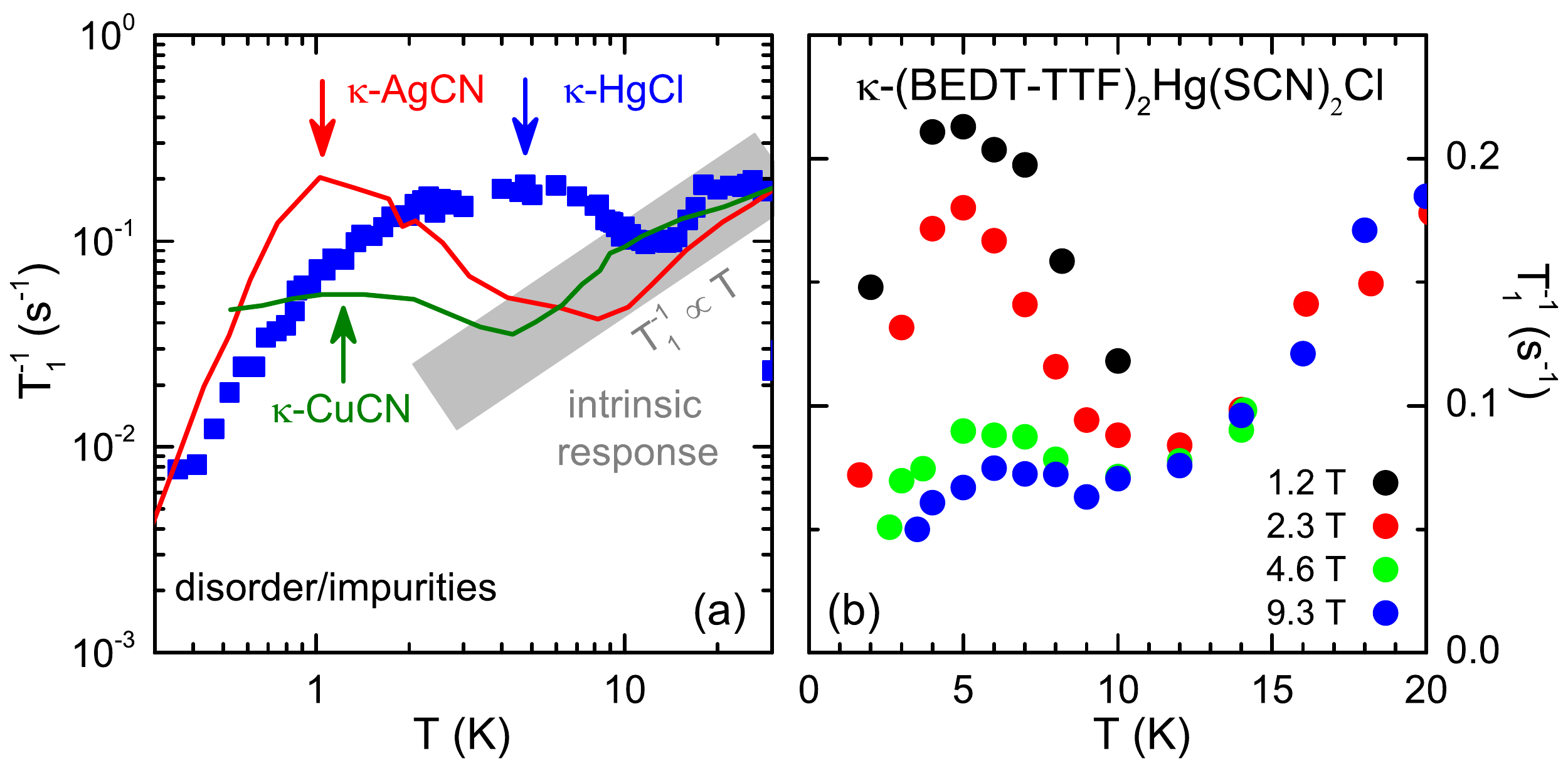}
	\caption{\label{fig:HgCl_NMR1}
		(a) Temperature dependence of the $^{1}$H-NMR relaxation rate of several spin liquid candidates. At temperatures above the maximum, the $T^{-1}_1$ data in the insulating state of \hgcl\ (indicated by $\kappa$-HgCl) \cite{Pustogow19b} coincide with the paradigmatic quantum spin liquid compounds \etcn\ \cite{Shimizu03} and \agcn\ \cite{Shimizu16}. Here, $T^{-1}_1$ follows a field-independent approximately linear $T$  dependence suggesting that this is the intrinsic response with $J \approx 200$~K. (b) Upon increasing $B_0$ the maximum is strongly suppressed and shifts to higher temperatures. This behavior is observed for the other compounds, too \cite{Shimizu03,Shimizu06,Shimizu16} (after \cite{Pustogow19b}). }
\end{figure}

Recently, the absence of long-range magnetic order is also concluded from $^{1}$H nuclear magnetic resonance data on \hgcl\ \cite{Pustogow19b}. This study reveals a classical Korringa temperature dependence in the metallic state; in other words  $(T_1T)^{-1} \approx 1000$~Ks remains constant from ambient temperatures down to $T_{\rm CO} =30$~K.
There the relaxation rate is strongly enhanced due to charge order, but otherwise the NMR spectra do not change when passing the phase transition; in fact no signature of long-range magnetic order was observed down to 25~mK. A maximum in the relaxation rate $T_1^{-1}$ occurs around 5~K, similar to previous observations in the spin liquid candidates \etcn\ and \agcn\ around 1~K, as plotted in Figure~\ref{fig:HgCl_NMR1}(a) \cite{Shimizu03,Shimizu06,Shimizu16}.
The strong field dependence of $T_1^{-1}$ observed at that temperature [(Figure~\ref{fig:HgCl_NMR1}(b)] is taken as indication of contributions for dynamic inhomogeneities.
It seems that the low-temperature NMR properties of all these frustrated materials are dominated by extrinsic magnetic contributions rather than by instrinsic spin degrees of freedom.

While antiferromagnetism is quite common in charge transfer salts, there has been only one report on indications of weak ferromagnetic order: shortly after the discovery of superconductivity in slightly pressurized \etcl\ at a record temperature of $T_c=12.8$~K by the Argonne group \cite{Williams90},
they observed an antiferromagnetic transition near $T_N=45$~K followed by a weak ferromagnetic hysteresis below 22~K \cite{Welp92}. As depicted in Figure~\ref{fig:kappa_phasediagram}(b), NMR, ESR und SQUID experiments confirm the antiferromagnetic ground state below 26~K; the weak ferromagnetic moment results from a canting of the ordered spins at lower temperatures \cite{Miyagawa95,Pinteric99,Yasin11}.

In this context, the recent discovery of weak ferromagnetic order in \hgbr\ below $T=20$~K
was fairly surprising \cite{Hemmida18}.
As discussed in Sections \ref{sec:COHgCl} and ~\ref{sec:hgbr}, the material is known to undergo a smooth metal-insulator transition at 80~K,
which is neither a Mott nor a charge-order transition and also not associated with structural changes  \cite{Aldoshina93,Ivek17}.
In the spin sector, the heat capacity shows a finite linear term $\gamma_e$ \cite{Hassan18},
while the spin susceptibility and ESR spectroscopy \cite{Hemmida18} suggest that frustrated spins in the molecular dimers suppress long-range antiferromagnetic order, forming a spin-glass-type ground state of the triangular lattice below $40$~K, which locally contains ferromagnetic polarons.
Interestingly, this estimate of $T_g$ corresponds to the energy of low-frequency Raman mode located at 40~\cm, which attains maximum intensity around $T=40$~K.
Moreover, most recent $^{13}$C-NMR investigation finds weak NMR line broadening starting right below $T_{\rm MIT}$ and strongly developing below 40\,K. The spin-lattice relaxation time forms a broad maximum centered around 5~K indicating a dynamical slowing down of magnetic fluctuations. Overall results are interpreted as disordered charge disproportionation and antiferromagnetism developed at short-range scales \cite{Le20}.
Considering both, the charge and spin sectors, we suggest for \hgbr\ an exotic quantum liquid state with glassy nature, that consists of entangled fluctuating electric dipoles and spins. Below $T_{\rm MIT}=80$~K, composite charge-spin clusters develop upon cooling, the dynamics gradually slows down and freezes at low temperatures. More efforts in future research, experimental as well theoretical, are needed in order to further elaborate this scenario.


\section{Summary and Prospects}
    \label{sec:summary}
Organic solids including charge-transfer, radical cation and anion salts and single-molecular materials represent a fascinating class of strongly correlated electrons systems in two dimensions.
These materials host a broad range of quantum phenomena, which
stay at the center stage of condensed-matter physics today: electronic ferroelectricity, massless Dirac fermions, quantum criticality, correlation-driven metal-insulator phase transition, quantum liquid states of spin and electric dipoles.

\subsection{Electronic ferroelectricity}
\label{sec:Summary_ferroelectricity}

Organic charge transfer salts attract attention because some of them exhibit ferroelectricity of electronic origin that arises from charge order instead of cooperative displacements of ions observed in conventional ferroelectrics.
The charge ordering is controlled by strong on-site and inter-site Coulomb interactions and is found  in a number of weakly dimerized materials with effectively quarter-filled bands. However, ferroelectricity is established only if the van den Brink and Khomski requirement  for the coexistence of non-equivalent sites with different charge density and non-equivalent bonds is fulfilled in the charge-ordered state.

The most prominent example is certainly \aeti, for which charge-order-driven ferroelectricity below the metal-insulator phase transition at $T_{\rm CO} = 135$~K is demonstrated unambiguously in numerous experiments including optical and THz-wave second-harmonic-generation, x-ray scattering, Raman and infrared vibrational spectroscopy, electron-spin and nuclear magnetic resonance and polarization switching. The phase transition is of first order as evidenced by an abrupt onset of second-harmonic-generation signal, charge imbalance and charge and spin gaps. Observation of near-field images of its spatial evolution corroborates these findings. The dynamic response of ferroelectric domains subjected to ultrafast external stimuli has been extensively studied by time-resolved measurements of electrical conductivity and femtosecond pump-probe spectroscopy and by now a rather fair understanding is achieved. New emergent phases, memory effect, photo-induced metallic state; overall, these observations strongly support a purely electronic mechanism of ferroelectricity with only minor involvement of electron-phonon interaction.

At longer time scales, the non-linear response observed by time- and field-dependent transport exhibits two distinct regimes, both of which are in line with domain wall motion. At low temperatures the behavior involves thermally excited topological defects over an electric field dependent potential barrier, whereas at temperatures closer to the phase transition the negative differential resistance behavior can be simulated consistently by a two-state model of excited charge carriers with the high mobility. The dynamic response in kHz-MHz frequency range studied by dielectric spectroscopy displays the two-mode response deeper in the charge ordered state than expected from a Curie-like peak. The anomalous response indicates the presence of disorder arising due to intrinsic heterogeneity but
no consensus has been reached yet how to explain these observations.
A plausible mechanism may involve the motion of two types of domain walls.
No doubt, a thorough study of the topology of domain structure is highly desired in order to reach a consistent understanding of the dielectric response as well as of the low level of switchable polarization.

For the related compound, {\it i.e.} the slowly-cooled \tetrz, charge order and ultrafast dynamics has been clearly identified,
while the experimental evidence for polar order is not complete and its recognition is mostly based
on a Curie-like dielectric peak. From a unifying view of experimental data and current theoretical approaches, it becomes rather clear that the stabilization of charge-order driven ferroelectricity takes place by a cooperative action between Coulomb repulsion --~both on-site and inter-site~-- and coupled molecular-anion subsystems.
There are also examples, such as $\beta^{\prime\prime}$-(BEDT-TTF)$_2$\-SF$_5$CHFCF$_2$SO$_3$, where some degree of charge order exists at all temperatures, but it is not associated with a metal-insulator phase transition, and the van den Brink and Khomski requirement is not fulfilled.

Glassy phases in the vicinity of charge order have been identified in two charge transfer salts with a frustrated triangular lattice of molecular units. Charge order can be avoided by rapid cooling in \tetrz\ or by replacing Rb by Cs in \tetcz. Slow charge dynamics and glassy freezing are demonstrated by resistance fluctuation spectroscopy: upon reducing the temperature, a broad peak shifts to lower frequency and its linewidth strongly increases. The development of slow dynamics is correlated with the growth of two-dimensional charge clusters observed by x-ray diffuse scattering.
Metastability of the charge glass phase in \tetrz\ is demonstrated by the time-evolution of the resistivity and nuclear magnetic resonance spectra at a fixed temperature; the obtained data show time-temperature-transformation curves commonly observed for crystallization of structural, ionic glasses and metallic alloys.
Additional fingerprints of glassy state --~charge vitrification and non-equilibrium aging phenomena~-- are demonstrated in \tetcz\ by means of the resistivity and resistance fluctuations spectroscopy: faster cooling results in a higher glass transition temperature, the resistivity behavior in time can be described by the stretched exponential function with the relaxation time obeying the gradual slowing down law; the equilibrium states at high temperature and non-equilibrium below the glass transition follow the same dynamics.

Overall, the data show that the interplay of long-range interactions and geometric frustration takes a primary role in the formation of charge glass; theoretical considerations confirm these observations and suggest \tetrz\ and \tetcz\ as prominent examples of a self-generated Coulomb glass.
However, in real materials the lattice degrees of freedom are involved in the creation of the electronic crystals states, both the charge glass as well as the charge order; the extent to which they influence the crystallization mechanism remains to be clarified.

Finally, among those two-dimensional organic materials, in which organic molecules are paired in dimers and organized on an anisotropic triangular lattice, charge ordering presents a prominent topic because a link to the magnetic degrees of freedom was proposed.
Until now, charge ordering below the metal-insulator transition at $T_{\rm CO} = 30$~K and a Curie-like non-dispersive dielectric peak
are detected only in \hgcl. There remain some open issues concerning the potentially polar nature of the charge-ordered state. The charge pattern and the symmetry space-group changes are not identified yet; this fact, together with indications that charge order melts below 15~K, calls for more efforts in x-ray diffraction measurements.
The dielectric investigations were conducted only for an electric field applied perpendicular to molecular layers where anionic contribution cannot be avoided; certainly, in-plane measurements are needed.


\subsection{Dirac electrons}
\label{sec:Summary_Dirac}

The Dirac electron state in the two-dimensional organic solid \aeti\ emerges under high pressure as evidenced by the electrical conductivity, Hall effect, specific heat, nuclear magnetic resonance, magnetotrasport and optical conductivity measurements. These measurements also indicate that low-mobility massive holes coexist with high-mobility massless Dirac carriers.
Theoretical considerations reveal that with increasing pressure, a topological transition occurs from a charge-ordered insulator to a zero-gap state with a pair of Dirac electrons of finite mass; their existence is characterized by a special structure of the Berry curvature inside the Brillouin zone. Angle-resolved photoemission spectroscopy is certainly the most desirable tool to verify modifications of the energy spectrum associated with the topological transition; unfortunately the material and parameter range constitutes a real challenge. Here molecular engineering might provide the possibility to enlarge the orbitals towards Dirac electrons at ambient conditions.

The outstanding feature of the Dirac state in \aeti\ is that the Dirac cones are fixed at the Fermi energy and are shifted away from the high crystallographic symmetry points in the first Brillouin zone, they are strongly anisotropic and tilted in the wavevector-energy space, very much in contrast to graphene with the isotropic Dirac cones at the corners of the first Brillouin zone. Consequently, the low-energy electronic and magnetic response show significant deviations  from theoretical expectations for simple Dirac systems.
The electronic correlations, long-range and short-range Coulomb interactions, induce non-uniform cone re-shaping and bandwidth reduction together with an emergent ferrimagnetic spin polarization. The long-range part is responsible for anomalous spin dynamics and excitonic fluctuations in the vicinity of the charge-order state. Effects  of correlations are also observed in dc resistivity and in optical studies; interaction among Dirac electrons can be modified  by temperature and pressure.
Numerical studies around the critical region of the charge ordering and Dirac state show how
the system passes through a phase change when increasing the intersite Coulomb interaction, which varies upon changing the external pressure: from massless Dirac via massive Dirac state coexistent with the charge order into the charge-ordered state with no Dirac cones. The much smaller gap extracted from dc resistivity compared to the one found in optics, is a result of the conduction of domain walls between ferroelectric domains with opposite polarization.

It remains unclear whether a true zero-gap Dirac state is realized in \aeti; spin-orbit coupling is suggested to may be responsible for that it is not the case.

\subsection{Mott metal-insulator phase transition}
\label{sec:Summary_Mott}

Organic charge-transfer salts turn out to constitute ideal model systems for studying the quantum-critical nature of the Mott transition and to verify different scenarios occurring in strongly correlated systems.
For $\kappa$-(BEDT-TTF)$_2$$X$ compounds a first-order Mott transition is observed up to the critical endpoints between $T_{\rm crit}=20$ and 40~K. Introducing disorder blurs the discontinuity and a smeared first-order transition remains.

The bandwidth-controlled Mott criticality involves fluctuations in electron and lattice degrees of freedom. A material-independent quantum critical scaling of the dc resisitivity, bifurcating into a Fermi liquid metal or Mott insulator, is identified regardless of the ground state; the findings reveal the incoherent charge transport in the crossover region at high temperatures as predicted by dynamical mean-field theory calculations. Hook's law of elasticity breaks down close to the critical endpoint due the coupling of the critical electronic system to the lattice;  critical elasticity shows the universal properties of an isostructural solid-solid endpoint with mean-field critical exponents.

The coexistence region of insulating and metallic phases below the critical endpoint, that is a result of the first-order Mott transition, is experimentally determined by dc conductivity measurements under varying pressure. Nuclear magnetic resonance, ac susceptibility and spatially resolved magneto-optical spectroscopy studies uncover the regime where antiferromagnetic and superconducting phases spatially coexist evidencing percolative superconductivity, whereas their competition deep in the Fermi liquid part of the phase diagram is revealed by ultrasonic velocity and attenuation measurements. The phase separation in the critical region of the phase diagram is also supported by slowing down of electron dynamics demonstrated by fluctuation spectroscopy.

Calculations by dynamical mean field theory find a first-order phase transition with a coexistence region on both sides limited by spinodal lines that end at the critical point at finite temperatures. In the high-temperature region above the critical point quantum critical transport spreads out, following the quantum Widom line; it separates the more-insulating from the more metallic features. The predicted behavior is successfully verified experimentally combining optical and transport measurements of three
highly-frustrated materials with no magnetic order \dmit, \etcn\ and \agcn. The genuine phase diagram of Mott insulators is established: quantum Widom line, the back-bending towards the critical endpoint and metallic fluctuations close to the first-order phase boundary are found.

Visualization of the real-space phase coexistence in the critical region is still limited to about 10~$\mu$m, disabling a direct proof that metallic regions coexist in the insulating background; scanning near-field microscopy at cryogenic temperatures is a desirable tool in future studies. Hence, at present indirect approaches must be employed. Tuning the bandwidth by hydrostatic pressure and chemical substitution enables us to follow the
temperature-dependent dc transport and dielectric constant from the strongly correlated Mott insulator via the range of phase coexistence into the metallic regime;
no hysteresis is observed.
The spatial coexistence of correlated insulating and metallic regions is successfully testified
by a strongly enhanced dielectric constant, when approaching the first-order transition; this behavior uncovers percolative nature of the first-order Mott transition. The experimental findings are supported by dynamical mean-field theory calculations including spatial inhomogeneities in a hybrid approach.

When the system is tuned into the metallic state, coherent electronic charge transport emerges, this is highlighted via the Fermi-liquid behavior including quadratic temperature dependence of dc resistivity and $\omega$-$T$ scaling in the optical conductivity. As a matter of fact, in these organic conductors
Fermi-liquid behavior is observed over a wider temperature and frequency range than in any inorganic compound, owing to the fact that the intrinsic energy scales (bandwidth, Fermi energy) are fairly low.
The upper limit of the Fermi liquid regime is given by deviations of the scattering rate from the $T^2$ and $\omega^2$-lines.

At low temperatures, in the Fermi-liquid regime, the data exhibit a gradual increase of the effective mass indicating that the electronic correlations become stronger when approaching the Mott transition from the metallic side; the findings are in agreement with the Brinkman-Rice theory and dynamical mean-field calculations. The comprehensive set of optical data evidences the universality of Landau's Fermi liquid concept upon varying the correlation strength.

At higher temperatures above the Fermi liquid range, a bad metal regime is found as commonly observed for strongly correlated materials; and resistivity even exceeds the Ioffe-Regel-Mott limit. Reduction of the low-frequency spectral weight is in agreement with theoretical calculations. The displacement and finally disappearance of the Drude peak is accompanied by a transfer of spectral weight to energies above 1\,eV. The phenomenon was ascribed to strong correlations, but arguments were raised that this might not be the main cause for all correlated organic materials.

Effects of disorder on the Mott transition were studied for different interaction and disorder strength by dynamical mean-field theory; the resulting Anderson-Hubbard phase diagram displays a correlated disordered metallic phase, surrounded by Anderson and Mott insulating phases. With increasing disorder the coexistence region gets smaller; when disorder becomes large enough (larger than twice the bandwidth), the critical temperature abruptly goes to zero and the coexistence region disappears.
In this regime, the effects of interaction and disorder are found to be comparably important for charge localization.

Experimentally, x-ray irradiation of the Mott insulator \etcl\ results in a shift of spectral weight from the interband transition to low energies where the Drude-like behavior is present. The result that disorder can make the system more metallic agrees with theoretical considerations; recent calculations find that the random potential broadens the bandwidth and moves the system away from the Mott transition. In addition, recently it was found
that x-ray irradiation can induce slow electronic fluctuations caused by a synergistic effect between the Mott boundary and randomness indicating the formation of electronic Griffiths phase. On the other hand,
\etcn\ is much less susceptible to x-ray irradiation;
the resistivity decreases but even for large doses the effect is not significant. However, effects of inherent disorder on the transport properties are clearly observed for several $\kappa$-(BEDT-TTF)$_2$$X$ quantum-spin-disordered Mott insulators and can be explained within the Mott-Anderson localization theory. While at high temperatures the transport properties is dominated by the correlation strength, upon lowering temperature variable-range hopping transport takes over; the most disordered material, \etcn, exhibits the lowest dc resistivity and the highest charge-carrier density,
indicating that it is located closest to the insulator-metal transition.

\subsection{Quantum spin liquid versus magnetic order}
\label{sec:Summary_QSL}

Quantum spin liquids are quantum disordered ground states of strongly correlated spins, in which the ground state is a superposition of multiple coonfigurations and quantum fluctuations are important enough to prevent conventional magnetic long-range order.
At this point, there are four prominent triangular organic solids, which are considered candidates for the realization of a quantum spin liquid state: charge-transfer salts \etcn, \agcn, \dmit\ and the hydrogen-bonded single-molecular compound \cat.

These materials are Mott insulators, in which --~despite the relatively large exchange coupling~-- there is no experimental indication of magnetic ordering: the susceptibility, specific heat and nuclear magnetic resonance show no singular features related to phase transitions. Initially, it was proposed that the key variable is frustration -- spatial anisotropy -- measured by the ratio of the next-nearest-neighbor and nearest-neighbor transfer integral, which is close to unity for \etcn, \agcn\ and \dmit, when estimated by the extended H{\"u}ckel method. Recent evidence challenges this idea because calculations by density functional theory yield values closer to 0.8, as well as the more one-dimensional anisotropy value of about 1.3 for \cat.

Low-energy excitations are probed by thermal and optical properties: good evidence supporting spin liquid and gapless spin excitations --~spinons~-- comes from specific heat displaying a large linear term for all four candidates. The Wilson ratio is slightly above unity, the value expected for free fermions; this result thus indicates that the same degrees of freedom determine the behavior of both the susceptibility and the specific heat.
However, very recent electron-spin resonance measurements on \etcn\ observe a drop of the spin susceptibility, which can only be explained by the opening of a gap due to the formation of spin singlets. Interestingly, a finite residual linear term of thermal conductivity is only observed in \cat, while it is fully suppressed in \etcn\ leading to the suggestion of a gap opening. In \dmit, in contrast to initial findings, most recent reports exclude the magnetic thermal conductivity as well. The absence of reproducible finite residual thermal conductivity questions the presence of delocalized gapless low-energy excitations in these spin-liquid candidates; spinons may become localized due to the spin-lattice decoupling, or to disorder.
On the other hand, the spinon contribution to the optical conductivity is successfully verified in \dmit\  in agreement with theory predicting a power-law absorption at low frequencies; similar experiments failed in the case of \etcn\ due to effects of metallic fluctuations in the vicinity of the Mott transition.


The legendary 6~K-anomaly in \etcn\ is omnipresent; effects observed in nuclear magnetic resonance, electron and muon spin resonance, in specific heat and thermal conductivity, in thermal expansion and ultrasonic velocity, and in microwave and dielectric response indicate coupled spin, lattice and charge degrees of freedom are involved.
Various scenarios are suggested for the explanation of this anomaly,
such as spin-chirality ordering, a spinon-pairing transition, or formation of an exciton condensate.
A large sample-to-sample variations in the size of 6~K anomaly as well as in the microwave and dielectric anomalies are consistent with inhomogeneity and disorder. Theoretical considerations suggest disorder-induced spin defects may provide a comprehensive explanation of the low-temperature properties of \etcn\ and maybe other compounds, too.
In this disorder-based approach, the 6~K anomaly can be interpreted as the formation  of a valence bond solid. The formation of singlets yields the drop in susceptibility but leaves sufficient defects spins randomly placed due to disorder also induced from the anion layer. At this point, a comprehensive explanation of the low-temperature anomaly and its possible relation to a spin-liquid state in all quantum spin liquid candidates is still lacking.

It is intriguing that randomness appears to support the quantum spin liquid state: a nuclear magnetic resonance experiment on \etcl\ found that the antiferromagnetic ordering disappears, when crystals are irradiated by x-rays while spin-liquid properties emerge. Alloying \etcn\ with BEDT-STF molecules results in no change of magnetic properties suggesting inherent disorder is already present in \etcn.
The theoretically developed randomness-induced quantum spin-liquid models show disorder enhance cooperative action of quantum fluctuations and triangular frustration.
Two inherent sources of randomness are anticipated: the intrinsic randomness is suggested to originate in the charge sector and to act via charge-spin coupling, while the extrinsic quenched randomness is due to disorder in the anion layers.

It remains a challenging task to verify the quantum spin liquid state experimentally.
Certainly, a crucial open issue is the extent and influence of disorder and inhomogeneity.
Whether any of these phenomena are also present in \cat, and whether they are important for the formation of the quantum spin liquid state in organics remains to be clarified in future.

\subsection{Quantum states of electric and magnetic dipoles}
\label{sec:Summary_Dipoles}

Although the spin liquid phase is insulating, anomalous charge dynamics in \etcn\ and \agcn\ are suggested for explaining the observed dielectric and low-energy optical responses. Several theoretical approaches have been developed to study the interplay of spins with charges on frustrated triangular lattice with dimerized sites. The approach based on extended Hubbard model considers quantum electric dipoles on organic dimers treating the inter-dimer interactions pertubatively; it results in the phase diagram, in which charge-spin liquid shares a common boundary with charge-ordered spin liquid and with spin-ordered charge liquid. An exact diagonalization study including electron-lattice coupling finds paired-electron crystal phase adjacent to the antiferromagnetic, spin gap and Wigner crystal phases.

Despite considerable efforts, no experimental evidence exists for sizeable electric dipoles; a negligibly small charge imbalance is found in vibrational spectroscopy and x-ray diffraction. But it has been shown that the symmetry of the non-polar mean P2$_1$/c structure is broken, {\it i.e.} that non-equivalent crystallographic sites exist. Whereas static electric dipoles are excluded, fluctuating dipoles may be concluded based on the unusually broad charge-sensitive molecular Raman and infrared modes. While these fast fluctuations cannot be invoked to explain the anomalous dielectric response in kHz-MHz range, suggestions were raised that they manifest in the microwave response;
the terahertz response, however, was proven to be due to coupled dimer-anion vibrations instead.

Since no sizeable static electric dipoles are present, the dielectric anomaly in quantum spin liquids is attributed to the cooperative motion of charged domain walls. The walls are created within random domain structure induced by quenched disorder originating in the anion layer, as identified by density functional theory calculations.
The domain wall scenario is also supported by an approach based on the one-dimensional tight binding model with on-site and inter-site Coulomb repulsion indicating that charge fluctuations occur in the boundary region between Mott and charge-order insulators. At low frequencies, oscillations occur in the double-well potential corresponding to a small charge disproportionation of opposite polarity; spatially extended domain walls connecting two respective domains may give rise to the observed dielectric response.

In \etcl\ an anomalous dielectric response shows up, which coincides with an antiferromagnetic state with signatures ranging from multiferroicity to ferroelectricity next to superconductivity; but relaxor-like response prevails in the majority of single crystals studied. Further experimental efforts are indispensable for the search for structural inversion-symmetry breaking in order to clarify the ferroelectric-like signatures in this antiferromagnetic system.

The low-temperature behavior of the hydrogen-bonded single-molecular compound \cat\ is unique among organics in the way that it demonstrates a quantum liquid of electric dipoles established simultaneously with a quantum spin liquid. The dielectric constant follows the Barrett behavior revealing strong quantum fluctuations originating in the zero-point motion of hydrogen atoms; density functional theory suggests that proton fluctuations coupled to charges and spins give rise to a quantum liquid of electric and magnetic dipoles. The crucial importance of the fluctuating proton-bond is evidenced in the behavior of deuterated sister system \dcat\ showing no signatures of liquid phases; instead a charge-ordered state with spin gap is established. Under strong dc electric fields, the negative differential resistance is observed and attributed to the deuterium dynamics coupled to the electron system. A negative differential resistance behavior is also seen in \cat. This surprising result may be explained
by density functional theory, which finds a quasi-degenerate electronic state implying random domains in the real space; the domain wall motion may be responsible for the nonlinear effects.

Another candidate for a quantum electric dipole liquid is given by \hgbr\ because no evidence for charge ordering is detected. The
$\nu_{2}(a_g)$ Raman band broadens on cooling and its shape is well described by Kubo’s two-states-jump model
assuming charge fluctuations with an exchange frequency $\omega_{\rm ex} \approx 30-40$\cm.
A broad feature of non-phonon origin around 40~\cm\ in the $A_{1g}$ Raman channel is taken as a fingerprint of these very fluctuations; arguments are given that the energy of the mode is lower than expected for magnetic excitations.
However, a quantum paraelectric behavior is not observed; dielectric response instead displays features characteristic of glassy dynamics indicating that the low-frequency Raman mode may be more appropriately interpreted as a Boson peak. The suggestion is supported by the significant non-Debye behavior found in the specific heat measurements; an excess of low-energy vibrational states, whose entropy is significantly larger than what is expected
for the pure magnetic entropy, indicates heterogeneity and glassy properties mostly in the charge sector of the liquid state. But spins are coupled to fluctuating charges as also indicated by a finite linear term observed in the specific heat measurements consistent with a spin liquid behavior; a spin-glass-like phase develops at lower temperatures as implied by the spin susceptibility and electron spin resonance measurements. An idea of exotic quantum liquid state with glassy nature that consists of entangled fluctuating electric dipoles and spins is worth to be verified in future.

Theoretical considerations of spin-charge coupling suggest possible mechanisms of charge-driven instabilities towards a long-range magnetic state. Initially the charge-ordered \hgcl\ was thought to tend to a magnetic transition at low temperatures. But the recent specific heat, electron spin resonance and nuclear magnetic resonance measurements give strong indications that magnetic order is absent and defect states dominate the properties at low temperatures. Nevertheless more efforts in this direction is needed to clarify the ground state in the spin sector and enable verification of current theoretical models based on dipolar-spin coupling.

\subsection{Outlook}
\label{sec:outlook}

We live in a three-dimensional world, but physics in smaller number of dimensions reveals a qualitative change in the system’s properties.  Despite not being atomically thin as graphene, quasi-one- and two-dimensional organic materials have many characteristics linked to their reduced dimensionality. It is associated with the overlap of molecular orbitals provides the cornerstone of organic metallicity and modifies the way quantum fluctuations compete with long-range order. They developed as a workbench of real materials to discover novel aspects of matter such as Peierls distortion and charge and spin density waves in one-dimensional materials, electronic ferroelectricity and quantum spin liquids in two dimensions. Unconventional non-phonon-mediated superconductivity was first discovered and studied in low-dimensional organics; the knowledge gained helped a lot in the quest for high $T_c$   {}superconductors since their discovery at the end of eighties.

While conducting polymers and organic semiconductors are implemented in a broad range of electronic devices, actual applications of crystalline two-dimensional organic conductors did not surface until today; and one should not make any strong claims or serious predictions in this regard. Whether or not these materials eventually find useful applications, they keep expanding our understanding of fundamental science and offer paths to designing other materials with practical use.

The family of molecular quantum materials draws its strength from the versatility of the compounds and selectivity of their physical properties. It allows us to tune the interplay of various degrees of freedom in an unprecedented way, to reach unexplored areas of the phase diagram and to shift the frontier towards new states of matter. In roughly ten years, research in two-dimensional organic solids has made spectacular advances: it is amazing to see how the collaboration between theory and experiment enables us to realize what has been suggested in the past, but also to understand what has been observed years ago.
Organic conductors always have been a niche in condensed matter science, the general interest is soaring since molecular solids have been realized as superior model compounds and often prime examples for investigating quantum phenomena.  There is a historical record that supports this and justifies believing this should continue to be the case.

But the paradigm of successful science is shifting nowadays. Accomplishing the bright future of the field will require a stronger effort in materials synthesis and design. To this end important advances can be achieved by integrating machine learning applications. Experimentally, the organic materials are attractive because the sample can be prepared from commercially low-cost components we find in nature. The clever performance of making the targeted material will vitalize the field together with up-to-date experimental methods center on imaging, scattering and spectroscopy tools with increasing spatial and time resolution, including instrumentation for the individual investigator and the scientific user facilities. Reaching the goal equally requires the use of state-of-the-art computational facilities and synergy of numerical calculations and microscopic theory.

Only time will tell, but the potential for the continued growth of the field is here. We would be gratified if this paper turns out to be useful to this end and succeeds to drive enthusiasm of coming generations in physical and chemical sciences.

\section{Acknowledgements}
Working in the field of low-dimensional organic conductors for decades, we have enjoyed fruitful collaborations and illuminating discussions with numerous colleagues and students whom we all want to thank very much. In this context we would like to mention in particular  M. Basleti{\'c}, K. Biljakovi{\'c}, S. Brown, E. Canadell, R.T. Clay, M. \v{C}ulo, V. Dobrosavljevi{\'c}, N. Do\v{s}li{\'c}, N. Drichko, M. Dumm, P. Foury-Leylekian, H. Fukuyama, S. Fratini, A. Girlando, B. Gorshunov, B. Gumhalter, C. Hotta, V. Ilakovac, S. Ishihara, T. Ivek, S. Kaiser, K. Kanoda, R. Kato, B. Korin-Hamzi{\'c}, M. Lang, P. Lazi{\'c}, W. Li, P. Littlewood, S. Mazumdar, J. Merino, O. Milat, J. M{\"u}ller, M. Pinteri{\'c}, J.P. Pouget, B. Powell, A. Pustogow, R. R{\"o}sslhuber, G. Saito, Y. Saito, T. Sasaki, J.A. Schlueter, D. Schweitzer, R. Valent{\'i}, S. Winter, Y. Yoshida.
The project was financially supported by the Deutsche Forschungsgemeinschaft (DFG), the Deutscher Akademischer Auslandsdienst (DAAD), the Croatian Ministry of Science and Education and by the Croatian Science Foundation. We acknowledge B.  Gumhalter, T. Ivek, M. Pinteri{\'c}, M. Prester, A. Pustogow and W. Strohmaier for their support and assistance during the preparation of this Review.

\bibliographystyle{tADP}
\bibliography{ADP}


\label{lastpage}

\end{document}